\documentclass[aps,floats]{revtex4-2}
\usepackage{amsmath,amssymb}
\usepackage{graphicx,epsfig}
\usepackage[greek,english]{babel}
\usepackage{bbold}

\begin{document}
\bibliographystyle {plain}

\pdfoutput=1
\def\oppropto{\mathop{\propto}} 
\def\opsimeq{\mathop{\simeq}}
\def\opoverderline{\mathop{\overline}}
\def\operarrow{\mathop{\longrightarrow}}
\def\opsim{\mathop{\sim}}

\def\opmin{\mathop{\min}} 
\def\opmax{\mathop{\max}} 
\def\oplim{\mathop{\lim}}

\title{ Classification of diffusion processes in dimension $d$ via the Carleman approach \\
with applications to models involving additive, multiplicative or square-root noises} 


\author{C\'ecile Monthus}
\affiliation{Universit\'e Paris-Saclay, CNRS, CEA, Institut de Physique Th\'eorique, 91191 Gif-sur-Yvette, France}


\begin{abstract}

The Carleman approach is well-known in the field of deterministic classical dynamics as a method to replace a finite number $d$ of non-linear differential equations by an infinite-dimensional linear system. Here this approach is applied to a system of $d$ stochastic differential equations for $[x_1(t),..,x_d(t)]$ when the forces and the diffusion-matrix elements are polynomials, in order to write the linear system governing the dynamics of the averaged values ${\mathbb E} ( x_1^{n_1}(t) x_2^{n_2}(t) ... x_d^{n_d}(t) )$ labelled by the $d$ integers $(n_1,..,n_d)$. The natural decomposition of the Carleman matrix into blocks associated to the global degree $n=n_1+n_2+..+n_d$ is useful to identify the models that have the simplest spectral decompositions in the bi-orthogonal basis of right and left eigenvectors. This analysis is then applied to models with a single noise per coordinate, that can be either additive or multiplicative or square-root, or with two types of noises per coordinate, with many examples in dimensions $d=1,2$. In $d=1$, the Carleman matrix governing the dynamics of the moments ${\mathbb E} ( x^{n}(t) )$ is diagonal for the Geometric Brownian motion, while it is lower-triangular for the family of Pearson diffusions containing the Ornstein-Uhlenbeck and the Square-Root processes, as well as the Kesten, the Fisher-Snedecor and the Student processes that converge towards steady states with power-law-tails. In dimension $d=2$, the Carleman matrix governing the dynamics of the correlations ${\mathbb E} ( x_1^{n_1}(t) x_2^{n_2}(t) )$ has a natural decomposition into blocks associated to the global degree $n=n_1+n_2$, and we discuss the simplest models where the Carleman matrix is either block-diagonal or block-lower-triangular or block-upper-triangular.

\end{abstract}

\maketitle

\section{ Introduction}

The Carleman Embedding (see the book \cite{carlemanBook} and references therein)
 has been introduced in 1932 by Carleman \cite{Carleman1932}
as a method to replace a finite number $d$ of non-linear differential equations
by an infinite-dimensional linear system.
At the conceptual level, this approach is related to the framework 
introduced at the same period by Koopman and von Neumann\cite{Koopman,KoopmanVonN,VonN},
where classical mechanics in dimension $d$ is reformulated as a linear dynamics in an infinite Hilbert space,
in order to take advantage of the technical tools developed in the field of quantum mechanics.
Note that these linear representations of non-linear classical models 
have attracted a lot of interest recently in relation with the field of quantum computing 
(see \cite{Koopman_vonNeumann,ThreeLinearEmbeddings,LinearRepresentations}
and references therein).
Since its introduction, 
the Carleman method has been applied to many classical deterministic models
(see the book \cite{carlemanBook} and references therein),
in particular to deterministic Lotka-Volterra models \cite{Steeb1980,Lotka-Volterra_Combinatorics}
and to the 3D chaotic Lorenz model \cite{Andrade1981,Andrade1982}.

In their pioneering work \cite{CarlemanStochastic} in 1982, 
Graham and Schenzle have extended the Carleman method to study some one-dimensional diffusions with multiplicative noise, but to the best of our knowledge, the Carleman approach has not spread afterwards in the field of stochastic processes.
The goal of the present paper is thus to give a self-contained pedestrian presentation of the Carleman perspective
for diffusion processes in dimension $d$, before focusing on specific examples involving 
either a single noise per coordinate, that can be either additive or multiplicative or square-root,
  or with two different types of these noises per coordinate.
  Indeed these types of noises are the most relevant from the physical point of view.
  For instance in population models involving positive variables (see \cite{KardarPopulation,MultiplicativeGrowth,RandomGrowth}
  and references therein),
  the most studied noises are the square-root noise (where it is sometimes called demographic noise) 
  and the multiplicative noise (where it is sometimes called seascape noise),
  while in the 3D Lorenz model, the most considered noises are the additive noise and the multiplicative noise 
\cite{LorenzMultiplicative,LorenzAddMulti,LorenzAdditive}.

The paper is organized as follows.
In section \ref{sec_general}, we introduce the notations that will be useful in the whole paper
to analyze diffusion processes $\vec x(t) \in {\mathbb R}^d$ in dimension $d$,
and we mention the link with deterministic processes when the noises vanish.
In section \ref{sec_Carleman}, we describe how the Carleman approach can be applied
 to diffusion processes in dimension $d$
when the Ito forces and the diffusion matrix elements are polynomials, focusing on the case
of polynomials of degree two to be more concrete.
In section \ref{sec_DfullyDiag}, we describe the simplifications of the Carleman matrix for models 
that involve a single noise 
  per coordinate, that can be 
  either additive or multiplicative or square-root,
  or that involve two different types of noises per coordinate,
  in order to give in section \ref{sec_list}
  the list of these models that have the simplest properties from the Carleman point of view.
  The application to the dimension $d=1$ is described in section \ref{sec_1D},
  while the application to the dimension $d=2$ is described
  in the three sections \ref{sec_2D} \ref{sec_BlockDiag2D} and \ref{sec_BlockLower2D}.  
   Our conclusions are summarized in section \ref{sec_conclusion},
   while appendices contain more technical computations.


\section{ Diffusion processes $\vec x(t) \in {\mathbb R}^d$ in dimension $d$
with their observables $O[\vec x(t) ]$ }

\label{sec_general}

In this section, we introduce the notations that will be useful in the whole paper
to analyze diffusion processes $\vec x(t) = [x_1(t),x_2(t),...,x_d(t) ]\in {\mathbb R}^d$ in dimension $d$,
and we mention the link with deterministic processes when the noises vanish.


\subsection{ Notations for diffusion processes in dimension $d$}

\subsubsection{ Ito Stochastic Differential Equations involving forces $F_j(\vec x) $ and 
amplitudes $G_{j \alpha}(\vec x) $ in front of the noises $dB_{\alpha}(t) $}

 The stochastic trajectories $\vec x(t)$ are generated by the
Ito Stochastic differential equations for the $d$ components $x_j(t)$ with $i=1,..,d$
\begin{eqnarray}
dx_j(t) \equiv x_j(t+dt)-x_j(t)  = F_j(\vec x(t)) dt 
+ \sum_{\alpha=1}^{\alpha_{max}} G_{j \alpha}(\vec x(t)) dB_{\alpha}(t) \ \ \ \ \ \ \ \ \ \ \ \ \ \ 
\text{[Ito]}
\label{ItoSDE}
\end{eqnarray}
 that involve $\alpha_{max}$ independent Brownian increments $dB_{\alpha}(t) =B_{\alpha}(t+dt)-B_{\alpha}(t)$ satisfying
 \begin{eqnarray}
{\mathbb E} [dB_{\alpha}(t) ] && =0 
\nonumber \\
{\mathbb E} [dB_{\alpha}(t) dB_{\beta}(t)] && = dt \delta_{\alpha,\beta}
\label{Brownian}
\end{eqnarray}
and that are independent of the trajectory $\vec x(0 \leq \tau \leq t )$ up to time $t$.
As a consequence, the averaged value of the increment $dx_j(t) $ of Eq. \ref{ItoSDE} 
reduces to the contribution of the Ito force $F_j(.)$
\begin{eqnarray}
{\mathbb E} [dx_j(t) ] && = {\mathbb E} [ F_j(\vec x(t)) ]dt 
\label{ItoSDEav}
\end{eqnarray}
while the averaged value of the connected correlations of the increments
\begin{eqnarray}
{\mathbb E} [ (dx_j(t)-F_j(\vec x(t)) dt ) (dx_i(t)-F_i(\vec x(t)) dt ) ] && 
= {\mathbb E} \left[ \sum_{\alpha=1}^{\alpha_{max}} G_{j \alpha}(\vec x(t)) dB_{\alpha}(t)
\sum_{\beta=1}^{\beta_{max}} G_{i \beta}(\vec x(t)) dB_{\beta}(t)\right] 
\nonumber \\
&& =  {\mathbb E} \left[ \sum_{\alpha=1}^{\alpha_{max}} G_{j \alpha}(\vec x(t)) 
 G_{i \alpha}(\vec x(t)) \right] dt
 \equiv 2 {\mathbb E} \left[ D_{ji}(\vec x(t)) \right] dt
\label{ItoSDEcorre}
\end{eqnarray}
involve the positive symmetric diffusion matrix constructed from the amplitudes $ G_{j \alpha}(\vec x)$
\begin{eqnarray}
D_{ji}(\vec x) 
\equiv \frac{1}{2} \sum_{\alpha=1}^{\alpha_{max}} G_{j \alpha}(\vec x)       G_{i \alpha}(\vec x)
= D_{ij}(\vec x)
\label{Dij}
\end{eqnarray}


\subsubsection{ Ito formula for the dynamics of an arbitrary observable $O(\vec x(t) )$}

The Ito SDE for the observable $O(\vec x(t) )$ can be obtained 
from the Ito SDE of Eq. \ref{ItoSDE} for the components $x_j(t) $
and from the Taylor expansion of $O(\vec x+\vec \epsilon) $ up to second order in $\vec \epsilon$
\begin{eqnarray}
 O(\vec x+\vec \epsilon )  &&  =O(\vec x) 
 + \sum_{j=1}^d \epsilon_j \partial_j O(\vec x  ) 
  +\frac{1}{2} \sum_{j=1}^d \sum_{i=1}^d \epsilon_j\epsilon_i 
  \partial_j  \partial_i O (\vec x  )+...
 \label{TaylorO}
\end{eqnarray}
Using the properties of Eq. \ref{Brownian} for the noises
in order to keep all the relevant terms up to order $dt$, one obtains that the SDE for the observable $O(\vec x(t) )$
reads
\begin{eqnarray}
d O(\vec x(t) )  &&  \equiv  
O(\vec x(t) + d \vec x(t)) - O(\vec x(t) )
  =   \sum_{j=1}^d dx_j(t) \partial_j O(\vec x(t)  ) 
  +\frac{1}{2} \sum_{j=1}^d \sum_{i=1}^d dx_j(t) dx_i(t) 
  \partial_j  \partial_i O (\vec x(t)  ) +...
  \nonumber \\
  && =   \sum_{j=1}^d 
  \left[ F_j(\vec x(t)) dt + \sum_{\alpha=1}^{\alpha_{max}} G_{j \alpha}(\vec x(t)) dB_{\alpha}(t)\right] 
  \partial_j O(\vec x(t)  ) 
\nonumber \\
&&  +\frac{1}{2} \sum_{j=1}^d \sum_{i=1}^d 
   \left[ F_j(\vec x(t)) dt + \sum_{\alpha=1}^{\alpha_{max}} G_{j \alpha}(\vec x(t)) dB_{\alpha}(t)\right]  
    \left[ F_i(\vec x(t)) dt + \sum_{\beta=1}^{\alpha_{max}} G_{i \beta}(\vec x(t)) dB_{\beta}(t)\right] 
  \partial_j  \partial_i O (\vec x(t)  ) +...
    \nonumber \\
  && \equiv  F^{[O]} (\vec x(t)) dt +  \sum_{\alpha=1}^{\alpha_{max}} G^{[O]}_{ \alpha}(\vec x(t)) dB_{\alpha}(t)
  \label{ItoFormula}
\end{eqnarray}
where the coefficient $G^{[O]}_{ \alpha}(\vec x(t)) $ in front of the noise $dB_{\alpha}(t) $ reads
\begin{eqnarray}
G^{[O]}_{ \alpha}(\vec x) =
  \sum_{j=1}^d    G_{j \alpha}(\vec x)  \partial_j O(\vec x  ) 
\label{GOalphax}
\end{eqnarray}
while the Ito force $F^{[O]} (\vec x(t)) $ in factor of $dt$ involves the Ito forces $F_j(\vec x) $
and the diffusion matrix $D_{ji}(\vec x) $ of Eq. \ref{Dij}
\begin{eqnarray}
F^{[O]}(\vec x) =  \sum_{j=1}^d   F_j(\vec x)   \partial_j O(\vec x  ) 
  + \sum_{j=1}^d \sum_{i=1}^d 
  \left( \frac{1}{2} \sum_{\alpha=1}^{\alpha_{max}} G_{j \alpha}(\vec x)       G_{i \alpha}(\vec x) \right)
  \partial_j  \partial_i O (\vec x  ) 
  \nonumber \\
=  \sum_{j=1}^d   F_j(\vec x)   \partial_j O(\vec x  ) 
  + \sum_{j=1}^d \sum_{i=1}^d 
  D_{ji}(\vec x)
  \partial_j  \partial_i O (\vec x  )   \equiv {\cal L} O (\vec x  )
\label{ItoFOx}
\end{eqnarray}
where ${\cal L} $ represents the second-order differential generator
\begin{eqnarray}
{\cal L}
&& \equiv  \sum_{j=1}^d   F_j(\vec x)   \frac{ \partial }{\partial x_j}
  + \sum_{j=1}^d \sum_{i=1}^d   D_{ji}(\vec x)   \frac{ \partial^2 }{\partial x_j \partial x_i}
  \nonumber \\
&& =  \sum_{j=1}^d   F_j(\vec x)   \frac{ \partial }{\partial x_j}
+  \sum_{j=1}^d    D_{jj}(\vec x)   \frac{ \partial^2 }{\partial x_j^2}  
  + 2 \sum_{j=1}^{d-1} \sum_{i=j+1}^{d}   D_{ji}(\vec x)   \frac{ \partial^2 }{\partial x_j \partial x_i}  
\label{Generator}
\end{eqnarray}
that governs the dynamics of the averaged values ${\mathbb E} [ O(\vec x(t) ) ] $ of observables 
\begin{eqnarray}
\partial_t {\mathbb E} [ O(\vec x(t) ) ]= {\mathbb E} [ F^{[O]} (\vec x(t)) ]  
=  {\mathbb E} [ ({\cal L} O ) (\vec x(t)) ]  
  \label{ItoFormulaAv}
\end{eqnarray}


\subsubsection{Corresponding Stratonovich interpretation : same amplitudes $G_{j \alpha}( \vec x) $ 
but modified forces $f_j(\vec x) $}

The Ito SDE of Eq. \ref{ItoSDE} can be translated into
\begin{eqnarray}
dx_j(t)  && = f_j(\vec x(t)) dt + \sum_{\alpha=1}^{\alpha_{max}} G_{j \alpha}(\vec x(t)) dB_{\alpha}(t) 
\ \ \ \ \ \ \ \ \ \ \ \ \ \ 
\text{in the Stratonovich interpretation}
\label{StratoSDE}
\end{eqnarray}
that involves the same amplitudes $G_{j \alpha}( \vec x) $ in front of the noises $dB_{\alpha}(t) $,
while the Stratonovich forces $ f_j(\vec x)  $ can be computed in terms of the Ito force $F_j(\vec x)  $  
and in terms of the amplitudes $G_{l \alpha} $ via
 \begin{eqnarray}
f_j(\vec x) = F_j(\vec x) 
- \frac{1}{2} \sum_{i=1}^d \sum_{\alpha=1}^{\alpha_{max}} G_{i \alpha}(\vec x) \frac{\partial G_{j \alpha}(\vec x)}{\partial x_i} 
\label{StratoItoCorrespondance}
\end{eqnarray}
so that they are different from the Ito forces $F_j(\vec x)  $ when
the coefficients $G_{j \alpha} (\vec x)$ explicitly depend on the position $\vec x$.

Using this correspondance of Eq. \ref{StratoItoCorrespondance}
and the diffusion matrix of Eq. \ref{Dij},
the differential generator ${\cal L} $ of Eq. \ref{Generator} reads in terms of the Stratonovich force $\vec f(\vec x)  $
\begin{eqnarray}
{\cal L}
&& = \sum_{j=1}^d   f_j(\vec x)  \frac{ \partial }{\partial x_j}
+ \frac{1}{2} \sum_{j=1}^d  \sum_{i=1}^d \sum_{\alpha=1}^{\alpha_{max}} 
\left( G_{i \alpha}(\vec x) \frac{\partial}{\partial x_i} \right)
\left( G_{j \alpha}(\vec x)   \frac{ \partial }{\partial x_j} \right)
\label{GeneratorStrato}
\end{eqnarray}

In the present paper, we will use both perspectives :
the Ito perspective with the Ito force $\vec F(\vec x) $ will be more convenient when one wishes 
to write the differential generator ${\cal L} $ with all the derivatives on the right as in Eq. \ref{Generator},
while the Stratonovich perspective with the Stratonovich force $f(\vec x)  $ will be more convenient 
when one wishes to make changes of variables in the SDE, since one can use the standard rules of calculus
(instead of the specific Ito rules of calculus if one uses the Ito forces).


\subsubsection{ Link with the Fokker-Planck dynamics for the probability density $\rho_t( \vec x ) $ }

The probability density $\rho_t( \vec x ) $ to be at configuration $x$ at time $t$
\begin{eqnarray}
  \rho_t( \vec x )     \equiv {\mathbb E} [ \delta( \vec x(t) - \vec x) ] 
\label{rhotasav}
\end{eqnarray}
can be used to rewrite 
the averaged value ${\mathbb E} [ O(\vec x(t) ) ] $  of the observable as
\begin{eqnarray}
 {\mathbb E} [O(\vec x(t)) ]   \equiv \int d^d \vec x \rho_t(\vec x)  O(\vec x) 
\label{EavObsdef}
\end{eqnarray}
As a consequence, the dynamics of Eq. \ref{ItoFormulaAv} that involves the differential generator 
${\cal L} $ of Eq. \ref{Generator} can be translated into
\begin{eqnarray}
\int d^d \vec x  O(\vec x) \partial_t \rho_t(\vec x) 
= \int d^d \vec x \rho_t(\vec x) \left(  {\cal L} O(\vec x) \right)
= \int d^d \vec x O(\vec x) \left(  {\cal L}^{\dagger} \rho_t(\vec x) \right)
  \label{ItoFormulaAvrho}
\end{eqnarray}
where the last expression obtained via integrations by parts involves the adjoint operator 
\begin{eqnarray}
{\cal L}^{\dagger}
\equiv  - \sum_{j=1}^d     \frac{ \partial }{\partial x_j}  F_j(\vec x)
  + \sum_{j=1}^d \sum_{i=1}^d   \frac{ \partial^2 }{\partial x_j \partial x_i}  D_{ji}(\vec x)  
\label{GeneratorAdjoint}
\end{eqnarray}
so that the Fokker-Planck dynamics obtained from Eq. \ref{ItoFormulaAvrho} reads
\begin{eqnarray}
 \partial_t \rho_t(\vec x) 
&& =   {\cal L}^{\dagger} \rho_t(\vec x) 
\nonumber \\
&& =- \sum_{j=1}^d     \frac{ \partial }{\partial x_j} \bigg(  F_j(\vec x)  \rho_t(\vec x)  \bigg)
  + \sum_{j=1}^d \sum_{i=1}^d   \frac{ \partial^2 }{\partial x_j \partial x_i} \bigg( D_{ji}(\vec x)  \rho_t(\vec x)  \bigg)
  \label{fokkerPlanck}
\end{eqnarray}


\subsection{ Deterministic processes $\vec x(t) \in {\mathbb R}^d$
in the limit of vanishing noises $G_{j\alpha}(\vec x)=0$ and $D_{ji}(\vec x)=0$}

In the limit of vanishing noises-amplitudes $G_{j\alpha}(\vec x)=0$ leading to the vanishing
of the diffusion matrix $D_{ji}(\vec x)=0$ of Eq. \ref{Dij},
the Ito SDE of Eq. \ref{ItoSDE} and the Stratonovich SDE of Eq. \ref{StratoSDE}
 become the deterministic equations
\begin{eqnarray}
\frac{ dx_j(t)}{dt}  = F_j(\vec x(t)) = f_j(\vec x(t)) 
\label{deterministe}
\end{eqnarray}
so that the only randomness comes from the probability distribution $\rho_{t=0}(\vec x)$
of the initial conditions $\vec x(t=0) $.
 The dynamics of Eq. \ref{ItoFormulaAv} for the averaged values of observables
 becomes the Koopman dynamics 
 \begin{eqnarray}
\partial_t {\mathbb E} [ O(\vec x(t) ) ]  
=  {\mathbb E} [ ({\cal L}_{deter} O ) (\vec x(t)) ]  
  \label{Koopman}
\end{eqnarray}
 where the Koopman differential generator ${\cal L}_{deter} $
 only contains the first-order differential contributions of Eq. \ref{Generator}
 and Eq. \ref{GeneratorStrato}
 \begin{eqnarray}
{\cal L}_{deter}
=  \sum_{j=1}^d   F_j(\vec x)   \frac{ \partial }{\partial x_j}
=  \sum_{j=1}^d   f_j(\vec x)   \frac{ \partial }{\partial x_j}
\label{GeneratorK}
\end{eqnarray}
Similarly, the Fokker-Planck dynamics of Eq. \ref{fokkerPlanck}
for the probability density $\rho_t(\vec x)$
becomes the Perron-Froebenius dynamics  
\begin{eqnarray}
 \partial_t \rho_t( \vec x )    =   {\cal L}_{deter}^{\dagger} \rho_t(\vec x) 
 =  - \sum_{j=1}^d \frac{\partial}{\partial x_j} \left[ F_j(\vec x)  \rho_t(\vec x)    \right]
\label{perronFroebenius}
\end{eqnarray}
where the Perron-Froebenius generator ${\cal L}_{deter}^{\dagger} $ is the adjoint of the Koopman generator ${\cal L}_{deter} $.

In the present paper, the goal is to focus on diffusion processes, 
but deterministic processes can always be recovered in the limit of vanishing noises as explained above.


\section{ Carleman dynamics for the moments ${\mathbb E} ( x_1^{n_1}(t) x_2^{n_2}(t) ... x_d^{n_d}(t) )$ 
with $n_j \in {\mathbb N}$  }

\label{sec_Carleman}

As recalled in the Introduction, the Carleman approach that is well-known 
in the field of deterministic dynamics (see also Appendix \ref{app_deterministic} for the simplest example)
has been applied 
to some one-dimensional processes with multiplicative noise in the pioneering work \cite{CarlemanStochastic}.
In the present section, we describe how the Carleman approach  
can be applied to diffusion processes in dimension $d$ 
when the Ito forces and the diffusion-matrix elements are polynomials.


\subsection{ Parametrization of the polynomial forces $F_j(\vec x) $ and of the polynomial diffusion-matrix elements $ D_{ji}(\vec x)$  }

The Carleman approach can be applied whenever one can write 
Taylor expansions for the Ito forces $F_j(\vec x)$
and for the diffusion matrix elements $D_{ji}(\vec x) $.
Here to be more concrete and to simplify the notations,
we will focus only on the cases where the Ito forces $F_j(\vec x)$  
and the diffusion matrix elements $D_{ji}(\vec x) $
are polynomials of degree two in the $d$ coordinates $(x_1,..,x_d)$
with the following bra-ket notations :

(i) the Taylor coefficients of the Ito forces
\begin{eqnarray}
F_j(\vec x) && 
= \langle j \vert {\bold F}^{[0]}  \rangle + \sum_{i=1}^d \langle j \vert {\bold F}^{[1]} \vert i \rangle x_i
+  \sum_{i_1=1}^d \sum_{i_2=1}^{d} \langle j \vert {\bold F}^{[2]} \vert i_1,i_2 \rangle x_{i_1} x_{i_2}
\nonumber \\
&& = \langle j \vert {\bold F}^{[0]}  \rangle + \sum_{i=1}^d \langle j \vert {\bold F}^{[1]} \vert i \rangle x_i
+  \sum_{i=1}^d  \langle j \vert {\bold F}^{[2]} \vert i,i \rangle x_{i}^2
+ 2 \sum_{i_1=1}^{d-1} \sum_{i_2=i_1+1}^d \langle j \vert {\bold F}^{[2]} \vert i_1,i_2 \rangle x_{i_1} x_{i_2}
\label{TaylorF}
\end{eqnarray}
  are gathered into the ket $ \vert {\bold F}^{[0]} \rangle $
and into the matrices ${\bold F}^{[1]} $ and ${\bold F}^{[2]} $ with the following matrix elements
\begin{eqnarray}
 \langle j \vert {\bold F}^{[0]}  \rangle && = F_j(\vec x=\vec 0)
 \nonumber \\
 \langle j \vert {\bold F}^{[1]} \vert i \rangle && 
 = \left[ \frac{\partial F_j(\vec x)}{\partial x_i}  \right]_{\vec x=\vec 0}
 \nonumber \\
  \langle j \vert {\bold F}^{[2]} \vert i_1,i_2 \rangle && 
  = \frac{1}{2} \left[ \frac{\partial^2 F_j(\vec x)}{\partial x_{i_1}\partial x_{i_2}}  \right]_{\vec x=\vec 0}
  =  \langle j \vert {\bold F}^{[2]} \vert i_2,i_1 \rangle
\label{TaylorFcoefs}
\end{eqnarray}

(ii) the Taylor coefficients of the diffusion-matrix elements
\begin{eqnarray}
D_{j_1 j_2}(\vec x) && = 
\langle j_1 j_2 \vert {\bold D}^{[0]}  \rangle + \sum_{i=1}^d \langle  j_1 j_2 \vert {\bold D}^{[1]} \vert i \rangle x_i
+ \sum_{i_1=1}^d \sum_{i_2=1}^d \langle  j_1 j_2 \vert {\bold D}^{[2]} \vert i_1,i_2 \rangle x_{i_1} x_{i_2}
\nonumber \\
&&= 
\langle j_1 j_2 \vert {\bold D}^{[0]}  \rangle + \sum_{i=1}^d \langle  j_1 j_2 \vert {\bold D}^{[1]} \vert i \rangle x_i
+ \sum_{i=1}^d  \langle  j_1 j_2 \vert {\bold D}^{[2]} \vert i,i \rangle x_i^2
+ 2 \sum_{i_1=1}^{d-1} \sum_{i_2=i_1+1}^d \langle  j_1 j_2 \vert {\bold D}^{[2]} \vert i_1,i_2 \rangle x_{i_1} x_{i_2}
\label{TaylorD}
\end{eqnarray}
are gathered into the ket $\vert {\bold D}^{[0]}  \rangle$, and into the matrices
${\bold D}^{[1]} $ and ${\bold D}^{[2]} $ with the following matrix elements
\begin{eqnarray}
\langle j_1 j_2 \vert {\bold D}^{[0]}  \rangle && = D_{j_1,j_2}(\vec x=\vec 0)
 \nonumber \\
\langle j_1 j_2 \vert {\bold D}^{[1]}  \vert i \rangle && 
 = \left[ \frac{\partial D_{j_1,j_2}(\vec x)}{\partial x_i}  \right]_{\vec x=\vec 0}
 \nonumber \\
 \langle j_1 j_2 \vert {\bold D}^{[2]} \vert i_1,i_2 \rangle && 
  = \frac{1}{2} \left[ \frac{\partial^2 D_{j_1,j_2}(\vec x)}{\partial x_{i_1}\partial x_{i_2}}  \right]_{\vec x=\vec 0}
  =  \langle j_1 j_2 \vert {\bold D}^{[2]} \vert i_2,i_1 \rangle
\label{TaylorDcoefs}
\end{eqnarray}

Note that if one in interested in models where the forces and the diffusion-matrix elements are polynomials of higher orders,
one just needs to include the appropriate higher matrices like $ {\bold F}^{[3]}, {\bold F}^{[4]}...$  and $ {\bold D}^{[3]}, {\bold D}^{[4]}...$ etc.


\subsection{ Decomposition of the generator ${\cal L} = {\cal L}^{[-2]} + {\cal L}^{[-1]}+{\cal L}^{[0]} + {\cal L}^{[1]}$ 
with respect to the global scaling dimension  }

Plugging the expressions of Eqs \ref{TaylorF} and \ref{TaylorD} into the differential generator ${\cal L} $ of Eq. \ref{Generator}
\begin{eqnarray}
{\cal L}
&& \equiv  \sum_{j=1}^d \left[ \langle j \vert {\bold F}^{[0]}  \rangle 
+ \sum_{i=1}^d \langle j \vert {\bold F}^{[1]} \vert i \rangle x_i
+  \sum_{i_1=1}^d \sum_{i_2=1}^{d} \langle j \vert {\bold F}^{[2]} \vert i_1,i_2 \rangle x_{i_1} x_{i_2}
\right]  \frac{ \partial }{\partial x_j}
 \nonumber \\
&&  + \sum_{j_1=1}^d \sum_{j_2=1}^d  \left[ \langle j_1 j_2 \vert {\bold D}^{[0]}  \rangle + \sum_{i=1}^d \langle  j_1 j_2 \vert {\bold D}^{[1]} \vert i \rangle x_i
 + \sum_{i_1=1}^d \sum_{i_2=1}^d \langle  j_1 j_2 \vert {\bold D}^{[2]} \vert i_1,i_2 \rangle x_{i_1} x_{i_2}
\right]   \frac{ \partial^2 }{\partial x_{j_1} \partial x_{j_2} }
  \nonumber \\
&& =  {\cal L}^{[-2]} + {\cal L}^{[-1]}+{\cal L}^{[0]} + {\cal L}^{[1]}
\label{GeneratorClassDegree}
\end{eqnarray}
leads to a natural decomposition into four contributions ${\cal L}^{[k=-2,-1,0,1]} $
where $k=-2,-1,0,1$ represents the global scaling dimension, i.e. 
the difference between the number of powers of $x_.$ and the number of derivatives $\frac{\partial}{\partial x_.} $
\begin{eqnarray}
{\cal L}^{[-2]}
&& \equiv   \sum_{j_1=1}^d \sum_{j_2=1}^d  \langle j_1 j_2 \vert {\bold D}^{[0]}  \rangle   \frac{ \partial^2 }{\partial x_{j_1} \partial x_{j_2} }
  \nonumber \\
{\cal L}^{[-1]}
&& \equiv  \sum_{j=1}^d \langle j \vert {\bold F}^{[0]}  \rangle   \frac{ \partial }{\partial x_j}
  + \sum_{j_1=1}^d \sum_{j_2=1}^d    \sum_{i=1}^d \langle  j_1 j_2 \vert {\bold D}^{[1]} \vert i \rangle x_i
\frac{ \partial^2 }{\partial x_{j_1} \partial x_{j_2} }
  \nonumber \\
    {\cal L}^{[0]}
&& \equiv  
\sum_{j=1}^d  \sum_{i=1}^d \langle j \vert {\bold F}^{[1]} \vert i \rangle x_i
  \frac{ \partial }{\partial x_j}
  + \sum_{j_1=1}^d \sum_{j_2=1}^d   \sum_{i_1=1}^d \sum_{i_2=1}^d \langle  j_1 j_2 \vert {\bold D}^{[2]} \vert i_1,i_2 \rangle x_{i_1} x_{i_2}   \frac{ \partial^2 }{\partial x_{j_1} \partial x_{j_2} }
  \nonumber \\
{\cal L}^{[1]}
&& \equiv  
 \sum_{j=1}^d  \sum_{i_1=1}^d \sum_{i_2=1}^{d} \langle j \vert {\bold F}^{[2]} \vert i_1,i_2 \rangle x_{i_1} x_{i_2}  \frac{ \partial }{\partial x_j}
\label{GeneratorClassDegree4}
\end{eqnarray}

Note that if one in interested in models where the forces and the diffusion matrix are polynomials of higher orders,
parametrized by higher-order matrices $( {\bold F}^{[k]}, {\bold D}^{[k]})$,
one needs to add the term containing ${\bold D}^{[3]} $ in $ {\cal L}^{[1]}$
and to write the higher contributions ${\cal L}^{[k]} $ of global dimension $k \geq 2$
 containing the terms associated to the matrices 
${\bold F}^{[k+1]} $ and ${\bold D}^{[k+2]} $.


\subsection{ Linear system governing the dynamics of the moments $m_t(n_1,..,n_d) \equiv {\mathbb E} ( x_1^{n_1}(t) x_2^{n_2}(t) ... x_d^{n_d}(t) )$ 
labelled by the $d$ integers $(n_1,n_2,..,n_d) \in {\mathbb N}^d$  }

In the Carleman approach, one focuses on the basis of observables corresponding to the monomials
\begin{eqnarray}
O_{n_1,..,n_d}(\vec x) \equiv x_1^{n_1} x_2^{n_2} ... x_d^{n_d}  =\prod_{j=1}^d x_j^{n_j} 
\label{MonomialsCarleman}
\end{eqnarray}
labelled by the $d$ integers $(n_1,n_2,..,n_d) \in {\mathbb N}^d$
and on their averaged values
\begin{eqnarray}
m_t(n_1,..,n_d) \equiv  {\mathbb E} ( O_{n_1,..,n_d}(\vec x(t)) )
= {\mathbb E} ( x_1^{n_1}(t) x_2^{n_2}(t) ... x_d^{n_d}(t) )
\label{Moments}
\end{eqnarray}
corresponding to the moments and correlations of arbitrary order of the $d$ components $x_j(t)$,
while the case where all the integers vanish $n_j=0$ correspond to the trivial value unity
\begin{eqnarray}
m_t(0,0,..,0,0) \equiv {\mathbb E} ( 1 ) =1
\label{Moment01}
\end{eqnarray}

The action of the differential generator ${\cal L}$ of Eq. \ref{Generator}
on the monomial of Eq. \ref{MonomialsCarleman}
\begin{eqnarray}
{\cal L} O_{n_1,..,n_d}(\vec x)
&& =  \sum_{j=1}^d   F_j(\vec x) n_j  x_j^{n_j-1}  \prod_{l \ne j}^d x_l^{n_l} 
  + \sum_{j=1}^d    D_{jj}(\vec x)   n_j (n_j-1) x_j^{n_j-2}  \prod_{l \ne j}^d x_l^{n_l} 
\nonumber \\
&&  + 2 \sum_{j_1=1}^d \sum_{j_2=j_1+1}^d   D_{j_1j_2}(\vec x)    n_{j_1}  x_{j_1}^{n_{j_1}-1}  
n_{j_2} x_{j_2}^{n_{j_2}-1} \prod_{l \ne j_1,j_2}^d x_l^{n_l}
\label{GeneratorO}
\end{eqnarray}
can be rewritten using the expressions of Eqs \ref{TaylorF} and \ref{TaylorD} for the forces and for the diffusion matrix elements
\begin{footnotesize}
\begin{eqnarray}
&& {\cal L} O_{n_1,..,n_d}(\vec x)
 =  \sum_{j=1}^d  
\left[  \langle j \vert {\bold F}^{[0]}  \rangle + \sum_{i=1}^d \langle j \vert {\bold F}^{[1]} \vert i \rangle x_i
+  \sum_{i=1}^d  \langle j \vert {\bold F}^{[2]} \vert i,i \rangle x_{i}^2
+ 2 \sum_{i_1=1}^{d-1} \sum_{i_2=i_1+1}^d \langle j \vert {\bold F}^{[2]} \vert i_1,i_2 \rangle x_{i_1} x_{i_2}
\right]
 n_j  x_j^{n_j-1}  \prod_{l \ne j}^d x_l^{n_l} 
\nonumber \\
&&  + \sum_{j=1}^d 
 \left[ \langle j j \vert {\bold D}^{[0]}  \rangle + \sum_{i=1}^d \langle  j j \vert {\bold D}^{[1]} \vert i \rangle x_i
 + \sum_{i=1}^d  \langle  j j \vert {\bold D}^{[2]} \vert i,i \rangle x_i^2
+ 2 \sum_{i_1=1}^{d-1} \sum_{i_2=i_1+1}^d \langle  j j \vert {\bold D}^{[2]} \vert i_1,i_2 \rangle x_{i_1} x_{i_2}
\right]
  n_j (n_j-1) x_j^{n_j-2}  \prod_{l \ne j}^d x_l^{n_l} 
\nonumber \\
&&  + 2 \sum_{j_1=1}^{d-1} \sum_{j_2=j_1+1}^d 
 \left[ \langle j_1 j_2 \vert {\bold D}^{[0]}  \rangle + \sum_{i=1}^d \langle  j_1 j_2 \vert {\bold D}^{[1]} \vert i \rangle x_i
 + \sum_{i=1}^d  \langle  j_1 j_2 \vert {\bold D}^{[2]} \vert i,i \rangle x_i^2
+2  \sum_{i_1=1}^{d-1} \sum_{i_2=i_1+1}^d \langle  j_1 j_2 \vert {\bold D}^{[2]} \vert i_1,i_2 \rangle x_{i_1} x_{i_2}
\right]
     n_{j_1}  x_{j_1}^{n_{j_1}-1}  
n_{j_2} x_{j_2}^{n_{j_2}-1} \prod_{l \ne j_1,j_2}^d x_l^{n_l}
\nonumber \\
&& \equiv 
 \sum_{q_1=0}^{+\infty} ...  \sum_{q_d=0}^{+\infty} M(n_1,..,n_d \vert q_1,...,q_d) O_{q_1,..,q_d}(\vec x)
\label{GeneratorOTaylor}
\end{eqnarray}
\end{footnotesize}
to obtain a linear combination of the monomials $O_{q_1,..,q_d}(\vec x) \equiv x_1^{q_1} x_2^{q_2} ... x_d^{q_d} $
with the matrix elements $M(n_1,..,n_d \vert q_1,...,q_d) $.

The application of Eq. \ref{ItoFormulaAv} to the monomials $O_{n_1,..,n_d}(\vec x) $
yields that the dynamics of the moments 
$m_t(n_1,..,n_d) $ of Eq. \ref{Moments}
\begin{eqnarray}
\partial_t m_t(n_1,..,n_d) 
&& = \partial_t  {\mathbb E} ( O_{n_1,..,n_d}(\vec x(t)) )
=  {\mathbb E} ( {\cal L}O_{n_1,..,n_d}(\vec x(t)) )
\nonumber \\
&& =  \sum_{q_1=0}^{+\infty} ...  \sum_{q_d=0}^{+\infty} M(n_1,..,n_d \vert q_1,...,q_d)  m_t(q_1,..,q_d) 
\label{DynMoments}
\end{eqnarray}
correspond to a linear system governed by the matrix elements $M(n_1,..,n_d \vert q_1,...,q_d) $
of the matrix ${\bold M}$ that will be called the Carleman matrix from now on.

It is useful to introduce bra-ket notations for the orthonormalized basis $\vert q_1,...,q_d \rangle $
satisfying
\begin{eqnarray}
  \langle n_1,..,n_d  \vert  q_1,...,q_d \rangle && = \prod_{i=1}^d \delta_{n_i,q_i}
  \nonumber \\
  \sum_{q_1=0}^{+\infty} ...  \sum_{q_d=0}^{+\infty}\vert q_1,...,q_d \rangle \langle q_1,..,q_d  \vert
 && = {\bold 1}
\label{BraKetBasis}
\end{eqnarray}
in order to rewrite the moment $ m_t(n_1,..,n_d)$ as the component of the ket $\vert m_t \rangle $ onto the bra $ \langle n_1,..,n_d  \vert$
\begin{eqnarray}
 m_t(n_1,..,n_d) = \langle n_1,..,n_d  \vert m_t \rangle 
\label{KetMoments}
\end{eqnarray}
and to rewrite the matrix elements of the Carleman matrix ${\bold M } $ as 
\begin{eqnarray}
\langle n_1,..,n_d  \vert {\bold M } \vert q_1,...,q_d \rangle = M(n_1,..,n_d \vert q_1,...,q_d)   
\label{BraKetMatrixM}
\end{eqnarray}
so that the linear dynamics of Eq. \ref{DynMoments} becomes at the matrix level
\begin{eqnarray}
\partial_t \vert m_t \rangle =  {\bold M } \vert m_t \rangle
\label{DynMomentsKet}
\end{eqnarray}
while the formal solution reads in terms of the initial condition $\vert m_{t=0} \rangle $
\begin{eqnarray}
 \vert m_t \rangle = e^{t   {\bold M } } \vert m_0 \rangle
\label{SolMomentsKet}
\end{eqnarray}

Once the dynamics of the moments $m_t(n_1,..,n_d)$ has been recast into the linear system of Eqs
\ref{DynMoments} or \ref{DynMomentsKet}, any method appropriate to solve such linear dynamics can be applied.

\subsubsection{ Standard method in the Carleman literature : via the time-Laplace-transforms ${\hat m}_s (n_1,..,n_d)$ of the moments $m_t(n_1,..,n_d) $}

In the literature on the Carleman method and in particular in \cite{CarlemanStochastic} concerning one-dimensional diffusion processes, the standard method involves the 
introduction of the time-Laplace-transforms
${\hat m}_s (n_1,..,n_d)$ of the moments $m_t(n_1,..,n_d) $
\begin{eqnarray}
{\hat m}_s(n_1,..,n_d) \equiv \int_0^{+\infty} dt e^{-s t} m_t(n_1,..,n_d)
\label{mlaplace}
\end{eqnarray}
The Laplace-transforms of their time-derivatives $\partial_t m_t(n_1,..,n_d) $
can be evaluated via integration by parts
\begin{eqnarray}
 \int_0^{+\infty} dt e^{-s t} \partial_t m_t(n_1,..,n_d) 
 && = \left[ e^{-s t}  m_t(n_1,..,n_d)\right]_{t=0}^{t=+\infty} + s  \int_0^{+\infty} dt m_t(n_1,..,n_d) e^{-s t}
 \nonumber \\
 && = - m_0(n_1,..,n_d) +s {\hat m}_s(n_1,..,n_d)
\label{mlaplacederi}
\end{eqnarray}
in terms of the initial moments $m_{t=0}(n_1,..,n_d) $ at time $t=0$.

So the linear dynamics of Eq. \ref{DynMoments} reads for the Laplace transforms of the moments
\begin{eqnarray}
s {\hat m}_s(n_1,..,n_d) 
-   \sum_{q_1=0}^{+\infty} ...  \sum_{q_d=0}^{+\infty} M(n_1,..,n_d \vert q_1,...,q_d)  {\hat m}_s(q_1,..,q_d) 
= m_0(n_1,..,n_d)
\label{dynmomentslaplace}
\end{eqnarray}

With the bra-ket notations of Eq. \ref{KetMoments}, the Laplace transforms ${\hat m}_s (n_1,..,n_d)$
are the components of the ket $\vert {\hat m}_s \rangle $ onto the bras $ \langle n_1,..,n_d  \vert$
\begin{eqnarray}
 {\hat m}_s(n_1,..,n_d) = \langle n_1,..,n_d  \vert {\hat m}_s \rangle 
\label{KetMomentsLaplace}
\end{eqnarray}
while the system of Eq. \ref{dynmomentslaplace}
becomes
\begin{eqnarray}
(s{\bold 1} - {\bold M }) \vert {\hat m}_s \rangle = \vert m_0 \rangle
\label{dynmomentslaplaceMatrix}
\end{eqnarray}
The solution for the Laplace-transform $\vert {\hat m}_s \rangle $ thus requires the application of the inverse 
$\frac{1}{s{\bold 1} - {\bold M }} $ to the initial condition $\vert m_0 \rangle $
\begin{eqnarray}
 \vert {\hat m}_s \rangle =\frac{1}{s{\bold 1} - {\bold M }} \vert m_0 \rangle
\label{Solmomentslaplace}
\end{eqnarray}
in agreement with the direct application of the Laplace transform to the formal solution of Eq. \ref{SolMomentsKet}
\begin{eqnarray}
 \vert {\hat m}_s \rangle=  \int_0^{+\infty} dt e^{-s t} \vert m_t \rangle 
 =  \int_0^{+\infty} dt e^{-s t} e^{t   {\bold M } } \vert m_0 \rangle  =\frac{1}{s{\bold 1} - {\bold M }} \vert m_0 \rangle
\label{SolLaplaceKet}
\end{eqnarray}
while the solution for the real-time moments $\vert m_t \rangle $
requires the computation of the inverse-Laplace-transformation of this solution $\vert {\hat m}_s \rangle $.
As a consequence in the present paper, we will instead focus on the direct solution in real-time
via the spectral analysisi of the Carleman matrix as described in the next subsection.


\subsubsection{ Direct solution in real-time via the spectral decomposition of the matrix ${\bold M}$ into its eigenvalues $E$ and its corresponding left and right eigenvectors  }

The spectral decomposition of the operator $ e^{t   {\bold M } }$
\begin{eqnarray}
 e^{t   {\bold M } }=\sum_{E} e^{t E} \vert R_E \rangle \langle L_E \vert
\label{Mspectral}
\end{eqnarray}
involves the (possibly complex) eigenvalues $E$ of the Carleman matrix ${\bold M }$,
while the corresponding right eigenvectors $\vert R_E \rangle $ and left eigenvectors
$\langle L_E \vert $
\begin{eqnarray}
{\bold M} \vert R_E \rangle= E \vert R_E \rangle 
\nonumber \\
\langle L_E \vert {\bold M} = E \langle L_E \vert
\label{MEigen}
\end{eqnarray}
satisfy the bi-orthogonalization
\begin{eqnarray}
 \langle L_E \vert R_{E'} \rangle = \delta_{E,E'}
\label{Mspectralbiorthog}
\end{eqnarray}
and the resolution of the identity
\begin{eqnarray}
{\bold 1}=\sum_{E}  \vert R_E \rangle \langle L_E \vert
\label{MspectralResolutionUnity}
\end{eqnarray}

Then the solution of Eq. \ref{SolMomentsKet} for the moments can be decomposed into the contributions of the eigenvalues $E$
\begin{eqnarray}
 \vert m_t \rangle = e^{t   {\bold M } } \vert m_0 \rangle
 = \sum_{E} e^{t E} \vert R_E \rangle \langle L_E \vert m_0 \rangle
\label{SolMomentsSpectral}
\end{eqnarray}
while the link with the Laplace-transform $ \vert {\hat m}_s \rangle $ of Eq \ref{SolLaplaceKet} reads
\begin{eqnarray}
 \vert {\hat m}_s \rangle 
 =\int_0^{+\infty} dt e^{-s t} \left( \sum_{E} e^{t E} \vert R_E \rangle \langle L_E \vert m_0 \rangle \right) 
  =  \sum_{E} \frac{1}{s-E } \vert R_E \rangle \langle L_E \vert m_0 \rangle  
\label{Solmomentslaplacespectral}
\end{eqnarray}

The property $m_t(0,0,..,0,0) \equiv {\mathbb E} ( 1 ) =1 $ of Eq. \ref{Moment01}
means that the projection of Eq. \ref{DynMomentsKet} on the bra $\langle 0,0, ..,0,0 \vert  $
vanish for any ket $\vert m_t \rangle $
\begin{eqnarray}
0= \partial_t m_t(0,0,..,0,0) = \langle 0,0, ..,0,0 \vert \bigg( \partial_t \vert m_t \rangle \bigg) 
= \langle 0,0, ..,0,0 \vert {\bold M } \vert m_t \rangle
\label{Moment01derit}
\end{eqnarray}
 so that the bra $\langle 0,0, ..,0,0 \vert $
 is the left eigenvector of the Carleman matrix ${\bold M } $ associated to the eigenvalue $E=0$
\begin{eqnarray}
 \langle L_{E=0} \vert  = \langle 0,0, ..,0,0 \vert
\label{leftMzero}
\end{eqnarray}
In cases where the Fokker-Planck dynamics of Eq. \ref{fokkerPlanck} for the probability density $\rho_t(\vec x) $ converges towards a normalizable steady state $\rho_{st}(\vec x) $ satisfying
\begin{eqnarray}
0 = \partial_t \rho_{st}(\vec x) 
&& =   {\cal L}^{\dagger} \rho_{st}(\vec x) 
  \label{fokkerPlancksteady}
\end{eqnarray}
then the corresponding steady moments $m_{st}(n_1,..,n_d) =\langle n_1,..,n_d  \vert m_t \rangle $ should be stable via the dynamics of Eq. \ref{DynMomentsKet}
\begin{eqnarray}
0= \partial_t \vert m_{st} \rangle =  {\bold M } \vert m_{st} \rangle
\label{SteadyMoments}
\end{eqnarray}
i.e. the steady ket $\vert m_{st} \rangle$ should be the right eigenvector $\vert R_{E=0} \rangle $ for the matrix ${\bold M} $
associated to the eigenvalue $E=0$
\begin{eqnarray}
\vert R_{E=0} \rangle=  \vert m_{st} \rangle
\label{SteadyMomentsMatrix}
\end{eqnarray}
Note that even when there is a normalizable steady state $\rho_{st}(\vec x) $, some of its moments may diverge,
as will be described in specific examples.

In order to better understand the spectral properties of the Carleman matrix ${\bold M}  $,
it is useful to introduce its decompositions into blocks as described in the next subsection.


\subsection{ Decomposition of the Carleman matrix $ {\bold M} =  {\bold M}^{[-2]} + {\bold M}^{[-1]}+{\bold M}^{[0]} + {\bold M}^{[1]}$ into blocks associated to the difference of global degrees }

The Carleman matrix $ {\bold M}$ introduced in Eq. \ref{GeneratorOTaylor}
can be decomposed 
into blocks ${\bold M}^{[k\equiv q-n]} $ according to the possible difference $k=-2,-1,0,1$ 
between the global degrees $q=q_1+q_2+.. +q_d$ and $n=n_1+n_2+.. +n_d$  
\begin{eqnarray}
{\bold M} =  {\bold M}^{[-2]} + {\bold M}^{[-1]}+{\bold M}^{[0]} + {\bold M}^{[1]}
\label{MClassDegree}
\end{eqnarray}
in correspondence with the decomposition of Eq. \ref{GeneratorClassDegree} for the generator ${\cal L} $
\begin{eqnarray}
{\cal L}^{[k]}  O_{n_1,..,n_d}(\vec x) =  \sum_{q_1=0}^{+\infty} ...  \sum_{q_d=0}^{+\infty} M^{[k]}(n_1,..,n_d \vert q_1,...,q_d) O_{q_1,..,q_d}(\vec x)
\label{Mnqcomponents}
\end{eqnarray}
with the following contributions for $k=-2,-1,0,1$ :

$ \bullet $ The contribution $ {\cal L}^{[-2]}$ associated to the matrix ${\bold D}^{[0]} $ 
lowers the global degree of $O_{n_1,..,n_d}(\vec x) =x_1^{n_1} x_2^{n_2} ... x_d^{n_d}$ by two
\begin{eqnarray}
{\cal L}^{[-2]}O_{n_1,..,n_d}(\vec x)
&& \equiv   \sum_{j=1}^d 
  \langle j j \vert {\bold D}^{[0]}  \rangle 
  n_j (n_j-1) x_j^{n_j-2}  \prod_{l \ne j} x_l^{n_l} 
  + 2 \sum_{j_1=1}^d \sum_{j_2=j_1+1}^d 
 \langle j_1 j_2 \vert {\bold D}^{[0]}  \rangle 
     n_{j_1}  x_{j_1}^{n_{j_1}-1}  
n_{j_2} x_{j_2}^{n_{j_2}-1} \prod_{l \ne j_1,j_2} x_l^{n_l}
\nonumber \\
&& \equiv 
 \sum_{q_1=0}^{+\infty} ...  \sum_{q_d=0}^{+\infty} M^{[-2]}(n_1,..,n_d \vert q_1,...,q_d) O_{q_1,..,q_d}(\vec x)
\label{Lm2}
\end{eqnarray}
with the matrix elements
\begin{eqnarray}
M^{[-2]}(n_1,..,n_d \vert q_1,...,q_d)
&& \equiv   \sum_{j=1}^d 
  \langle j j \vert {\bold D}^{[0]}  \rangle 
  n_j (n_j-1) \delta_{q_j,n_j-2}  \prod_{l \ne j} \delta_{q_l,n_l} 
 \nonumber \\ &&
 + 2 \sum_{j_1=1}^d \sum_{j_2=j_1+1}^d 
 \langle j_1 j_2 \vert {\bold D}^{[0]}  \rangle 
     n_{j_1}  x_{j_1}^{n_{j_1}-1}  
n_{j_2} x_{j_2}^{n_{j_2}-1} \prod_{l \ne j_1,j_2} \delta_{q_l,n_l} 
\label{Mm2}
\end{eqnarray}

$ \bullet $ The contribution $ {\cal L}^{[-1]}$ associated to the matrices $ {\bold F}^{[0]}$
and ${\bold D}^{[1]} $ 
lowers the global degree of $O_{n_1,..,n_d}(\vec x) =x_1^{n_1} x_2^{n_2} ... x_d^{n_d}$ by one
\begin{eqnarray}
{\cal L}^{[-1]} O_{n_1,..,n_d}(\vec x)
&& =  \sum_{j=1}^d  
 \langle j \vert {\bold F}^{[0]}  \rangle 
 n_j  x_j^{n_j-1}  \prod_{l \ne j}^d x_l^{n_l} 
  + \sum_{j=1}^d 
  \sum_{i=1}^d \langle  j j \vert {\bold D}^{[1]} \vert i \rangle 
  n_j (n_j-1) x_i x_j^{n_j-2}  \prod_{l \ne j} x_l^{n_l} 
\nonumber \\
&&  + 2 \sum_{j_1=1}^d \sum_{j_2=j_1+1}^d 
 \sum_{i=1}^d \langle  j_1 j_2 \vert {\bold D}^{[1]} \vert i \rangle 
     n_{j_1} n_{j_2} 
     x_i x_{j_1}^{n_{j_1}-1}  
 x_{j_2}^{n_{j_2}-1} \prod_{l \ne j_1,j_2} x_l^{n_l}
\nonumber \\
&& \equiv 
 \sum_{q_1=0}^{+\infty} ...  \sum_{q_d=0}^{+\infty} M^{[-1]}(n_1,..,n_d \vert q_1,...,q_d) O_{q_1,..,q_d}(\vec x)
\label{Lm1}
\end{eqnarray}

$ \bullet $ The contribution $ {\cal L}^{[0]}$ associated to the matrices $ {\bold F}^{[1]}$
and ${\bold D}^{[2]} $ 
conserves the global degree of $O_{n_1,..,n_d}(\vec x) =x_1^{n_1} x_2^{n_2} ... x_d^{n_d}$ 
\begin{eqnarray}
&& {\cal L}^{[0]} O_{n_1,..,n_d}(\vec x)
 =  \sum_{j=1}^d  
 \sum_{i=1}^d \langle j \vert {\bold F}^{[1]} \vert i \rangle x_i
 n_j  x_j^{n_j-1}  \prod_{l \ne j}^d x_l^{n_l} 
\nonumber \\
&&  + \sum_{j=1}^d 
 \left[  \sum_{i=1}^d  \langle  j j \vert {\bold D}^{[2]} \vert i,i \rangle x_i^2
+ 2 \sum_{i_1=1}^{d-1} \sum_{i_2=i_1+1}^d \langle  j j \vert {\bold D}^{[2]} \vert i_1,i_2 \rangle x_{i_1} x_{i_2}
\right]
  n_j (n_j-1) x_j^{n_j-2}  \prod_{l \ne j} x_l^{n_l} 
\nonumber \\
&&  + 2 \sum_{j_1=1}^d \sum_{j_2=j_1+1}^d 
 \left[  \sum_{i=1}^d  \langle  j_1 j_2 \vert {\bold D}^{[2]} \vert i,i \rangle x_i^2
+2  \sum_{i_1=1}^{d-1} \sum_{i_2=i_1+1}^d \langle  j_1 j_2 \vert {\bold D}^{[2]} \vert i_1,i_2 \rangle x_{i_1} x_{i_2}
\right]
     n_{j_1}  x_{j_1}^{n_{j_1}-1}  
n_{j_2} x_{j_2}^{n_{j_2}-1} \prod_{l \ne j_1,j_2} x_l^{n_l}
\nonumber \\
&& \equiv 
 \sum_{q_1=0}^{+\infty} ...  \sum_{q_d=0}^{+\infty} M^{[0]}(n_1,..,n_d \vert q_1,...,q_d) O_{q_1,..,q_d}(\vec x)
\label{Lm0}
\end{eqnarray}

$ \bullet $ The contribution $ {\cal L}^{[1]}$ associated to the matrix $ {\bold F}^{[2]}$
raises the global degree of $O_{n_1,..,n_d}(\vec x) =x_1^{n_1} x_2^{n_2} ... x_d^{n_d}$ by one
\begin{eqnarray}
&& {\cal L}^{[1]} O_{n_1,..,n_d}(\vec x)
 =  \sum_{j=1}^d  
\left[   \sum_{i=1}^d  \langle j \vert {\bold F}^{[2]} \vert i,i \rangle x_{i}^2
+ 2 \sum_{i_1=1}^{d-1} \sum_{i_2=i_1+1}^d \langle j \vert {\bold F}^{[2]} \vert i_1,i_2 \rangle x_{i_1} x_{i_2}
\right]
 n_j  x_j^{n_j-1}  \prod_{l \ne j} x_l^{n_l} 
\nonumber \\
&& = \sum_{j=1}^d  
\langle j \vert {\bold F}^{[2]} \vert j,j \rangle 
  n_j  x_j^{n_j+1}  \prod_{l \ne j} x_l^{n_l} 
+ \sum_{j=1}^d  \sum_{i\ne j}  \langle j \vert {\bold F}^{[2]} \vert i,i \rangle n_j
x_{i}^{n_i+2}    x_j^{n_j-1}  \prod_{l \ne j,i} x_l^{n_l} 
\nonumber \\
&&+ 2 \sum_{i_1=1}^{d-1} \sum_{i_2=i_1+1}^d 
\left[\langle i_1 \vert {\bold F}^{[2]} \vert i_1,i_2 \rangle n_{i_1} 
     x_{i_2}^{n_{i_2}+1} \prod_{l \ne i_2} x_l^{n_l} 
+\langle i_2 \vert {\bold F}^{[2]} \vert i_1,i_2 \rangle n_{i_2}
x_{i_1}^{n_{i_1}+1}     \prod_{l \ne i_1} x_l^{n_l} 
\right]
\nonumber \\
&&+ 2 \sum_{i_1=1}^{d-1} \sum_{i_2=i_1+1}^d 
\sum_{j\ne i_1,i_2} \langle j \vert {\bold F}^{[2]} \vert i_1,i_2 \rangle n_j
x_{i_1}^{n_{i_1}+1} x_{i_2}^{n_{i_2}+1}  x_j^{n_j-1}  \prod_{l \ne j,i_1,i_2} x_l^{n_l} 
\nonumber \\
&& \equiv 
 \sum_{q_1=0}^{+\infty} ...  \sum_{q_d=0}^{+\infty} M^{[1]}(n_1,..,n_d \vert q_1,...,q_d) O_{q_1,..,q_d}(\vec x)
\label{L1}
\end{eqnarray}
with the matrix elements
\begin{eqnarray}
M^{[1]}(n_1,..,n_d \vert q_1,...,q_d)
&& \equiv  
\sum_{j=1}^d  
\langle j \vert {\bold F}^{[2]} \vert j,j \rangle 
  n_j  \delta_{q_j,n_j+1}  \prod_{l \ne j} \delta_{q_l,n_l} 
+ \sum_{j=1}^d  \sum_{i\ne j}  \langle j \vert {\bold F}^{[2]} \vert i,i \rangle n_j
\delta_{q_i,n_i+2}    \delta_{q_j,n_j-1} \prod_{l \ne j,i} \delta_{q_l,n_l} 
\nonumber \\
&&+ 2 \sum_{i_1=1}^{d-1} \sum_{i_2=i_1+1}^d 
\left[\langle i_1 \vert {\bold F}^{[2]} \vert i_1,i_2 \rangle n_{i_1} 
     \delta_{q_{i_2},n_{i_2}+1} \prod_{l \ne i_2} \delta_{q_l,n_l} 
+\langle i_2 \vert {\bold F}^{[2]} \vert i_1,i_2 \rangle n_{i_2}
\delta_{q_{i_1},n_{i_1}+1}     \prod_{l \ne i_1} \delta_{q_l,n_l} 
\right]
\nonumber \\
&&+ 2 \sum_{i_1=1}^{d-1} \sum_{i_2=i_1+1}^d 
\sum_{j\ne i_1,i_2} \langle j \vert {\bold F}^{[2]} \vert i_1,i_2 \rangle n_j
\delta_{q_{i_1},n_{i_1}+1}  \delta_{q_{i_2},n_{i_2}+1}  \delta_{q_j,n_j-1}  \prod_{l \ne j,i_1,i_2} \delta_{q_l,n_l} 
\label{M1}
\end{eqnarray}

As examples we have written the explicit matrix elements of ${\bold M}^{[-2]} $ in Eq. \ref{Mm2}
and of ${\bold M}^{[1]} $ in Eq. \ref{M1},
while the matrix elements of ${\bold M}^{[k=-1,0]} $
can be also deduced from Eqs \ref{Lm1} \ref{Lm0},
but one needs to distinguish whether some indices $(i,i_1,i_2)$ coincide with some indices $(j,j_1,j_2)$ or not.
As a consequence, the fully general expressions are somewhat heavy
when all the Taylor coefficients of Eq \ref{TaylorDcoefs}
for the diffusion matrix elements $D_{j_1,j_2}(\vec x) $
are present.
However in practice, one is often interested in models with various simplifications,
that may occur either at the block-level, as described in the following subsection,
or in the internal structure of some blocks, as will be described in the further section \ref{sec_DfullyDiag}.


\subsection{ Simplifications when the Carleman matrix ${\bold M}$ is block-diagonal or block-triangular }

\label{subsec_BlockTriangular}

In the present paper, we will be interested in three types of simplifications
when some blocks of Carleman matrix ${\bold M}
 =  {\bold M}^{[-2]} + {\bold M}^{[-1]}+{\bold M}^{[0]} + {\bold M}^{[1]}$ vanish,
 as described in the following three subsections.

\subsubsection{ Simplifications when the Carleman matrix is block-diagonal ${\bold M} =  {\bold M}^{[0]} $ }
 
 \label{subsec_blockDiag}

 The Carleman matrix ${\bold M}  $ reduces to its block-diagonal component $  {\bold M}^{[0]} $ of Eq. \ref{Lm0}
when the generator ${\cal L}$ reduces to the contribution ${\cal L}^{[0]} $ of Eq. \ref{GeneratorClassDegree4}
i.e. the forces $F_j(\vec x) $ of Eq. \ref{TaylorF} should only contain the linear terms in the variables $x_i$ parametrized by the matrix $ {\bold F}^{[1]}$
\begin{eqnarray}
F_j(\vec x) && 
=  \sum_{i=1}^d \langle j \vert {\bold F}^{[1]} \vert i \rangle x_i
\label{TaylorFLinear}
\end{eqnarray}
while the diffusion matrix elements of Eq. \ref{TaylorD}
should only contain the quadratic terms parametrized by the matrix $ {\bold D}^{[2]}$
\begin{eqnarray}
D_{j_1 j_2}(\vec x) = 
\sum_{i_1=1}^d \sum_{i_2=1}^d \langle  j_1 j_2 \vert {\bold D}^{[2]} \vert i_1,i_2 \rangle x_{i_1} x_{i_2}
\label{TaylorDQuadratic}
\end{eqnarray}

Since the matrix $  {\bold M}^{[0]} $ can be decomposed into blocks $  {\bold M}^{[0]}_{[n,n]} $
associated to a given degree $n=n_1+..+n_d$, 
then the eigenvalues of the full Carleman matrix $  {\bold M}^{[0]} $ can be obtained from
the diagonalizations of the individual blocks $ {\bold M}^{[0]}_{[n,n]} $,
while the dynamics of of Eq. \ref{DynMoments} for the moment $m_t(n_1,..,n_d) $ of degree $n=n_1+..+n_d$
only involves moments $m_t(q_1,..,q_d) $ whose total degree $q= q_1+..+q_d $
 is the same $q=n$.


\subsubsection{ Simplifications when the Carleman matrix is block-lower-triangular ${\bold M}
 = {\bold M}^{[-2]} + {\bold M}^{[-1]}+ {\bold M}^{[0]} $ }
 
 When the Carleman matrix is block-lower-triangular ${\bold M}
 = {\bold M}^{[-2]} + {\bold M}^{[-1]}+ {\bold M}^{[0]} $,
 i.e. when the matrix $  {\bold F}^{[2]}=0$ 
 vanishes in the parametrization of Eqs \ref{TaylorF} for the forces,
 then the eigenvalues of the full Carleman matrix $  {\bold M} $ are still given
by the eigenvalues of the individual blocks $ {\bold M}^{[0]}_{[n,n]} $,
while the dynamics of Eq. \ref{DynMoments} for the moment $m_t(n_1,..,n_d) $  of degree $n= n_1+..+n_d$
now involves moments $m_t(q_1,..,q_d) $ whose total degree $q= q_1+..+q_d $
 is either the same $q=n$ or smaller $q=n-1,n-2$ by one or two.
 So the solutions for moments can be then constructed iteratively starting from the lower ones.

 
 \subsubsection{ Simplifications when the Carleman matrix is block-upper-triangular ${\bold M}
 =  {\bold M}^{[0]} + {\bold M}^{[1]}$ }
 
 When the Carleman matrix is block-upper-triangular ${\bold M}
 ={\bold M}^{[0]} + {\bold M}^{[1]} $,
 i.e. when the matrices $  {\bold F}^{[0]}=0$ and ${\bold D}^{[0]}=0= {\bold D}^{[1]}$ vanish in 
 in the parametrization of Eqs \ref{TaylorF} and \ref{TaylorD} for the Ito forces and the diffusion matrix elements,
then the eigenvalues of the full Carleman matrix $  {\bold M} $ are still given
by the eigenvalues of the individual blocks $ {\bold M}^{[0]}_{[n,n]} $,
while the dynamics of Eq. \ref{DynMoments} for the moment $m_t(n_1,..,n_d) $  of degree $n= n_1+..+n_d$
now involves moments $m_t(q_1,..,q_d) $ whose total degree $q= q_1+..+q_d $
 is either the same $q=n$ or higher $q=n+1$ by one.
 The simplest example concerning the one-dimensional deterministic model is described in Appendix \ref{app_deterministic}.


\subsection{  Models where ${\bold D}^{[2]} =0$ : linear change of variables to diagonalize the matrix $F^{[1]}$
and thus ${\cal L}^{[0]} $  
 before applying the Carleman approach  }

\label{subsec_D2vanishDiagoF1}

In models where the matrix ${\bold D}^{[2]} $ vanishes ${\bold D}^{[2]} =0$ in the parametrization of Eqs \ref{TaylorD} for the diffusion matrix,
then the contribution $ {\cal L}^{[0]}$ of the generator
that determines the diagonal block $  {\bold M}^{[0]} $ of the Carleman matrix ${\bold M}$
only involves the matrix ${\bold F}^{[1]} $ parametrizing the linear forces
\begin{eqnarray}
  {\cal L}^{[0]}
 =  \sum_{j=1}^d  \sum_{i=1}^d \langle j \vert {\bold F}^{[1]} \vert i \rangle x_i  \frac{ \partial }{\partial x_j}
 \ \ \ \ \text{ if } \ \ {\bold D}^{[2]} =0
\label{LzeroOnlyF1}
\end{eqnarray}
Before applying the Carleman approach, it can be then useful 
 to make the linear change of variables from the coordinates $(x_1,..x_d)$
 towards the new coordinates $(\xi_1,..,\xi_d)$ that will make diagonal
  the operator ${\cal L}^{[0]} $.
  The spectral decomposition of the $d \times d$ matrix ${\bold F}^{[1]} $
  involves its $d$ eigenvalues $\lambda_{\nu=1,..,d}$ (that may be complex)
  \begin{eqnarray}
{\bold F}^{[1]} = \sum_{\nu=1}^d \lambda_{\nu}  \vert r_{\nu} \rangle \langle l_{\nu} \vert
\label{DiagoF1}
\end{eqnarray}
while the corresponding right eigenvectors $\vert r_{\nu} \rangle $ and left eigenvectors $\langle l_{\nu} \vert $
form a bi-orthogonal basis satisfying
\begin{eqnarray}
 \langle l_{\nu} \vert r_{\mu} \rangle = \delta_{\nu,\mu}
\nonumber \\
{\bold 1}=\sum_{\nu=1}^d   \vert r_{\nu} \rangle \langle l_{\nu} \vert
\label{F1biorthog}
\end{eqnarray}

The change of variables between the initial coordinates $x_i \equiv \langle i \vert x\rangle$
 and the new coordinates $\xi_{\nu} \equiv \langle l_{\nu} \vert x\rangle$ 
 involves these left and right eigenvectors
  \begin{eqnarray}
  \xi_{\nu} && = \langle l_{\nu} \vert x\rangle  
  = \sum_{i=1}^d  \langle l_{\nu} \vert i\rangle  \langle i \vert x\rangle
  =  \sum_{i=1}^d  \langle l_{\nu} \vert i\rangle  x_i
  \nonumber \\
x_i  && = \langle i \vert x \rangle
  = \sum_{\nu=1}^d \langle i  \vert r_{\nu} \rangle \langle l_{\nu} \vert x \rangle
 =  \sum_{\nu=1}^d \langle i  \vert r_{\nu} \rangle  \xi_{\nu}
\label{ChangrVarF1}
\end{eqnarray} 
and leads to the corresponding change of variables for the partial derivatives 
\begin{eqnarray}
\frac{ \partial }{\partial x_j} 
= \sum_{\nu=1}^d \frac{ \partial  \xi_{\nu}}{\partial x_j}  \frac{ \partial }{\partial  \xi_{\nu}} 
=  \sum_{\nu=1}^d \langle l_{\nu} \vert j \rangle   \frac{ \partial }{\partial  \xi_{\nu}} 
\label{ChangrVarF1deri}
\end{eqnarray}
Then the contribution $ {\cal L}^{[0]}$ of Eq. \ref{LzeroOnlyF1} becomes diagonal in the new variables $\xi_{\nu} $
\begin{eqnarray}
  {\cal L}^{[0]}
&& =  \sum_{j=1}^d  \sum_{i=1}^d \langle j \vert {\bold F}^{[1]} \vert i \rangle \langle i \vert x \rangle  \frac{ \partial }{\partial x_j}
=  \sum_{j=1}^d   \langle j \vert \left( \sum_{\nu=1}^d \lambda_{\nu}  \vert r_{\nu} \rangle \langle l_{\nu} \vert \right)  \vert x \rangle  \left(  \sum_{\mu=1}^d \langle l_{\mu} \vert j \rangle   \frac{ \partial }{\partial  \xi_{\mu}} \right)
\nonumber \\
&& =    \sum_{\nu=1}^d \sum_{\mu=1}^d
\lambda_{\nu} \left(  \sum_{j=1}^d   \langle l_{\mu} \vert j \rangle \langle j \vert r_{\nu} \rangle\right)  \langle l_{\nu}   \vert x \rangle       \frac{ \partial }{\partial  \xi_{\mu}} 
=   \sum_{\nu=1}^d \sum_{\mu=1}^d
\lambda_{\nu} \left(  \delta_{\mu,\nu} \right)  \langle l_{\nu}   \vert x \rangle       \frac{ \partial }{\partial  \xi_{\mu}} 
=  \sum_{\nu=1}^d 
\lambda_{\nu}  \left( \xi_{\nu}    \frac{ \partial }{\partial  \xi_{\nu}} \right)
\label{LzeroOnlyF1diag}
\end{eqnarray}

The Carleman approach can be now applied to 
the monomials in the new variables $\xi_{\nu}$
\begin{eqnarray}
\Omega_{\kappa_1,..,\kappa_d}(\xi_1,..,\xi_d) \equiv \xi_1^{\kappa_1} \xi_2^{\kappa_2} ... \xi_d^{\kappa_d}  =\prod_{\mu=1}^d \xi_{\mu}^{\kappa_{\mu}} 
\ \ \ \text{ with the $d$ integers $\kappa_{\nu} \in {\mathbb N}$}
\label{MonomialsCarlemannu}
\end{eqnarray}
instead of the monomials $O_{n_1,..,n_d}(\vec x) = x_1^{n_1} x_2^{n_2} ... x_d^{n_d} $ of Eq. \ref{MonomialsCarleman}
involving the initial variables $x_i$.
The application of ${\cal L}^{[0]} $ of Eq. \ref{LzeroOnlyF1diag}
to Eq. \ref{MonomialsCarlemannu} reduces to
\begin{eqnarray}
  {\cal L}^{[0]}\Omega_{\kappa_1,..,\kappa_d}(\xi_1,..,\xi_d) && =  \left( \sum_{\nu=1}^d 
\lambda_{\nu}  \  \xi_{\nu}    \frac{ \partial }{\partial  \xi_{\nu}}  \right)\prod_{\mu=1}^d \xi_{\mu}^{\kappa_{\mu}}
= \left( \sum_{\nu=1}^d \lambda_{\nu}   \kappa_{\nu} \right)   \prod_{\mu=1}^d \xi_{\mu}^{\kappa_{\mu}}
 \nonumber \\ &&
 = \left( \sum_{\nu=1}^d \lambda_{\nu}   \kappa_{\nu} \right) \Omega_{\kappa_1,..,\kappa_d}(\xi_1,..,\xi_d)
 \equiv E_{\kappa_1,..,\kappa_d} \Omega_{\kappa_1,..,\kappa_d}(\xi_1,..,\xi_d)
\label{L0MonomialsCarlemannu}
\end{eqnarray}
i.e. the monomials $\Omega_{\kappa_1,..,\kappa_d}(\xi_1,..,\xi_d) $ 
are directly the eigenvectors of the operator ${\cal L}^{[0]} $
and of the corresponding diagonal-blocks $  {\bold M}^{[0]} $ of the Carleman matrix ${\bold M}$,
while the corresponding eigenvalues of ${\cal L}^{[0]} $ and $ {\bold M}^{[0]} $
\begin{eqnarray}
E_{\kappa_1,..,\kappa_d} =  \sum_{\nu=1}^d   \kappa_{\nu}  \lambda_{\nu} 
\label{EigenCLEigenF1}
\end{eqnarray}
reduce to linear combinations of the $d$ eigenvalues $(\lambda_1,\lambda_2,...,\lambda_d ) $ of the $d \times d$ matrix ${\bold F}^{[1]} $
of Eq. \ref{DiagoF1} with the integers coefficients $(\kappa_1,\kappa_2,...,\kappa_d ) \in {\mathbb N}^d $.

Let us summarize the conclusion for models with ${\bold D}^{[2]} =0$ : whenever the eigenvalues of the full Carleman matrix ${\bold M} $
are given by the eigenvalues of the diagonal block $  {\bold M}^{[0]} $,
i.e. when the full Carleman matrix ${\bold M} $ 
is either block-diagonal or block-lower-triangular or block-upper-triangular
as discussed in the previous subsection \ref{subsec_BlockTriangular},
then the eigenvalues of the full Carleman matrix ${\bold M} $ are given by Eq. \ref{EigenCLEigenF1}.
For deterministic models,
this result is usually derived via a tensor-representation in the initial variables $x_i$
\cite{Steeb1980}
 that is redundant since the eigenvalue $ E_{\kappa_1,..,\kappa_d}$ then appears with the artificial multiplicity
$ \frac {(\kappa_1+..+\kappa_d)!}{ \kappa_1! \kappa_2! .. \kappa_d!}$ (see the discussion in \cite{Andrade1981}).
So the change described above from the initial variables $x_i$ towards the new variables $\xi_{\nu}$ 
 seems clearer to obtain directly the eigenvalues 
of Eq. \ref{EigenCLEigenF1} without artificial multiplicities.
 Then one should of course translate 
the other contributions ${\cal L}^{[k \ne 0]} $ of Eq. \ref{GeneratorClassDegree}
 into the new variables
in order to obtain the corresponding off-diagonal-blocks of the Carleman matrix ${\bold M}^{[k \ne 0]} $
associated to the monomials $\Omega_{\kappa_1,..,\kappa_d}(\xi_1,..,\xi_d) \equiv \xi_1^{\kappa_1} \xi_2^{\kappa_2} ... \xi_d^{\kappa_d} $ in the new variables using Eqs \ref{ChangrVarF1} and \ref{ChangrVarF1deri}:

$\bullet$ The contribution ${\cal L}^{[-2]} $
\begin{eqnarray}
{\cal L}^{[-2]}
&& =   \sum_{j_1=1}^d \sum_{j_2=1}^d  \langle j_1 j_2 \vert {\bold D}^{[0]}  \rangle 
 \left( \sum_{\nu_1=1}^d \langle l_{\nu_1} \vert j_1 \rangle   \frac{ \partial }{\partial  \xi_{\nu_1}}  \right)
  \left( \sum_{\nu_2=1}^d \langle l_{\nu_2} \vert j_2 \rangle   \frac{ \partial }{\partial  \xi_{\nu_2}}  \right)
\nonumber \\
&&  =  \sum_{\nu_1=1}^d \sum_{\nu_2=1}^d   
 \left( \sum_{j_1=1}^d \sum_{j_2=1}^d \langle l_{\nu_1} \vert j_1 \rangle   
   \langle l_{\nu_2} \vert j_2 \rangle \langle j_1 j_2 \vert {\bold D}^{[0]}  \rangle  \right)
  \frac{ \partial^2 }{\partial  \xi_{\nu_1}\partial  \xi_{\nu_2}}
\nonumber \\
&& \equiv \sum_{\nu_1=1}^d \sum_{\nu_2=1}^d   
  \langle l_{\nu_1} l_{\nu_2} \vert {\bold D}^{[0]}  \rangle  
  \frac{ \partial^2 }{\partial  \xi_{\nu_1}\partial  \xi_{\nu_2}}
\label{GeneratorClassDegree4xim2}
\end{eqnarray}
involves the new matrix elements $\langle l_{\nu_1} l_{\nu_2} \vert {\bold D}^{[0]}  \rangle $ instead
of the old ones $ \langle j_1 j_2 \vert {\bold D}^{[0]}  \rangle $.

$\bullet$ The contribution ${\cal L}^{[-1]} $
\begin{eqnarray}
{\cal L}^{[-1]}
&& = \sum_{j=1}^d \langle j \vert {\bold F}^{[0]}  \rangle 
\left(  \sum_{\nu=1}^d \langle l_{\nu} \vert j \rangle   \frac{ \partial }{\partial  \xi_{\nu}} \right)
  + \sum_{j_1=1}^d \sum_{j_2=1}^d    \sum_{i=1}^d \langle  j_1 j_2 \vert {\bold D}^{[1]} \vert i \rangle 
  \left(  \sum_{\nu=1}^d \langle i  \vert r_{\nu} \rangle  \xi_{\nu} \right)
\left( \sum_{\nu_1=1}^d \langle l_{\nu_1} \vert j_1 \rangle   \frac{ \partial }{\partial  \xi_{\nu_1}}  \right)
  \left( \sum_{\nu_2=1}^d \langle l_{\nu_2} \vert j_2 \rangle   \frac{ \partial }{\partial  \xi_{\nu_2}}  \right)  
  \nonumber \\
&&  =\sum_{\nu=1}^d  
\left(  \sum_{j=1}^d \langle l_{\nu} \vert j \rangle \langle j \vert {\bold F}^{[0]}  \rangle   \right)
\frac{ \partial }{\partial  \xi_{\nu}}
  + \sum_{\nu_1=1}^d \sum_{\nu_2=1}^d    \sum_{\nu=1}^d 
    \left(   
   \sum_{j_1=1}^d    
    \sum_{j_2=1}^d  \sum_{j=1}^d \langle l_{\nu_1} \vert j_1 \rangle
      \langle   l_{\nu_2} \vert j_2 \rangle  \langle  j_1 j_2 \vert {\bold D}^{[1]} \vert i \rangle  \langle i  \vert r_{\nu} \rangle  \right)
  \xi_{\nu}
   \frac{ \partial^2 }{\partial  \xi_{\nu_1}\partial  \xi_{\nu_2}}
   \nonumber \\
&&  = \sum_{\nu=1}^d  
 \langle l_{\nu} \vert {\bold F}^{[0]}  \rangle  
\frac{ \partial }{\partial  \xi_{\nu}}
  + \sum_{\nu_1=1}^d \sum_{\nu_2=1}^d    \sum_{\nu=1}^d 
     \langle l_{\nu_1} l_{\nu_2} \vert {\bold D}^{[1]} \vert r_{\nu} \rangle 
  \xi_{\nu}
   \frac{ \partial^2 }{\partial  \xi_{\nu_1}\partial  \xi_{\nu_2}}
\label{GeneratorClassDegree4xim1}
\end{eqnarray}
involves the new matrix elements $\langle l_{\nu} \vert {\bold F}^{[0]}  \rangle $ and $\langle l_{\nu_1} l_{\nu_2} \vert {\bold D}^{[1]} \vert r_{\nu} \rangle $ instead
of the old ones $\langle j \vert {\bold F}^{[0]}  \rangle  $ and $\langle  j_1 j_2 \vert {\bold D}^{[1]} \vert i \rangle $.

$\bullet$ The contribution ${\cal L}^{[1]} $
\begin{eqnarray}
{\cal L}^{[1]}
&& =  
 \sum_{j=1}^d  \sum_{i_1=1}^d \sum_{i_2=1}^{d} \langle j \vert {\bold F}^{[2]} \vert i_1,i_2 \rangle
  \left(  \sum_{\nu_1=1}^d \langle i_1  \vert r_{\nu_1} \rangle  \xi_{\nu_1} \right)
 \left(  \sum_{\nu_2=1}^d \langle i_2  \vert r_{\nu_2} \rangle  \xi_{\nu_2} \right)
  \left( \sum_{\nu=1}^d \langle l_{\nu} \vert j \rangle   \frac{ \partial }{\partial  \xi_{\nu}}  \right)
     \nonumber \\
&& =
 \sum_{\nu=1}^d  \sum_{\nu_1=1}^d \sum_{\nu_2=1}^{d} 
  \left(  \sum_{i_1=1}^d  
    \sum_{i_2=1}^d  
    \sum_{j=1}^d \langle l_{\nu} \vert j \rangle  \langle j \vert {\bold F}^{[2]} \vert i_1,i_2 \rangle
     \langle i_1  \vert r_{\nu_1} \rangle \langle i_2  \vert r_{\nu_2} \rangle
     \right)
  \xi_{\nu_1} \xi_{\nu_2} \frac{ \partial }{\partial  \xi_{\nu}}
   \nonumber \\
&& =
 \sum_{\nu=1}^d  \sum_{\nu_1=1}^d \sum_{\nu_2=1}^{d} 
 \langle l_{\nu}  \vert {\bold F}^{[2]} \vert  r_{\nu_1}  r_{\nu_2} \rangle 
  \xi_{\nu_1} \xi_{\nu_2} \frac{ \partial }{\partial  \xi_{\nu}}
\label{GeneratorClassDegree4xip1}
\end{eqnarray}
involves the new matrix elements $\langle l_{\nu}  \vert {\bold F}^{[2]} \vert  r_{\nu_1}  r_{\nu_2} \rangle $ instead
of the old ones $\langle j \vert {\bold F}^{[2]} \vert i_1,i_2 \rangle $.

In summary, all the matrices ${\bold F}^{[k]} $ and ${\bold D}^{[k]} $ now appear via their matrix elements
in the bi-orthogonal basis of ${\bold F}^{[1]} $ with the left eigenvectors for all indices on the left 
and the right eigenvectors for all indices on the right.


\subsection{  Discussion  }

In this section we have described how the Carleman approach can be applied to stochastic processes
in dimension $d$ whenever the Ito forces $F_j(\vec x)$ and the diffusion matrix elements $D_{j_1 j_2}(\vec x)$
can be expanded in Taylor series with respect to the $d$ coordinates $(x_1,x_2,..,x_d)$
via Eqs \ref{TaylorF} and \ref{TaylorD}.
In the remaining sections, we will focus on models where 
the diffusion matrices ${\bold D}^{[k=0,1,2]} $ appearing in Eq. \ref{TaylorD} 
have a simpler structure called fully-diagonal in the next section.


\section{ Simplifications when the diffusion matrices ${\bold D}^{[k=0,1,2]} $ are fully-diagonal  }

\label{sec_DfullyDiag}

In this section, the Carleman approach described in the previous section
is applied to models where the diffusion matrices ${\bold D}^{[k=0,1,2]} $ appearing in Eq. \ref{TaylorD}
have non-vanishing matrix elements only if all indices coincide.

\subsection{ Models where the diffusion matrices ${\bold D}^{[k=0,1,2]} $ are fully-diagonal}

In many models of interest,
 the diffusion matrix is diagonal $D_{j,i}(\vec x) =\delta_{j,i} D_{j,j}(\vec x)$
and in addition, the diagonal element $D_{j,j}(\vec x) $ is either independent of $\vec x$ 
or only depends on the coordinate $x_{j}$,
so that the Taylor expression of Eq. \ref{TaylorD} 
reduces to
\begin{eqnarray}
 D_{j,i}(\vec x) && = \delta_{j,i} D_{j,j}(x_{j})
\nonumber \\
D_{j,j}(x_j) && = \langle j j \vert {\bold D}^{[0]}  \rangle 
+  \langle  j j \vert {\bold D}^{[1]} \vert j \rangle x_{j}
+  \langle  j j \vert {\bold D}^{[2]} \vert j,j \rangle x_{j}^2
\nonumber \\
&&\equiv D^{[0]}_j+D^{[1]}_j x_j + D^{[2]}_j x_j^2
\label{DdiagoEtSingle}
\end{eqnarray}
i.e. the only non-vanishing elements of the matrices
${\bold D}^{[k=0,1,2]} $ appearing in Eq. \ref{TaylorD}
are $\langle j j \vert {\bold D}^{[0]}  \rangle $, $\langle  j j \vert {\bold D}^{[1]} \vert j \rangle $ and $\langle  j j \vert {\bold D}^{[2]} \vert j,j \rangle $
corresponding to all-coinciding-indices and will be denoted by the simplified notations $D^{[k=0,1,2]}_j$ from now on.

Then the generator of Eqs \ref{Generator}  and \ref{GeneratorClassDegree}
reduces to
\begin{eqnarray}
{\cal L} && = \sum_{j=1}^d   F_j(\vec x)   \frac{ \partial }{\partial x_j}
  + \sum_{j=1}^d    D_{jj}(\vec x)   \frac{ \partial^2 }{\partial x_j^2 }
\nonumber \\
&& =  \sum_{j=1}^d \left[ F^{[0]}_j 
+ \sum_{i=1}^d F^{[1]}_{ji} x_i
+   \sum_{i=1}^d  F^{[2]}_{j;ii}  x_{i}^2
+ 2 \sum_{i_1=1}^{d-1} \sum_{i_2=i_1+1}^d F^{[2]}_{j;i_1i_2} x_{i_1} x_{i_2}
\right]  \frac{ \partial }{\partial x_j}
 \nonumber \\
&&  + \sum_{j=1}^d   \left[ 
D^{[0]}_j+D^{[1]}_j x_j + D^{[2]}_j x_j^2
\right]   \frac{ \partial^2 }{\partial x_j^2 }
\label{GeneratorDiag}
\end{eqnarray}


\subsection{ Simplifications for the Carleman matrix ${\bold M}$  }

The structure of Eq. \ref{DdiagoEtSingle} for the diffusion matrix elements
 leads to the following simplifications for the three contributions $({\bold M}^{[-2]},{\bold M}^{[-1]},{\bold M}^{[0]})$ 
 of the Carleman matrix ${\bold M}$
(while the fourth contribution $ {\bold M}^{[1]}$ of Eq. \ref{M1} is unchanged since it only depends on the matrix ${\bold F}^{[2]} $) :

$ \bullet $ The action of $ {\cal L}^{[-2]}$ in Eq. \ref{Lm2} becomes
\begin{eqnarray}
{\cal L}^{[-2]}O_{n_1,..,n_d}(\vec x)
&& \equiv   \sum_{j=1}^d 
  D^{[0]}_j 
  n_j (n_j-1) x_j^{n_j-2}  \prod_{l \ne j} x_l^{n_l} 
\equiv 
 \sum_{q_1=0}^{+\infty} ...  \sum_{q_d=0}^{+\infty} M^{[-2]}(n_1,..,n_d \vert q_1,...,q_d) O_{q_1,..,q_d}(\vec x)
\label{Lm2simpli}
\end{eqnarray}
and leads to the simplified matrix elements with respect to Eq. \ref{Mm2}
\begin{eqnarray}
M^{[-2]}(n_1,..,n_d \vert q_1,...,q_d)
&& \equiv   \sum_{j=1}^d 
  D^{[0]}_j 
  n_j (n_j-1) \delta_{q_j,n_j-2}  \prod_{l \ne j} \delta_{q_l,n_l} 
\label{Mm2simpli}
\end{eqnarray}
that involves only the $d$ possibilities where one integer $q_j=n_j-2$ is lower by two units 
while all the other $l \ne j$ coincide $q_l=n_l$.

$ \bullet $ The action of $ {\cal L}^{[-1]}$ in Eq. \ref{Lm1} becomes
\begin{eqnarray}
{\cal L}^{[-1]} O_{n_1,..,n_d}(\vec x)
&& =   \sum_{j=1}^d  
 \langle j \vert {\bold F}^{[0]}  \rangle 
 n_j  x_j^{n_j-1}  \prod_{l \ne j} x_l^{n_l} 
  + \sum_{j=1}^d 
   \langle  j j \vert {\bold D}^{[1]} \vert j \rangle 
  n_j (n_j-1)  x_j^{n_j-1}  \prod_{l \ne j} x_l^{n_l} 
\nonumber \\
&& \equiv 
 \sum_{q_1=0}^{+\infty} ...  \sum_{q_d=0}^{+\infty} M^{[-1]}(n_1,..,n_d \vert q_1,...,q_d) O_{q_1,..,q_d}(\vec x)
\label{Lm1simpli}
\end{eqnarray}
and leads to the matrix elements
\begin{eqnarray}
M^{[-1]}(n_1,..,n_d \vert q_1,...,q_d)
&& =   
  \sum_{j=1}^d  
\left[  F^{[0]}_j    n_j   +    D^{[1]}_j   n_j (n_j-1) \right] 
\delta_{q_j,n_j-1}  \prod_{l \ne j} \delta_{q_l,n_l} 
\label{m1simpli}
\end{eqnarray}
that involves only the $d$ possibilities where one integer $q_j=n_j-1$ is lower by one unit 
while all the other $l \ne j$ coincide $q_l=n_l$.

$ \bullet $ The action of $ {\cal L}^{[0]}$ in Eq. \ref{Lm0} becomes
\begin{eqnarray}
&& {\cal L}^{[0]} O_{n_1,..,n_d}(\vec x)
 = \left( \sum_{j=1}^d  
\left[  F^{[1]}_{jj}   n_j   
+ \sum_{j=1}^d    D^{[2]}_j   n_j (n_j-1) \right]  \right)
\prod_{l =1}^d x_l^{n_l}    
 + \sum_{j=1}^d  
 \sum_{i\ne j} F^{[1]}_{ji} 
 n_j x_i^{n_i+1} x_j^{n_j-1}  \prod_{l \ne i, j}^d x_l^{n_l} 
\nonumber \\
&& \equiv 
 \sum_{q_1=0}^{+\infty} ...  \sum_{q_d=0}^{+\infty} M^{[0]}(n_1,..,n_d \vert q_1,...,q_d) O_{q_1,..,q_d}(\vec x)
\label{Lm0simpli}
\end{eqnarray}
and leads to the matrix elements
\begin{eqnarray}
M^{[0]}(n_1,..,n_d \vert q_1,...,q_d)
&& =  \left(  \sum_{j=1}^d  
\left[  F^{[1]}_{jj}  n_j   
+     D^{[2]}_j   n_j (n_j-1) \right]  \right)
\prod_{l =1}^d  \delta_{q_l,n_l}     
\nonumber \\
&& + \sum_{j=1}^d  
 \sum_{i\ne j} F^{[1]}_{ji} 
 n_j \delta_{q_i,n_i+1} \delta_{q_j,n_j-1}  \prod_{l \ne j,i}^d  \delta_{q_l,n_l} 
\label{m0simpli}
\end{eqnarray}
where the first line corresponds to the diagonal contribution where the $d$ integers coincide $q_l=n_l$
\begin{eqnarray}
M^{[0]}(n_1,..,n_d \vert n_1,...,n_d)
&& =  \sum_{j=1}^d  
\left[  F^{[1]}_{jj}  n_j   
+     D^{[2]}_j   n_j (n_j-1) \right]  
\label{m0simplidiagnq}
\end{eqnarray}
while the second line of Eq. \ref{m0simpli}
corresponds to the terms where one integer is bigger $q_i=n_i+1$
and another integer is lower $q_j=n_j-1$, while the other $l \ne (j,i )$ coincide.


\subsection{ Interpretations of the 'fully-diagonal' diffusion matrix of Eq. \ref{DdiagoEtSingle} in terms of SDE  }

The interpretation in terms of Ito SDE of Eq. \ref{ItoSDE} involving $\alpha_{max}$ noises $dB_{\alpha}(t)$
require that the amplitudes $G_{j \alpha}(\vec x) $ are compatible 
with the diffusion matrix of Eq. \ref{Dij} satisfying the fully-diagonal structure of Eq. \ref{DdiagoEtSingle}
\begin{eqnarray}
 \frac{1}{2} \sum_{\alpha=1}^{\alpha_{max}} G_{j \alpha}(\vec x)       G_{i \alpha}(\vec x)
=
 D_{j,i}(\vec x) = \delta_{ji} D_{jj}(x_j)
 = \delta_{j,i} \left[ D^{[0]}_j+D^{[1]}_j x_j + D^{[2]}_j x_j^2 \right]
\label{Dijsingle}
\end{eqnarray}


  \subsubsection{ Interpretation with a single noise $dB_j(t)$ for each coordinate $x_j(t)$}
  
  If there is a single noise $dB_j(t)$ in the Ito SDE of Eq. \ref{ItoSDE} each coordinate $x_j(t)$,
  then $\alpha_{max}=d$ and the matrix $G_{j \alpha}(\vec x)$ is diagonal 
    \begin{eqnarray}
 G_{j \alpha}(\vec x)    =\delta_{j,\alpha} G_{j j}(\vec x)
 \label{Gdiagonal}
\end{eqnarray}
   while the diagonal elements $G_{j j}(\vec x) $ determined by Eq. \ref{Dijsingle} only depend on $x_j$
  \begin{eqnarray}
 G_{j j}(\vec x)
= \sqrt{  2 \left[ D^{[0]}_j+D^{[1]}_j x_j + D^{[2]}_j x_j^2
\right] } \equiv \sqrt{ 2 D_{jj}(x_j)} \equiv G_{j j}(x_j)
\label{Dijsinglediagonal}
\end{eqnarray}

The Stratonovich forces $ f_j(\vec x)  $ of Eq. \ref{StratoItoCorrespondance} then reads
 \begin{eqnarray}
f_j(\vec x) && = F_j(\vec x) 
- \frac{1}{2} \sum_{i=1}^d \left( \sum_{\alpha=1}^{\alpha_{max}} \delta_{i,\alpha} \delta_{j,\alpha} \right) 
G_{ii}(x_i) 
\frac{\partial G_{j j}(x_j)}{\partial x_i}  
=  F_j(\vec x) - \frac{1}{2}  G_{j j}(x_j) \frac{\partial }{\partial x_j}  G_{j j}(x_j)
\nonumber \\
&&  
= F_j(\vec x) - \frac{1}{4}   \frac{\partial }{\partial x_j}  G^2_{j j}(x_j)
=  F_j(\vec x) - \frac{1}{2}   \frac{\partial }{\partial x_j} \left[ 
 D^{[0]}_j+D^{[1]}_j x_j + D^{[2]}_j x_j^2
\right]
\nonumber \\
&& = F_j(\vec x) - \frac{D^{[1]}_j }{2}  
-  D^{[2]}_j  x_{j}
\label{StratoItoCorrespondanceSingle}
\end{eqnarray}

So for the present models where the Ito forces $F_j(\vec x) $ are given by the polynomials of degree two of
Eq. \ref{TaylorF}, the Stratonovich forces $f_j(\vec x) $ 
\begin{eqnarray}
f_j(\vec x)  && 
= \left( F^{[0]}_j  - \frac{D^{[1]}_j}{2} \right)
+ \left(  F^{[1]}_{jj} - D^{[2]}_j \right) x_j
+ \sum_{i\ne j}^d F^{[1]}_{ji} x_i
+  \sum_{i_1=1}^d \sum_{i_2=1}^{d} F^{[2]}_{j;i_1,i_2} x_{i_1} x_{i_2}
\nonumber \\
&& \equiv f^{[0]}_j + \sum_{i=1}^d f^{[1]}_{ji} x_i
+  \sum_{i_1=1}^d \sum_{i_2=1}^{d} f^{[2]}_{j;  i_1,i_2 } x_{i_1} x_{i_2}
\label{TaylorFStrato}
\end{eqnarray}
are also polynomials of degree two 
with the following coefficients
  \begin{eqnarray}
 f^{[0]}_j && =F^{[0]}_j  - \frac{D^{[1]}_j}{2} 
\nonumber \\
f^{[1]}_{ji} && =  F^{[1]}_{ji} - \delta_{i,j} D^{[2]}_j
\nonumber \\
f^{[2]}_{j;  i_1,i_2 } && =F^{[2]}_{j;i_1,i_2}
\label{DfullyStratoSingle}
\end{eqnarray}

In conclusion,  the SDE involving a single noise $dB_j(t) $ per coordinate $x_j(t)$ 
have similar forms in the Ito and Stratonovich interpretations
  \begin{eqnarray}
dx_j(t)   && =  \left( F^{[0]}_j + \sum_{i=1}^d F^{[1]}_{ji} x_i(t)
+  \sum_{i_1=1}^d \sum_{i_2=1}^{d} F^{[2]}_{j;  i_1,i_2 } x_{i_1}(t) x_{i_2}(t)\right) dt + \sqrt{ 2 \left[D^{[0]}_j+D^{[1]}_j x_j(t) + D^{[2]}_j x_j^2(t) \right]} dB_j(t) 
\text{[Ito]}
\nonumber \\
dx_j(t)   && = \left( f^{[0]}_j + \sum_{i=1}^d f^{[1]}_{ji} x_i(t)
+  \sum_{i_1=1}^d \sum_{i_2=1}^{d} f^{[2]}_{j;  i_1,i_2 } x_{i_1}(t) x_{i_2}(t)\right) dt 
+\sqrt{ 2 \left[D^{[0]}_j+D^{[1]}_j x_j(t) + D^{[2]}_j x_j^2(t) \right]} dB_j(t) 
\text{[Stratonovich]}
\nonumber \\
\label{ItoSDEsinglegj}
\end{eqnarray}
and only some coefficients are different and related via Eq. \ref{DfullyStratoSingle}.
  
  In practice, one is often interested into the three special cases where $D_{jj}(x_j)$ involves a single monomial
instead of the second-order polynomial $D_{jj}(x_j) = D^{[0]}_j+D^{[1]}_j x_j + D^{[2]}_j x_j^2 $
of Eq. \ref{DdiagoEtSingle}
\begin{eqnarray}
  \begin{cases}
  \text{ $D_{jj}(x_j) = D^{[0]}_j$ corresponding to the additive-noise term $\sqrt{ 2 D^{[0]}_j} dB_j(t)$   } 
\\
\text{ $D_{jj}(x_j) =  D^{[2]}_j x_j^2 $ corresponding to the multiplicative-noise term $\sqrt{ 2 D^{[2]}_j} \ x_j(t) dB_j(t)$     } 
\\
\text{ $D_{jj}(x_j) = D^{[1]}_j x_j $ corresponding to the square-root-noise term $\sqrt{ 2 D^{[1]}_j x_j(t) } dB_j(t)$ for positive processes $x_j \in ]0,+\infty[$   } 
\end{cases}
\label{SingleNoisesCases}
\end{eqnarray}
Since these three types of noises are considered as the most relevant from a physical point of view,
one may wish to reinterpret the cases where $D_{jj}(x_j)$ involves two monomials
as the result of two different types of noises for each coordinate $x_j(t)$
as described in the two following subsections.

  
    \subsubsection{ Case $D_{jj}(x_j) = D^{[0]}_j + D^{[2]}_j x_j^2 $  : alternative interpretation with two noises for each coordinate $x_j(t) \in ]-\infty,+\infty[$}

When  $D_{jj}(x_j) = D^{[0]}_j + D^{[2]}_j x_j^2 $, the interpretation of Eq. \ref{ItoSDEsinglegj} 
involving a single noise $B_j(t)$ per coordinate $x_j(t)$  
    can be replaced by the alternative interpretation 
 with an additive noise $dB_j^{additive}(t) $ 
and a multiplicative noise $dB_j^{multiplicative}(t) $ for each coordinate $x_j(t)$
via the equivalence
\begin{eqnarray}
\sqrt{ 2[ D^{[0]}_j + D^{[2]}_j x_j^2(t)  ]} dB_j(t) = \sqrt{ 2 D^{[0]}_j }dB_j^{additive}(t) +  \sqrt{ 2 D^{[2]}_j } x_j(t) dB_j^{multiplicative}(t)
\label{ItoAddAndMulti}
\end{eqnarray}
that produces the same diffusion matrix.


    \subsubsection{ Case $D_{jj}(x_j) = D^{[1]}_j x_j+ D^{[2]}_j x_j^2 $  : alternative interpretation with two noises for each coordinate $x_j(t) \in ]0,+\infty[$}

Similarly when $D_{jj}(x_j) = D^{[1]}_j x_j+ D^{[2]}_j x_j^2 $, the interpretation of Eq. \ref{ItoSDEsinglegj} 
involving a single noise $B_j(t)$ per coordinate $x_j(t)$  
    can be replaced by the alternative interpretation 
 with a square-root noise $dB_j^{square-root}(t) $ 
and a multiplicative noise $dB_j^{multiplicative}(t) $ for each coordinate $x_j(t)$
via the equivalence
\begin{eqnarray}
\sqrt{ 2[ D^{[1]}_j x_j+ D^{[2]}_j x_j^2(t)  ]} dB_j(t) = \sqrt{ 2 D^{[1]}_j x_j }dB_j^{square-root}(t) +  \sqrt{ 2 D^{[2]}_j } x_j(t) dB_j^{multiplicative}(t)
\label{ItoSquareRootAndMulti}
\end{eqnarray}
that produces the same diffusion matrix.


    \subsection{ Discussion }

In summary, the models with fully-diagonal diffusion matrices $D^{[k=0,1,2]}$ of Eq. \ref{DdiagoEtSingle}
considered in the present section
  already contain a lot of interesting diffusion processes, 
  and it is thus useful in the following section to recapitulate 
  the models with one or two noises per coordinate
  that are the simplest from the Carleman point of view.
  
  
  \section{ List of the simplest models involving one or two noises per coordinate}
  
  \label{sec_list}
  
  In this section, we focus on the models discussed in the previous section
  that involve either a single noise $dB_j(t)$
  per coordinate $x_j(t)$, that can be 
  either additive or multiplicative or square-root as described in Eq. \ref{SingleNoisesCases},
  or with two different types of noises per coordinate $x_j(t)$ 
  as described in Eqs \ref{ItoAddAndMulti} and \ref{ItoSquareRootAndMulti}.
  For each of these five types of models, we list the corresponding Ito forces that
  lead to Carleman matrices ${\bold M}$ that are are either block-diagonal or block-triangular
  in order to have the simplifications described in subsection \ref{subsec_BlockTriangular}.

 \subsection{ Models with only one additive-noise $dB_j^{additive}(t) $ for each $x_j$ 
 having  
 ${\bold M}={\bold M}^{[-2]} + {\bold M}^{[-1]}+{\bold M}^{[0]} $ }

Since the additive noises $dB_j^{additive}(t) $ produce the diffusion matrix ${\bold D}^{[0]}$ that appears in  the Carleman block ${\bold M}^{[-2]}$, the Carleman matrix will be block-lower-triangular  
 ${\bold M}={\bold M}^{[-2]} + {\bold M}^{[-1]}+{\bold M}^{[0]} $ when the Ito forces involve 
 only the matrices ${\bold F}^{[0]}$ and ${\bold F}^{[1]}$ that appear in the Carleman blocks ${\bold M}^{[-1]}$ and ${\bold M}^{[0]} $ respectively.
 
 The SDE of Eq. \ref{ItoSDEsinglegj} 
  \begin{eqnarray}
dx_j(t)   && =  \left( F^{[0]}_j + \sum_{i=1}^d F^{[1]}_{ji} x_i(t)
\right) dt + \sqrt{ 2 D^{[0]}_j } dB_j^{additive}(t) 
\label{ItoSDEsinglegjonlyadd}
\end{eqnarray}
then correspond to the Ornstein-Uhlenbeck processes in dimension $d$
that are well-known for their explicit Gaussian finite-time propagators and steady states
described in textbooks \cite{gardiner,vankampen,risken}, 
and that   
have thus been reconsidered in various contexts over the years to 
obtain explicit solutions for many specific problems. In particular, they are vey useful
to address the new issues raised by the progresses in the field of nonequilibrium processes
in arbitrary dimension $d$ \cite{Thouless,CGetJML,us_Gyrator,duBuisson_gyrator,duBuisson_thesis},
and the special case $d=2$ has attracted a lot of interest recently  \cite{Gyr_vanWijland,Gyr_2013,Gyr_elec,Gyr_exp,Gyr_Tryphon_Harvesting,Gyr_Tryphon_Engine,Gyr_Tryphon_Geometry,Gyr_Inferring,Gyr_Inference,cerasoli,duBuisson_stochasticArea}.
As a consequence, these Ornstein-Uhlenbeck processes will not be rediscussed in the present paper.


\subsection{ Models with only one square-root-noise $dB_j^{square-root}(t) $ for each $x_j >0$ having  
 ${\bold M}= {\bold M}^{[-1]}+{\bold M}^{[0]} $}
 
 \label{subsec_onlysquareroot}
 
 Since the square-root-noises $dB_j^{square-root}(t) $ produce the diffusion matrix ${\bold D}^{[1]}$ that appears in  the Carleman block ${\bold M}^{[-1]}$, the Carleman matrix will be block-lower-bidiagonal  
 ${\bold M}= {\bold M}^{[-1]}+{\bold M}^{[0]} $ when the Ito forces involve 
 only the matrices ${\bold F}^{[0]}$ and ${\bold F}^{[1]}$ that appear in the Carleman blocks ${\bold M}^{[-1]}$ and ${\bold M}^{[0]} $ respectively.
 
 The SDE of Eq. \ref{ItoSDEsinglegj} then read
   \begin{eqnarray}
dx_j(t)   && =  \left( F^{[0]}_j + \sum_{i=1}^d F^{[1]}_{ji} x_i(t)
\right) dt + \sqrt{ 2 D^{[1]}_j x_j(t)} dB_j^{square-root}(t) 
\ \ \ \ \ \text{[Ito]}
\nonumber \\
dx_j(t)   && = \left( f^{[0]}_j + \sum_{i=1}^d f^{[1]}_{ji} x_i(t)
\right) dt 
+\sqrt{ 2 D^{[1]}_j x_j(t) } dB_j^{square-root}(t) 
\ \ \ \ \ \text{[Stratonovich]}
\label{ItoSDEsinglegjonlysquareroot}
\end{eqnarray}
In order to remain in the domain $x_j \in ]0,+\infty[$ of validity of the square-root-noises,
 the Ito increment $dx_j(t)  $ should remain positive when the coordinate $x_j(t)$ vanishes
 $x_j(t)=0$  
    \begin{eqnarray}
0 \leq dx_j(t) \bigg\vert_{x_j(t)=0}  && = \left( F^{[0]}_j + \sum_{i \ne j}^d F^{[1]}_{ji} x_i(t)
\right) dt 
\label{ItoSDEsinglegjonlysquarerootpositive}
\end{eqnarray}
for any positive values of the $(d-1)$ other coordinates $x_{i \ne j}(t) \in ]0,+\infty[$,
so that one obtains that the $d$ coefficients $F^{[0]}_j $ and the $(d^2-d)$ off-diagonal coefficients $F^{[1]}_{ji}  $
should be positive
    \begin{eqnarray}
 F^{[0]}_j && \geq 0 \ \ \ \text{ for $j=1,2,..,d$ }
 \nonumber \\
  F^{[1]}_{ji} && \geq 0 \ \ \ \text{ for $j=1,2,..,d$  and $i \ne j$}
\label{ConditionsF1forSquareRootNoise}
\end{eqnarray}
while the diagonal coefficients $F^{[1]}_{jj} $ have no constraints.

Since ${\bold D}^{[2]} =0$, the discussion of subsection \ref{subsec_D2vanishDiagoF1}
can be applied to these models: the eigenvalues of the full Carleman matrix ${\bold M}= {\bold M}^{[-1]}+{\bold M}^{[0]} $ coincide with the eigenvalues of the diagonal-block ${\bold M}^{[0]} $
that only involves the matrix ${\bold F}^{[1]}$ and that 
are directly given by Eq. \ref{EigenCLEigenF1}
\begin{eqnarray}
E_{\kappa_1,..,\kappa_d} =  \sum_{\nu=1}^d   \kappa_{\nu}  \lambda_{\nu} 
\label{EigenCLEigenF1Laguerre}
\end{eqnarray}
as linear combinations of the $d$ eigenvalues $\lambda_{\nu} $ of the $d \times d$ matrix ${\bold F}^{[1]} $
of Eq. \ref{DiagoF1} with the integers coefficients $\kappa_{\nu} \in {\mathbb N} $.

The diffusion processes of Eq. \ref{ItoSDEsinglegjonlysquareroot} will be discussed for the dimensions $d=1$
and $d=2$ in subsections  \ref{subsec_onlySQ1D}
 and \ref{subsec_onlySQ2D} respectively.


  \subsection{ Models with only one multiplicative noise $dB_j^{multiplicative}(t) $ for each $x_j$ having
   ${\bold M}={\bold M}^{[0]} $ }
  
  \label{subsec_onlymultiplicativediag}
  
  Since the multiplicative noises $dB_j^{multiplicative}(t) $ produce the diffusion matrix ${\bold D}^{[2]}$ that appears in  the Carleman block ${\bold M}^{[0]}$, one can have
 a block-diagonal  Carleman matrix
 ${\bold M}= {\bold M}^{[0]} $ when the Ito forces involve 
 only the matrix ${\bold F}^{[1]}$, 
 so that the SDE of Eq. \ref{ItoSDEsinglegj} reduce to
  \begin{eqnarray}
dx_j(t)   && =  \left( \sum_{i=1}^d F^{[1]}_{ji} x_i(t)
\right) dt + \sqrt{ 2  D^{[2]}_j } \  x_j(t) dB_j^{multiplicative}(t) 
\ \ \ \ \ \text{[Ito]}
\nonumber \\
dx_j(t)   && =  \left( \sum_{i=1}^d f^{[1]}_{ji} x_i(t)
\right) dt + \sqrt{ 2  D^{[2]}_j } \  x_j(t) dB_j^{multiplicative}(t) 
\ \ \ \ \ \text{[Stratonovich]}
\label{ItoSDEsinglegjmultiplicativediag}
\end{eqnarray}
 The special cases of the dimensions $d=1$ and $d=2$
 will be considered in Eqs \ref{ItoSDE1dGBM} and \ref{StratoSDE2DF1D2}
 respectively, while the generalisation to $d>2$ is discussed in Appendix \ref{app_dimensiond}
 with the example $d=3$.
 The mean-field case of large dimension $d \to + \infty$ has attracted a lot of interest over the years
 with many applications in various fields 
(see the recent work \cite{RandomGrowth} and references therein).


 \subsection{ Models with only one multiplicative noise $dB_j^{multiplicative}(t) $ for each $x_j$
 having ${\bold M}={\bold M}^{[-1]}+{\bold M}^{[0]} $ }
 
  \label{subsec_onlymultiplicativelower}
 
In the presence of only one multiplicative noise $dB_j^{multiplicative}(t) $ for each $x_j$, 
the Carleman matrix will be block-lower-bidiagonal  
 ${\bold M}={\bold M}^{[-1]}+ {\bold M}^{[0]} $ when the Ito forces involve 
 only the matrices ${\bold F}^{[0]}$ and ${\bold F}^{[1]}$, so that the SDE of Eq. \ref{ItoSDEsinglegj} reduce to
  \begin{eqnarray}
dx_j(t)   && =  \left( F^{[0]}_j + \sum_{i=1}^d F^{[1]}_{ji} x_i(t))\right) dt 
+ \sqrt{ 2  D^{[2]}_j } \  x_j(t) dB_j^{multiplicative}(t) 
\ \ \ \text{[Ito]}
\nonumber \\
dx_j(t)   && = \left( f^{[0]}_j + \sum_{i=1}^d f^{[1]}_{ji} x_i(t)\right) dt 
+ \sqrt{ 2  D^{[2]}_j } \  x_j(t) dB_j^{multiplicative}(t) 
\ \ \ \text{[Stratonovich]}
\label{ItoSDEsinglegjmultiplicativelower}
\end{eqnarray}
The special cases of the dimensions $d=1$ and $d=2$
 will be discussed in subsections \ref{subsec_Kesten} and \ref{subsec_Kesten2D}
 respectively, while 
 related models formulated in discrete time are discussed in detail in the recent works 
 \cite{Sornette_EigenvectorGeometry,Sornette_DiscreteTimeKesten} both for $d=2$ and for large dimension $d\to + \infty$.



 \subsection{ Models with only one multiplicative noise $dB_j^{multiplicative}(t) $ for each $x_j$ having 
  ${\bold M}={\bold M}^{[0]}+{\bold M}^{[1]} $ }
 
  \label{subsec_onlymultiplicativeupper}
 
  In the presence of only one multiplicative noise $dB_j^{multiplicative}(t) $ for each $x_j$, the Carleman matrix will be block-upper-triangular 
 ${\bold M}= {\bold M}^{[0]} +{\bold M}^{[1]}$ when the Ito forces involve 
 only the matrices ${\bold F}^{[1]}$ and $ {\bold F}^{[2]}$, so that the SDE of Eq. \ref{ItoSDEsinglegj} reduce to
 \begin{small}
  \begin{eqnarray}
dx_j(t)   && =  \left(  \sum_{i=1}^d F^{[1]}_{ji} x_i(t)
+   \sum_{i=1}^d  F^{[2]}_{j;  i,i } x_i^2(t) 
+ 2 \sum_{i_1=1}^{d-1} \sum_{i_2=i_1+1}^{d} F^{[2]}_{j;  i_1,i_2 } x_{i_1}(t) x_{i_2}(t)\right) dt 
+ \sqrt{ 2  D^{[2]}_j } \  x_j(t) dB_j^{multiplicative}(t) 
\ \ \ \text{[Ito]}
\nonumber \\
dx_j(t)   && = \left(  \sum_{i=1}^d f^{[1]}_{ji} x_i(t)
+   \sum_{i=1}^d  f^{[2]}_{j;  i,i } x_i^2(t) 
+ 2 \sum_{i_1=1}^{d-1} \sum_{i_2=i_1+1}^{d} f^{[2]}_{j;  i_1,i_2 } x_{i_1}(t) x_{i_2}(t)\right) dt 
+ \sqrt{ 2  D^{[2]}_j } \  x_j(t) dB_j^{multiplicative}(t) 
\ \ \ \text{[Stratonovich]}
\nonumber \\
\label{ItoSDEsinglegjmultiplicativeupper}
\end{eqnarray}
\end{small}
In this most general form, there are $d^2$ different matrix elements for $F^{[1]}$ or $f^{[1]}$, 
while there are $d \times \left( d+ \frac{d (d-1)}{2} \right)= \frac{d^2(d+1)}{2}$
different matrix elements for $F^{[2]}=f^{[2]}$, but in practice one is often interested in models with less coefficients
as discussed in the examples of the two next subsections.


 \subsubsection{ Example of Lotka-Volterra models for positive variables $x_j>0$ with multiplicative noises having
  ${\bold M}={\bold M}^{[0]}+{\bold M}^{[1]} $ }

 In Lotka-Volterra models for positive variables, the matrices $F^{[1]}$ or $f^{[1]}$ are diagonal
with only $d$ diagonal matrix elements $F^{[1]}_{jj} $, and the matrix elements of $F^{[2]}=f^{[2]}$ 
are non-vanishing only for $j=i_1$ or $j=i_2$ so that there are only $d^2$ distinct matrix elements
 \begin{small}
  \begin{eqnarray}
dx_j(t)   && = x_j(t)  \left[  \left( F^{[1]}_{jj} 
+ F^{[2]}_{j;  j ,j }  x_j(t)
+ 2  \sum_{i_1=1}^{j-1} F^{[2]}_{j;  i_1,j }  x_{i_1}(t)
+2  \sum_{i_2=j+1}^{d} F^{[2]}_{j;  j,i_2 }  x_{i_2}(t) \right) dt 
+ \sqrt{ 2  D^{[2]}_j }  dB_j^{multiplicative}(t) \right]
\ \ \ \text{[Ito]}
\nonumber \\
dx_j(t)   && =  x_j(t)  \left[ \left(  f^{[1]}_{jj} 
+ f^{[2]}_{j;  j ,j }  x_j(t)
+ 2  \sum_{i_1=1}^{j-1} f^{[2]}_{j;  i_1,j }  x_{i_1}(t)
+2  \sum_{i_2=j+1}^{d} f^{[2]}_{j;  j,i_2 }  x_{i_2}(t)\right) dt 
+ \sqrt{ 2  D^{[2]}_j }  dB_j^{multiplicative}(t) \right]
\ \ \ \text{[Stratonovich]}
\nonumber \\
\label{ItoSDEsinglegjmultiplicativeupperLV}
\end{eqnarray}
\end{small}

This leads to the following simplifications for the Carleman matrix ${\bold M}={\bold M}^{[0]}+{\bold M}^{[1]}  $ :

$\bullet$ The matrix ${\bold M}^{[0]} $ of Eq. \ref{m0simpli} is diagonal 
\begin{eqnarray}
M^{[0]}(n_1,..,n_d \vert q_1,...,q_d)
&& = M^{[0]}(n_1,..,n_d \vert n_1,...,n_d)
\prod_{l =1}^d  \delta_{q_l,n_l}     
\label{m0simplidiagnqLV}
\end{eqnarray}
and the diagonal elements directly give the eigenvalues of the Carleman matrix
\begin{eqnarray}
E_{n_1,..,n_d} = M^{[0]}(n_1,..,n_d \vert n_1,...,n_d)
 =  \sum_{j=1}^d  
\left[  F^{[1]}_{jj}  n_j   
+     D^{[2]}_j   n_j (n_j-1) \right]  
\label{m0simplidiagnqLVeigen}
\end{eqnarray}

$ \bullet$ The matrix elements of Eq. \ref{M1} for ${\bold M}^{[1]} $ simplify into
\begin{eqnarray}
M^{[1]}(n_1,..,n_d \vert q_1,...,q_d)
&& =
\sum_{j=1}^d  
\langle j \vert {\bold F}^{[2]} \vert j,j \rangle 
  n_j  \delta_{q_j,n_j+1}  \prod_{l \ne j} \delta_{q_l,n_l} 
\nonumber \\
&&+ 2 \sum_{i_1=1}^{d-1} \sum_{i_2=i_1+1}^d 
\left[\langle i_1 \vert {\bold F}^{[2]} \vert i_1,i_2 \rangle n_{i_1} 
     \delta_{q_{i_2},n_{i_2}+1} \prod_{l \ne i_2} \delta_{q_l,n_l} 
+\langle i_2 \vert {\bold F}^{[2]} \vert i_1,i_2 \rangle n_{i_2}
\delta_{q_{i_1},n_{i_1}+1}     \prod_{l \ne i_1} \delta_{q_l,n_l} 
\right] 
\nonumber \\
&& = \sum_{i=1}^d  
\langle i \vert {\bold F}^{[2]} \vert i,i \rangle 
  n_i  \delta_{q_i,n_i+1}  \prod_{l \ne i} \delta_{q_l,n_l} 
\nonumber \\
&&+ 2 \sum_{i=2}^d \sum_{i_1=1}^{i-1} 
\langle i_1 \vert {\bold F}^{[2]} \vert i_1,i \rangle n_{i_1} 
     \delta_{q_{i},n_{i}+1} \prod_{l \ne i} \delta_{q_l,n_l} 
     + 2 \sum_{i=1}^{d-1} \sum_{i_2=i+1}^d 
\langle i_2 \vert {\bold F}^{[2]} \vert i,i_2 \rangle n_{i_2}
\delta_{q_{i},n_{i}+1}     \prod_{l \ne i} \delta_{q_l,n_l} 
\nonumber \\
&&= \sum_{i=1}^d  \left[ \delta_{q_i,n_i+1}  \prod_{l \ne i} \delta_{q_l,n_l} \right]
\left( \langle i \vert {\bold F}^{[2]} \vert i,i \rangle 
  n_i   
+ 2  \sum_{i_1=1}^{i-1} 
\langle i_1 \vert {\bold F}^{[2]} \vert i_1,i \rangle n_{i_1} 
   + 2  \sum_{i_2=i+1}^d 
\langle i_2 \vert {\bold F}^{[2]} \vert i,i_2 \rangle n_{i_2}
\right)
\nonumber \\
\label{M1LV}
\end{eqnarray}
i.e. the only non-vanishing matrix elements are when only one integer $q_i=n_i+1$ is higher by one unit
while all the other coincide $q_l=n_l $ for $l \ne i$.

The Lotka-Volterra processes of Eq. \ref{ItoSDEsinglegjmultiplicativeupperLV} will be discussed in dimensions $d=1$
and $d=2$ in subsections  \ref{subsec_upperbidiag1D}
 and \ref{subsec_2Dupper} respectively.

 
  \subsubsection{ Example of Lorenz-type models in $d=3$  with multiplicative noises having
  ${\bold M}={\bold M}^{[0]}+{\bold M}^{[1]} $ }
 
 In another type of models, the matrix elements of $F^{[2]}=f^{[2]}$ 
are non-vanishing only if the three indices $(j,i_1, i_2)$ are all distinct,
so the smallest dimension where this is possible is $d=3$ with only three distinct 
matrix elements for $F^{[2]}=f^{[2]}$ that appear in the Ito-SDE system
   \begin{eqnarray}
dx_1(t)   && = \left(  \sum_{i=1}^3 F^{[1]}_{1i} x_i(t)
+  2 F^{[2]}_{1;  2,3 } x_2(t) x_3(t)\right) dt 
+ \sqrt{ 2  D^{[2]}_1 } \  x_1(t) dB_1^{multiplicative}(t) 
\nonumber \\
dx_2(t)   && = \left(  \sum_{i=1}^3 F^{[1]}_{2i} x_i(t)
+ 2  F^{[2]}_{2;  1,3 } x_1(t) x_3(t)\right) dt 
+ \sqrt{ 2  D^{[2]}_2 } \  x_2(t) dB_2^{multiplicative}(t) 
\nonumber \\
dx_3(t)   && = \left(  \sum_{i=1}^3 F^{[1]}_{3i} x_i(t)
+ 2  F^{[2]}_{3;  1,2 } x_1(t) x_2(t)\right) dt 
+ \sqrt{ 2  D^{[2]}_3 } \  x_3(t) dB_3^{multiplicative}(t) 
\label{ItoSDEsinglegjmultiplicativeupperDistinct}
\end{eqnarray}
that includes the Lorenz model in the presence of multiplicative noises.

In these models where the three indices $(j,i_1, i_2)$ are all distinct, the matrix elements of Eq. \ref{M1}
for ${\bold M}^{[1]}$ reduce to the last contribution
\begin{eqnarray}
M^{[1]}(n_1,..,n_d \vert q_1,...,q_d)
&& = 2 \sum_{i_1=1}^{d-1} \sum_{i_2=i_1+1}^d 
\sum_{j\ne i_1,i_2} \langle j \vert {\bold F}^{[2]} \vert i_1,i_2 \rangle n_j
\delta_{q_{i_1},n_{i_1}+1}  \delta_{q_{i_2},n_{i_2}+1}  \delta_{q_j,n_j-1}  \prod_{l \ne j,i_1,i_2} \delta_{q_l,n_l} 
\label{M1Lorenztype}
\end{eqnarray}
i.e. in dimension $d=3$ for the example of Eq. \ref{ItoSDEsinglegjmultiplicativeupperDistinct}
\begin{eqnarray}
M^{[1]}(n_1,n_2,n_3 \vert q_1,q_2,q_3)
&& = 2 n_3 
  F^{[2]}_{3;12}  
\delta_{q_{1},n_{1}+1}  \delta_{q_{2},n_{2}+1}  \delta_{q_3,n_3-1}  
\nonumber \\
&&  +2 n_2 F^{[2]}_{2;13}
\delta_{q_{1},n_{1}+1}  \delta_{q_{3},n_{3}+1}  \delta_{q_2,n_2-1} 
+ 2 n_1 F^{[2]}_{1;23} 
\delta_{q_{2},n_{2}+1}  \delta_{q_{3},n_{3}+1}  \delta_{q_1,n_1-1} 
\label{M1Lorenz3D}
\end{eqnarray}


 \subsection{ Models with one additive-noise $dB_j^{additive}(t) $ and one multiplicative-noise $dB_j^{multiplicative}(t) $ for each $x_j$ having  
 ${\bold M}={\bold M}^{[-2]} + {\bold M}^{[-1]}+{\bold M}^{[0]} $ }
 
  \label{subsec_twoaddmulti}
 
 Since the additive noises $dB_j^{additive}(t) $ and the multiplicative-noises $dB_j^{multiplicative}(t) $
 produce the diffusion matrices $D^{[0]}$ and $D^{[2]}$
 that enter the Carleman blocks ${\bold M}^{[-2]}$ and ${\bold M}^{[0]}$, the Carleman matrix will be block-lower-triangular  
 ${\bold M}={\bold M}^{[-2]} + {\bold M}^{[-1]}+{\bold M}^{[0]} $ when the Ito forces involve 
 only the matrices $F^{[0]}$ and $F^{[1]}$.
 
The corresponding SDE of Eq. \ref{ItoSDEsinglegj} read using Eq. \ref{ItoAddAndMulti}
  \begin{eqnarray}
dx_j(t)   && =  \left( F^{[0]}_j + \sum_{i=1}^d F^{[1]}_{ji} x_i(t)\right) dt 
+\sqrt{ 2 D^{[0]}_j }dB_j^{additive}(t) +  \sqrt{ 2 D^{[2]}_j } x_j(t) dB_j^{multiplicative}(t)
\ \ \ \text{[Ito]}
\nonumber \\
dx_j(t)   && = \left( f^{[0]}_j + \sum_{i=1}^d f^{[1]}_{ji} x_i(t)
\right) dt 
+\sqrt{ 2 D^{[0]}_j }dB_j^{additive}(t) +  \sqrt{ 2 D^{[2]}_j } x_j(t) dB_j^{multiplicative}(t)
\ \ \ \text{[Stratonovich]}
\nonumber \\
\label{ItoSDEsinglegjtwoaddmulti}
\end{eqnarray}

The Carleman matrix will only involve the two blocks ${\bold M}={\bold M}^{[-2]} +{\bold M}^{[0]} $
when the constant forces $F^{[0]}_j=0=f^{[0]}_j$ vanish corresponding to
   \begin{eqnarray}
dx_j(t)   && =  \left(  \sum_{i=1}^d F^{[1]}_{ji} x_i(t)\right) dt 
+\sqrt{ 2 D^{[0]}_j }dB_j^{additive}(t) +  \sqrt{ 2 D^{[2]}_j } x_j(t) dB_j^{multiplicative}(t)
\ \ \ \text{[Ito]}
\nonumber \\
dx_j(t)   && = \left(  \sum_{i=1}^d f^{[1]}_{ji} x_i(t)
\right) dt 
+\sqrt{ 2 D^{[0]}_j }dB_j^{additive}(t) +  \sqrt{ 2 D^{[2]}_j } x_j(t) dB_j^{multiplicative}(t)
\ \ \ \text{[Stratonovich]}
\nonumber \\
\label{ItoSDEsinglegjtwoaddmultiBiblock}
\end{eqnarray}
The special cases of the dimensions $d=1$ and $d=2$
 will be considered in subsections \ref{subsec_twoaddmulti1D} and \ref{subsec_twoaddmulti1D}
 respectively.

 
  \subsection{ Models with one square-root noise $dB_j^{square-root}(t) $ 
  and one multiplicative noise $dB_j^{multiplicative}(t) $ for each coordinate $x_j>0$ having  
 ${\bold M}= {\bold M}^{[-1]}+{\bold M}^{[0]} $}
 
  \label{subsec_twosquarerootmulti}
 
 Since the square-root-noises $dB_j^{additive}(t) $ and the multiplicative-noises $dB_j^{multiplicative}(t) $
 produce the diffusion matrices $D^{[1]}$ and $D^{[2]}$
 that appear in the Carleman blocks ${\bold M}^{[-1]}$ and ${\bold M}^{[0]}$, the Carleman matrix will be block-lower-bidiagonal 
 ${\bold M}= {\bold M}^{[-1]}+{\bold M}^{[0]} $ when the Ito forces involve 
 only the matrices $F^{[0]}$ and $F^{[1]}$.
 
The corresponding SDE of Eq. \ref{ItoSDEsinglegj} read using Eq. \ref{ItoSquareRootAndMulti}
  \begin{eqnarray}
dx_j(t)   && =  \left( F^{[0]}_j + \sum_{i=1}^d F^{[1]}_{ji} x_i(t)\right) dt 
+  \sqrt{ 2 D^{[1]}_j x_j }dB_j^{square-root}(t) +  \sqrt{ 2 D^{[2]}_j } x_j(t) dB_j^{multiplicative}(t)
\ \ \text{[Ito]}
\nonumber \\
dx_j(t)   && = \left( f^{[0]}_j + \sum_{i=1}^d f^{[1]}_{ji} x_i(t)\right) dt 
+ \sqrt{ 2 D^{[1]}_j x_j }dB_j^{square-root}(t) +  \sqrt{ 2 D^{[2]}_j } x_j(t) dB_j^{multiplicative}(t)
\ \ \text{[Stratonovich]}
\nonumber \\
\label{ItoSDEsinglegjtwoSquareRootMulti}
\end{eqnarray}
In order to remain in the domain $x_j \in ]0,+\infty[$ of validity of the square-root-noises,
one has the same conditions as in Eq. \ref{ConditionsF1forSquareRootNoise}.
The special cases of the dimensions $d=1$ and $d=2$
 will be considered in subsections \ref{subsec_twosquarerootmulti1D} and \ref{subsec_twosquarerootmulti2D}
 respectively.


  \subsection{ Discussion }

In the remaining sections, we focus on the dimensions $d=1$ and $d=2$ 
in order to describe the results of the Carleman approach
for the above list of models.


 \section{ Application of the Carleman approach to diffusion processes in $d=1$ } 
 
 \label{sec_1D}

 In this section, the Carleman approach is applied to diffusion processes in $d=1$
 in order to generalize the pionnering work \cite{CarlemanStochastic} concerning 
 some diffusion processes with multiplicative noise.

 
  \subsection{ Notations for diffusions in dimension $d=1$ involving the Ito force $F(x)$ and the diffusion coefficient $D(x)$}
 
 In dimension $d=1$, the differential generator of Eq. \ref{Generator}
 reduces to
\begin{eqnarray}
{\cal L}
=    F( x)   \frac{ \partial }{\partial x}  +    D( x)   \frac{ \partial^2 }{\partial x^2}
\label{Generator1D}
\end{eqnarray}
 where the Taylor expansions 
 of Eqs \ref{TaylorF} and \ref{TaylorD} for the Ito force $F( x)$ and for the diffusion coefficient $D( x)$
\begin{eqnarray}
 F(  x) && = F^{[0]}+  F^{[1]} x + F^{[2]} x^2
  \nonumber \\
 D(  x) && = D^{[0]}+  D^{[1]} x + D^{[2]} x^2
\label{TaylorFD1d}
\end{eqnarray}
only involve the numbers $(F^{[0]},F^{[1]} ,F^{[2]},D^{[0]}, D^{[1]},D^{[2]})$ instead of matrices,
while the Stratonovich force $f(x)$ of Eq. \ref{TaylorFStrato}
is also a polynomial of degree two
\begin{eqnarray}
f(x) &&  = \left( F^{[0]} - \frac{D^{[1]}  }{2}\right) + \left( F^{[1]}- D^{[2]}\right)+ F^{[2]} x^2
\nonumber \\
&& \equiv f^{[0]}+  f^{[1]} x + f^{[2]} x^2
\label{StratoItoCorrespondance1dx}
\end{eqnarray}
with the coefficients of Eq. \ref{DfullyStratoSingle}
\begin{eqnarray}
 f^{[0]} &&  =  F^{[0]} - \frac{D^{[1]}  }{2}
\nonumber \\ 
 f^{[1]} && =  F^{[1]}- D^{[2]}
\nonumber \\
 f^{[2]} && = F^{[2]}
\label{StratoItoCorrespondance1d}
\end{eqnarray}

 The SDE of Eq. \ref{ItoSDEsinglegj}
 involving a single noise $dB(t) $ read
\begin{eqnarray}
dx(t)  && = \left(F^{[0]}+  F^{[1]} x(t) + F^{[2]} x^2(t) \right)  dt 
+ \sqrt{ 2 \left[ D^{[0]}+  D^{[1]} x(t) + D^{[2]} x^2(t)\right]} dB(t) 
\ \ \ \ \ 
\text{[Ito]}
\nonumber \\
dx(t)  && =  \left(f^{[0]}+  f^{[1]} x(t) + f^{[2]} x^2(t) \right)  dt  + 
+ \sqrt{ 2 \left[ D^{[0]}+  D^{[1]} x(t) + D^{[2]} x^2(t)\right]} dB(t)
\ \ \ \ \text{[Stratonovich]}
\label{ItoSDE1d}
\end{eqnarray}
with the three special cases of additive/multiplicative/square-root noises of Eq. \ref{SingleNoisesCases},
or with the re-interpretation in terms of two noises of Eqs \ref{ItoAddAndMulti} and \ref{ItoSquareRootAndMulti}.

 
  \subsection{Carleman dynamics for the moments $m_t( n) \equiv  {\mathbb E} ( x^n(t) )   $ labelled by the integer 
  $n \in {\mathbb N}$ }

In dimension $d=1$, the dynamics of Eq. \ref{DynMoments}
for the moments $m_t( n) \equiv  {\mathbb E} ( x^n(t) )   $ labelled by the integer 
  $n \in {\mathbb N}$
 \begin{eqnarray}
\partial_t m_t(n) = \sum_{q=n-2}^{n+1} M(n,q) m_t(q)
\label{dynmomentsd1}
\end{eqnarray}
involves the matrix elements of Eqs \ref{Mm2simpli} \ref{m1simpli} \ref{m0simpli} \ref{M1}
 \begin{eqnarray}
M(n, n-2)  && =      n (n-1)   D^{[0]}
\nonumber \\
M(n, n-1)  && =   n    F^{[0]}+   n (n-1)   D^{[1]}
\nonumber \\
M(n, n)  && =   n    F^{[1]}+   n (n-1)   D^{[2]} =  n    f^{[1]}+   n^2   D^{[2]}
\nonumber \\
M(n, n+1)  && =   n    F^{[2]}
 \label{MCarlemand1detailed}
\end{eqnarray}

As a consequence, the Carleman system of Eq. \ref{dynmomentsd1}
 reads starting with $n=0$ associated to conservation of $m_0(t)={\mathbb E} ( 1 )=1 $
\begin{eqnarray}
\partial_t m_t(0) && = 0 
\nonumber \\
\partial_t m_t(1) && 
= M(1,0) m_t(0) 
+ M(1,1) m_t(1)
+ M(1,2) m_t(2)
\nonumber \\
\partial_t m_t(2) && 
= M(2,0) m_t(0) 
+ M(2,1) m_t(1)
+ M(2,2) m_t(2)
+ M(2,3) m_t(3)
\nonumber \\
\partial_t m_t(3) && 
= M(3,1) m_t(1)
+ M(3,2) m_t(2)
+ M(3,3)m_t(3)
+ M(3,4) m_t(4) 
\nonumber \\
...
\nonumber \\
\partial_t m_t(n) && 
= M(n,n-2) m_t(n-2) 
+ M(n,n-1) m_t(n-1)
+ M(n,n) m_t(n)
+ M(n,n+1) m_t(n+1) + ...
\nonumber \\
...
\label{dynotxexpli}
\end{eqnarray}

In the three following subsections, we describe the three simplifications 
where the Carleman matrix ${\bold M}$ is diagonal, or lower-triangular or upper-triangular,
in relation with the general discussion of subsection \ref{subsec_BlockTriangular}.


\subsection{ Cases where the Carleman matrix is ${\bold M}$ diagonal : 
$F(  x)  = x  F^{[1]} $ and $D(  x)  =   x^2  D^{[2]} $  }

\label{subsec_GBM1D}

The Carleman matrix ${\bold M} $ of Eq. \ref{MCarlemand1detailed} will be diagonal
with its eigenvalues $E_n$ given by the diagonal matrix elements $M(n,n)$
 \begin{eqnarray}
E_n= M(n, n)  && =   n    F^{[1]}+   n (n-1)   D^{[2]}  = n    f^{[1]}+n^2 D^{[2]} 
 \label{MTdiag}
\end{eqnarray}
governing the exponential behaviors
of the moments $m_t(n) $ of Eq. \ref{dynotxexpli}
 \begin{eqnarray}
m_t(n) =e^{t E_n } m_0(n) =e^{t M(n, n) } m_0(n) = e^{t \left( n    f^{[1]}+n^2 D^{[2]}  \right) } m_0(n)
 \label{MTdiagsol}
\end{eqnarray}
when the only non-vanishing Taylor coefficients in Eq. \ref{TaylorFD1d}
are $F^{[1]} $ and $D^{[2]} $, 
i.e. when the diffusion coefficient $D(x)$ involves only the quadratic term (multiplicative noise)
while the Ito force $F(x)$ and the Stratonovich force $f(x)$ involve only
 the linear terms
\begin{eqnarray}
D(  x) && =   x^2  D^{[2]} 
  \nonumber \\
F(  x) && = x  F^{[1]} 
\nonumber \\
f(x) &&=    x f^{[1]} \ \ \text{ with } \ \ f^{[1]}= F^{[1]} -  D^{[2]}
\label{TaylorFD1dcasdiag}
\end{eqnarray}
The corresponding differential generator of Eq. \ref{Generator1D}
\begin{eqnarray}
{\cal L}
=      F^{[1]}  x  \frac{ \partial }{\partial x}  +     D^{[2]}  x^2   \frac{ \partial^2 }{\partial x^2}
= f^{[1]} \left( x  \frac{ \partial }{\partial x}\right)  +     D^{[2]}  \left( x  \frac{ \partial }{\partial x}\right) \left( x  \frac{ \partial }{\partial x}\right) 
\label{Generator1DGBM}
\end{eqnarray}
only involves the scale-invariant operator $ x  \frac{ \partial }{\partial x}  $ 
that acts on the power $x^n$ as
\begin{eqnarray}
  \left( x  \frac{ \partial }{\partial x}\right) x^n = n x^n
\label{scaleinvxn}
\end{eqnarray}
while the associated SDEs of Eq. \ref{ItoSDE1d} 
correspond to the case $d=1$ of the $d$-dimensional case of Eq. \ref{ItoSDEsinglegjmultiplicativediag}
\begin{eqnarray}
dx(t)  && =   F^{[1]} x(t) dt + \sqrt{ 2 D^{[2]}} x(t) dB(t) \ \ \ \ \ \ \ \ \ \ \ \ \ \ 
\text{[Ito]}
\nonumber \\
dx(t)  && =   f^{[1]} x(t) dt + \sqrt{ 2 D^{[2]}} x(t) dB(t) \ \ \ \ \ \ \ \ \ \ \ \ \ \ 
\text{[Stratonovich]}
\label{ItoSDE1dGBM}
\end{eqnarray}
where one recognizes the Geometric Brownian motion on $x \in ]0,+\infty[$ as recalled in the next subsection.


\subsubsection{ Reminder on the Geometric Brownian motion } 

The Stratonovich SDE of Eq. \ref{ItoSDE1dGBM} with multiplicative noise for $x(t)$
translates for the logarithmic variable
\begin{eqnarray}
y(t) \equiv \ln x(t) 
\label{YlogGBM}
\end{eqnarray}
into the following SDE 
\begin{eqnarray}
dy(t)  =  f^{[1]} +  \sqrt{ 2 D^{[2]} } \ dB(t) 
\label{ItoSDEGBMY}
\end{eqnarray}
involving the constant drift $f^{[1]}$ 
and the additive noise of amplitude $\sqrt{ 2 D^{[2]} } $
(here the Ito and the Stratonovich forms coincide).

The solution $y(t)$ is simply a Brownian motion with drift
\begin{eqnarray}
y(t)=  y(0)+f^{[1]} t+  \sqrt{ 2 D^{[2]} } \ B(t)
\label{ItoSDEGBMYinteg}
\end{eqnarray}
and its exponential gives the following representation of the initial process $x(t)$ 
in terms of the noise $B(t)$
\begin{eqnarray}
x(t) \equiv e^{ y(t)} = x(0) e^{f^{[1]} t+  \sqrt{ 2 D^{[2]} } \ B(t)}
\label{YlogGBMinverse}
\end{eqnarray}

The probability density $ p_t(y)$ of the variable $y$ at $t$
 involves the well-known Gaussian propagator 
\begin{eqnarray}
p_t(y) = \int_{-\infty}^{+\infty} dy_0 \frac{1 }{ \sqrt{ 4 \pi D^{[2]} t }}e^{ - \frac{(y-y_0- f^{[1]} t)^2 }{ 4 D^{[2]} t }} p_0(y_0)
\label{GaussianPropay}
\end{eqnarray}
and can be translated
for the probability density $ \rho_t(x)$ of the variable $x=e^y$ at $t$
\begin{eqnarray}
\rho_t(x) = p_t(y) \frac{dy}{dx} = \frac{p_t \left(y=\ln x \right) }{x}
= \int_{0}^{+\infty} dx_0 \frac{1 }{ x \sqrt{ 4 \pi D^{[2]} t }}e^{ - \frac{(\ln x -\ln x_0- f^{[1]} t)^2 }{ 4 D^{[2]} t}} \rho_0(x_0)
\label{LognormalPropax}
\end{eqnarray}
where the propagator is log-normal.

The moments $m_t(n) $ computed from this solution $ \rho_t(x) $
\begin{eqnarray}
m_t(n) && = \int_0^{+\infty} dx \rho_t(x) \ x^n= \int_{-\infty}^{+\infty} dy p_t(y) \ e^{ n y } 
= \int_{-\infty}^{+\infty} dy_0 p_0(y_0)
\int_{-\infty}^{+\infty} dy  e^{ n y }  \frac{1 }{ \sqrt{ 4 \pi D^{[2]} t }}e^{ - \frac{(y-y_0- f^{[1]} t)^2 }{ 4 D^{[2]} t } }
\nonumber \\
&& = e^{ t ( n^2 D^{[2]} + n f^{[1]} )}  \int_{-\infty}^{+\infty} dy_0 p_0(y_0) e^{ n y_0}
= e^{t M(n, n) } m_0(n)
\label{MomentsLognormalPropax}
\end{eqnarray}
are in agreement with the results of Eqs \ref{MTdiag} \ref{MTdiagsol}
of the Carleman approach, as it should for consistency.

The Fourier transform
of Eq. \ref{GaussianPropay} can be obtained via the same computation as in Eq. \ref{MomentsLognormalPropax}
with the replacement $n \to -ik$
\begin{eqnarray}
{\hat p_t}(k) \equiv \int_{-\infty}^{+\infty} dy p_t(y) \ e^{ -ik y } 
=e^{ t ( -k^2 D^{[2]} - ik f^{[1]} )}  \int_{-\infty}^{+\infty} dy_0 p_0(y_0) e^{- ik y_0}
= e^{ t ( -k^2 D^{[2]} - ik f^{[1]})} {\hat p_0}(k)
\label{GaussianPropayFourier}
\end{eqnarray}
showing that each Fourier mode $k$ evolves independently.
The inverse-Fourier-transform yields the spectral decomposition of $p_t(y) $
\begin{eqnarray}
p_t(y) =  \int_{-\infty}^{+\infty} \frac{dk}{2 \pi} {\hat p_t}(k) e^{ ik y }
=\int_{-\infty}^{+\infty} \frac{dk}{2 \pi}  e^{ ik y } e^{t ( -k^2 D^{[2]} - ik f^{[1]} )} {\hat p_0}(k)
\label{GaussianPropayFourierInverse}
\end{eqnarray}
that can be translated for the probability density $\rho_t(x) $ of Eq. \ref{LognormalPropax} 
\begin{eqnarray}
\rho_t(x) = \frac{p_t \left(y=\ln x \right) }{x}
&& =  \int_{-\infty}^{+\infty} \frac{dk}{2 \pi}  x^{ik -1} e^{t ( -k^2 D^{[2]} - ik f^{[1]} )}
\int_{-\infty}^{+\infty} dy_0 p_0(y_0) e^{- ik y_0}
\nonumber \\
&& = \int_{-\infty}^{+\infty} \frac{dk}{2 \pi}  x^{ik -1} e^{t ( -k^2 D^{[2]} - ik f^{[1]} )}
\int_{0}^{+\infty} dx_0 \rho_0(x_0) x_0^{- ik }
\label{LognormalPropaxspectral}
\end{eqnarray}
to obtain its own spectral decomposition in terms of the continuous spectrum labelled by $k \in ]-\infty,+\infty[$.


\subsubsection{ Discussion of the asymptotic behaviors for large time $t \to + \infty$ }

At large time $t \to + \infty$, 
there is of course no steady state 
for the drifted Brownian motion $y(t)$ of Eq. \ref{ItoSDEGBMYinteg}
and its exponential $x(t) \equiv e^{ y(t)} $,
but it is useful to discuss the various possibilities in terms of the sign 
$ f^{[1]}  =  \left(  F^{[1]} -  D^{[2]}\right)$ 
of the drift of the process $y(t)$ satisfying the SDE of Eq. \ref{ItoSDEGBMY}:

$\bullet$  when the drift is positive $f^{[1]}>0$, then the drifted Brownian motion $y(t)$ flows towards $(+\infty)$, 
and the Geometric-Brownian $x(t)=e^{y(t)}$ also flows towards 
$(+\infty)$, while all the moments $m_n(t) $ of Eq. \ref{MTdiagsol} diverge
exponentially with 
the eigenvalues $ E_n= M(n, n)   = n    f^{[1]}+n^2 D^{[2]} $.

$\bullet$ when the drift vanishes $f^{[1]}=0$, then $y(t)= \ln x(t)$ is a Brownian motion without drift, and all the moments $m_n(t) $ of Eq. \ref{MTdiagsol} diverge exponentially with 
the eigenvalues $E_n=M(n,n) = n^2 D^{[2]} $.

$\bullet$ when the drift is negative $f^{[1]}<0$, 
then the drifted Brownian motion $y(t)$ flows towards $(-\infty)$, and the Geometric Brownian motion
$x(t)=e^{y(t)}$ flows towards $0$.
However the moments $m_n(t) $ of Eq. \ref{MTdiagsol} will converge towards zero only if
the eigenvalue $E_n =M(n,n) = n^2 D^{[2]}+nf^{[1]}$ is strictly negative, i.e. only if the order $n$ is small enough 
$n  < \frac{(-f^{[1]})}{D^{[2]} } $,
while the higher orders $n  > \frac{(-f^{[1]})}{D^{[2]} } $ diverge towards $(+\infty)$
\begin{eqnarray}
m_t(n)  =e^{t M(n, n) } m_0(n) = e^{t \left( n    f^{[1]}+n^2 D^{[2]}  \right) } m_0(n)
  \begin{cases}
\text{ converges towards $0$    if } \ \ n  < \frac{(-f^{[1]})}{D^{[2]} } \equiv \mu
\\
\text{ diverges towards $+ \infty$     if } \ \  n  > \frac{(-f^{[1]})}{D^{[2]} } \equiv \mu
\end{cases}
\label{DVmoments}
\end{eqnarray}
where it is convenient to introduce the notation
\begin{eqnarray}
\mu \equiv \frac{(-f^{[1]})}{D^{[2]} } = \frac{(D^{[2]}-F^{[1]})}{D^{[2]} }
\label{defmu}
\end{eqnarray}
for the transition between the convergent moments $n<\mu$ and the diverging moments $n>\mu$,
while the eigenvalues $E_n= M(n, n) $ of the Carleman matrix become
 \begin{eqnarray}
E_n= M(n, n)  && =   n    F^{[1]}+   n (n-1)   D^{[2]}  = D^{[2]} n  (n - \mu) 
 \label{MTdiagmu}
\end{eqnarray}
This change of sign at the non-integer value $n=\mu$ when $\mu>0$
also plays an essential role for the other cases where the Carleman matrix is triangular
with the same eigenvalues $E_n$ as will be described in the following subsections.

As a final remark, let us mention that for the drifted Brownian motion $y(t)$,
the moments of $x(t)= e^{y(t)}$ 
\begin{eqnarray}
m_t(n)  = {\mathbb E } \left( x^n(t) \right) = {\mathbb E } \left( e^{n y(t) } \right) 
\label{MomentsxnGeneydef}
\end{eqnarray}
correspond to the generating function of $y(t)$ when $n$ is considered as a continuous parameter.
In particular, the exponential behavior for large time $t$
\begin{eqnarray}
 {\mathbb E } \left( e^{n y(t) } \right)  \opsimeq_{t \to + \infty} e^{t E_n }
\label{MomentsxnGeney}
\end{eqnarray}
yields that the eigenvalue $E_n=M(n,n)$ of the diagonal Carleman matrix $M$ as a function of 
the continuous variable $n$
can be interpreted as the Scaled Cumulant Generating Function (SCGF) of the drifted Brownian motion $y(t)$
in the language
of large deviations (see the reviews \cite{oono,ellis,review_touchette} and references therein).


\subsection{ Cases where the Carleman matrix ${\bold M}$ is lower-triangular : $F(  x) = F^{[0]}+x F^{[1]}$
and $D( x) = D^{[0]}+x D^{[1]} +x^2 D^{[2]}$  }

\label{subsec_1Dlower}

\subsubsection{Pearson diffusions
with forces $F(  x) = F^{[0]}+x F^{[1]}$
and diffusion coefficients $D( x) = D^{[0]}+x D^{[1]} +x^2 D^{[2]}$ }
 
 \label{subsec_pearson}

When the diffusion coefficient $D(x)$ is a polynomial of degree two,
while the Ito force and the Stratonovich forces are polynomials of order one 
\begin{eqnarray}
D(  x) && = D^{[0]}+x D^{[1]} +  x^2  D^{[2]} 
\nonumber \\
F(  x) && =F^{[0]} + x  F^{[1]} 
  \nonumber \\
f(x) && = \left( F^{[0]} - \frac{ D^{[1]}  }{2}\right) + x  \big(F^{[1]} - D^{[2]} \big) 
\equiv f^{[0]} + x  f^{[1]}
\label{FDPearson}
\end{eqnarray}
then the Carleman matrix $M(n,q)$ of Eq. \ref{MCarlemand1detailed}
 is lower-triangular, and the only non-vanishing matrix elements
are the diagonal $q=n$ and the two lower diagonals $q=n-1,n-2$ of Eq. \ref{MCarlemand1detailed}
(while the upper diagonal $M(n, n+1)  =   n    F^{[2]} =0$ vanishes).

The triangular property of the Carleman matrix ${\bold M}$
yields that its eigenvalues $E_n$ are directly given by the diagonal matrix elements $M(n,n)$ as 
in Eq. \ref{MTdiag} of the previous subsection concerning the diagonal case
 \begin{eqnarray}
E_n = M(n,n)= n    F^{[1]}+   n (n-1)   D^{[2]} = n    f^{[1]}+   n^2   D^{[2]}
 \label{EigenMPearson}
\end{eqnarray}
while the Carleman dynamics of Eq. \ref{dynotxexpli}
 for the moments 
\begin{eqnarray}
\partial_t m_t(0) && = 0 
\nonumber \\
\partial_t m_t(1) && 
= M(1,0) m_t(0) 
+ M(1,1) m_t(1)
\nonumber \\
\partial_t m_t(2) && 
= M(2,0) m_t(0) 
+ M(2,1) m_t(1)
+ M(2,2) m_t(2)
\nonumber \\
\partial_t m_t(3) && 
= M(3,1) m_t(1)
+ M(3,2) m_t(2)
+ M(3,3)m_t(3)
\nonumber \\
...
\nonumber \\
\partial_t m_t(n) && 
= M(n,n-2) m_t(n-2) 
+ M(n,n-1) m_t(n-1)
+ M(n,n) m_t(n)
\nonumber \\
...
\label{dynotxexpliPearson}
\end{eqnarray}
can be solved iteratively in the order $n=0,1,..$ as explained in detail in \cite{c_Pearson}
with the comparison with various other methods that can be used to solve Pearson diffusions
characterized by Eq. \ref{FDPearson}
(see  \cite{pearson_wong,diaconis,autocorrelation,pearson_class,pearson2012,PearsonHeavyTailed,pearson2018}
and references therein). 

From the point of view of the spectral decomposition of Eq. \ref{SolMomentsSpectral},
this means that the moment $m_t(k) $ actually involves only the eigenvalues
$E_n=M(n,n)$ of Eq. \ref{EigenMPearson} in the interval $0 \leq n \leq k$, 
\begin{eqnarray}
 m_t(k) && 
 =   \sum_{n=0}^{k}  e^{t E_n} \langle k  \vert R_n \rangle 
 \langle L_n \vert  m_0 \rangle 
 = \langle k  \vert R_0 \rangle 
   +   \sum_{n=1}^{k}  e^{t E_n} \langle k  \vert R_n \rangle 
  \langle L_n \vert  m_0 \rangle 
\label{SolMomentsSpectralPearson}
\end{eqnarray}
So the moment $m_t(k)  $ will converge towards the steady value $m_{st}(k)=  \langle k  \vert R_0 \rangle$
corresponding to the right eigenvector $\vert R_{E=0} \rangle $ associated to the eigenvalue $E=0$ discussed in Eq. \ref{SteadyMomentsMatrix} (while 
the trivial left eigenvector is $\langle L_{E=0} \vert =\langle q=0 \vert$ as discussed in Eq \ref{leftMzero})
only if the $(k-1)$ exponentials $e^{t E_n} $ involve negative eigenvalues $E_n=M(n,n)= n    f^{[1]}+n^2 D^{[2]}
= D^{[2]} n  (n - \mu) <0$
for $n=1,2,..,k$
\begin{eqnarray}
 m_t(k)     \opsimeq_{t \to +\infty}
m_{st}(k)=  \langle k  \vert R_0 \rangle \ \ \ 
&& \text{only if $E_n=M(n,n)=D^{[2]} n  (n - \mu)<0$ for $1 \leq n \leq k$} 
\nonumber \\
&& \text{i.e. only if $f^{[1]} <0 $ and $k< \mu =\frac{(-f^{[1]})}{D^{[2]} } $} 
 \label{OavdynPearsonkm2integ}
\end{eqnarray}

In conclusion,, when the multiplicative noise is present $D^{[2]}>0$
and when $f^{[1]} <0 $, then the steady moments $ m_t(k) $ are finite only for $k<\mu$
and the corresponding steady state $\rho_{st}(x) $
will display the power-law tail
\begin{eqnarray}
\rho_{st}(x) \oppropto_{x \to + \infty} \frac{1}{x^{1+\mu}}
\label{powerlawmu}
\end{eqnarray}
as will be described in more details in various examples below,
namely in the Kesten distribution of Eq. \ref{kesten},
in the Fisher-Snedecor distribution of Eq. \ref{fisher},
and in the Student distribution of \ref{student}.

After this general discussion concerning the whole Pearson family of diffusions,
we will focus in the next subsections
on special examples 
in order to make the link with the models listed in section \ref{sec_list}.


\subsubsection{ Cases where the Carleman matrix ${\bold M}$ is lower-bidiagonal :  $D(x)=D^{[1]} x+D^{[2]} x^2$ and $F(x)=F^{[0]} + x  F^{[1]}  $ }

Let us focus on the case $D^{[0]}=0 $ in Eq. \ref{FDPearson} 
\begin{eqnarray}
D(  x) && = D^{[1]} x+  x^2  D^{[2]} 
  \nonumber \\
F(  x) && = F^{[0]}+x  F^{[1]} 
\nonumber \\
f(x)&& = \left( F^{[0]} - \frac{ D^{[1]}  }{2}\right) + x  \big(F^{[1]} - D^{[2]} \big) \equiv f^{[0]} + x  f^{[1]} 
\label{Kestenaugmented}
\end{eqnarray}
so that the Carleman matrix is lower-bidiagonal with the matrix elements of Eq. \ref{MCarlemand1detailed}
 \begin{eqnarray}
M(n, n-1)  && =   n    F^{[0]}+   n (n-1)   D^{[1]} = n (F^{[0]} - D^{[1]}) + n^2 D^{[1]}
\nonumber \\
M(n,n) && = n    F^{[1]}+   n (n-1)   D^{[2]} =n f ^{[1]}+ n^2 D^{[2]} = E_n
 \label{MCarlemand1detailedPearsonlower}
\end{eqnarray}

For this bidiagonal matrix, the 
spectral decomposition of Eq. \ref{SolMomentsSpectralPearson}
\begin{eqnarray}
 m_t(k)
&& = \langle k  \vert R_0 \rangle 
   +   \sum_{n=1}^{k}  e^{t E_n} \langle k  \vert R_n \rangle 
 \left( \sum_{q=0}^{n}  \langle L_n \vert  q \rangle \langle q \vert  m_0 \rangle \right)
  \nonumber \\
  && = m_{st}(k) +   \sum_{n=1}^{k}  e^{t E_n} \langle k  \vert R_n \rangle 
  \sum_{q=0}^{n}  \langle L_n \vert  q \rangle m_0(q)
\label{SolMomentsSpectralPearsonBidiag}
\end{eqnarray}
can be written explicitly as described in subsection \ref {app_lowerBidiag} of Appendix \ref{app_SpectralMatriceBiDiag}
and involves the following eigenvectors :

(i) the left eigenvectors $\langle L_n \vert $ of Eq. \ref{RightUpperBiT}
read using Eq. \ref{MCarlemand1detailedPearsonlower}
\begin{eqnarray}
\langle L_n \vert q \rangle 
= \begin{cases}
\displaystyle  \prod_{j=q}^{n-1} \frac{ M_{j+1,j} }{E_n-E_j }
=   \prod_{j=q}^{n-1} \frac{  (j+1)    F^{[0]}+   (j+1) j   D^{[1]}
 }{(n-j)  f ^{[1]}+ (n^2 -j^2)  D^{[2]} }
 \ \ \ \text{  for } \ \ 0 \leq q \leq n-1
\\
1   \text{ for } \ \ q=n
\\
0   \text{  for } \ \ q>n 
\end{cases}
\label{LeftPearsonlower}
\end{eqnarray}

(ii) the right eigenvectors $ \vert R_n \rangle$ of Eq. \ref{xquadLrecT}
read using Eq. \ref{MCarlemand1detailedPearsonlower}
\begin{eqnarray}
\langle k \vert R_n \rangle 
 = \begin{cases}
 0   \text{ for } \ \ k<n
\\
1   \text{  for } \ \ k=n 
\\
 \displaystyle   \prod_{j=n+1}^k  \frac{ M_{j,j-1} }{E_n-E_j}
 = \prod_{j=n+1}^k \left( -  \frac{ j (F^{[0]} - D^{[1]}) + j^2 D^{[1]} }{ (j-n)  f ^{[1]}+ (j^2-n^2)  D^{[2]}} \right)
 \ \ \text{ for } \ \ k>n
 \end{cases}
\label{RightPearsonlower}
\end{eqnarray}
with the special case $n=0$ giving the steady moments
\begin{eqnarray}
m_{st}(k) = \langle k \vert R_0 \rangle 
 = \begin{cases}
1   \text{  for } \ \ k=0 
\\
 \displaystyle \prod_{j=1}^k \left( -  \frac{  (F^{[0]} - D^{[1]}) + j D^{[1]} }{  f ^{[1]}+ j  D^{[2]}} \right)
 \ \ \text{ for } \ \ k>0
 \end{cases}
\label{msteadyPearsonlower}
\end{eqnarray}
that will produce expressions involving the Gamma function $\Gamma(.)$,
but one should distinguish the three cases described in the three next subsections, 
depending on whether one of the two coefficients $(D^{[1]},D^{[2]})$ vanishes or not.


\subsubsection{ Case $F(x)=F^{[0]} + x  F^{[1]}  $ with only the square-root noise $D(x)=D^{[1]} x$  }

\label{subsec_onlySQ1D}

When $D^{[1]}>0$ and $D^{[2]} =0$, the model corresponds to the case $d=1$ in Eq. \ref{ItoSDEsinglegjonlysquareroot}
where only the square-root noise is present
   \begin{eqnarray}
dx(t)   && =  \left( F^{[0]} +  F^{[1]} x(t)
\right) dt + \sqrt{ 2 D^{[1]} x(t)} dB^{square-root}(t) 
\ \ \ \ \ \text{[Ito]}
\nonumber \\
dx(t)   && =  \left( f^{[0]} +  f^{[1]} x(t)
\right) dt + \sqrt{ 2 D^{[1]} x(t)} dB^{square-root}(t) 
\ \ \ \ \ \text{[Stratonovich]}
\label{ItoSDEsinglegjonlysquareroot1S}
\end{eqnarray}
with the condition $F^{[0]} >0 $ of Eq. \ref{ConditionsF1forSquareRootNoise}.

The linear eigenvalues of Eq. \ref{MCarlemand1detailedPearsonlower} 
 \begin{eqnarray}
E_n=M(n,n)=  n    F^{[1]}= n    f^{[1]}
 \label{EnLinear}
\end{eqnarray}
yield that one should distinguish two cases :

(i) When $ F^{[1]}=     f^{[1]} >0$, 
 then the leading behavior of the moments $m_t(k) $ of Eq. \ref{SolMomentsSpectralPearsonBidiag} for $k>0$
at large time $t \to + \infty $ is the exponential growth  
\begin{eqnarray}
 m_t(k) \oppropto_{t \to + \infty}     e^{t k f^{[1]}} 
\label{mtkLaguerreDV}
\end{eqnarray}
and the process $x(t)$ of Eq. \ref{ItoSDEsinglegjonlysquareroot1S} flows towards $(+\infty)$.

(ii) When $ F^{[1]}=     f^{[1]} <0$, 
then the the moments $m_t(k) $ of Eq. \ref{SolMomentsSpectralPearsonBidiag} 
converge towards
the steady moments of Eq. \ref{msteadyPearsonlower} that read for $k>0$
\begin{eqnarray}
m_{st}(k) =  \prod_{j=1}^k \left( -  \frac{  (F^{[0]} - D^{[1]}) + j D^{[1]} }{  f ^{[1]}} \right)
=  \prod_{j=1}^k   \frac{  \frac{F^{[0]} }{ D^{[1]}}-1 + j  }{  \left( -\frac{ f ^{[1]}}{D^{[1]}} \right) } 
\equiv \prod_{j=1}^k    \frac{      (\alpha-1) +   j   }{    \gamma} 
  = \frac{ \Gamma(k+ \alpha) }{ \gamma^k \Gamma(\alpha)}
\label{SteadyMomentsLaguerre}
\end{eqnarray}
in terms of the two parameters
\begin{eqnarray}
 \alpha && \equiv \frac{ F^{[0]}}{ D^{[1]}} >0
 \nonumber \\
 \gamma && \equiv  \frac{ (- f^{[1]})}{D^{[1]}} >0
\label{NotationsLaguerre}
\end{eqnarray}
while the corresponding normalizable steady state
is given by the Gamma-distribution \cite{c_Pearson,pearson_wong,diaconis,autocorrelation,pearson_class,pearson2012,PearsonHeavyTailed,pearson2018} 
\begin{eqnarray}
\rho^{Gamma}_{st}(x) = \frac{\gamma^{\alpha}}{ \Gamma(\alpha)} x^{\alpha-1}e^{- \gamma x} \ \ \ \ \ {\rm for } \ \ x \in ]0,+\infty[
\label{gamma}
\end{eqnarray}
All the integer moments for $k=1,2,..,+\infty$ are finite, and the expression of Eq. \ref{SteadyMomentsLaguerre}
is actually valid even for non-integer $k \in ]-\alpha,+\infty[ $.

The corresponding diffusion process is
called either the Square-Root process \cite{dufresne} or the Cox-Ingersoll-Ross process \cite{CIR},
while the spectral decomposition of the propagator involves the Laguerre orthogonal polynomials 
\cite{pearson_wong,pearson_class,pearson2012,PearsonHeavyTailed,pearson2018}.


\subsubsection{ Case $F(x)=F^{[0]} + x  F^{[1]}  $ with only the multiplicative noise $D(x)=D^{[2]} x^2$   }

\label{subsec_Kesten}

When $D^{[1]}=0$ and $D^{[2]} >0$, the model corresponds to the case $d=1$ in Eq. \ref{ItoSDEsinglegjonlysquareroot} where only the multiplicative noise is present
 \begin{eqnarray}
dx(t)   && =  \left( F^{[0]} +  F^{[1]}x(t))\right) dt 
+ \sqrt{ 2  D^{[2]} } \  x(t) dB^{multiplicative}(t) 
\ \ \ \text{[Ito]}
\nonumber \\
dx(t)   && =  \left( f^{[0]} +  f^{[1]}x(t))\right) dt 
+ \sqrt{ 2  D^{[2]} } \  x(t) dB^{multiplicative}(t) 
\ \ \ \text{[Stratonovich]}
\label{ItoSDEsinglegjmultiplicativelower1D}
\end{eqnarray}

The quadratic eigenvalues of Eq. \ref{MCarlemand1detailedPearsonlower} 
 \begin{eqnarray}
E_n=M(n,n)= n    F^{[1]}+   n (n-1)   D^{[2]} =n f ^{[1]}+ n^2 D^{[2]} \equiv D^{[2]} n ( n - \mu )
\ \ \ \text{ with } \ \ \mu \equiv \frac{(-f^{[1]})}{D^{[2]} } = \frac{(D^{[2]}-F^{[1]})}{D^{[2]} }
 \label{EnQuadratic}
\end{eqnarray}
yields the following criterion for the convergence or divergence at large time $t \to + \infty $
of the moments $m_t(k) $ of Eq. \ref{SolMomentsSpectralPearsonBidiag} for $k>0$
\begin{eqnarray}
m_t(k)  
  \begin{cases}
\text{ converges towards some finite steady value $m_{st}(k)$    if } \ \ k  <  \mu
\\
\text{ diverges as $e^{ t D^{[2]} k ( k - \mu )}  $ towards $+ \infty$     if } \ \  k  >  \mu
\end{cases}
\label{CVDVmoments}
\end{eqnarray}
with the following discussion :

(i) If $f^{[1]} > 0 $ corresponding to $\mu<0$, then 
 the process $x(t)$ of Eq. \ref{ItoSDEsinglegjmultiplicativelower1D} flows towards $(+\infty)$.

(ii) If $f^{[1]} < 0 $ corresponding to $\mu>0$, then the moments of order $0<k<\mu$
converge towards the steady values given by Eq. \ref{msteadyPearsonlower} 
\begin{eqnarray}
m_{st}(k) =  \prod_{j=1}^k \left( -  \frac{  F^{[0]}   }{  f ^{[1]}+ j  D^{[2]}} \right)
= \prod_{j=1}^k \left(   \frac{  \lambda   }{ \mu - j} \right)
  =  \lambda^k \frac{ \Gamma(\mu-k) }{  \Gamma(\mu)}
\label{msteadyPearsonlowerKesten}
\end{eqnarray}
in terms of the notation
\begin{eqnarray}
 \lambda && \equiv \frac{   F^{[0]}}{D^{[2]}} =\frac{   f^{[0]}}{D^{[2]}} >0
\label{NotationsKesten}
\end{eqnarray}
while the corresponding normalizable steady state
is given by the Inverse-Gamma-distribution
\cite{c_Pearson,pearson_wong,diaconis,autocorrelation,pearson_class,pearson2012,PearsonHeavyTailed,pearson2018} 
\begin{eqnarray}
\rho^{Kesten}_{st}(x) = \frac{\lambda^{\mu}}{\Gamma(\mu) x^{1+\mu} } e^{- \frac{\lambda}{x}} \ \ \ {\rm } \ \ {\rm for } \ \ x \in ]0,+\infty[
\label{kesten}
\end{eqnarray}
displaying the power-law decay of Eq. \ref{powerlawmu} as $\frac{1}{ x^{1+\mu} } $ at $x \to + \infty$.
 The expression of Eq. \ref{SteadyMomentsLaguerre}
is actually valid for $k \in ]-\infty,\mu[ $.

The Statonovich SDE of Eq. \ref{ItoSDEsinglegjmultiplicativelower1D}
 can be considered as an inhomogeneous linear differential
equation that can be integrated via the method of variation of constant 
\begin{eqnarray}
x(t)  && = F^{[0]} e^{f^{[1]} t + \sqrt{2 D^{[2]} } B(t)  }   \int_0^t ds \ e^{- f^{[1]} s - \sqrt{2 D^{[2]} } B(s) }  
\nonumber \\
&& = F^{[0]} \int_0^t ds \ e^{f^{[1]} (t-s) + \sqrt{2 D^{[2]} } (B(t) -B(s)) }  
\label{xexpfunctionalBrownian}
\end{eqnarray}
to obtain that the process $x(t)$ is an exponential functional of the Brownian motion $B(t)$
that has been much studied \cite{c_flux,c_these,yor}.
In particular, it corresponds to the continuous limit 
of the Kesten random variables 
that appears in many disordered systems
\cite{Kesten,Solomon,sinai,jpb_review,annphys90,Der_Hil,Cal,strong_review,c_microcano,c_watermelon,c_mblcayley,c_reset}.

As a final remark, let us mention that the logarithmic variable $y(t) \equiv \ln x(t)  $
satisfies the following SDE with additive noise of amplitude $\sqrt{ 2 D^{[2]} } $
(here the Ito and the Stratonovich forms coincide)
\begin{eqnarray}
dy(t)  = \left( F^{[0]}  e^{ - y(t) }+  f^{[1]} \right) dt+  \sqrt{ 2 D^{[2]} } \ dB(t) 
\label{ItoSDEKesten}
\end{eqnarray}
instead of the drifted-Brownian SDE of Eq. \ref{ItoSDEGBMY} when $F^{[0]} =0 $.


\subsubsection{ Case $F(x)=F^{[0]} + x  F^{[1]}  $ with both square-root noise  $D^{[1]}>0$
  and multiplicative noise $D^{[2]} >0$ }

\label{subsec_twosquarerootmulti1D}

When $D^{[1]}>0$ and $D^{[2]} >0$, 
the model corresponds to the case $d=1$ in Eq. \ref{ItoSDEsinglegjtwoSquareRootMulti} where
 both the square-root noise and the multiplicative noise are present
   \begin{eqnarray}
dx(t)   && =  \left( F^{[0]} + F^{[1]}x(t)\right) dt 
+  \sqrt{ 2 D^{[1]} x }dB^{square-root}(t) +  \sqrt{ 2 D^{[2]} } x(t) dB^{multiplicative}(t)
\ \ \text{[Ito]}
\nonumber \\
dx(t)   && = \left( f^{[0]} + f^{[1]}x(t)\right) dt 
+  \sqrt{ 2 D^{[1]} x }dB^{square-root}(t) +  \sqrt{ 2 D^{[2]} } x(t) dB^{multiplicative}(t)
\ \ \text{[Stratonovich]}
\nonumber \\
\label{ItoSDEsinglegjtwoSquareRootMulti1D}
\end{eqnarray}
with the condition $F^{[0]}>0 $ of Eq. \ref{ConditionsF1forSquareRootNoise}.
 
 The quadratic eigenvalues of Eq. \ref{EnQuadratic} 
 leads to the same criterion of Eq. \ref{CVDVmoments}
 with the similar discussion :

(i) If $f^{[1]} > 0 $ corresponding to $\mu<0$, then 
 the process $x(t)$ of Eq. \ref{ItoSDEsinglegjtwoSquareRootMulti1D} flows towards $(+\infty)$.

(ii) If $f^{[1]} < 0 $ corresponding to $\mu= - \frac{  f^{[1]}) }{D^{[2]} }>0$, then the moments of order $0<k<\mu$
converge towards the steady values given by Eq. \ref{msteadyPearsonlower} 
 \begin{eqnarray}
m_{st}(k) = \prod_{j=1}^k \left( -  \frac{  (F^{[0]} - D^{[1]}) + j D^{[1]} }{  f ^{[1]}+ j  D^{[2]}} \right)
  = \prod_{j=1}^k   \frac{   D^{[1]} \left(   \frac{ F^{[0]}}{ D^{[1]}} -1 +   j \right) }
  {   D^{[2]}  \left( - \frac{  f^{[1]}) }{D^{[2]} } -   j \right) } 
  = \prod_{n=1}^k   \frac{  c \left(   \alpha  +   n \right) }  {    \left(  \mu -   n \right) } 
  =  c^k \frac{ \Gamma(k+ \alpha) \Gamma(\mu-k) }{ \Gamma( \alpha) \Gamma(\mu)}
\label{SteadyMomentsSnedeckor}
\end{eqnarray}
in terms of the notation $\alpha \equiv \frac{ F^{[0]}}{ D^{[1]}}>0$ of Eq. \ref{NotationsLaguerre}
and
\begin{eqnarray}
c && \equiv \frac{   D^{[1]}}{D^{[2]}} 
\label{cratiod12}
\end{eqnarray}
while the corresponding normalizable steady state
is given by the Fisher-Snedecor-distribution 
\cite{c_Pearson,pearson_wong,diaconis,autocorrelation,pearson_class,pearson2012,PearsonHeavyTailed,pearson2018} 
\begin{eqnarray}
\rho^{Fisher-Snedecor}_{st}(x) = c^{\mu} \frac{\Gamma(\alpha+\mu)}{ \Gamma(\alpha) \Gamma(\mu) } \frac{ x^{\alpha-1} }{(c+x)^{\alpha+\mu} }
 \ \ \ {\rm for } \ x \in ]0,+\infty[
\label{fisher}
\end{eqnarray}
displaying the power-law decay of Eq. \ref{powerlawmu} as $\frac{1}{ x^{1+\mu} } $ at $x \to + \infty$.

The corresponding process is called the Fisher-Snedecor diffusion 
(see  \cite{c_Pearson,pearson_wong,pearson_class,pearson2012,PearsonHeavyTailed,pearson2018,pearson_fisher,pearson_fisherSnedecor}
and references therein).


\subsubsection{ Case $F(x)= x  F^{[1]}  $ with both additive noise  $D^{[0]}>0$
  and multiplicative noise $D^{[2]} >0$ }

\label{subsec_twoaddmulti1D}

The case $d=1$ in Eq. \ref{ItoSDEsinglegjtwoaddmultiBiblock} where
 both the additive noise and the multiplicative noise are present
 while the force reduces to $F(x)= x  F^{[1]} $ (i.e. the constant term vanishes $F^{[0]}=0$)
  \begin{eqnarray}
dx(t)   && =   F^{[1]} x(t) dt 
+\sqrt{ 2 D^{[0]} }dB^{additive}(t) +  \sqrt{ 2 D^{[2]}} x(t) dB^{multiplicative}(t)
\ \ \ \text{[Ito]}
\nonumber \\
dx(t)   && =   f^{[1]} x(t) dt 
+\sqrt{ 2 D^{[0]} }dB^{additive}(t) +  \sqrt{ 2 D^{[2]}} x(t) dB^{multiplicative}(t)
\ \ \ \text{[Stratonovich]}
\nonumber \\
\label{ItoSDEsinglegjtwoaddmulti1D}
\end{eqnarray}
corresponds to the Carleman matrix ${\bold M}$ 
where the only non-vanishing matrix elements are on the diagonal $q=n$ 
and on the second-lower-diagonal $q=n-2$ in Eq. \ref{MCarlemand1detailed}
 \begin{eqnarray}
M(n, n-2)  && =      n (n-1)   D^{[0]}
\nonumber \\
M(n, n)  && =   n    F^{[1]}+   n (n-1)   D^{[2]} =  n    f^{[1]}+   n^2   D^{[2]} = E_n
 \label{MCarlemand1detailedsecondlower}
\end{eqnarray}
so that one can adapt the computations of Appendix \ref{app_SpectralMatriceBiDiag}
to obtain its explicit spectral decomposition.
In particular, the even steady moments $m_{st}(k=2l) $
can be computed from the recurrence obtained from Eq. \ref{dynmomentsd1}
 \begin{eqnarray}
m_{st}(2l) && = -\frac{ M(2l,2l-2) }{M(2l,2l)} m_{st}(2l-2)   
= \prod_{j=1}^{l} \left(  -\frac{ M(2j,2j-2) }{M(2j,2j)}\right)
= \prod_{j=1}^{l} \left(  -\frac{   (2j-1)   D^{[0]} }
{    f^{[1]}+   2 j   D^{[2]}}\right)
\nonumber \\
&& = \prod_{j=1}^{l}   \frac{   \left(j-\frac{1}{2}\right)   \frac{D^{[0]}}{D^{[2]}}  }
{     \frac{\mu}{2} -    j   }
= \left( \frac{D^{[0]}}{D^{[2]}} \right)^l \frac{\Gamma(\frac{1}{2}+l)\Gamma(\frac{\mu}{2}-l)}{\Gamma(\frac{1}{2})\Gamma(\frac{\mu}{2}) }
 \ \ {\rm for } \ \ \ \ 2l < \mu
\label{studentmomentsd1rec}
\end{eqnarray}
while the corresponding normalizable steady state
is given by the Student-distribution 
\cite{c_Pearson,pearson_wong,diaconis,autocorrelation,pearson_class,pearson2012,PearsonHeavyTailed,pearson2018,pearson_student_processes,pearson_student} 
\begin{eqnarray}
\rho^{Student}_{st}(x) = \sqrt{ \frac{D^{[2]}}{D^{[0]}}}  \frac{\Gamma(\frac{\mu+1}{2}) }
{\Gamma(\frac{1}{2})\Gamma(\frac{\mu}{2}) \left( 1+\frac{D^{[2]}}{D^{[0]}} x^2 \right)^{\frac{1+\mu}{2}}}
  \ \ \ {\rm } \ \ {\rm for } \ \ x \in ]-\infty,+\infty[
\label{student}
\end{eqnarray}
displaying the power-law decay of Eq. \ref{powerlawmu} as $\frac{1}{ x^{1+\mu} } $ at $x \to + \infty$.
Note that the Cauchy distribution appears for 
the special case $\mu=1$
\begin{eqnarray}
\text{ Case $\mu=1$ } \ : \ \ \rho^{Cauchy}_{st}(x) = \sqrt{ \frac{D^{[2]}}{D^{[0]}}}  \frac{1 }
{ \pi \left( 1+\frac{D^{[2]}}{D^{[0]}} x^2 \right)}
  \ \ \ {\rm } \ \ {\rm for } \ \ x \in ]-\infty,+\infty[
\label{cauchy}
\end{eqnarray}



\subsection{ Cases where the Carleman matrix ${\bold M}$ is upper-triangular :  $F^{[0]}=0=D^{[0]}=D^{[1]}$ }

\label{subsec_upperbidiag1D}

When the three Taylor coefficients $F^{[0]}=0=D^{[0]}=D^{[1]}$ vanish in Eq. \ref{TaylorFD1d}, 
the Carleman matrix ${\bold M}$ is upper-bidiagonal
where the non-vanishing matrix elements of Eq. \ref{MCarlemand1detailed} are
 \begin{eqnarray}
M(n, n)  && =   n    F^{[1]}+   n (n-1)   D^{[2]} =  n    f^{[1]}+   n^2   D^{[2]} =E_n
\nonumber \\
M(n, n+1)  && =   n    F^{[2]}
 \label{MCarlemand1detailedupper}
\end{eqnarray}
So one can write its explicit spectral decomposition of Eq. \ref{Uupperbidiagspectral}
derived in Appendix \ref{app_SpectralMatriceBiDiag},
as already done for the deterministic version without noise $D^{[2]}=0 $ in Appendix \ref{app_deterministic}.

The SDE correspond to the case $d=1$ in Eqs \ref{ItoSDEsinglegjmultiplicativeupper}
\begin{eqnarray}
dx(t)  && =  \left( F^{[1]} x(t) + F^{[2]} x^2(t)\right) dt + \sqrt{ 2 D^{[2]}} x(t) dB^{multiplicative}(t) 
\ \ \ \text{[Ito]}
\nonumber \\
dx(t)  && =  \left( f^{[1]} x(t)  + F^{[2]} x^2(t)\right) dt + \sqrt{ 2 D^{[2]}} x(t) dB^{multiplicative}(t) 
\ \ \ \text{[Stratonovich]}
\label{ItoSDE1dGBMupper}
\end{eqnarray}

The change of variables
\begin{eqnarray}
{\breve x}(t) = \frac{1}{x(t) }
\label{Zinvx}
\end{eqnarray}
that is already useful in the deterministic case (see Eq. \ref{zinversex} in Appendix \ref{app_deterministic})
leads here to the Stratonovich SDE 
\begin{eqnarray}
d{\breve x}(t)  && = - \frac{\left( f^{[1]} x(t)  + F^{[2]} x^2(t)\right) dt + \sqrt{ 2 D^{[2]}} x(t) dB^{multiplicative}(t)}{x^2(t)}
\nonumber \\
&& =  -  \left( f^{[1]} {\breve x}(t)  + F^{[2]} \right) dt - \sqrt{ 2 D^{[2]}} {\breve x}(t) dB^{multiplicative}(t)
\nonumber \\
&& \equiv \left(  {\breve f}^{[1]} {\breve x}(t) + {\breve F}^{[0]} \right) dt   - \sqrt{ 2 D^{[2]}} {\breve x}(t) dB^{multiplicative}(t)
\label{SDEZstratoupper}
\end{eqnarray}
where the noise remains multiplicative with respect to the new variable $ {\breve x}(t)$
while the Stratonovich force ${\cal F}({\breve x})=(  {\breve F}^{[0]} {\breve x} +{\breve F}^{[0]}) $
 involves the two Taylor coefficients
\begin{eqnarray}
{\breve F}^{[0]}&& = - F^{[2]}
 \nonumber \\
{\breve f}^{[1]} && = -  f^{[1]}
\label{coefsupperlower}
\end{eqnarray}

So the process ${\breve x}(t) $ corresponds to
the process discussed in the previous subsection \ref{subsec_Kesten} 
with the parameters
\begin{eqnarray}
{\breve \mu}&& = - \frac{{\breve f}^{[1]}}{D^{[2]} } = \frac{f^{[1]}}{D^{[2]} } 
 \nonumber \\
{\breve \lambda}&& = \frac{   {\breve F}^{[0]}}{D^{[2]}} = - \frac{F^{[2]}}{D^{[2]}}
\label{coefsupperlowerkesten}
\end{eqnarray}
and one can directly translate the results as follows for the present process $x(t) = \frac{1}{{\breve x}(t) } $ 
in the region $ f^{[1]} >0$ and $F^{[2]} <0 $ corresponding to ${\breve \mu}>0$ and to ${\breve \lambda}>0$ :

$\bullet$ The steady state distribution ${\breve \rho}_{st}({\breve x}) $
given by the Inverse-Gamma-distribution of Eq. \ref{kesten}
of parameters ${\breve \mu} $ and ${\breve \lambda} $ 
translates into the Gamma distribution for the variable $x=\frac{1}{ {\breve x}}$
\begin{eqnarray}
\rho_{st}(x) = \frac{{\breve \rho}_{st}({\breve x}=\frac{1}{x} ) }{x^2}
= \frac{{\breve \lambda}^{{\breve \mu}}}{\Gamma({\breve \mu})  } x^{ {\breve \mu}-1} e^{- {\breve \lambda}  x} 
\label{kestenInverse}
\end{eqnarray}

$\bullet$ The functional representation of Eq. \ref{xexpfunctionalBrownian} for the process ${\breve x}(t) $ 
\begin{eqnarray}
{\breve x}(t)   && = {\breve F}^{[0]}   \int_0^t ds \ e^{ {\breve f}^{[1]} (t-s) - \sqrt{2 D^{[2]} } (B(t) -B(s)) }  
\label{xexpfunctionalBrownianbreve}
\end{eqnarray}
translates for the process $x(t)=\frac{1}{ {\breve x}(t)}$
into
\begin{eqnarray}
x(t)=\frac{1}{ {\breve x}(t)}   && =
\frac{1}{ [- F^{[2]} ]   \int_0^t ds \ e^{  -  f^{[1]} (t-s) - \sqrt{2 D^{[2]} } (B(t) -B(s)) }   }
\label{xexpfunctionalBrownianbreveInverse}
\end{eqnarray}

As a final remark, let us mention that the present analysis 
concerning the quadratic non-linearity of coefficient $F^{[2]}$
can be done for any other degree $p=3,4,..$ of non-linearity of coefficient $F^{[p]}$
in the presence of multiplicative noise
as described in \cite{CarlemanStochastic}.
The corresponding Carleman matrix elements $M(n,q)$ 
are then non-vanishing only on the diagonal $q=n$ and on the upper-diagonal $q=n+p-1$.


 \subsection{ Discussion }

In conclusion, the Carleman perspective is very useful in dimension $d=1$ 
to recover the well-known simplest soluble diffusions,
namely :

(i) the case where the Carleman matrix is diagonal corresponds to the Geometric Brownian motion;

(ii) the cases where the Carleman matrix is lower-triangular correspond to the soluble 
family of Pearson diffusions;

(iii) the cases where the Carleman matrix is upper-triangular
include the cases with only multiplicative noise $D(x)=D^{[2]} x^2$,
while the the force may contain any degree $p=2,3,4,..$ of non-linearity of coefficient $F^{[p]}$
besides the linear contribution parametrized by $F^{[1]}$.

In the remaining sections, we focus on the dimension $d=2$ 
to analyze the analogous simplest models from the Carleman perspective.


 \section{ Application of the Carleman approach to diffusion processes in $d=2$ } 

 \label{sec_2D}

 In this section, the Carleman approach is applied to the diffusion processes of section \ref{sec_DfullyDiag}
  in the dimension $d=2$. To be more concrete, let us first recapitulate explicitly
  the various notations in the following subsection.

  \subsection{ Notations for the diffusion processes in dimension $d=2$ }
  
  \label{subsec_Nota2D}
  
In dimension $d=2$, the diffusion matrix of Eq. \ref{DdiagoEtSingle}
reads
\begin{eqnarray}
 D_{11}(x_1,x_2) 
&& = D^{[0]}_1+D^{[1]}_1 x_1 + D^{[2]}_1 x_1^2
\nonumber \\
 D_{22}(x_1,x_2) 
&& = D^{[0]}_2+D^{[1]}_2 x_2 + D^{[2]}_2 x_2^2
\nonumber \\
 D_{1,2}(x_1,x_2) && = 0
\label{DdiagoEtSingle2D}
\end{eqnarray}
while the Ito forces 
 of Eq. \ref{TaylorF}  read
  \begin{eqnarray}
F_1(x_1,x_2) && 
 = F^{[0]}_1   
 +  F^{[1]}_{11} x_1
+ F^{[1]}_{12} x_2
+   F^{[2]}_{1;11} x_1^2
+    F^{[2]}_{1;22} x_2^2
+ 2  F^{[2]}_{1;12} x_1x_2
\nonumber \\
F_2(x_1,x_2) && 
 = F^{[0]}_2   
 +  F^{[1]}_{21} x_1
+ F^{[1]}_{22} x_2
+   F^{[2]}_{2;11} x_1^2
+    F^{[2]}_{2;22} x_2^2
+ 2  F^{[2]}_{2;12} x_1x_2
\label{TaylorF2D}
\end{eqnarray}

The Stratonovich forces of Eq. \ref{TaylorFStrato} are also polynomials of degree two
 \begin{eqnarray}
f_1(x_1,x_2) &&  = \left(F^{[0]}_2 -  \frac{D^{[1]}_1}{2} \right) 
+  \bigg( F^{[1]}_{11}-  D^{[2]}_1  \bigg) x_1
+  F^{[1]}_{12} x_2
+   F^{[2]}_{1;11} x_1^2
+    F^{[2]}_{1;22} x_2^2
+ 2  F^{[2]}_{1;12} x_1x_2
\nonumber \\
&& \equiv f^{[0]}_1   
 +  f^{[1]}_{11} x_1
+ f^{[1]}_{12} x_2
+   f^{[2]}_{1;11} x_1^2
+    f^{[2]}_{1;22} x_2^2
+ 2  f^{[2]}_{1;12} x_1x_2
\nonumber \\ 
f_2(x_1,x_2) && = \left( F^{[0]}_2 - \frac{ D^{[1]}_2}{2} \right)
+  F^{[1]}_{21} x_1
+ \bigg( F^{[1]}_{22} -  D^{[2]}_2 \bigg) x_2
+   F^{[2]}_{2;11} x_1^2
+    F^{[2]}_{2;22} x_2^2
+ 2  F^{[2]}_{2;12} x_1x_2
\nonumber \\
&& \equiv f^{[0]}_2   
 +  f^{[1]}_{21} x_1
+ f^{[1]}_{22} x_2
+   f^{[2]}_{2;11} x_1^2
+    f^{[2]}_{2;22} x_2^2
+ 2  f^{[2]}_{2;12} x_1x_2
\label{StratoItoCorrespondance2D}
\end{eqnarray}
and the only different coefficients of Eq. \ref{DfullyStratoSingle} are
\begin{eqnarray}
 f^{[0]}_1 &&  =  F^{[0]}_1 - \frac{D^{[1]}_1  }{2}
\nonumber \\ 
f^{[0]}_2 &&  =  F^{[0]}_2 - \frac{D^{[1]}_2  }{2}
\nonumber \\ 
 f^{[1]}_{11} && =  F^{[1]}_{11}- D^{[2]}_1
\nonumber \\
 f^{[1]}_{22} && =  F^{[1]}_{22}- D^{[2]}_{22}
\label{DfullyStratoSingle2D}
\end{eqnarray}

  The Stratonovich system of Eq. \ref{ItoSDEsinglegj} involving a single noise $dB_j(t) $ for each components $x_j(t)$
  reads
  \begin{small}
  \begin{eqnarray}
dx_1(t)   && = \left(f^{[0]}_1   
 +  f^{[1]}_{11} x_1(t)
+ f^{[1]}_{12} x_2(t)
+   f^{[2]}_{1;11} x_1^2(t)
+    f^{[2]}_{1;22} x_2^2(t)
+ 2  f^{[2]}_{1;12} x_1(t)x_2(t) \right) dt 
+ \sqrt{ 2 \left[ D^{[0]}_1+D^{[1]}_1 x_1(t) + D^{[2]}_1 x_1^2(t) \right] }  dB_1(t) 
\nonumber \\
dx_2(t)   && = \left(f^{[0]}_2   
 +  f^{[1]}_{21} x_1(t)
+ f^{[1]}_{22} x_2(t)
+   f^{[2]}_{2;11} x_1^2(t)
+    f^{[2]}_{2;22} x_2^2(t)
+ 2  f^{[2]}_{2;12} x_1(t)x_2(t) \right) dt 
+ \sqrt{ 2  \left[ D^{[0]}_2+D^{[1]}_2 x_2(t) + D^{[2]}_2 x_2^2(t) \right]} dB_2(t) 
\nonumber \\
\label{StratoSDE2D}
\end{eqnarray} 
\end{small}
 with the three special cases of additive/multiplicative/square-root noises of Eq. \ref{SingleNoisesCases},
or the re-interpretation in terms of two noises of Eqs \ref{ItoAddAndMulti} and \ref{ItoSquareRootAndMulti}.

 The differential generator of Eq. \ref{GeneratorDiag} reads
\begin{eqnarray}
{\cal L}
&& =    F_1( x_1,x_2)   \frac{ \partial }{\partial x_1} 
 + F_2( x_1,x_2)   \frac{ \partial }{\partial x_2}  
+    D_{11}( x_1,x_2)   \frac{ \partial^2 }{\partial x_1^2}
+     D_{22}( x_1,x_2)   \frac{ \partial^2 }{\partial x_2^2}
\nonumber \\
&& =   \left[ 
F^{[0]}_1   
 +  F^{[1]}_{11} x_1
+ F^{[1]}_{12} x_2
+   F^{[2]}_{1;11} x_1^2
+    F^{[2]}_{1;22} x_2^2
+ 2  F^{[2]}_{1;12} x_1x_2
\right]  \frac{ \partial }{\partial x_1} 
\nonumber \\
&& +   \left[ 
 F^{[0]}_2   
 +  F^{[1]}_{21} x_1
+ F^{[1]}_{22} x_2
+   F^{[2]}_{2;11} x_1^2
+    F^{[2]}_{2;22} x_2^2
+ 2  F^{[2]}_{2;12} x_1x_2
\right]   \frac{ \partial }{\partial x_2}  
\nonumber \\
&&+    \left[ D^{[0]}_1+D^{[1]}_1 x_1 + D^{[2]}_1 x_1^2 \right] 
   \frac{ \partial^2 }{\partial x_1^2}
+    \left[ D^{[0]}_2+D^{[1]}_2 x_2 + D^{[2]}_2 x_2^2 \right]   
   \frac{ \partial^2 }{\partial x_2^2}
     \nonumber \\
&& =  {\cal L}^{[-2]} + {\cal L}^{[-1]}+{\cal L}^{[0]} + {\cal L}^{[1]}
\label{Generator2D}
\end{eqnarray}
while the natural decomposition into the four contributions ${\cal L}^{[k=-2,-1,0,1]} $
of Eq. \ref{GeneratorClassDegree4} involves the operators
\begin{eqnarray}
{\cal L}^{[-2]}
&& \equiv     D^{[0]}_1   \frac{ \partial^2 }{\partial x_1^2}
+   D^{[0]}_2   \frac{ \partial^2 }{\partial x_2^2}
  \nonumber \\
{\cal L}^{[-1]}
&& \equiv  F^{[0]}_1  \frac{ \partial }{\partial x_1} 
 +   F^{[0]}_2     \frac{ \partial }{\partial x_2}  
 +    D^{[1]}_1 x_1     \frac{ \partial^2 }{\partial x_1^2}
+    D^{[1]}_2 x_2    \frac{ \partial^2 }{\partial x_2^2}
  \nonumber \\
    {\cal L}^{[0]}
&& \equiv  
  \left[   F^{[1]}_{11} x_1+ F^{[1]}_{12} x_2 \right]  \frac{ \partial }{\partial x_1} 
 +   \left[   F^{[1]}_{21} x_1+ F^{[1]}_{22} x_2\right]   \frac{ \partial }{\partial x_2}  
+    D^{[2]}_1 x_1^2    \frac{ \partial^2 }{\partial x_1^2}
+   D^{[2]}_2 x_2^2    \frac{ \partial^2 }{\partial x_2^2}
  \nonumber \\
{\cal L}^{[1]}
&& \equiv  
 \left[    F^{[2]}_{1;11} x_1^2
+    F^{[2]}_{1;22} x_2^2
+ 2  F^{[2]}_{1;12} x_1x_2
\right]  \frac{ \partial }{\partial x_1} 
 +   \left[   F^{[2]}_{2;11} x_1^2
+    F^{[2]}_{2;22} x_2^2
+ 2  F^{[2]}_{2;12} x_1x_2
\right]   \frac{ \partial }{\partial x_2}  
\label{GeneratorClassDegree4in2D}
\end{eqnarray}

 
  \subsection{Carleman dynamics for the moments $m_t( n_1,n_2) \equiv  {\mathbb E} ( x_1^{n_1}(t) x_2^{n_2}(t))   $ labelled by the two integers 
  $(n_1,n_2) \in {\mathbb N}^2$ }

In dimension $d=2$, the dynamics of Eq. \ref{DynMoments}
for the moments $m_t( n_1,n_2) \equiv  {\mathbb E} ( x_1^{n_1}(t) x_2^{n_2}(t))   $ labelled by the two integers 
  $(n_1,n_2) \in {\mathbb N}^2$
\begin{eqnarray}
\partial_t m_t(n_1,n_2) 
 =  \sum_{q_1=0}^{+\infty}  \sum_{q_2=0}^{+\infty} M(n_1,n_2 \vert q_1,q_2)  m_t(q_1,q_2) 
\label{DynMoments2D}
\end{eqnarray}
  involves the Carleman matrix ${\bold M}=  {\bold M}^{[-2]} + {\bold M}^{[-1]}+{\bold M}^{[0]} + {\bold M}^{[1]}$ 
  that can be decomposed into blocks ${\bold M}^{[k\equiv q-n]} $ according to the possible difference $k=-2,-1,0,1$ 
between the global degrees $q=q_1+q_2$ and $n=n_1+n_2$ as discussed around Eq. \ref{MClassDegree}.
 Their matrix elements are given by
  Eqs \ref{Mm2simpli} \ref{m1simpli} \ref{m0simpli}
    respectively  
  \begin{eqnarray}
M^{[-2]}(n_1,n_2 \vert q_1,q_2)
&& \equiv  
  D^{[0]}_1  
  n_1 (n_1-1) \delta_{q_1,n_1-2}   \delta_{q_2,n_2}   
 +   D^{[0]}_2 
  n_2 (n_2-1) \delta_{q_2,n_2-2}  \delta_{q_1,n_1} 
\nonumber \\
M^{[-1]}(n_1,n_2 \vert q_1,q_2)
&& =   
\left[  F^{[0]}_1    n_1   +    D^{[1]}_1   n_1 (n_1-1) \right] 
\delta_{q_1,n_1-1}  \delta_{q_2,n_2}   
  + \left[  F^{[0]}_1  n_2   +    D^{[1]}_1   n_2 (n_2-1) \right] 
\delta_{q_2,n_2-1}   \delta_{q_1,n_1} 
\nonumber \\
M^{[0]}(n_1,n_2 \vert q_1,q_2)
&& =  
\left[  F^{[1]}_{11}  n_1   +   D^{[2]}_1   n_1 (n_1-1) 
+ F^{[1]}_{22}  n_2   +     D^{[2]}_2   n_2 (n_2-1) 
\right]  
 \delta_{q_1,n_1}   \delta_{q_2,n_2} 
\nonumber \\
&& + F^{[1]}_{12} 
 n_1 \delta_{q_1,n_1-1}  \delta_{q_2,n_2+1} 
+   F^{[1]}_{21} 
 n_2 \delta_{q_1,n_1+1} \delta_{q_2,n_2-1} 
\label{m0simpli2D}
\end{eqnarray}
   while the matrix ${\bold M}^{[1]} $ of Eq. \ref{M1} reads
  \begin{eqnarray}
M^{[1]}(n_1,n_2 \vert q_1,q_2)
&& = 
\left( F^{[2]}_{1;11}   n_1  +2 F^{[2]}_{2;12} n_2 \right) 
\delta_{q_1,n_1+1}     \delta_{q_2,n_2} 
+ \left( F^{[2]}_{2;22}    n_2  + 2 F^{[2]}_{1;12}  n_1  \right) 
  \delta_{q_1,n_1}    \delta_{q_2,n_2+1}
\nonumber \\
&&+ F^{[2]}_{1;22} n_1  \delta_{q_1,n_1-1}  \delta_{q_2,n_2+2}   
+ F^{[2]}_{2;11} n_2 \delta_{q_1,n_1+2}    \delta_{q_2,n_2-1} 
\label{M1in2D}
\end{eqnarray}

Besides this formulation on the 2D integrer lattice $(n_1,n_2) \in {\mathbb N}^2$,
it is useful to consider the reformulation into blocks associated to the total degree $n=n_1+n_2$
as described in the next subsection.

 
  \subsection{ Reshaping the Carleman matrix $M(n_1,n_2 \vert q_1,q_2)$  
    into blocks ${\bold M}^{[k\equiv q-n]}_{[n,q]} $ of sizes $(n+1) \times (q+1)$ 
  associated to the total degrees $n=n_1+n_2$ and $q=q_1+q_2$
  }
  
  The moments $m_t( n_1,n_2) = {\mathbb E} ( x_1^{n_1}(t) x_2^{n_2}(t))   $ labelled by the two integers   $(n_1,n_2) \in {\mathbb N}^2$ can be gathered into kets $\vert {\cal M}_t^{[n]} \rangle $ 
 associated to the total degree $n=n_1+n_2$ 
  containing $(n+1)$ moments labelled by $n_2=0,1,..,n$ while $n_1=n-n_2$ 
  so that the correspondence with the previous integer-lattice formulation $m_t(n_1,n_2)$ reads
    \begin{eqnarray}
\langle n_2 \vert {\cal M}_t^{[n]} \rangle  =m_t(n-n_2,n_2)   \ \ \ \text{ for } \ \ n_2=0,1,..,n
\label{kYn}
\end{eqnarray}  
   or more explicitly starting with the first blocks $n=0,1,2..$
   \begin{eqnarray}
\vert {\cal M}_t^{[n=0]} \rangle  = \begin{pmatrix} 
 m_t(0,0)
  \end{pmatrix} \ \ ;    
\vert {\cal M}_t^{[n=1]} \rangle  = \begin{pmatrix} 
 m_t(1,0)
\\ m_t(0,1)
  \end{pmatrix} \ \ ;  \ \
\vert {\cal M}_t^{[n=2]} \rangle  = \begin{pmatrix} 
 m_t(2,0)
\\ m_t(1,1)
\\ m_t(0,2)
  \end{pmatrix}  
 \ \ ...  \ \ 
\vert {\cal M}_t^{[n]} \rangle  = \begin{pmatrix} 
m_t(n,0)
\\ m_t(n-1,1)
\\ m_t(n-2,2)
\\ ...
\\ m_t(1,n-1)
\\ m_t(0,n)
 \end{pmatrix}  
\label{ketmomentsdegren}
\end{eqnarray}

  Then the dynamics of Eq. \ref{DynMoments2D} for the moments $m_t(n_1,n_2) $
  can be translated in terms of these blocks as
 \begin{eqnarray}
\partial_t \vert {\cal M}_t^{[n]} \rangle 
  = {\bold M}^{[-2]}_{[n,n-2]}   \vert {\cal M}_t^{[n-2]} \rangle
 + {\bold M}^{[-1]}_{[n,n-1]}   \vert {\cal M}_t^{[n-1]} \rangle
 +  {\bold M}^{[0]}_{[n,n]}    \vert {\cal M}_t^{[n]} \rangle 
  +  {\bold M}^{[1]}_{[n,n+1]}   \vert {\cal M}_t^{[n+1]} \rangle
\label{DynMoments2DBlocs}
\end{eqnarray} 
 where the matrices ${\bold M}^{[k=-2,-1,0,1]}_{[n,q=n+k]}$ of size $(n+1) \times (q+1)$
 are the restrictions of the matrices ${\bold M}^{[k=-2,-1,0,1]} $ to the degree $n$ on the left and the degree $q=n+k$ 
 on the right with the matrix elements for $n_2=0,..,n$ and $q_2=0,..,q$
     \begin{eqnarray}
\langle n_2 \vert  {\bold M}_{[n,q]}   \vert q_2 \rangle = M^{[k=q-n]}(n-n_2,n_2 \vert q-q_2,q_2 ) 
\label{Matrix2DBlocs}
\end{eqnarray}
 So one can translate the 2D integer-lattice matrix elements $M^{[k=-2,-1,0,1]}(n_1,n_2 \vert q_1,q_2 ) $
 of Eqs \ref{m0simpli2D} and \ref{M1in2D}
 as follows:

$\bullet$ The block ${\bold M}^{[-2]}_{[n,n-2]}  $ has 
matrix elements on the diagonal $q_2=n_2$ and on the second-lower-diagonal $q_2=n_2-2$
given by
\begin{eqnarray}
\langle n_2 \vert {\bold M}^{[-2]}_{[n,n-2]}  \vert q_2 \rangle
&& = M^{[-2]}(n-n_2,n_2 \vert n-2-q_2,q_2 )
\nonumber \\
&&=  D^{[0]}_{1}    (n-n_2) (n-n_2-1)    \delta_{q_2,n_2}   
 +   D^{[0]}_{2}   n_2 (n_2-1) \delta_{q_2,n_2-2}  
\label{mm2Blocs}
\end{eqnarray}

$\bullet$ The block ${\bold M}^{[-1]}_{[n,n-1]} $ has matrix elements
on the diagonal $q_2=n_2$ and on the first-lower-diagonal $q_2=n_2-1$
 \begin{eqnarray}
\langle n_2 \vert {\bold M}^{[-1]}_{[n,n-1]}  \vert q_2 \rangle
&& = M^{[-1]}(n-n_2,n_2 \vert n-1-q_2,q_2 )
\nonumber \\
&&=  \left[  F^{[0]}_1   (n-n_2)   +    D^{[1]}_1    (n-n_2) (n-n_2-1) \right]
   \delta_{q_2,n_2}   
\nonumber \\
&&  + \left[  F^{[0]}_1  n_2   +    D^{[1]}_2   n_2 (n_2-1) \right] 
\delta_{q_2,n_2-1} 
\label{mm1Blocs}
\end{eqnarray}

$\bullet$ The block ${\bold M}^{[0]}_{[n,n]} $ 
has matrix elements on the diagonal $q_2=n_2$, on the first-upper-diagonal $q_2=n_2+1$
and on the first-lower-diagonal $q_2=n_2-1$
 \begin{eqnarray}
\langle n_2 \vert {\bold M}^{[0]}_{[n,n]}  \vert q_2 \rangle
&& = M^{[0]}(n-n_2,n_2 \vert n-q_2,q_2 )
\nonumber \\
&&= \left[  F^{[1]}_{11}  (n-n_2)
   +     D^{[2]}_1  (n-n_2) (n-n_2-1) 
+ F^{[1]}_{22}  n_2  
 +    D^{[2]}_1  n_2 (n_2-1) 
\right]    \delta_{q_2,n_2} 
\nonumber \\
&& + F^{[1]}_{12}  (n-n_2)  \delta_{q_2,n_2+1} 
+   F^{[1]}_{21}  n_2  \delta_{q_2,n_2-1} 
\label{mm0Blocs}
\end{eqnarray}

$\bullet$ The block ${\bold M}^{[1]}_{[n,n+1]} $ 
has matrix elements on the diagonal $q_2=n_2$, on the first-lower-diagonal $q_2=n_2-1$
and on the two first-upper-diagonals $q_2=n_2+1$ and $q_2=n_2+2$
 \begin{eqnarray}
\langle n_2 \vert {\bold M}^{[1]}_{[n,n+1]}  \vert q_2 \rangle
&& = M^{[1]}(n-n_2,n_2 \vert n+1-q_2,q_2 )
\nonumber \\
&&= \left( F^{[2]}_{1;11}   (n-n_2)  
+2 F^{[2]}_{2;12}  n_2 \right)     \delta_{q_2,n_2} 
+ \left( F^{[2]}_{2;22}    n_2  
+ 2 F^{[2]}_{1;12}  (n-n_2)  \right)      \delta_{q_2,n_2+1}
\nonumber \\
&&+ F^{[2]}_{1;22}  (n-n_2)   \delta_{q_2,n_2+2}   
+ F^{[2]}_{2;11}  n_2     \delta_{q_2,n_2-1} 
\label{m1Blocs}
\end{eqnarray}


\subsection{ Discussion   }

 This block-structure of the Carleman matrix
is thus very useful to identify the simplifications that can occur 
depending on the Taylor coefficients of the forces and of the diffusion coefficients 
that are non-vanishing in any given model one is interested in :

(i) one should first determine what blocks $ {\bold M}^{[k]}_{[n,n+k]} $  are vanishing or non-vanishing,
in particular to see if the Carleman matrix ${\bold M}$ happens to be block-diagonal ${\bold M}= {\bold M}^{[0]}$ , or happens to be block-lower-triangular ${\bold M}= {\bold M}^{[-2]}+ {\bold M}^{[-1]}+ {\bold M}^{[0]}$
or block-upper-triangular ${\bold M}= {\bold M}^{[0]}+ {\bold M}^{[1]}+..$.

(ii) one should then analyze the internal structure of each non-vanishing block ${\bold M}^{[k]}_{[n,q=n+k]}$
with its matrix elements $\langle n_2 \vert {\bold M}^{[k]}_{[n,q=n+k]} \vert q_2 \rangle $ written above,
in order to see if the internal structure of the block ${\bold M}^{[k]}_{[n,q=n+k]}$ is diagonal or bidiagonal or tridiagonal or quadridiagonal.

Let us describe the example of models with
block-upper-bidiagonal Carleman matrices ${\bold M}= {\bold M}^{[0]}+ {\bold M}^{[1]}$
in the next subsection,
while models with
block-diagonal Carleman matrices ${\bold M}= {\bold M}^{[0]}$ are analyzed in section \ref{sec_BlockDiag2D}
and models with block-lower-triangular Carleman matrices ${\bold M}= {\bold M}^{[-2]}+ {\bold M}^{[-1]}+ {\bold M}^{[0]}$ are discussed in section \ref{sec_BlockLower2D}.


\subsection{ Example with models whose Carleman matrix is block-upper-bidiagonal  ${\bold M} = {\bold M}^{[0]} + {\bold M}^{[1]} $ }

\label{subsec_2Dupper}

Let us apply the previous discussion to 
models whose Carleman matrix is block-upper-bidiagonal  ${\bold M} = {\bold M}^{[0]} + {\bold M}^{[1]} $:

$\bullet $ (i) The simplifications associated to the block-upper-bidiagonal  ${\bold M} = {\bold M}^{[0]} + {\bold M}^{[1]} $
structure are that the eigenvalues of the Carleman matrix ${\bold M} $
are the same as the eigenvalues of ${\bold M}^{[0]} $
that can be found as the eigenvalues of the individual diagonal blocks ${\cal M}^{[0]}_{[n,n]} $
of size $(n+1) \times (n+1)$, with their internal tridiagonal structure of Eq. \ref{mm0Blocs},
and that the spectral decomposition of ${\bold M} = {\bold M}^{[0]} + {\bold M}^{[1]}  $
can be explicitly written as described in subsection \ref{app_upperBiBLOCK}
of Appendix \ref{app_SpectralMatriceBiBLOCK}.

 The associated Ito system correspond to the case $d=2$ of Eq. \ref{ItoSDEsinglegjmultiplicativeupper}
  \begin{eqnarray}
dx_1(t)   && =  \left(   F^{[1]}_{11} x_1(t) + F^{[1]}_{12} x_2(t)
+     F^{[2]}_{1;  11 } x_1^2(t) + F^{[2]}_{1;  22 } x_1^2(t
+ 2  F^{[2]}_{1;  12 } x_{1}(t) x_{2}(t)\right) dt 
+ \sqrt{ 2  D^{[2]}_1 } \  x_1(t) dB_1^{multiplicative}(t) 
\nonumber \\
dx_2(t)   && = \left(   F^{[1]}_{21} x_1(t) + F^{[1]}_{22} x_2(t)
+     F^{[2]}_{2;  11 } x_1^2(t) + F^{[2]}_{2;  22 } x_1^2(t
+ 2  F^{[2]}_{2;  12 } x_{1}(t) x_{2}(t)\right) dt 
+ \sqrt{ 2  D^{[2]}_2 } \  x_2(t) dB_2^{multiplicative}(t) 
\nonumber \\
\label{ItoSDEsinglegjmultiplicativeupper2D}
\end{eqnarray}
For models that do not involve all the coefficients present in Eq. \ref{ItoSDEsinglegjmultiplicativeupper2D},
there can be further simplifications in the internal structures of $ {\bold M}^{[0]} $ and $ {\bold M}^{[1]} $,
as discussed in the examples below.

$\bullet $ (ii) For the Lotka-Volterra models for positive variables of Eq. \ref{ItoSDEsinglegjmultiplicativeupperLV}
in dimension $d=2$
  \begin{eqnarray}
dx_1(t)   && = x_1(t)  \left[  \left( F^{[1]}_{11} 
+ F^{[2]}_{1; 11 }  x_1(t)
+2   F^{[2]}_{1; 12 }  x_{2}(t) \right) dt 
+ \sqrt{ 2  D^{[2]}_1 }  dB_1^{multiplicative}(t) \right]
\nonumber \\
dx_2(t)   && = x_2(t)  \left[  \left( F^{[1]}_{22} 
+ F^{[2]}_{2; 22 }  x_2(t)
+2   F^{[2]}_{2; 12 }  x_1(t) \right) dt 
+ \sqrt{ 2  D^{[2]}_1 }  dB_1^{multiplicative}(t) \right]
\label{ItoSDEsinglegjmultiplicativeupperLV2D}
\end{eqnarray}
there are several simplifications :

(ii-a) the vanishing of the off-diagonal elements $F^{[1]}_{12} =0 =F^{[1]}_{21}  $ of $F^{[1]}$
yields that ${\bold M}^{[0]}$ is diagonal both the the integer-lattice formulation 
$M^{[0]}(n_1,n_2 \vert q_1,q_2) $ of Eqs \ref{m0simplidiagnqLV} and \ref{m0simpli2D}
or in the reshaped formulation $\langle n_2 \vert {\bold M}^{[0]}_{[n,n]}  \vert q_2 \rangle
 = M^{[0]}(n-n_2,n_2 \vert n-q_2,q_2 ) $ of Eq. \ref{mm0Blocs}
so that the diagonal elements directly give the eigenvalues of Eq. \ref{m0simplidiagnqLVeigen} for the Carleman matrix ${\bold M}$
\begin{eqnarray}
E_{n_1,n_2} 
&& =  
  F^{[1]}_{11}  n_1   +   D^{[2]}_1   n_1 (n_1-1) 
+ F^{[1]}_{22}  n_2   +     D^{[2]}_2   n_2 (n_2-1) 
\nonumber \\
&& =  F^{[1]}_{11}  (n-n_2)
   +     D^{[2]}_1  (n-n_2) (n-n_2-1) 
+ F^{[1]}_{22}  n_2  
 +    D^{[2]}_1  n_2 (n_2-1)  = E^{[n]}_{n_2}
\label{EigenLV2D}
\end{eqnarray}

(ii-b) the vanishing of the matrix elelements $F^{[2]}_{1;22}=0=F_{2;11} $ of $F^{[2]}$ 
yields that the matrix elements of $M^{[1]} $ simplify into Eq. \ref{M1LV}
for $d=2$ in the integer-lattice formulation
\begin{eqnarray}
M^{[1]}(n_1,n_2 \vert q_1,q_2)
&& =   \delta_{q_1,n_1+1}   \delta_{q_2,n_2} 
\left( n_1 F^{[2]}_{1;11}
   + 2  n_2F^{[2]}_{2;12} 
\right)
+   \delta_{q_1,n_1}  \delta_{q_2,n_2+1} 
\left(  n_2 F^{[2]}_{2;22}
+ 2 n_1  F^{[2]}_{1;12} 
\right)
\label{M1LVd2}
\end{eqnarray}
or in the following upper-bidiagonal form in the reshaped formulation of Eq. \ref{m1Blocs}
 \begin{eqnarray}
\langle n_2 \vert {\bold M}^{[1]}_{[n,n+1]}  \vert q_2 \rangle
&& = M^{[1]}(n-n_2,n_2 \vert n+1-q_2,q_2 )
\nonumber \\
&&= \left( (n-n_2) F^{[2]}_{1;11}     
+2 n_2 F^{[2]}_{2;12}   \right)     \delta_{q_2,n_2} 
+ \left( n_2 F^{[2]}_{2;22}    
+ 2  (n-n_2) F^{[2]}_{1;12}   \right)      \delta_{q_2,n_2+1}
\label{m1BlocsLV}
\end{eqnarray}


 \section{ Models in dimension $d=2$ with block-diagonal Carleman matrices ${\bold M}= {\bold M}^{[0]}$   }

\label{sec_BlockDiag2D}

In this section, we analyze the properties of diffusion processes of section \ref{sec_2D}  
whose Carleman matrices are block-diagonal ${\bold M}= {\bold M}^{[0]}$,
that correspond to the case $d=2$ of Eq \ref{ItoSDEsinglegjmultiplicativediag}.

 \subsection{ Notations for models in dimension $d=2$ when the Carleman matrix is block-diagonal   ${\bold M}= {\bold M}^{[0]}$   }

 The discussion of subsection \ref{subsec_blockDiag} concerning the dimension $d$
  can be applied to the present models in dimension $d=2$ as follows :
  when the differential generator of Eq. \ref{Generator2D} reduces to the contribution ${\cal L}^{[0]}$
  of Eq. \ref{GeneratorClassDegree4in2D}
\begin{eqnarray}
{\cal L} = {\cal L}^{[0]}
&&  =   \left[   F^{[1]}_{11} x_1+ F^{[1]}_{12} x_2 \right]  \frac{ \partial }{\partial x_1} 
 +   \left[    F^{[1]}_{21} x_1+ F^{[1]}_{22} x_2
\right]   \frac{ \partial }{\partial x_2}  
+     D^{[2]}_1 x_1^2 
   \frac{ \partial^2 }{\partial x_1^2}
+  D^{[2]}_2 x_2^2  
   \frac{ \partial^2 }{\partial x_2^2}
\label{Generator2DBlockDiag}
\end{eqnarray}
then the Carleman matrix ${\bold M}  $ reduces to its block-diagonal component $  {\bold M}^{[0]} $
 with its matrix elements $M^{[0]}(n_1,n_2 \vert q_1,q_2)$ of Eq. \ref{m0simpli2D}
  or equivalently its diagonal blocks ${\bold M}^{[0]}_{[n,n]}$ of Eq. \ref{mm0Blocs}
  with their internal tridiagonal structure
   \begin{eqnarray}
\langle n_2 \vert {\bold M}^{[0]}_{[n,n]}  \vert q_2 \rangle
&& = M^{[0]}(n-n_2,n_2 \vert n-q_2,q_2 )
\nonumber \\
&&=\langle n_2 \vert {\bold M}^{[0]}_{[n,n]}  \vert n_2 \rangle   \delta_{q_2,n_2} 
 + F^{[1]}_{12}  (n-n_2)  \delta_{q_2,n_2+1} 
+   F^{[1]}_{21}  n_2  \delta_{q_2,n_2-1} 
\label{mm0BlocsDiag}
\end{eqnarray} 
with the diagonal elements for $q_2=n_2$
  \begin{eqnarray}
\langle n_2 \vert {\bold M}^{[0]}_{[n,n]}  \vert n_2 \rangle 
&& = F^{[1]}_{11}  (n-n_2)
   +     D^{[2]}_1  (n-n_2) (n-n_2-1) 
+ F^{[1]}_{22}  n_2  
 +    D^{[2]}_1  n_2 (n_2-1) 
\label{mm0BlocsDiagel}
\end{eqnarray} 

This tridiagonal matrix ${\bold M}^{[0]}_{[n,n]} $ of size $(n+1) \times (n+1)$ governs
  the dynamics of Eq. \ref{DynMoments2DBlocs} for the ket $\vert {\cal M}_t^{[n]} \rangle $ of Eq. \ref{ketmomentsdegren} gathering the $(n+1)$ moments $m_t( n-n_2,n_2)  $ of the given global degree $n$ 
  labeled by $n_2=0,1,..,n$
 \begin{eqnarray}
\partial_t m_t( n-n_2,n_2)
&& = \partial_t \langle n_2 \vert {\cal M}_t^{[n]} \rangle 
  = \sum_{q_2=n_2-1}^{n_2+1} \langle n_2 \vert {\bold M}^{[0]}_{[n,n]} \vert q_2 \rangle  \langle q_2 \vert {\cal M}_t^{[n]} \rangle 
  \nonumber \\
  && = \langle n_2 \vert {\bold M}^{[0]}_{[n,n]}  \vert n_2 \rangle m_t( n-n_2,n_2)
 + F^{[1]}_{21}  n_2 m_t( n-n_2+1,n_2-1) 
  + F^{[1]}_{12}  (n-n_2) m_t( n-n_2-1,n_2+1)
   \nonumber \\
\label{DynMoments2DBlocscasl0}
\end{eqnarray}

In the models governed by the generator of Eq. \ref{Generator2DBlockDiag},  
the diffusion coefficients involve only the quadratic terms 
\begin{eqnarray}
 D_{11}(x_1,x_2) 
&& =  D^{[2]}_1 x_1^2
\nonumber \\
 D_{22}(x_1,x_2) 
&& =   D^{[2]}_2 x_1^2
\label{DdiagoEtSingle2DonlyD2}
\end{eqnarray}
while the Ito forces involve only
 the linear terms parametrized by the $2 \times 2$ matrix $ {\bold F}^{[1]}$
  \begin{eqnarray}
F_1(x_1,x_2) && = F^{[1]}_{11} x_1 +  F^{[1]}_{12} x_2
\nonumber \\
F_2(x_1,x_2) &&  =  F^{[1]}_{21} x_1 +  F^{[1]}_{22} x_2
\label{TaylorF2DonlyF1}
\end{eqnarray}
The Stratonovich forces of Eq. \ref{StratoItoCorrespondance2D} also involve only
 the linear terms
  \begin{eqnarray}
f_1(x_1,x_2) &&  =   \bigg( F^{[1]}_{11}-  D^{[2]}_1  \bigg) x_1+  F^{[1]}_{12} x_2
\equiv   f^{[1]}_{11} x_1+ f^{[1]}_{12} x_2
\nonumber \\ 
f_2(x_1,x_2) && =   F^{[1]}_{21} x_1+ \bigg( F^{[1]}_{22} -  D^{[2]}_2 \bigg) x_2
 \equiv  f^{[1]}_{21} x_1+ f^{[1]}_{22} x_2
\label{StratoItoCorrespondance2DF1}
\end{eqnarray}
so that the generator ${\cal L} ={\cal L}^{[0]} $ of Eq. \ref{Generator2DBlockDiag}
becomes in terms of the Stratonovich forces
\begin{eqnarray}
{\cal L} = {\cal L}^{[0]}
&&  =   f^{[1]}_{11} \left( x_1  \frac{ \partial }{\partial x_1} \right)  
+ f^{[1]}_{12} \left( x_2  \frac{ \partial }{\partial x_1} \right)
 +    f^{[1]}_{21} \left( x_1  \frac{ \partial }{\partial x_2} \right)
 + f^{[1]}_{22}   \left( x_2  \frac{ \partial }{\partial x_2} \right)
\nonumber \\
&& +     D^{[2]}_1  \left( x_1  \frac{ \partial }{\partial x_1} \right)  \left( x_1  \frac{ \partial }{\partial x_1} \right)
 +  D^{[2]}_2   \left( x_2  \frac{ \partial }{\partial x_2} \right)  \left( x_2  \frac{ \partial }{\partial x_2} \right)   
\label{Generator2DBlockDiagStrato}
\end{eqnarray}
while the Stratonovich system of Eq. \ref{StratoSDE2D} involving multiplicative noises
corresponds to the case $d=2$ of Eq. \ref{ItoSDEsinglegjmultiplicativediag}
  \begin{eqnarray}
dx_1(t)   && = \left(f^{[1]}_{11} x_1(t)+ f^{[1]}_{12} x_2(t) \right) dt 
+ \sqrt{ 2   D^{[2]}_1  }  \ x_1(t) dB_1(t) 
\nonumber \\
dx_2(t)   && = \left( f^{[1]}_{21} x_1(t)+ f^{[1]}_{22} x_2(t) \right) dt 
+ \sqrt{ 2   D^{[2]}_2  } \ x_2(t) dB_2(t) 
\label{StratoSDE2DF1D2}
\end{eqnarray}

 As already discussed around Eqs \ref{MomentsxnGeney} 
 concerning the dimension $d=1$, the Carleman moments have a direct interpretation in terms of large deviations.
 For the present models in dimension $d=2$, 
 it is useful to introduce 
 the finite-time Lyapunov exponents $\lambda_j(T)$ 
 in order to characterize the exponential growths of the absolute-values of $x_j(t)$ 
over the time-window $t \in [0,T]$ 
 \begin{eqnarray}
\lambda_j(T)  \equiv  \frac{1}{T} \ln \left\vert \frac{x_j(T)} {x_j(0) } \right\vert    
= \frac{1}{T} \int_0^T \frac{ dx_j(t)}{x_j(t) }
\label{lambdajT}
\end{eqnarray}

Then the Carleman moments $m_T(n_1,n_2) $ in the integer lattice representation $(n_1,n_2) \in {\mathbb N}^2$
\begin{eqnarray}
m_T(n_1,n_2) = {\mathbb E} \left( x_1^{n_1}(T) x_2^{n_2}(T)\right) 
&& = {\mathbb E} \left( x_1^{n_1}(0) x_2^{n_2}(0) e^{ T (n_1 \lambda_1(T)+n_2 \lambda_2(T) )} \right)
\label{LyapunovTgenerating12}
\end{eqnarray}
can be reinterpreted as the joint generating function of the two finite-time Lyapunov exponents
$(\lambda_1(T) , \lambda_2(T))$ when 
$(n_1,n_2) $ are considered as real parameters.
The leading exponential behavior for large $T$
\begin{eqnarray}
m_T(n_1,n_2) \opsimeq_{T \to + \infty} e^{ T {\cal E}(n_1,n_2)}
\label{LyapunovTgenerating12largedev}
\end{eqnarray}
is governed by their joint Scaled Cumulant Generating Function ${\cal E}(n_1,n_2) $   
in the language
of large deviations (see the reviews \cite{oono,ellis,review_touchette} and references therein).

Since the physical properties of the 2D process of Eq. \ref{StratoSDE2DF1D2}
depend on the vanishing or non-vanishing of the off-diagonal coefficients $f^{[1]}_{12} $ and $f^{[1]}_{21} $ 
that mediate the interactions between the two coordinates $x_1(t)$ and $x_2(t)$,
let us discuss the three possible cases in the three following subsections.


 \subsection{ When $f^{[1]}_{12}=0$ and $ f^{[1]}_{21} =0$ : $x_1(t)$ and $x_2(t)$ are two independent positive Geometric Brownian motions  }

When the two off-diagonal matrix coefficients $F^{[1]}_{12}=f^{[1]}_{12} $ and $F^{[1]}_{21}=f^{[1]}_{21}  $ 
vanish, then $x_1(t)$ and $x_2(t)$ are two independent positive Geometric Brownian motions
as in Eq. \ref{YlogGBMinverse} discussed in the previous section concerning the dimension $d=1$
\begin{eqnarray}
\text{ If $f^{[1]}_{12}=0 = f^{[1]}_{21} $ } : \ \ \ x_1(t) &&= x_1(0) e^{f^{[1]}_{11} t+  \sqrt{ 2 D^{[2]}_1 } \ B_1(t)}
\nonumber \\
x_2(t) &&= x_2(0) e^{f^{[1]}_{22} t+  \sqrt{ 2 D^{[2]}_2 } \ B_2(t)}
\label{YlogGBM2Dindep}
\end{eqnarray}
and their joint moments $m_t(n_1,n_2) $ reduce to the products of their moments 
given by Eq. \ref{MTdiagsol} concerning the one-dimensional case
 \begin{eqnarray}
m_t(n_1,n_2) =e^{t E(n_1,n_2) } m_0(n_1,n_2) 
 \label{MTdiagsol2D}
\end{eqnarray}
The eigenvalues $E(n_1=n-n_2,n_2) $ of the Carleman block ${\bold M}^{[0]}_{[n,n]} $
of Eq. \ref{mm0BlocsDiagel},
 that is diagonal when the two off-diagonal matrix coefficients $F^{[1]}_{12}=f^{[1]}_{12} $ and $F^{[1]}_{21}=f^{[1]}_{21}  $ vanish, are given by the diagonal matrix elements of Eq. \ref{mm0BlocsDiag}
 \begin{eqnarray}
E(n-n_2,n_2)  = 
\langle n_2 \vert {\bold M}^{[0]}_{[n,n]}  \vert n_2 \rangle 
&& = F^{[1]}_{11}  (n-n_2)
   +     D^{[2]}_1  (n-n_2) (n-n_2-1) 
+ F^{[1]}_{22}  n_2  
 +    D^{[2]}_1  n_2 (n_2-1) 
 \label{Ediagsol2D}
\end{eqnarray}
that reduce to the sum of the two one-dimensional-eigenvalues of Eq. \ref{MTdiag} with $n-n_2=n_1$.

In order to stress the difference with the next subsection concerning the interacting case,
let us stress that here the finite-time Lyapunov exponents $\lambda_j(T) $ of Eq. \ref{lambdajT}
reduce to
 \begin{eqnarray}
\lambda_j(T)  \equiv  \frac{1}{T} \ln \left\vert \frac{x_j(T)} {x_j(0) } \right\vert   
&& = f^{[1]}_{jj} +  \sqrt{ 2 D^{[2]}_j } \frac{ B_j(T) }{T}
\label{LyapunovGBMj}
\end{eqnarray}
while the Brownian scalings $B_j(T) \propto \sqrt{T} $ lead to the asymptotic values for $T \to + \infty$
 \begin{eqnarray}
\lambda_j(T=\infty)  = f^{[1]}_{jj} 
\label{LyapunovGBMjinfty}
\end{eqnarray}

 
 \subsection{ Case $f^{[1]}_{12} \ne 0$ and $ f^{[1]}_{21} \ne 0$ : 
 closed dynamics for the ratio $R(t) = \frac{ x_2(t) }{x_1(t) }$ converging towards an explicit steady state    }
 
 \label{subsec_Ricatti}
 
 \subsubsection{ Replacing the second coordinate $x_2(t)$ by the ratio $R(t) = \frac{ x_2(t) }{x_1(t) }$ and keeping the first coordinate $x_1(t)$    }

 Instead of the cartesian coordinates $(x_1,x_2) \in ]-\infty,+\infty[^2$, 
 one could consider the polar coordinates  
 \begin{eqnarray}
x_1(t) && = w(t)  \cos \theta(t)
\nonumber \\
 x_2(t) && = w(t)  \sin \theta(t)
\label{Polar}
\end{eqnarray}
where the radial coordinate $w(t) = \sqrt{ x_1^2(t)+x_2^2(t) } $ lives on the semi-infinite line $]0,+\infty[$,
while the polar angle $ \theta(t) $ lives on the finite interval $  [0,2 \pi[$
with periodic boundary conditions $\theta(0)=\theta(2 \pi)$.
However as discussed in detail in \cite{c_Lyapunov} concerning similar models,
 it is technically more convenient to consider the tangent of the polar angle $\theta(t)$
 that corresponds to the ratio of the two cartesian coordinates
 \begin{eqnarray}
R(t) = \frac{ x_2(t) }{x_1(t) } = \tan \theta(t) \in ]-\infty,+\infty[ 
\label{Ricatti}
\end{eqnarray} 
and to keep the variable $x_1(t)$ as the other independent variable instead of the radial coordinate. 
Indeed, in the reshaping of the Carleman matrix where $n_1=n-n_2$, the moments 
$m_t(n_1=n-n_2,n_2) $
associated to a given global degree $n$ while $n_2=0,..,n$
\begin{eqnarray}
m_t(n-n_2,n_2) = {\mathbb E} \left( x_1^{n-n_2}(t) x_2^{n_2}(t)\right) 
&& = {\mathbb E} \left( x_1^{n}(t) \left( \frac{x_2(t)}{x_1(t)}\right)^{n_2} \right)
= {\mathbb E} \left( x_1^{n}(t) R^{n_2}(t) \right)
\label{LyapunovTgenerating12reshape}
\end{eqnarray}
correspond to the moments of the two variables $x_1(t)$ and $R(t)$,
while the difference between 
the two finite-time Lyapunov exponents $\lambda_j(T) $ of Eq. \ref{lambdajT}
 \begin{eqnarray}
\lambda_2(T) - \lambda_1(T) =  \frac{1}{T} \left( \ln \left\vert \frac{x_2(T)} {x_2(0) } \right\vert  
- \ln \left\vert \frac{x_1(T)} {x_1(0) } \right\vert    \right)
= \frac{1}{T}  \ln \left\vert \frac{R(T)} {R(0) } \right\vert 
\label{lambdaTdifference21}
\end{eqnarray}
represents the exponential growth of the absolute-values of $R(t)$ 
over the time-window $t \in [0,T]$.

When $x_2(t)$ is replaced by the ratio $R(t) = \frac{ x_2(t) }{x_1(t) } $,
 the Stratonovich system of Eq. \ref{StratoSDE2DF1D2} is replaced by 
the dynamics of $x_1(t)$ that depends now on $R(t)$
  \begin{eqnarray}
dx_1(t)   && = \left(f^{[1]}_{11} x_1(t)+ f^{[1]}_{12} R(t) x_1(t) \right) dt 
+ \sqrt{ 2   D^{[2]}_1  }  \ x_1(t) dB_1(t) 
\label{StratoSDE2DF1D2dynx1R}
\end{eqnarray} 
and by the dynamics of $R(t)$ that does not depend on $x_1(t)$
  \begin{eqnarray}
d R(t) && = \frac{ \left( f^{[1]}_{21} x_1(t)+ f^{[1]}_{22} x_2(t) \right) dt 
+ \sqrt{ 2   D^{[2]}_2  } \ x_2(t) dB_2(t) }{x_1(t) } 
- x_2(t) \frac{ \left(f^{[1]}_{11} x_1(t)+ f^{[1]}_{12} x_2(t) \right) dt 
+ \sqrt{ 2   D^{[2]}_1  }  \ x_1(t) dB_1(t)   }{ x_1^2(t)}
\nonumber \\
&& = \left[  f^{[1]}_{21} 
+ ( f^{[1]}_{22} - f^{[1]}_{11})  R(t)
- f^{[1]}_{12} R^2(t) \right] dt 
+  R(t) \left[ \sqrt{ 2 D^{[2]}_2 } dB_2(t) 
 -    \sqrt{ 2 D^{[2]}_1}  dB_1(t)  \right]
\label{RicattiSDE}
\end{eqnarray} 
So 
we will first analyze the dynamics of $R(t)$ alone before returning to the properties of $x_1(t)$.


\subsubsection{ Properties of the dynamics of the ratio $R(t) = \frac{ x_2(t) }{x_1(t) }$ alone    }

The single-variable Stratonovich SDE of Eq. \ref{RicattiSDE} can be rewritten as
  \begin{eqnarray}
d R(t) &&  \equiv  f(R(t)) dt +  R(t) \sqrt{ 2 D^{[2]} } dB_R(t) 
\label{RicattiDeri}
\end{eqnarray} 
where  the Stratonovich force 
  \begin{eqnarray}
f(R) \equiv f^{[0]} 
+  f^{[1]} R
+ f^{[2]} R^2
\label{fRstrato}
\end{eqnarray} 
 is a polynomial of degree two in $R$ with the coefficients
   \begin{eqnarray}
 f^{[0]} && = f^{[1]}_{21} 
  \nonumber \\
  f^{[1]} && = f^{[1]}_{22} - f^{[1]}_{11}
   \nonumber \\
 f^{[2]} && =- f^{[1]}_{12}
\label{fRstratocoef}
\end{eqnarray} 
while the effective multiplicative Brownian noise defined via
  \begin{eqnarray}
  \sqrt{ 2 D^{[2]} } dB_R(t) \equiv \sqrt{ 2 D^{[2]}_2 } dB_2(t)  -    \sqrt{ 2 D^{[2]}_1}  dB_1(t)
 \label{DReff}
\end{eqnarray} 
  involves the diffusion coefficient 
  \begin{eqnarray}
 D^{[2]} \equiv D^{[2]}_1+D^{[2]}_2
\label{DReffnorm}
\end{eqnarray}

The generator of Eq. \ref{GeneratorStrato} associated to the single-variable Stratonovich SDE of Eq. \ref{RicattiDeri}
\begin{eqnarray}
{\cal L}
&& = f(R) \frac{ \partial }{\partial R}
+  D^{[2]}
\left( R \frac{\partial}{\partial R} \right)
\left( R  \frac{ \partial }{\partial R} \right)
\label{GeneratorStratoR}
\end{eqnarray}
leads to the adjoint operator
\begin{eqnarray}
{\cal L}^{\dagger}
&& = -  \frac{ \partial }{\partial R} f(R)
+  D^{[2]}
\left(  \frac{\partial}{\partial R}  R  \frac{ \partial }{\partial R} R \right)
\label{GeneratorStratoRdagger}
\end{eqnarray}
that governs the Fokker-Planck dynamics of Eq. \ref{fokkerPlanck}
for the probability density $\rho_t( R ) $ of $R$ at time $t$
\begin{eqnarray}
 \partial_t \rho_t(R) 
&& =   {\cal L}^{\dagger} \rho_t(R) 
\nonumber \\
&& = -  \frac{ \partial }{\partial R} \left[  f(R) \rho_t(R)
-  D^{[2]}  R  \frac{ \partial }{\partial R} \left( R \rho_t(R) \right)
\right] \equiv -  \frac{ \partial }{\partial R} J_t(R)
  \label{fokkerPlanckR}
\end{eqnarray}
that can be interpreted as a continuity equation involving the current
\begin{eqnarray}
J_t(R) && \equiv   f(R) \rho_t(R)
-  D^{[2]}  R  \frac{ \partial }{\partial R} \left( R \rho_t(R) \right)
\nonumber \\
&& = \left[ f(R) - D^{[2]}  R \right] \rho_t(R)
-  D^{[2]}  R^2  \frac{ \partial  \rho_t(R) }{\partial R} 
  \label{fokkerPlanckcurrentR}
\end{eqnarray}


\subsubsection{ Computation of the explicit steady state $\rho_{st}(R) $ for the ratio $R(t) = \frac{ x_2(t) }{x_1(t) }$ alone    }

The steady solution $\rho_{st}(R) $ of the Fokker-Planck dynamics of Eq. \ref{fokkerPlanckR}
is associated to the steady current $J_{st}(R) = J_{st}$ that cannot depend on $R$
but that can be non-vanishing as a consequence of the periodic boundary conditions at $ R=\pm \infty$
inherited from the polar-angle interpretation of Eq. \ref{Ricatti}
\begin{eqnarray}
J_{st} && = \left[ f(R) - D^{[2]}  R \right] \rho_{st}(R)
-  D^{[2]}  R^2  \frac{ d  \rho_{st}(R) }{\ dR} 
\nonumber \\
&& = \left[ f^{[0]} 
+  (f^{[1]} - D^{[2]}) R
+ f^{[2]} R^2    \right] \rho_{st}(R)
-  D^{[2]}  R^2  \frac{ d  \rho_{st}(R) }{\ dR} 
\nonumber \\
&& \equiv  D^{[2]}  R^2 \left( - U'(R) \rho_{st}(R) - \frac{ d  \rho_{st}(R) }{\ dR} \right)
  \label{SteadycurrentR}
\end{eqnarray}
where we have used the polynomial form $f(R) = f^{[0]} 
+  f^{[1]} R
+ f^{[2]} R^2 $ of Eq. \ref{fRstrato}
for the Stratonovich force 
 and where on the last line we have introduced the effective potential $U(R)$ via its derivative 
 \begin{eqnarray}
U'(R)  && \equiv - \frac{ \left[ f^{[0]} 
+  (f^{[1]} - D^{[2]}) R
+ f^{[2]} R^2    \right] }{   D^{[2]}  R^2  }
=- \frac{  f^{[0]}  }{   D^{[2]}  R^2  }
+ \frac{ D^{[2]} - f^{[1]}  
 }{   D^{[2]}  R  }
- \frac{  f^{[2]}  }{   D^{[2]}    }
  \label{URderi}
\end{eqnarray}
The final result for the potential $U(R)$ itself reads using the coefficients of Eq. \ref{fRstratocoef}
 \begin{eqnarray}
U(R)  &&  =\frac{  f^{[0]}  }{   D^{[2]}  R  }
+ \frac{ D^{[2]} - f^{[1]}   }{   D^{[2]}    } \ln \vert R \vert
- \frac{  f^{[2]}  }{   D^{[2]}    } R
\nonumber \\
&& = \frac{  f^{[1]}_{21}   }{   D^{[2]}  R  }
+ \frac{ D^{[2]} + f^{[1]}_{11} - f^{[1]}_{22}   }{   D^{[2]}    } \ln \vert R \vert
+ \frac{ f^{[1]}_{12} }{   D^{[2]}    } R
  \label{UR}
\end{eqnarray}

The first term in $1/R$ of Eq. \ref{UR}
gives the leading diverging behaviors on both sides of the origin $R \to 0^{\pm}$
whose signs are determined by the sign of the off-diagonal coefficient $f^{[1]}_{21} $
 \begin{eqnarray}
U(R)  &&  \opsimeq_{R \to 0^+ }  \frac{  f^{[1]}_{21}   }{   D^{[2]}  R  } 
\opsimeq_{R \to 0^+ } \left[  \rm{sgn} ( f^{[1]}_{21}) \right] \infty
\nonumber \\
U(R)  &&  \opsimeq_{R \to 0^- } - \frac{  f^{[1]}_{21}   }{   D^{[2]}  (-R)  } 
\opsimeq_{R \to 0^- } \left[ - \rm{sgn} ( f^{[1]}_{21}) \right] \infty
  \label{URorigin}
\end{eqnarray}
The last term in $R$ of Eq. \ref{UR}
determines the leading asymptotic behaviors at the two infinities $R \to \pm \infty$
whose signs are determined by the sign of the off-diagonal coefficient $f^{[1]}_{12} $
 \begin{eqnarray}
U(R)  &&  \opsimeq_{R \to +\infty } \frac{ f^{[1]}_{12} }{   D^{[2]}    } R 
 \opsimeq_{R \to +\infty } \left[  \rm{sgn} ( f^{[1]}_{12}) \right] \infty
\nonumber \\
U(R)  &&  \opsimeq_{R \to - \infty } - \frac{ f^{[1]}_{12} }{   D^{[2]}    } (-R) 
\opsimeq_{R \to -\infty } \left[  - \rm{sgn} ( f^{[1]}_{12}) \right] \infty
  \label{URinfty}
\end{eqnarray}

 As discussed in detail in \cite{c_Lyapunov} concerning similar models,
 it is very important to distinguish whether the uniform steady current $J_{st} $ of Eq. \ref{SteadycurrentR}
 vanishes or not :
 
$\bullet$ (i) When the steady current vanishes $J_{st} =0$, then the dynamics of $R(t)$ is an equilibrium dynamics satisfying
 detailed-balance,
 and the corresponding steady state $\rho_{st}^{eq}(R)$ satisfying Eq. \ref{SteadycurrentR}
 \begin{eqnarray}
\frac{ d  \ln \rho_{st}^{eq}(R) }{\ dR} = - U'(R) 
  \label{SteadycurrentRzero}
\end{eqnarray}
 is the Boltzmann distribution associated to the potential $U(R)$ of Eq. \ref{UR}
 \begin{eqnarray}
 \rho_{st}^{eq}(R)  && = {\cal N} e^{- U(R) } 
 =  {\cal N} \ e^{ \displaystyle - \frac{  f^{[1]}_{21}   }{   D^{[2]}  R  }
- \frac{ D^{[2]} + f^{[1]}_{11} - f^{[1]}_{22}   }{   D^{[2]}    } \ln \vert R \vert
- \frac{ f^{[1]}_{12} }{   D^{[2]}    } R }
\nonumber \\
&& =   {\cal N}  
\ \vert R \vert^{-1 - \frac{  f^{[1]}_{11} - f^{[1]}_{22}   }{   D^{[2]}    }}  
\ e^{ \displaystyle - \frac{  f^{[1]}_{21}   }{   D^{[2]}  R  } - \frac{ f^{[1]}_{12} }{   D^{[2]}    } R }
  \label{SteadyEq}
\end{eqnarray}
that should be normalized on the appropriate interval 
 $R \in ]R_{min},R_{max} [$. Here the two possible cases are : 
 
 (i-a) When both off-diagonal coefficients are positive  $f^{[1]}_{12} >0$ and $f^{[1]}_{21} > 0$,
 then the equilibrium steady state of Eq. \ref{SteadyEq} is normalized on the positive half-line $R \in ]0,+\infty[$
  \begin{eqnarray}
\text{ Case $f^{[1]}_{12} >0$ and $f^{[1]}_{21} > 0$} : \ \  \rho_{st}^{eq}(R)  && = \frac{ e^{- U(R) } }
 { \int_0^{+\infty} dr e^{- U(r) } } \ \ \text{ for $R \in ]0,+\infty[$} 
  \label{SteadyEqpos}
\end{eqnarray}
 
 (i-b) When both off-diagonal coefficients are negative  $f^{[1]}_{12} <0$ and $f^{[1]}_{21} < 0$,
 then the equilibrium steady state of Eq. \ref{SteadyEq} is normalized on 
 the negative half-line $R \in ]-\infty,0[$
  \begin{eqnarray}
\text{ Case $f^{[1]}_{12} <0$ and $f^{[1]}_{21} < 0$} : \ \  \rho_{st}^{eq}(R)  && = \frac{ e^{- U(R) } }
 { \int_{-\infty}^0 dr e^{- U(r) } } \ \ \text{ for $R \in ]-\infty,0[$} 
  \label{SteadyEqneg}
\end{eqnarray}

$\bullet $ (ii) When the steady current does not vanish $J_{st} \ne 0$, then the dynamics of $R(t)$ is an out-of-equilibrium dynamics breaking
 detailed-balance, and the corresponding steady state $\rho_{st}^{neq}(R)$ satisfying Eq. \ref{SteadycurrentR}
\begin{eqnarray}
  \frac{ d  \rho_{st}^{neq}(R) }{\ dR} + U'(R) \rho_{st}^{neq}(R) = - \frac{J_{st} }{  D^{[2]}  R^2 } 
  \label{SteadycurrentRnoneq}
\end{eqnarray}
can be computed to obtain the two following cases (see \cite{c_Lyapunov} for very detailed discussions concerning similar models) :

(ii-a) When $f^{[1]}_{12} >0$ and $f^{[1]}_{21} < 0$, then the potential $U(R)$ grows from $U(R \to - \infty) =-\infty$
towards $U(R \to 0^-)=+\infty$ and from $U(R \to 0^+)=-\infty$ towards $U(R \to + \infty) =+\infty$,
so the steady current is negative $J_{st}<0$ with the corresponding steady state
\begin{eqnarray}
\text{ Case $f^{[1]}_{12} >0$ and $f^{[1]}_{21} < 0$} : \ \  
\rho^{neq}_{st}(R) =
 \begin{cases}
 \displaystyle 
  (- J_{st}) e^{ -U(R)}   \int_{-\infty}^R \frac{d r}{D^{[2]}  r^2} e^{ U(r) }  
   \text{ for } \ \ R \in ]-\infty,0[
\\
 \displaystyle 
 (- J_{st}) e^{ -U(R)}   \int_{0}^R \frac{d r}{D^{[2]}  r^2} e^{ U(r) }  
  \  \text{  for } \ \ R \in ]0,+\infty[ 
\end{cases}
\label{noneqsolRegplus}
\end{eqnarray}

(ii-b) When $f^{[1]}_{12} <0$ and $f^{[1]}_{21} > 0$, then the potential $U(R)$ decays from $U(R \to - \infty) =+\infty$
towards $U(R \to 0^-)=-\infty$ and from $U(R \to 0^+)=+\infty$ towards $U(R \to + \infty) =-\infty$,
so the steady current is positive $J_{st}>0$ with the corresponding steady state
\begin{eqnarray}
\text{ Case $f^{[1]}_{12} <0$ and $f^{[1]}_{21} > 0$} : \ \  \rho^{neq}_{st}(R) =
   \begin{cases}
    \displaystyle 
   J_{st} e^{ -U(R)}   \int_{R}^{0} \frac{d r}{D^{[2]}  r^2} e^{ U(r) }  
   \text{ for } \ \ R \in ]-\infty,0[
\\
 \displaystyle 
  J_{st} e^{ -U(R)}   \int_{R}^{+\infty} \frac{d r}{D^{[2]}  r^2} e^{ U(r) } 
   \text{  for } \ \ R \in ]0,+\infty[ 
\end{cases}
\label{noneqsolRegmoins}
\end{eqnarray}

In both cases of Eqs \ref{noneqsolRegplus} and \ref{noneqsolRegmoins}, it is the normalization 
of the non-equilibrium steady state $\rho^{neq}_{st}(R) $ on $R \in ]-\infty,+\infty[ $
\begin{eqnarray}
1= \int_{-\infty}^{+\infty} dR  \rho^{neq}_{st}(R) 
=  \int_{-\infty}^{0} dR  \rho^{neq}_{st}(R) 
+  \int_{0}^{+\infty} dR  \rho^{neq}_{st}(R) 
\label{jstplus}
\end{eqnarray}
that determines the value of steady-state current $J_{st}$
in terms of the parameters of the model.

$\bullet $ In summary, when $f^{[1]}_{12} \ne 0$ and $ f^{[1]}_{21} \ne 0$,
the ratio $R(t)=\frac{x_2(t)}{x_1(t)}$ satisfies the single-variable Stratonovich SDE of
Eq. \ref{RicattiDeri} that converges towards a steady state $\rho_{st}(R) $
that can be  
either an equilibrium steady state $\rho^{eq}_{st}(R) $ associated to a vanishing steady current $J^{eq}_{st} =0 $
when $ f^{[1]}_{12}  f^{[1]}_{21} >0$ with the explicit expressions of Eqs \ref{SteadyEqpos}
and \ref{SteadyEqneg},
or a non-equilibrium steady state $\rho^{neq}_{st}(R) $ associated to a non-vanishing steady current $J^{neq}_{st} \ne 0 $ when $f^{[1]}_{12}  f^{[1]}_{21} <0$, with the explicit expressions of Eqs \ref{noneqsolRegplus}
and \ref{noneqsolRegmoins}.


\subsubsection{ Analysis of the dynamics of $x_1(t)$ that depends on $R(t)$}

Now that the dynamics for $R(t)$ alone has been well understood in the two previous subsections,
let us return to the dynamics of Eq. \ref{StratoSDE2DF1D2dynx1R}
for the variable $x_1(t)$ that depends on $R(t)$.

Since we have introduced in Eqs \ref{DReff} \ref{DReffnorm}
the effective Brownian noise 
 \begin{eqnarray}
   dB_R(t) \equiv \frac{ \left(-    \sqrt{  D^{[2]}_1}  dB_1(t) + \sqrt{  D^{[2]}_2 } dB_2(t)   \right) }{\sqrt{  D^{[2]} }}
   =  \frac{ \left( -    \sqrt{  D^{[2]}_1}  dB_1(t) + \sqrt{  D^{[2]}_2 } dB_2(t) \right)}{\sqrt{  D^{[2]}_1+D^{[2]}_2}  }
 \label{DReffexpli}
\end{eqnarray} 
that appears in the single-variable Stratonovich SDE of Eq. \ref{RicattiDeri} for $R(t)$,
it is convenient to introduce the orthogonal Brownian noise
 \begin{eqnarray}
   dB_R^{\perp} (t) \equiv \frac{     \sqrt{  D^{[2]}_2}  dB_1(t) + \sqrt{  D^{[2]}_1 } dB_2(t) }{\sqrt{  D^{[2]} }}
   =  \frac{  \sqrt{  D^{[2]}_2}  dB_1(t) + \sqrt{  D^{[2]}_1 } dB_2(t) }{\sqrt{  D^{[2]}_1+D^{[2]}_2}}
 \label{DReffexpliperp}
\end{eqnarray} 
that is uncorrelated with $dB_R(t) $
 \begin{eqnarray}
{\mathbb E} \left( dB_R(t)  dB_R^{\perp}(t) \right) = 0
 \label{DReffexplicheckortho}
\end{eqnarray} 
The inversion yields the initial Brownian noise $ dB_1(t)  $ can be rewritten in terms of $dB_R(t)$ and $dB_R^{\perp}(t) $
 \begin{eqnarray}
 dB_1(t)  = \frac{ \sqrt{  D^{[2]}_2}   dB_R^{\perp} (t) -    \sqrt{  D^{[2]}_1}  dB_R(t) } { \sqrt{  D^{[2]} } }
 = \frac{ \sqrt{  D^{[2]}_2}   dB_R^{\perp} (t) -    \sqrt{  D^{[2]}_1}  dB_R(t) } {\sqrt{  D^{[2]}_1+D^{[2]}_2}}
   \label{B1inversion}
\end{eqnarray} 
and can be plugged into Eq. \ref{StratoSDE2DF1D2dynx1R} to obtain
  \begin{eqnarray}
\frac{ dx_1(t) }{x_1(t)}  && = \left(f^{[1]}_{11}+ f^{[1]}_{12} R(t)  \right) dt + \sqrt{ 2   D^{[2]}_1  }   dB_1(t) 
\nonumber \\
&&  = \left(f^{[1]}_{11}+ f^{[1]}_{12} R(t)  \right) dt 
-    \frac{   \sqrt{ 2     }    D^{[2]}_1   } { \sqrt{  D^{[2]} } } dB_R(t)
+    \frac{ \sqrt{ 2   D^{[2]}_1 D^{[2]}_2}    } { \sqrt{  D^{[2]} } } dB_R^{\perp} (t) 
\label{StratoSDE2DF1D2dynx1RBRetorthog}
\end{eqnarray} 
One can then use the single-variable Stratonovich SDE of Eq. \ref{RicattiDeri} to rewrite the noise $dB_R(t) $ in terms of $R(t)$ and its increment $dR(t)$
  \begin{eqnarray}
\sqrt{ 2 D^{[2]} } dB_R(t)  = \frac{ d R(t)  } { R(t)  } 
- \left[ \frac{ f^{[1]}_{21} }  { R(t)  }
+ ( f^{[1]}_{22} - f^{[1]}_{11})  
- f^{[1]}_{12} R(t) \right]   dt
\label{RicattiInversionBRnoise}
\end{eqnarray} 
and plug it into Eq. \ref{StratoSDE2DF1D2dynx1RBRetorthog}
  \begin{eqnarray}
\frac{ dx_1(t) }{x_1(t)}  && = 
\left(f^{[1]}_{11}+ f^{[1]}_{12} R(t)  \right) dt 
-    \frac{       D^{[2]}_1   } {  D^{[2]}  } 
\left( 
\frac{ d R(t)  } { R(t)  } 
- \left[ \frac{ f^{[1]}_{21} }  { R(t)  }+ ( f^{[1]}_{22} - f^{[1]}_{11})  - f^{[1]}_{12} R(t) \right]   dt 
\right)
+    \frac{ \sqrt{ 2   D^{[2]}_1 D^{[2]}_2}    } { \sqrt{  D^{[2]} } } dB_R^{\perp} (t) 
\nonumber \\
&&  =   
 \left[ 
 \left(f^{[1]}_{11} + \frac{       D^{[2]}_1   } {  D^{[2]}  } ( f^{[1]}_{22} - f^{[1]}_{11})  \right)
+\left( 1 - \frac{       D^{[2]}_1   } {  D^{[2]}  }  \right)  f^{[1]}_{12} R(t) 
+ \frac{       D^{[2]}_1  f^{[1]}_{21}   } {  D^{[2]}  R(t) } \right]   dt 
-    \frac{       D^{[2]}_1   } {  D^{[2]}  } \ \frac{ d R(t)  } { R(t)  } 
+    \frac{ \sqrt{ 2   D^{[2]}_1 D^{[2]}_2}    } { \sqrt{  D^{[2]} } } dB_R^{\perp} (t) 
\nonumber \\
\label{StratoSDE2DF1D2dynx1RBRetorthogR}
\end{eqnarray} 
Let us now replace
$ D^{[2]} = D^{[2]}_1+D^{[2]}_2 $ of Eq. \ref{DReffnorm} to obtain 
the final result in terms of the initial parameters
  \begin{eqnarray}
\frac{ dx_1(t) }{x_1(t)}  
  =   
 \left[ 
 \left( \frac{ f^{[1]}_{11} D^{[2]}_2   +        f^{[1]}_{22} D^{[2]}_1  } {  D^{[2]}_1+D^{[2]}_2  }   \right)
+\left(  \frac{        f^{[1]}_{12} D^{[2]}_2 R(t) + \frac{         f^{[1]}_{21} D^{[2]}_1  } {   R(t) } } {  D^{[2]}_1+D^{[2]}_2  }  \right)   
 \right]   dt 
-    \frac{       D^{[2]}_1   } {  D^{[2]}_1+D^{[2]}_2  } \ \frac{ d R(t)  } { R(t)  } 
+    \frac{ \sqrt{ 2   D^{[2]}_1 D^{[2]}_2}    } { \sqrt{  D^{[2]}_1+D^{[2]}_2 } } dB_R^{\perp} (t) 
\nonumber \\
\label{StratoSDE2DF1D2dynx1RBRetorthogRfinal}
\end{eqnarray}

This expression is useful to analyze the 
finite-time Lyapunov exponent $\lambda_1(T)$ as discussed in the next subsection.


\subsubsection{ Analysis of the two finite-time Lyapunov exponents $\lambda_1(T)$ and $\lambda_2(T)$
with their explicit common asymptotic value $\lambda_1(T=\infty) = \lambda_2(T=\infty)   $}

The integration of Eq. \ref{StratoSDE2DF1D2dynx1RBRetorthogRfinal}
over the time interval $t \in [0,T]$ contains fully-integrated terms involving $T$, $R(T)$
and the Brownian $B_R^{\perp} (T) $
  \begin{eqnarray}
\ln \left\vert \frac{x_1(T)} {x_1(0) } \right\vert   
&&  =  \left( \frac{ f^{[1]}_{11} D^{[2]}_2   +        f^{[1]}_{22} D^{[2]}_1  } {  D^{[2]}_1+D^{[2]}_2  }   \right) T
+    \frac{ \sqrt{ 2   D^{[2]}_1 D^{[2]}_2}    } { \sqrt{  D^{[2]}_1+D^{[2]}_2 } } B_R^{\perp} (t) 
-    \frac{       D^{[2]}_1   } {  D^{[2]}_1+D^{[2]}_2    } \ln \left\vert \frac{R(T)} {R(0) } \right\vert   
\nonumber \\
&& + \int_0^T dt  \left(  \frac{        f^{[1]}_{12} D^{[2]}_2 R(t) 
+ \frac{         f^{[1]}_{21} D^{[2]}_1  } {   R(t) } } {  D^{[2]}_1+D^{[2]}_2  }  \right) 
\label{StratoSDE2DF1D2dynx1RBRetorthogRinteg}
\end{eqnarray} 
while the remaining integral corresponds to an additive functional 
 of the diffusion process $R(t)$
over the time-window $t \in [0,T]$.
[Note that the integration of $\frac{ d R(t)  } { R(t)  }  $  into $\ln \left\vert \frac{R(T)} {R(0) } \right\vert $
is actually valid only when $R(t)$ is not able to use the periodic boundary conditions to go around the ring, 
i.e. here for the cases of Eqs \ref{SteadyEqpos} and \ref{SteadyEqneg},
and should be thus analyzed more precisely for the cases of Eqs 
\ref{noneqsolRegplus} and \ref{noneqsolRegmoins}
concerning non-equilibrium steady states with non-vanishing steady currents,
as discussed in details in \cite{c_Lyapunov} for analog models. However in the present case,
the results concerning the large deviations for large $T$
will not involve this term, as found in a similar example described in the last section of \cite{c_Lyapunov},
so that here we will continue the discussion with Eq. \ref{StratoSDE2DF1D2dynx1RBRetorthogRinteg}].

Let us divide Eq. \ref{StratoSDE2DF1D2dynx1RBRetorthogRinteg} by $T$
to obtain the finite-time Lyapunov exponent $\lambda_1(T)$ 
of Eq. \ref{lambdajT} 
 \begin{eqnarray}
\lambda_1(T)  =  \frac{1}{T} \ln \left\vert \frac{x_1(T)} {x_1(0) } \right\vert   
&&  =  \left( \frac{ f^{[1]}_{11} D^{[2]}_2   +        f^{[1]}_{22} D^{[2]}_1  } {  D^{[2]}_1+D^{[2]}_2  }   \right) 
+    \frac{ \sqrt{ 2   D^{[2]}_1 D^{[2]}_2}    } { \sqrt{  D^{[2]}_1+D^{[2]}_2 } } \frac{ B_R^{\perp} (t) }{T}
-    \frac{       D^{[2]}_1   } {  (D^{[2]}_1+D^{[2]}_2 ) T   } \ln \left\vert \frac{R(T)} {R(0) } \right\vert   
 \nonumber \\ &&
 + A_T  [R(0 \leq t \leq T)]
\label{StratoSDE2DF1D2dynx1RBRetorthogRintegLyapunov}
\end{eqnarray}
where the last contribution 
 \begin{eqnarray}
A_T[R(0 \leq t \leq T) ]&& \equiv  \frac{1}{T} \int_0^T dt 
   \left(  \frac{        f^{[1]}_{12} D^{[2]}_2 R(t) 
+ \frac{         f^{[1]}_{21} D^{[2]}_1  } {   R(t) } } {  D^{[2]}_1+D^{[2]}_2  }  \right)
\label{defTimeAverage}
\end{eqnarray}
is a functional of the stochastic trajectory $R(0 \leq t \leq T) $
that corresponds to a time-averaging over the time-window $[0,T]$
that can be thus evaluated for large $T \to + \infty$ by an average over the steady state $\rho_{st}(R) $
 \begin{eqnarray}
A[R(0 \leq t \leq T) \opsimeq_{T \to + \infty}
\int_{-\infty}^{+\infty} dR \rho_{st}(R)    \left(  \frac{        f^{[1]}_{12} D^{[2]}_2 R 
+ \frac{         f^{[1]}_{21} D^{[2]}_1  } {   R } } {  D^{[2]}_1+D^{[2]}_2  }  \right) \equiv A(T=\infty)
\label{ergodicForLyapunov}
\end{eqnarray}
Since the Brownian $B_R^{\perp} (T) $ scales as $\sqrt{T}$ and since $R(T)$ converges towards the steady state $
\rho_{st}(R)$, 
one obtains that
the asymptotic Lyapunov exponent $\lambda_1(T=\infty) $ 
\begin{eqnarray}
\lambda_1(T=\infty)  &&  =   \left( \frac{ f^{[1]}_{11} D^{[2]}_2   +        f^{[1]}_{22} D^{[2]}_1  } {  D^{[2]}_1+D^{[2]}_2  }   \right) + A(T=\infty)
\nonumber \\
&&  =   \left( \frac{ f^{[1]}_{11} D^{[2]}_2   +        f^{[1]}_{22} D^{[2]}_1  } {  D^{[2]}_1+D^{[2]}_2  }   \right) 
+ \int_{-\infty}^{+\infty} dR \rho_{st}(R)    \left(  \frac{        f^{[1]}_{12} D^{[2]}_2 R 
+ \frac{         f^{[1]}_{21} D^{[2]}_1  } {   R } } {  D^{[2]}_1+D^{[2]}_2  }  \right)
\label{Lyapunovinfinity}
\end{eqnarray}
can be computed from the averages of $R$ and of $\frac{1}{R}$ over the steady state $\rho_{st}(R) $
with the four possible explicit expressions of Eqs \ref{SteadyEqpos}, \ref{SteadyEqneg}, \ref{noneqsolRegplus}
and \ref{noneqsolRegmoins}.

The finite-time Lyapunov exponent $\lambda_2(T)$ of Eq. \ref{lambdajT} for the process $x_2(t)$
 \begin{eqnarray}
\lambda_2(T) && = \frac{1}{T} \ln \left\vert \frac{x_2(T)} {x_0(0) } \right\vert  
=  \frac{1}{T} \ln \left\vert \frac{x_1(T)} {x_1(0) } \right\vert  
+ \frac{1}{T} \ln \left\vert \frac{ \frac{x_2(T)}{x_1(T)}} {\frac{x_2(0)}{x_1(0)} } \right\vert  
\nonumber \\
&& = \lambda_1(T) + \frac{1}{T}  \ln \left\vert \frac{R(T)} {R(0) } \right\vert 
 = \lambda_1(T) + O \left(\frac{1}{T}  \right)
\label{lambdaTdiff21}
\end{eqnarray}
differs from the finite-time Lyapunov exponent $\lambda_1(T) $ for the process $x_1(t)$
only via a correction of order $\frac{1}{T}$ since $R(T)$ converges towards its steady state.
As a consequence, the asymptotic value $\lambda_2(T=\infty) $ coincides with the asymptotic $\lambda_1(T=\infty) $ computed in Eq. \ref{Lyapunovinfinity}
\begin{eqnarray}
\lambda_2(T=\infty) = \lambda_1(T=\infty)   
\label{Lyapunovinfinity12}
\end{eqnarray}


\subsubsection{ Large deviations properties of the two finite-time Lyapunov exponents $\lambda_1(T)$ and $\lambda_2(T)$
in relation with the Carleman moments $m_T(n_1,n_2) $}

\label{subsec_LargeDevLyapunov}

Besides their common asymptotic value of Eq. \ref{Lyapunovinfinity12},
the difference of Eq. \ref{lambdaTdiff21}
yields moreover that the leading exponential behavior for large $T$ of Eq. \ref{LyapunovTgenerating12largedev}
concerning the Carleman moment of Eq. \ref{LyapunovTgenerating12}
\begin{eqnarray}
m_T(n_1,n_2) && = {\mathbb E} \left( x_1^{n_1}(T) x_2^{n_2}(T)\right) 
 = {\mathbb E} \left( x_1^{n_1}(0) x_2^{n_2}(0) e^{ T (n_1 \lambda_1(T)+n_2 \lambda_2(T) )} \right)
\nonumber \\
&& \opsimeq_{T \to + \infty} e^{ T {\cal E}(n=n_1+n_2)}
\label{LyapunovTgeneratinglargedevn}
\end{eqnarray}
will involve the Scaled Cumulant Generating Function ${\cal E}(n=n_1+n_2) $
that depends only on the global degree $n=n_1+n_2$,
and that can be thus evaluated from the special case $(n_1=n,n_2=0)$ involving only $\lambda_1(T)$ of Eq. \ref{StratoSDE2DF1D2dynx1RBRetorthogRintegLyapunov}
\begin{eqnarray}
m_T(n,0) && = {\mathbb E} \left( x_1^{n}(T) \right) 
 = {\mathbb E} \left( x_1^{n}(0) e^{ T n \lambda_1(T) } \right)
\nonumber \\
&& = {\mathbb E} \left( x_1^{n}(0) e^{ T n \left[ 
  \left( \frac{ f^{[1]}_{11} D^{[2]}_2   +        f^{[1]}_{22} D^{[2]}_1  } {  D^{[2]}_1+D^{[2]}_2  }   \right) 
+    \frac{ \sqrt{ 2   D^{[2]}_1 D^{[2]}_2}    } { \sqrt{  D^{[2]}_1+D^{[2]}_2 } } \frac{ B_R^{\perp} (t) }{T}
-    \frac{       D^{[2]}_1   } {  (D^{[2]}_1+D^{[2]}_2 ) T   } \ln \left\vert \frac{R(T)} {R(0) } \right\vert   
 + A_T  [R(0 \leq t \leq T)] \right] }
 \right)  
\nonumber \\
&&  \opsimeq_{T \to + \infty}   e^{ T n
  \left( \frac{ f^{[1]}_{11} D^{[2]}_2   +        f^{[1]}_{22} D^{[2]}_1  } {  D^{[2]}_1+D^{[2]}_2  }   \right) }
  {\mathbb E} \left(  e^{  n  \frac{ \sqrt{ 2   D^{[2]}_1 D^{[2]}_2}    } { \sqrt{  D^{[2]}_1+D^{[2]}_2 } }  B_R^{\perp} (t) } \right)
  {\mathbb E} \left(  e^{ T n A_T  [R(0 \leq t \leq T)]  } \right)  
  \nonumber \\
&&  \opsimeq_{T \to + \infty}   e^{ T n
  \left( \frac{ f^{[1]}_{11} D^{[2]}_2   +        f^{[1]}_{22} D^{[2]}_1  } {  D^{[2]}_1+D^{[2]}_2  }   \right) }
    e^{ T  n^2 \left( \frac{  2   D^{[2]}_1 D^{[2]}_2    } {   D^{[2]}_1+D^{[2]}_2  } \right)  }
 e^{ \displaystyle T \Phi(n) }
\label{LyapunovTgeneratinglargedevnExpli}
\end{eqnarray}
where $\Phi(n)$ represents the Scaled Cumulant Generating Function 
of the additive observable $A_T  [R(0 \leq t \leq T)] $ of the one-dimensional diffusion process $R(t)$
\begin{eqnarray}
  {\mathbb E} \left(  e^{ T n A_T  [R(0 \leq t \leq T)]  } \right)  
&&  \opsimeq_{T \to + \infty}   e^{ T \Phi(n) }
\label{SCGFA}
\end{eqnarray}
that can be expanded in powers of $n$ around $n=0$
\begin{eqnarray}
 \Phi(n) = \sum_{k=1}^{+\infty} n^k \frac{\Phi^{(k)}(0)}{k!} =  n \Phi'(0) + n^2 \frac{ \Phi''(0)}{2}+...
\label{SCGFAn}
\end{eqnarray}
in order to obtains the cumulants $c_n$ of the additive observable $A_T  [R(0 \leq t \leq T)] $ via
\begin{eqnarray}
c_k = \frac{\Phi^{(k)}(0)}{T^{k-1}}
\label{SCGFAck}
\end{eqnarray}
In particular, the first cumulant $c_1$ coincides with the asymptotic value $A(T=\infty) $ evaluated in Eq. \ref{ergodicForLyapunov}
 \begin{eqnarray}
 c_1&& = \Phi'(0)
 =   {\mathbb E} \left(   A_T  [R(0 \leq t \leq T)]  \right)
 = {\mathbb E} \left[   \frac{1}{T} \int_0^T dt 
   \left(  \frac{        f^{[1]}_{12} D^{[2]}_2 R(t) 
+ \frac{         f^{[1]}_{21} D^{[2]}_1  } {   R(t) } } {  D^{[2]}_1+D^{[2]}_2  }  \right)  \right]
\nonumber \\
&&  =
\int_{-\infty}^{+\infty} dR \rho_{st}(R)    \left(  \frac{        f^{[1]}_{12} D^{[2]}_2 R 
+ \frac{         f^{[1]}_{21} D^{[2]}_1  } {   R } } {  D^{[2]}_1+D^{[2]}_2  }  \right) =  A(T=\infty)
\label{c1ergodicForLyapunov}
\end{eqnarray}
while the second cumulant $c_2$ corresponding to the variance
characterizes the typical fluctuations of order $\frac{1}{\sqrt{T}}$ 
\begin{eqnarray}
c_2 && = \frac{\Phi''(0)}{T} = 
 {\mathbb E} \left[   A_T^2  [R(0 \leq t \leq T)]  \right] - \left({\mathbb E} \left(   A_T  [R(0 \leq t \leq T)]  \right) \right)^2
 =  {\mathbb E} \left[   \left( A_T  [R(0 \leq t \leq T)] -c_1 \right)^2 \right]
\nonumber \\
&&  = {\mathbb E} \left[   \frac{1}{T^2} \int_0^T dt_1  \int_0^T dt_2 
  \left(  \frac{f^{[1]}_{12} D^{[2]}_2 R(t_1) + \frac{         f^{[1]}_{21} D^{[2]}_1  } {   R(t_1) } } {  D^{[2]}_1+D^{[2]}_2  }  \right) \left(  \frac{f^{[1]}_{12} D^{[2]}_2 R(t_2) + \frac{         f^{[1]}_{21} D^{[2]}_1  } {   R(t_2) } } {  D^{[2]}_1+D^{[2]}_2  }  \right) \right] - c_1^2
\label{SCGFAck2}
\end{eqnarray}
and involves the two-time correlations of the process $R(t)$.

If one wishes to analyze the full Scaled Cumulant Generating Function $\Phi(n)$ that contains
the cumulants $c_n$ of arbitrary order, i.e. if one wishes to study the whole large deviations 
properties of the additive observable $A_T  [R(0 \leq t \leq T)] $ 
of the one-dimensional diffusion process $R(t)$, one can use various methods
as discussed in detail in \cite{c_Lyapunov} concerning similar models.
One of them is the famous Feynman-Kac formula \cite{feynman,kac,review_maj}, where  
the Scaled Cumulant Generating Function $\Phi(n)$ represents 
the highest eigenvalue of the following tilted Fokker-Planck generator,
where one adds the contribution $ n \left(  \frac{        f^{[1]}_{12} D^{[2]}_2 R 
+ \frac{         f^{[1]}_{21} D^{[2]}_1  } {   R } } {  D^{[2]}_1+D^{[2]}_2  }  \right) $ 
to the generator of Eq \ref{GeneratorStratoR}
\begin{eqnarray}
{\cal L}_n && = {\cal L} + n \left(  \frac{        f^{[1]}_{12} D^{[2]}_2 R 
+ \frac{         f^{[1]}_{21} D^{[2]}_1  } {   R } } {  D^{[2]}_1+D^{[2]}_2  }  \right)
\nonumber \\
&& = f(R) \frac{ \partial }{\partial R}
+  D^{[2]}
\left( R \frac{\partial}{\partial R} \right)
\left( R  \frac{ \partial }{\partial R} \right) + n  \left(  \frac{        f^{[1]}_{12} D^{[2]}_2 R 
+ \frac{         f^{[1]}_{21} D^{[2]}_1  } {   R } } {  D^{[2]}_1+D^{[2]}_2  }  \right)
\label{GeneratorStratoRn}
\end{eqnarray}
or equivalently to its adjoint
\begin{eqnarray}
{\cal L}^{\dagger}_n && = {\cal L}^{\dagger} + n \left(  \frac{        f^{[1]}_{12} D^{[2]}_2 R 
+ \frac{         f^{[1]}_{21} D^{[2]}_1  } {   R } } {  D^{[2]}_1+D^{[2]}_2  }  \right)
\nonumber \\
&& = -  \frac{ \partial }{\partial R} f(R)
+  D^{[2]}
\left(  \frac{\partial}{\partial R}  R  \frac{ \partial }{\partial R} R \right)
+ n  \left(  \frac{        f^{[1]}_{12} D^{[2]}_2 R 
+ \frac{         f^{[1]}_{21} D^{[2]}_1  } {   R } } {  D^{[2]}_1+D^{[2]}_2  }  \right)
\label{GeneratorStratoRdaggern}
\end{eqnarray}
that will govern the following dynamics for the population $Z_t(R)  $
\begin{eqnarray}
 \partial_t Z_t(R) 
&& =   {\cal L}_n^{\dagger} Z_t(R) 
  \label{fokkerPlanckRtilted}
\end{eqnarray}
that replaces the conserved Fokker-Planck dynamics of Eq. \ref{fokkerPlanckR}
for the probability density $\rho_t( R ) $.
The physical interpretation is that the additional contribution $n \left(  \frac{        f^{[1]}_{12} D^{[2]}_2 R 
+ \frac{         f^{[1]}_{21} D^{[2]}_1  } {   R } } {  D^{[2]}_1+D^{[2]}_2  }  \right) $
can be interpreted as a replicating-rate if it is positive
 or as a killing rate if it is negative,
 and that the corresponding population $Z_t(R)  $ 
 \begin{eqnarray}
 Z_t(R) \oppropto_{t \to + \infty} e^{t \Phi(n) }
  \label{fokkerPlanckRtiltedinfty}
\end{eqnarray}
 will either grow exponentially in time if $\Phi(n) >0 $
 or decay exponentially in time  if $\Phi(n) <0 $.
 
 The quantitative analysis of the whole Scaled Cumulant Generating Function $\Phi(n) $
for the various possible diffusion processes $R(t)$
clearly goes beyond the goals of the present work and will not be discussed in the present paper
(see \cite{c_Lyapunov} for more details on the methods and on the results that can be obtained in examples).
However for the present perspective, 
the important conclusion of Eq. \ref{LyapunovTgeneratinglargedevnExpli}
is that the Scaled Cumulant Generating Function ${\cal E}(n) $
of the finite-time Lyapunov exponent $\lambda_1(T)$ 
that governs the Carleman moment $m_T(n,0) $ for large time $T$ reads
\begin{eqnarray}
{\cal E}(n)  = n
  \left( \frac{ f^{[1]}_{11} D^{[2]}_2   +        f^{[1]}_{22} D^{[2]}_1  } {  D^{[2]}_1+D^{[2]}_2  }   \right) 
    +  n^2 \left( \frac{  2   D^{[2]}_1 D^{[2]}_2    } {   D^{[2]}_1+D^{[2]}_2  } \right)  
 + \Phi(n) 
\label{EnPhin}
\end{eqnarray}
and is thus directly related, apart from the explicit terms involving $n$ and $n^2$,
to the Scaled Cumulant Generating Function $\Phi(n) $ of Eq. \ref{SCGFA}
for the additive observable $A_T  [R(0 \leq t \leq T)] $ of the one-dimensional diffusion process $R(t)$.

Let us now summarize the asymptotic behavior of the two couples processes $[x_1(t),x_2(t)]$ 
whose ratio $R(t)=\frac{x_2(t)}{x_1(t)} $ converges towards a steady state :

$\bullet$ If their common asymptotic Lyapunov exponent of Eq. \ref{Lyapunovinfinity}
is positive $\lambda_2(T=\infty) = \lambda_1(T=\infty)   >0$,
then they flow towards infinity.

$\bullet$ If their common asymptotic Lyapunov exponent of Eq. \ref{Lyapunovinfinity}
is negative $\lambda_2(T=\infty) = \lambda_1(T=\infty)   <0$,
then they flow towards zero. However,
as in Eq. \ref{DVmoments} concerning the one-dimensional case,
it is then useful to define the positive value $\mu>0$ where the Scaled Cumulant Generating Function 
 ${\cal E}(n) $ vanishes
\begin{eqnarray}
{\cal E}(n=\mu)  = 0
\label{Enmuzero}
\end{eqnarray}
in order to separate the moments of order $n<\mu$ converging towards zero 
and the moments of order $n>\mu$ diverging towards infinity 
\begin{eqnarray}
m_T(n,0)  \oppropto_{T \to + \infty} e^{T {\cal E}(n)}
  \begin{cases}
\text{ converges towards $0$    if } \ \ n  <  \mu
\\
\text{ diverges towards $+ \infty$     if } \ \  n  >  \mu
\end{cases}
\label{DVmoments2D}
\end{eqnarray}
This exponent $\mu$ plays also a very important role in the next section \ref{sec_BlockLower2D}
concerning models whose Carleman matrices are block-lower-triangular 
${\bold M} = {\bold M}^{[-2]} + {\bold M}^{[-1]}+ {\bold M}^{[0]} $
as will be discussed in subsection \ref{subsec_ArgumentMU}.


 \subsubsection{ Discussion   }

In summary, the non-vanishing coefficients $f^{[1]}_{12} \ne 0$ and $ f^{[1]}_{21} \ne 0$
produce the common asymptotic Lyapunov exponent $ \lambda_1(T=\infty) = \lambda_2(T=\infty)   $
of Eqs \ref{Lyapunovinfinity} \ref{Lyapunovinfinity12}
for the two components $x_1(t)$ and $x_2(t)$, instead of the two different values 
of Eq. \ref{LyapunovGBMj} for the case without interactions $f^{[1]}_{12}=0=f^{[1]}_{21} $.
It is thus interesting in the next subsection to consider the intermediate case
where only one of the two off-diagonal coefficients vanishes.


 \subsection{ When $f^{[1]}_{12}=0$ and $ f^{[1]}_{21} \ne 0$ : $x_1(t)$ is a Geometric Brownian motion 
 that acts as an external forcing for $x_2(t)$   }

 In the two previous subsections, we have discussed the cases where the two off-diagonal elements 
 $f^{[1]}_{12}$ and $ f^{[1]}_{21} $ are either both vanishing or both non-vanishing.
 So let us now analyze the cases where only one of them vanishes,
 and let us choose the
 case $f^{[1]}_{12}=0$ and $ f^{[1]}_{21} \ne 0$ 
 (since the other case $f^{[1]}_{12} \ne 0$ and $ f^{[1]}_{21} = 0$ can be obtained by exchanging the two variables $x_1 \leftrightarrow x_2$).
 
 When $f^{[1]}_{12}=0 $, then $x_1(t)$ remains the independent Geometric Brownian motion
of Eq. \ref{YlogGBM2Dindep} 
   \begin{eqnarray}
 x_1(t)   && =  x_1(0) e^{f^{[1]}_{11} t+  \sqrt{ 2 D^{[2]}_1 } \ B_1(t)}
\label{x1tGBMalone}
\end{eqnarray} 
with the properties of Eqs \ref{LyapunovGBMj}
\begin{eqnarray}
\lambda_1(T)  \equiv  \frac{1}{T} \ln \left\vert \frac{x_j(T)} {x_j(0) } \right\vert   
&& = f^{[1]}_{11} +  \sqrt{ 2 D^{[2]}_1 } \frac{ B_1(T) }{T}
\label{LyapunovGBMj1}
\end{eqnarray}
and \ref{LyapunovGBMj}
 \begin{eqnarray}
\lambda_1(T=\infty)  = f^{[1]}_{11} 
\label{LyapunovGBMj1infty}
\end{eqnarray}

One can plug the solution $x_1(t)$ of Eq. \ref{x1tGBMalone}
 into the SDE of Eq. \ref{StratoSDE2DF1D2}
 for $x_2(t)$
where it appears as an inhomogeneous forcing term
   \begin{eqnarray}
dx_2(t)   && =  f^{[1]}_{21} x_1(0) e^{f^{[1]}_{11} t+  \sqrt{ 2 D^{[2]}_1 } \ B_1(t)}+ x_2(t) \left( f^{[1]}_{22}  dt 
+ \sqrt{ 2   D^{[2]}_2  }  dB_2(t) \right)
\label{StratoSDE2DF1D2asym}
\end{eqnarray} 
so that one obtains 
the following integral representation for $x_2(t)$ in terms of the two Brownian motions $B_1(s)$ and $B_2(s)$
\begin{eqnarray}
x_2(t) &&=  e^{f^{[1]}_{22} t+  \sqrt{ 2 D^{[2]}_2 } \ B_2(t)}
\left[ x_2(0) + f^{[1]}_{21} x_1(0)
\int_0^t ds    e^{ ( f^{[1]}_{11} -f^{[1]}_{22} ) s+  \sqrt{ 2 D^{[2]}_1 } \ B_1(s)-  \sqrt{ 2 D^{[2]}_2 } \ B_2(s)}
\right] 
\label{YlogGBM2Dindep2ct}
\end{eqnarray}

To make the link with the previous subsection,
let us write the ratio $ R(t) = \frac{x_2(t)}{x_1(t) }$ 
in terms of the effective noise $\sqrt{ 2 D^{[2]} } B_R(t) = \sqrt{ 2 D^{[2]}_2 } B_2(t)  -    \sqrt{ 2 D^{[2]}_1}  B_1(t) $ introduced in Eq. \ref{DReff}

\begin{eqnarray}
R(t) = \frac{x_2(t)}{x_1(t) } &&
=  e^{(f^{[1]}_{22} - f^{[1]}_{11}) t+  \sqrt{ 2 D^{[2]}_2 } \ B_2(t)-  \sqrt{ 2 D^{[2]}_1 } \ B_1(t)}
\left[ \frac{x_2(0)}{x_1(0)} + f^{[1]}_{21} 
\int_0^t ds    e^{ ( f^{[1]}_{11} -f^{[1]}_{22} ) s+  \sqrt{ 2 D^{[2]}_1 } \ B_1(s)-  \sqrt{ 2 D^{[2]}_2 } \ B_2(s)}
\right] 
\nonumber \\
&& =e^{(f^{[1]}_{22} - f^{[1]}_{11}) t+  \sqrt{ 2 D^{[2]} } B_R(t)}
\left[  R(0) + f^{[1]}_{21} 
\int_0^t ds    e^{ ( f^{[1]}_{11} -f^{[1]}_{22} ) s - \sqrt{ 2 D^{[2]} } B_R(s)}
\right] 
\nonumber \\
&& =e^{(f^{[1]}_{22} - f^{[1]}_{11}) t+  \sqrt{ 2 D^{[2]} } B_R(t)}  R(0) 
+ f^{[1]}_{21} 
\int_0^t ds    e^{ - ( f^{[1]}_{11} -f^{[1]}_{22} ) (t-s) +  \sqrt{ 2 D^{[2]} } (B_R(t)- B_R(s))}
\label{Rratiof12vanish}
\end{eqnarray}

The asymptotic behavior of $R(t)$ for large time $t \to + \infty$
depends on the sign of the difference $( f^{[1]}_{11} -f^{[1]}_{22} ) $
between the two diagonal coefficients $f^{[1]}_{11} $ and $f^{[1]}_{22} $ :

$\bullet $ (i) If $ ( f^{[1]}_{11} -f^{[1]}_{22} ) >0 $, then the first contribution of Eq. \ref{Rratiof12vanish}
becomes exponentially small for large times $t \to +\infty$,
while the second contribution remains finite 
and thus dominates the ratio $R(t)$ of Eq. \ref{Rratiof12vanish}
 for large times $t \to + \infty$
\begin{eqnarray}
R(t) \opsimeq_{t \to + \infty}  f^{[1]}_{21} 
\int_0^t ds    e^{ - ( f^{[1]}_{11} -f^{[1]}_{22} ) (t-s) +  \sqrt{ 2 D^{[2]} } (B_R(t)- B_R(s))}
\label{Rratiof12vanishasympKesten}
\end{eqnarray}
This expression is similar to the exponential functional of the Brownian motion of Eq. \ref{xexpfunctionalBrownian},
whose Inverse-Gamma steady distribution of Eq. \ref{kesten}
is in agreement with the equilibrium distribution of Eq. \ref{SteadyEq}
for $f^{[1]}_{12}=0$
 \begin{eqnarray}
 \rho_{st}^{eq}(R)  = {\cal N} e^{- U(R) } =   {\cal N}  
\ \vert R \vert^{-1 - \frac{  f^{[1]}_{11} - f^{[1]}_{22}   }{   D^{[2]}    }}  
\ e^{ \displaystyle - \frac{  f^{[1]}_{21}   }{   D^{[2]}  R  }  }
  \label{SteadyEqf12vanish}
\end{eqnarray}
with the following possible steady states from Eqs \ref{SteadyEqpos} and \ref{SteadyEqneg}
depending of the sign of $f^{[1]}_{21} $
  \begin{eqnarray}
\text{ Case $f^{[1]}_{12} =0$ and $f^{[1]}_{11} > f^{[1]}_{22} $ and $f^{[1]}_{21} > 0$ } : \ \  \rho_{st}^{eq}(R)  && = \frac{ e^{- U(R) } }
 { \int_0^{+\infty} dr e^{- U(r) } } \ \ \text{ for $R \in ]0,+\infty[$} 
 \nonumber \\
\text{ Case $f^{[1]}_{12} =0$ and $f^{[1]}_{11} > f^{[1]}_{22} $ and $f^{[1]}_{21} < 0$ } : \ \  \rho_{st}^{eq}(R)  && = \frac{ e^{- U(R) } }
 { \int_{-\infty}^0 dr e^{- U(r) } } \ \ \text{ for $R \in ]-\infty,0[$} 
  \label{SteadyEqOneOffzero}
\end{eqnarray} 

The physical interpretation is that when $ f^{[1]}_{11} >f^{[1]}_{22}$,
i.e. when the self-growth of $x_1(t)$ is bigger than the self-growth of $x_2(t)$,
then $x_1(t)$ imposes its growth to $x_2(t)$ via the off-diagonal coefficient $f^{[1]}_{21} \ne 0 $, 
so that the ratio $R(t)$ remains finite
and converges towards the steady state of Eq. \ref{SteadyEqOneOffzero}.

$\bullet $ (ii) If $ ( f^{[1]}_{11} -f^{[1]}_{22} ) <0 $,
then both contributions of Eq. \ref{Rratiof12vanish}
diverge exponentially for large times $t \to +\infty$,
and
there is no normalizable steady state for the ratio $R(t)$ of Eq. \ref{Rratiof12vanish}
that flows towards infinity.
The physical interpretation is that when $ f^{[1]}_{11} <f^{[1]}_{22}$,
i.e. when the self-growth of $x_1(t)$ is smaller than the self-growth of $x_2(t)$,
then $x_1(t)$ becomes negligible with respect to $x_2(t)$ for large times,
so that the ratio $R(t)$ flows towards infinity.

As a final remark, let us mention that from the point of view of the Carleman matrix
${\bold M}  =  {\bold M}^{[0]} $ with the matrix elements of Eq. \ref{mm0BlocsDiag},
the vanishing of the coefficient $F^{[1]}_{12}=f^{[1]}_{12}=0 $
 yields that the diagonal blocks ${\bold M}^{[0]}_{[n,n]}$ of Eq. \ref{mm0Blocs}
 are bidiagonal instead of tridiagonal, so that their eigenvalues are given by the diagonal elements $q_2=n_2$
   of Eq. \ref{mm0BlocsDiag} and are thus the same as in Eq. \ref{Ediagsol2D}
   with $n_1=n-n_2$.


\subsection{ Discussion }

In summary, we have discussed in detail the effects of the interactions mediated by the two off-diagonal
coefficients $f^{[1]}_{12} $ and $f^{[1]}_{21} $ for the 2D stochastic system of Eq. \ref{StratoSDE2DF1D2}.
In particular when they are both non-vanishing $f^{[1]}_{12} \ne 0$ and $f^{[1]}_{21}\ne 0 $, we have explained why it was useful to keep $x_1(t)$
and to replace $x_2(t)$ by the ratio $R(t)=\frac{x_2(t)}{x_1(t)}$ whose dynamics converging towards an
explicit steady state determines the statistical properties of the finite-time Lyapunov exponent $\lambda_1(T)$.

In Appendix \ref{app_dimensiond}, 
we discuss what happens when one tries to apply the same strategy for the 
Stratonovich system of Eq. \ref{ItoSDEsinglegjmultiplicativediag} in dimension $d>2$ and in particular $d=3$.


\section{ Models in dimension $d=2$ with Carleman matrices ${\bold M}= {\bold M}^{[-2]} + {\bold M}^{[-1]}+ {\bold M}^{[0]}$   }

\label{sec_BlockLower2D}

In this section, we analyze the properties of diffusion processes of section \ref{sec_2D}  
whose Carleman matrices are block-lower-triangular 
${\bold M} = {\bold M}^{[-2]} + {\bold M}^{[-1]}+ {\bold M}^{[0]} $.
So with respect to the previous section where the Carleman matrix was block-diagonal 
$ {\bold M}^{[0]} $ and involved the matrices ${\bold F}^{[1]}$ (linear forces) and ${\bold D}^{[2]}$ (multiplicative noises), 
the block ${\bold M}^{[-2]} $ involves the matrix ${\bold D}^{[0]} $ parametrizing the additive noises,
while the matrix ${\bold M}^{[-1]}$ involves the matrices ${\bold F}^{[0]}$ (constant forces) 
and ${\bold D}^{[1]}$ (square-root noises).
The goal is thus to analyze the generalizations in dimension $d=2$
of the Pearson family of one-dimensional discussed in section \ref{subsec_1Dlower}.


\subsection{ General properties of the Carleman moments }

When ${\bold M}^{[1]}=0$, 
the dynamics of Eq. \ref{DynMoments2DBlocs} for the ket $\vert {\cal M}_t^{[n]} \rangle $
of Eq. \ref{ketmomentsdegren}
that gathers the $(n+1)$ moments $m_t(n_1=n-n_2,n_2)$ of a given degree $n$ with $n_2=0,..,n$ 
becomes
 \begin{eqnarray}
 \partial_t \vert {\cal M}_t^{[0]} \rangle  
 && = 0
 \nonumber \\
 \partial_t \vert {\cal M}_t^{[1]} \rangle 
&&  =  {\bold M}^{[-1]}_{[1,0]}   \vert {\cal M}_t^{[0]} \rangle
 +  {\bold M}^{[0]}_{[1,1]}    \vert {\cal M}_t^{[1]} \rangle
 \nonumber \\
 \partial_t \vert {\cal M}_t^{[2]} \rangle 
&&  = {\bold M}^{[-2]}_{[2,0]}   \vert {\cal M}_t^{[0]} \rangle
 + {\bold M}^{[-1]}_{[2,1]}   \vert {\cal M}_t^{[1]} \rangle
 +  {\bold M}^{[0]}_{[2,2]}    \vert {\cal M}_t^{[2]} \rangle
\nonumber \\
....
\nonumber \\
 \partial_t \vert {\cal M}_t^{[n]} \rangle 
&&  = {\bold M}^{[-2]}_{[n,n-2]}   \vert {\cal M}_t^{[n-2]} \rangle
 + {\bold M}^{[-1]}_{[n,n-1]}   \vert {\cal M}_t^{[n-1]} \rangle
 +  {\bold M}^{[0]}_{[n,n]}    \vert {\cal M}_t^{[n]} \rangle
 \nonumber \\
...
\label{DynMoments2DBlocsPearson}
\end{eqnarray} 
that generalizes the dynamics of Eq. \ref{dynotxexpliPearson}
concerning one-dimensional Pearson diffusions,
and that should also be solved recursively iteratively in the order $n=0,1,..$.

At each step $n$, one should thus diagonalise the  
${\bold M}^{[0]}_{[n,n]}  $ of size $(n+1) \times (n+1)$ with its tridiagonal structure of Eq. \ref{mm0Blocs}
in order to obtain in particular its $(n+1)$ eigenvalues $E_{n,\alpha} $ labelled by $\alpha=0,..,n$.

From the point of view of the spectral decomposition of Eq. \ref{SolMomentsSpectral},
this means that the moment 
$m_t(k_1=k-k_2,k_2) = \langle k_2 \vert {\cal M}_t^{[k]}\rangle $ can only involve
 the eigenvalues $E_{n,\alpha}$ of the blocks $0 \leq n \leq k$, 
\begin{eqnarray}
m_t(k_1=k-k_2,k_2) = \langle k_2 \vert {\cal M}_t^{[k]}\rangle && 
 =   \sum_{n=0}^{k} \sum_{\alpha=0}^n  e^{t E_{n,\alpha}} \langle k,k_2  \vert R_{n,\alpha} \rangle 
 \langle L_{n,\alpha} \vert  {\cal M}_{t=0} \rangle 
 \nonumber \\
 && =  \langle k,k_2  \vert R_{0} \rangle
 + \sum_{n=1}^{k} \sum_{\alpha=0}^n  e^{t E_{n,\alpha}} \langle k,k_2  \vert R_{n,\alpha} \rangle 
 \langle L_{n,\alpha} \vert  {\cal M}_{t=0} \rangle 
\label{SolMomentsSpectralPearson2D}
\end{eqnarray}
So the moment $ m_t(k_1=k-k_2,k_2)$ will converge towards its steady value 
$m_{st}(k_1=k-k_2,k_2)=  \langle k,k_2  \vert R_0 \rangle$
only if all the eigenvalues $E_{n,\alpha}$ that really appear in Eq. \ref{SolMomentsSpectralPearson2D}
have negative real parts.
If one considers all the moments $\vert {\cal M}_t^{[k]} \rangle $ of a given degree $k$,
the conclusion is that the convergence toward steady moments require
that all the eigenvalues $E_{n,\alpha}$ of the blocks $1 \leq k \leq n$ should have negative real parts
\begin{eqnarray}
\vert {\cal M}_t^{[k]} \rangle   \opsimeq_{t \to +\infty}
\vert {\cal M}_{st}^{[k]} \rangle \ \ \ 
&& \text{only if all the eigenvalues $E_{n,\alpha} $ labelled by $\alpha=0,..,n$ of the blocks $1 \leq n \leq k$ } 
\nonumber \\
&& \text{ have negative real parts $\text{ Re}(E_{n,\alpha} )<0$  } 
 \label{OavdynPearsonkm2integd2}
\end{eqnarray}

In subsection \ref{app_lowerBiBLOCK} of Appendix \ref{app_upperBiBLOCK}, 
we describe how the spectral decomposition
of block-lower-bidiagonal matrices ${\bold M}= {\bold M}^{[-1]} + {\bold M}^{[0]}    $
can be explicitly computed,
with the final results of Eqs \ref{RightLowerBlock}
and \ref{LeftLowerBlock}
for the right and left eigenvectors respectively.
Note that these calculations can be extended to the case ${\bold M}= {\bold M}^{[-2]} + {\bold M}^{[0]} $.
So in the following subsections, we describe the corresponding simplest examples of section \ref{sec_list}
in dimension $d=2$.


\subsection{ Models with only one square-root-noise $dB_j^{square-root}(t) $ for each $x_j >0$ having  
 ${\bold M}= {\bold M}^{[-1]}+{\bold M}^{[0]} $}
 
 \label{subsec_onlySQ2D}

The case $d=2$ of Eq. \ref{ItoSDEsinglegjonlysquareroot},
that generalizes the one-dimension case of Eq. \ref{ItoSDEsinglegjonlysquareroot1S},
correspond to the following Ito system
for the positive variables $x_1(t)>0$ and $x_2(t)$
   \begin{eqnarray}
dx_1(t)   && =  \left( F^{[0]}_1 +  F^{[1]}_{11} x_1(t)+  F^{[1]}_{12} x_2(t)
\right) dt + \sqrt{ 2 D^{[1]}_1 x_1(t)} dB_1^{square-root}(t) 
\nonumber \\
dx_2(t)   && =  \left( F^{[0]}_2 +  F^{[1]}_{21} x_1(t)+  F^{[1]}_{22} x_2(t)
\right) dt + \sqrt{ 2 D^{[1]}_2 x_2(t)} dB_2^{square-root}(t) 
\label{ItoSDEsinglegjonlysquareroot2D}
\end{eqnarray}
where the discussion of Eq. \ref{ConditionsF1forSquareRootNoise}
yield that the four following coefficients $(F^{[0]}_1,F^{[0]}_2,F^{[1]}_{12}, F^{[1]}_{21})$ should be positive
  \begin{eqnarray}
 F^{[0]}_1 && \geq 0 \ \ \ ; \ \ \ F^{[0]}_2 \geq 0
 \nonumber \\
  F^{[1]}_{12} && \geq 0 \ \ \ ; \ \ \ F^{[1]}_{21}  \geq 0
  \label{ConditionsF1forSquareRootNoise2D}
\end{eqnarray}
while $(F^{[1]}_{11},F^{[1]}_{22})$ are arbitrary.

The eigenvalues of the full Carleman matrix ${\bold M}= {\bold M}^{[-1]}+{\bold M}^{[0]} $ coincide with the eigenvalues of the diagonal-block ${\bold M}^{[0]} $
that only involves the matrix $F^{[1]}$
that are directly given by Eq. \ref{EigenCLEigenF1Laguerre}
\begin{eqnarray}
E_{\kappa_1,\kappa_2} =     \kappa_1  \lambda_1+  \kappa_2  \lambda_2
\label{EigenCLEigenF1Laguerre2D}
\end{eqnarray}
as linear combinations with integers coefficients $(\kappa_{1},\kappa_2) \in {\mathbb N}^2 $
of the two eigenvalues $\lambda_{1,2} $ of the $2 \times 2$ matrix ${\bold F}^{[1]} $
\begin{eqnarray}
\lambda_{1,2} && = \frac{ F^{[1]}_{11}+F^{[1]}_{22} \pm \sqrt{\left(F^{[1]}_{11}-F^{[1]}_{22}\right)^2 + 4   F^{[1]}_{12} F^{[1]}_{21} }  }{2}
\label{H2by2eigenvaluessol}
\end{eqnarray}


 \subsection{ Models with only one multiplicative noise $dB_j^{multiplicative}(t) $ for each $x_j$
 having ${\bold M}={\bold M}^{[-1]}+{\bold M}^{[0]} $ }
 
  \label{subsec_Kesten2D}
 
 The case $d=2$ of Eq. \ref{ItoSDEsinglegjmultiplicativelower}
 that generalizes the one-dimension case of Eq. \ref{ItoSDEsinglegjmultiplicativelower1D}
correspond to the following Stratonovich system
  \begin{eqnarray}
dx_1(t)   && = \left( f^{[0]}_1 + f^{[1]}_{11} x_1(t)+  f^{[1]}_{12} x_2(t)\right) dt 
+ \sqrt{ 2  D^{[2]}_1 } \  x_1(t) dB_1^{multiplicative}(t) 
\nonumber \\
dx_2(t)   && = \left( f^{[0]}_2 + f^{[1]}_{21} x_1(t)+  f^{[1]}_{22} x_2(t)\right) dt 
+ \sqrt{ 2  D^{[2]}_2 } \  x_2(t) dB_2^{multiplicative}(t) 
\label{ItoSDEsinglegjmultiplicativelower2D}
\end{eqnarray}

To get some intuition, let us discuss progressively the effects of the interactions
mediated by the off-diagonal coefficients $f^{[1]}_{12} $ and $f^{[1]}_{21} $ as follows :

$\bullet$ When $f^{[1]}_{12}=0$ and $ f^{[1]}_{21} =0$, then $x_1(t)$ and $x_2(t)$ are two independent 
Kesten processes that were discussed in subsection \ref{subsec_Kesten}.
In particular in the region of parameters
\begin{eqnarray}
\mu_j && \equiv \ \frac{(-f^{[1]}_{jj})}{D^{[2]}_j } >0
\nonumber \\
 \lambda_j && \equiv \frac{   f^{[0]}_j}{D^{[2]}_j} >0
\label{NotationsKesten2D}
\end{eqnarray}
the steady state $\rho_{st}(x_1,x_2)$ is factorized into two Inverse-Gamma distributions of Eq. \ref{kesten}
of parameters $(\mu_j,\lambda_j)$
\begin{eqnarray}
\rho_{st}(x_1,x_2) = \prod_{j=1}^2 \frac{\lambda_j^{\mu_j}}{\Gamma(\mu_j) x_j^{1+\mu_j} } e^{- \frac{\lambda_j}{x_j}} \ \ \ {\rm } \ \ {\rm for } \ \ (x_1,x_2)  \in (]0,+\infty[)^2
\label{kesten2D}
\end{eqnarray}
with possibly different power-laws for large $x_1\to + \infty$ and large $x_2\to + \infty$ when $\mu_1 \ne \mu_2$.

$\bullet$ When $f^{[1]}_{12}=0$ and $ f^{[1]}_{21} \ne 0$ : then $x_1(t)$ still converges towards
its own one-dimensional Inverse-Gamma distribution $\rho_{st}(x_1)$ of parameters $(\mu_1,\lambda_1)$
and appears as an external forcing in the SDE for $x_2(t)$ that would otherwise converge towards
its own one-dimensional Inverse-Gamma distribution of parameters $(\mu_2,\lambda_2)$ if it were alone.
As a consequence, one needs to distinguish two cases :

(i) if $\mu_2<\mu_1$, i.e. when the steady state for $x_1$ alone is less broad, 
then $x_1$ acting as external forcing will not change the exponent $\mu_2$
that $x_2(t)$ would produce alone, the moments of $x_2(t)$ of any order $n_2>\mu_2$
will remain infinite, so that there will be the two distincts exponents $\mu_1 \ne \mu_2$ as in Eq. \ref{kesten2D}
\begin{eqnarray}
\rho_{st}(x_1,x_2) \oppropto_{\vert x_j \vert \to + \infty}  \frac{1}{\vert x_j \vert^{1+\mu_j} }
\label{kesten2Dexternaldifferent}
\end{eqnarray}

(ii) if $\mu_2>\mu_1$, i.e. when the steady state for $x_1$ alone is broader, 
then $x_1$ acting as external forcing will impose to $x_2$ its smaller exponent $\mu_1$
and make infinite the moments of $x_2(t)$ of any order $n_2>\mu_1$,
so that the same exponent $\mu_1$ will govern the power-law for the two variables in contrast to
Eqs \ref{kesten2D} and \ref{kesten2Dexternaldifferent}
\begin{eqnarray}
\rho_{st}(x_1,x_2) \oppropto_{\vert x_j \vert \to + \infty}  \frac{1}{\vert x_j \vert^{1+\mu_1} }
\label{kesten2Dexternalsame}
\end{eqnarray}

$\bullet$ When $f^{[1]}_{12} \ne 0$ and $ f^{[1]}_{21} \ne 0$, then both processes acting on each other
will produce a common exponent $\mu$ that will govern the tails for both $x_1$ and $x_2$
\begin{eqnarray}
\rho_{st}(x_1,x_2) \oppropto_{\vert x_j \vert \to + \infty}  \frac{1}{\vert x_j \vert^{1+\mu} }
\label{kesten2Dcommon}
\end{eqnarray}
Then to make the link with subsection \ref{subsec_Ricatti} concerning the dynamics of Eq. \ref{ItoSDEsinglegjmultiplicativelower2D}
without the constant forces $f^{[0]}_1 $ and $f^{[0]}_1 $, it is useful to replace the process $x_2(t)$
by the ratio $R(t)= \frac{ x_2(t) }{x_1(t) }$ of Eq. \ref{Ricatti} 
to obtain the Stratonovich system obtained
from Eq. \ref{ItoSDEsinglegjmultiplicativelower2D}
 \begin{eqnarray}
 dx_1(t)   && = \left[ f^{[0]}_1 + \left( f^{[1]}_{11} +  f^{[1]}_{12} R(t) \right) x_1(t)  \right) dt 
+ \sqrt{ 2  D^{[2]}_1 } \  x_1(t) dB_1(t) 
\label{RicattiKesten}
 \\
dR(t) &&=  \left[ \left( f^{[1]}_{21} + \frac{ f^{[0]}_2 }{x_1(t) } \right)
+ \left( f^{[1]}_{22} - f^{[1]}_{11} - \frac{  f^{[0]}_1}{x_1(t) } \right)  R(t)
- f^{[1]}_{12} R^2(t)   \right] dt 
+  R(t) \left[ \sqrt{ 2 D^{[2]}_2 } dB_2(t) 
 -    \sqrt{ 2 D^{[2]}_1}  dB_1(t)  \right]
\nonumber
\end{eqnarray} 
In the dynamics of $R(t)$, the new terms with respect to single-variable SDE for $R(t)$ of Eq. \ref{RicattiSDE} 
discussed in detail in subsection \ref{subsec_Ricatti} are the contributions 
$\frac{ f^{[0]}_2 - R(t) f^{[0]}_1}{x_1(t) } $ to the Stratonovich force that involve $x_1(t)$,
but that are not expected to change the fact that all the moments of $R(t)$ will converge 
towards finite steady values, so that 
the joint steady state $\rho_{st}(x_1,R)$ will involve power-law only in the variable $x_1$
\begin{eqnarray}
\rho_{st}(x_1,R) \oppropto_{\vert x_1 \vert \to + \infty}  \frac{1}{\vert x_1 \vert^{1+\mu} }
\label{kesten2Dx1R}
\end{eqnarray}
as will be further discussed in subsection \ref{subsec_ArgumentMU}
after describing other examples involving analogous power-laws.

 
  \subsection{ Models with one square-root noise $dB_j^{square-root}(t) $ 
  and one multiplicative noise $dB_j^{multiplicative}(t) $ for each coordinate $x_j>0$ having  
 ${\bold M}= {\bold M}^{[-1]}+{\bold M}^{[0]} $}
 
  \label{subsec_twosquarerootmulti2D}
  
   The case $d=2$ of Eq. \ref{ItoSDEsinglegjtwoSquareRootMulti}
that generalizes the one-dimensional case of Eq. \ref{ItoSDEsinglegjtwoSquareRootMulti1D}
correspond to the following Ito system for the two positive processes $x_1(t)>0$ and $x_2(t)>0$
   \begin{eqnarray}
dx_1(t)   && =  \left( F^{[0]}_1 +  F^{[1]}_{11} x_1(t)+  F^{[1]}_{12} x_2(t)
\right) dt + \sqrt{ 2 D^{[1]}_1 x_1(t)} dB_1^{square-root}(t) 
+  \sqrt{ 2 D^{[2]}_1 } x_1(t) dB_1^{multiplicative}(t)
\nonumber \\
dx_2(t)   && =  \left( F^{[0]}_2 +  F^{[1]}_{21} x_1(t)+  F^{[1]}_{22} x_2(t)
\right) dt + \sqrt{ 2 D^{[1]}_2 x_2(t)} dB_2^{square-root}(t)    
   +  \sqrt{ 2 D^{[2]}_2 } x_2(t) dB_2^{multiplicative}(t)
\label{ItoSDEsinglegjtwoSquareRootMulti2D}
\end{eqnarray}
with the same conditions as in Eq. \ref{ConditionsF1forSquareRootNoise2D}.

 Here again to understand the effects of the interactions
mediated by the off-diagonal coefficients $F^{[1]}_{12} $ and $F^{[1]}_{21} $,
the discussion is as follows :

$\bullet$ When $F^{[1]}_{12}=0$ and $ F^{[1]}_{21} =0$, then $x_1(t)$ and $x_2(t)$ are two independent 
one-dimensional Fisher-Snedecor processes that were discussed in subsection \ref{subsec_twosquarerootmulti1D}.
In particular in the region of parameters
\begin{eqnarray}
\mu_j && \equiv \ \frac{(-f^{[1]}_{jj})}{D^{[2]}_j } >0
\nonumber \\
\alpha_j &&\equiv \frac{ F^{[0]}_j}{ D^{[1]}_j}>0
\nonumber \\
 c_j &&  \equiv \frac{   D^{[1]}_j}{D^{[2]}_j} >0
\label{Notationsfisher2D}
\end{eqnarray}
the steady state $\rho_{st}(x_1,x_2)$ is factorized into two Fisher-Snedecor distributions of Eq. \ref{fisher}
of parameters $(\mu_j,\alpha_j,c_j)$
\begin{eqnarray}
\rho_{st}(x_1,x_2) = \prod_{j=1}^2 
\left( c_j^{\mu} \frac{\Gamma(\alpha_j+\mu_j)}{ \Gamma(\alpha_j) \Gamma(\mu_j) } \frac{ x^{\alpha_j-1} }{(c_j+x_j)^{\alpha_j+\mu_j} } \right)
\ \ \ {\rm } \ \ {\rm for } \ \ (x_1,x_2)  \in (]0,+\infty[)^2
\label{fisher2D}
\end{eqnarray}
with possibly different power-laws for large $x_1 \to + \infty$ and large $x_2 \to + \infty$ when $\mu_1 \ne \mu_2$.

 $\bullet$  When $F^{[1]}_{12} \ne 0$ and $ F^{[1]}_{21} \ne 0$, then both processes acting on each other
will produce a common exponent $\mu$ that will govern the tails for both $x_1$ and $x_2$
\begin{eqnarray}
\rho_{st}(x_1,x_2) \oppropto_{\vert x_j \vert \to + \infty}  \frac{1}{\vert x_j \vert^{1+\mu} }
\label{fisher2Dcommon}
\end{eqnarray}
that will be discussed further in subsection \ref{subsec_ArgumentMU}.


 \subsection{ Models with one additive-noise $dB_j^{additive}(t) $ and one multiplicative-noise $dB_j^{multiplicative}(t) $ for each $x_j$ having  
 ${\bold M}={\bold M}^{[-2]} +{\bold M}^{[0]} $ }
 
  \label{subsec_twoaddmulti2D}

  The case $d=2$ of Eq. \ref{ItoSDEsinglegjtwoaddmultiBiblock}
that generalizes the one-dimensional case of Eq. \ref{ItoSDEsinglegjtwoaddmulti1D}
correspond to the following Stratonovich system
   \begin{eqnarray}
   dx_1(t)   && = \left(  f^{[1]}_{11} x_1(t)+  f^{[1]}_{12} x_2(t)\right) dt 
+\sqrt{ 2 D^{[0]}_1 }dB_1^{additive}(t) +  \sqrt{ 2 D^{[2]}_1 } x_1(t) dB_1^{multiplicative}(t)
\nonumber \\
dx_2(t)   && = \left(  f^{[1]}_{21} x_1(t)+  f^{[1]}_{22} x_2(t)
\right) dt 
+\sqrt{ 2 D^{[0]}_2 }dB_2^{additive}(t) +  \sqrt{ 2 D^{[2]}_2 } x_2(t) dB_2^{multiplicative}(t)
\label{ItoSDEsinglegjtwoaddmultiBiblock2D}
\end{eqnarray}

 Here again to understand the effects of the interactions
mediated by the off-diagonal coefficients $f^{[1]}_{12} $ and $f^{[1]}_{21} $,
the discussion is as follows :

$\bullet$ When $f^{[1]}_{12}=0$ and $ f^{[1]}_{21} =0$, then $x_1(t)$ and $x_2(t)$ are two independent 
one-dimensional Student processes that were discussed in subsection \ref{subsec_twoaddmulti1D}.
In particular in the region of parameters $
\mu_j = \ \frac{(-f^{[1]}_{jj})}{D^{[2]}_j } >0$,
the steady state $\rho_{st}(x_1,x_2)$ is factorized into two Student distributions of Eq. \ref{student}
\begin{eqnarray}
\rho_{st}(x_1,x_2) = \prod_{j=1}^2 
\left( \sqrt{ \frac{D_j^{[2]}}{D_j^{[0]}}}  \frac{\Gamma(\frac{\mu_j+1}{2}) }
{\Gamma(\frac{1}{2})\Gamma(\frac{\mu_j}{2}) \left( 1+\frac{D_j^{[2]}}{D_j^{[0]}} x^2 \right)^{\frac{1+\mu_j}{2}}} \right)
\ \ \ {\rm } \ \ {\rm for } \ \ (x_1,x_2)  \in (]-\infty,+\infty[)^2
\label{student2D}
\end{eqnarray}
with possibly different power-laws for large $x_1 \to + \infty$ and large $x_2 \to + \infty$ when $\mu_1 \ne \mu_2$. 
 
 $\bullet$ When $f^{[1]}_{12} \ne 0$ and $ f^{[1]}_{21} \ne 0$, then both processes acting on each other
will produce a common exponent $\mu$ that will govern the tails for both $x_1$ and $x_2$
\begin{eqnarray}
\rho_{st}(x_1,x_2) \oppropto_{\vert x_j \vert \to + \infty}  \frac{1}{\vert x_j \vert^{1+\mu} }
\label{student2Dcommon}
\end{eqnarray}
Then to make the link with subsection \ref{subsec_Ricatti} concerning the dynamics of Eq. \ref{ItoSDEsinglegjmultiplicativelower2D}
without the additive noises, it is useful to replace the process $x_2(t)$
by the ratio $R(t)= \frac{ x_2(t) }{x_1(t) }$ of Eq. \ref{Ricatti} 
to obtain the Stratonovich system obtained
from Eq. \ref{ItoSDEsinglegjtwoaddmultiBiblock2D}
 \begin{eqnarray}
 dx_1(t)   && = \left(  f^{[1]}_{11} x_1(t)+  f^{[1]}_{12} R(t) x_1(t)\right) dt 
+\sqrt{ 2 D^{[0]}_1 }dB_1^{additive}(t) +  \sqrt{ 2 D^{[2]}_1 } x_1(t) dB_1^{multiplicative}(t)
 \nonumber \\
dR(t) &&= \left[  f^{[1]}_{21} 
+ ( f^{[1]}_{22} - f^{[1]}_{11})  R(t)
- f^{[1]}_{12} R^2(t) \right] dt 
+  R(t) \left[ \sqrt{ 2 D^{[2]}_2 } dB^{multiplicative}_2(t) 
 -    \sqrt{ 2 D^{[2]}_1}  dB^{multiplicative}_1(t)  \right]
 \nonumber \\
 && + \frac{\sqrt{ 2 D^{[0]}_2 }dB_2^{additive}(t)}{x_1(t)} - R(t) \frac{\sqrt{ 2 D^{[0]}_1 }dB_1^{additive}(t)}{x_1(t)}
\label{RicattiStudent}
\end{eqnarray} 
In the dynamics of $R(t)$, the new terms with respect to single-variable SDE for $R(t)$ of Eq. \ref{RicattiSDE} 
discussed in detail in subsection \ref{subsec_Ricatti} are the contributions 
of the two additive noises with non-trivial amplitudes involving $x_1(t)$ in the denominator,
but that are not expected to change the fact that all the moments of $R(t)$ will converge 
towards finite steady values, so that 
the joint steady state $\rho_{st}(x_1,R)$ will involve power-law only in the variable $x_1$
\begin{eqnarray}
\rho_{st}(x_1,R) \oppropto_{\vert x_1 \vert \to + \infty}  \frac{1}{\vert x_1 \vert^{1+\mu} }
\label{student2Dx1R}
\end{eqnarray}
as will be further discussed in the next subsection.

 
 \subsection{ Relation between the power-law exponent $\mu$ and the Scaled Cumulant Generating Function ${\cal E}(n) $ of Eq. \ref{EnPhin}  }
 
 \label{subsec_ArgumentMU}
 
 As discussed in detail in subsection \ref{subsec_Ricatti} concerning the interacting case 
 $f^{[1]}_{12} \ne 0$ and $ f^{[1]}_{21} \ne 0$ when the Carleman matrix is block-diagonal 
 ${\bold M}={\bold M}^{[0]} $,
 in particular in subsection \ref{subsec_LargeDevLyapunov} concerning 
 the two finite-time Lyapunov exponents $\lambda_1(T)$ and $\lambda_2(T)$,
 the Carleman moments $m_T(n_1,n_2) $ of Eq. \ref{LyapunovTgeneratinglargedevn}
 are governed for large $T$
 \begin{eqnarray}
m_T(n_1,n_2)  \opsimeq_{T \to + \infty} e^{ T {\cal E}(n=n_1+n_2)}
\label{LyapunovTgeneratinglargedevnfor mu}
\end{eqnarray}
by the Scaled Cumulant Generating Function ${\cal E}(n=n_1+n_2) $
that depends only on the global degree $n=n_1+n_2$, 
and the value $n=\mu$ where 
${\cal E}(n=n_1+n_2) $ vanishes is useful in the case $\lambda_2(T=\infty) = \lambda_1(T=\infty)   <0 $
to separate the moments of order $n<\mu$ converging towards zero 
and the moments of order $n>\mu$ diverging towards infinity 
in Eq. \ref{DVmoments2D}

In the three models of the three previous subsections, the additional contributions
with respect to the block-diagonal ${\bold M}={\bold M}^{[0]} $ case
do not change the fact that the moments are diverging for $n>\mu$,
but they turn the vanishing moments $n<\mu$ of Eq. \ref{DVmoments2D} into finite steady values,
so that the parameter $\mu$ introduced via ${\cal E}(n=\mu)  = 0 $ in Eq. \ref{Enmuzero} of the previous section
is the power-law exponent that appear in the steady distributions 
of Eq. \ref{kesten2Dx1R} for the 2D Kesten process,
of Eq. \ref{fisher2Dcommon} for the 2D Fisher-Snedecor process,
of Eq. \ref{student2Dx1R} for the 2D Student process.

Related models formulated in discrete time are discussed in detail in the recent works 
 \cite{Sornette_EigenvectorGeometry,Sornette_DiscreteTimeKesten}
 both for $d=2$ and for large dimension $d \to + \infty$.


 \section{ Conclusion }
 
 \label{sec_conclusion}
 
  In summary, we have described how the Carleman approach can be applied to diffusion processes $[x_1(t),..,x_d(t)]$ in dimension $d$ when the forces and the diffusion-matrix elements are polynomials, in order to write the linear system governing the dynamics of the averaged values ${\mathbb E} ( x_1^{n_1}(t) x_2^{n_2}(t) ... x_d^{n_d}(t) )$ labelled by the $d$ integers $(n_1,..,n_d)$. We have explained why the natural decomposition of the Carleman matrix into blocks associated to the global degree $n=n_1+n_2+..+n_d$ is very useful to identify the models that have the simplest spectral decompositions in the bi-orthogonal basis of right and left eigenvectors. We have discussed in detail the application to models with a single noise per coordinate, that can be either additive or multiplicative or square-root, or with two types of noises per coordinate, with many examples in dimensions $d=1,2$. In dimension $d=1$, we have described how the Carleman dynamics of the moments ${\mathbb E} ( x^{n}(t) )$ 
  labelled by the integer $n$ is able  
to recover the well-known simplest soluble diffusions: 
 the Carleman matrix is diagonal for the Geometric Brownian motion, while it is lower-triangular for the family of Pearson diffusions containing the Ornstein-Uhlenbeck and the Square-Root processes, as well as the Kesten, the Fisher-Snedecor and the Student processes that converge towards steady states with power-law-tails. 
  We have then focused on the dimension $d=2$ where the Carleman matrix governing the dynamics of the correlations ${\mathbb E} ( x_1^{n_1}(t) x_2^{n_2}(t) )$ has a natural decomposition into blocks associated to the global degree $n=n_1+n_2$, and we have discussed in detail the simplest models where the Carleman matrix is either block-diagonal or block-lower-triangular or block-upper-triangular.


\appendix

\section{ Simplest deterministic example in dimension $d=1$ to illustrate the Carleman approach } 

\label{app_deterministic}

In this Appendix, the Carleman approach is illustrated with the one-dimensional deterministic dynamics 
\begin{eqnarray}
\frac{ dx(t)}{dt}  = F^{[1]}x(t) +F^{[2]}x^2(t) = x(t) \left[ F^{[1]} +F^{[2]}x(t) \right]
\label{xquad}
\end{eqnarray}
In the field of population dynamics, the linear coefficient $ F^{[1]} >0 $ represents the growth rate
that would lead to the exponential growth as $x(t) = e^{t F^{[1]} } x(0) $ if it were alone,
while the quadratic coefficient $F^{[2]} <0 $ represents the saturation mechanism
that will produce the finite fixed point
\begin{eqnarray}
x_*  =  - \frac{F^{[1]}}{F^{[2]}}
\label{fixedPoint}
\end{eqnarray}

Let us first write the direct solution before describing the Carleman method.

\subsection{ Explicit solution for $x(t)$ via direct integration}

\label{subsec_deterDirect}

The new variable
\begin{eqnarray}
 z(t)\equiv \frac{1}{x(t)}
\label{zinversex}
\end{eqnarray}
satisfies the linear dynamics
\begin{eqnarray}
\frac{ dz(t)}{dt}  =- \frac{ \frac{ dx(t)}{dt} }{x^2(t)} = - \frac{ F^{[1]}x(t) +F^{[2]}x^2(t) }{x^2(t)}  
 = -F^{[1]}z(t) - F^{[2]} 
\label{xquadz}
\end{eqnarray}
that can be integrated 
\begin{eqnarray}
z(t)=  e^{- F^{[1]}t} \left( z(0) +\frac{F^{[2]}}{F^{[1]}} \right) - \frac{F^{[2]}}{F^{[1]}}
\label{xquadzinteg}
\end{eqnarray}
to obtain that $z(t)$ converges towards its fixed point $z_*  =   \frac{1}{x_*} =- \frac{F^{[2]}}{F^{[1]}} $
exponentially in time with the relaxation rate $F^{[1]} $.

The solution for the variable $x(t)=\frac{1}{z(t)}$ reads in terms of the initial condition $x(0)=\frac{1}{z(0)} $
and in terms of the fixed point $x_*  =  - \frac{F^{[1]}}{F^{[2]}} $ of Eq. \ref{fixedPoint}
\begin{eqnarray}
x(t)= \frac{1}{z(t)}= \frac{1}{   \frac{1}{x_*} +  e^{- F^{[1]}t} \left( \frac{1}{x(0)} - \frac{1}{x_*} \right)  }
\label{xquadsolu}
\end{eqnarray}
Let us now write two different series expansions of this explicit solution in the two following subsections.


\subsubsection{ Series expansion of the explicit solution $x(t)$ of Eq. \ref{xquadsolu}
to analyze the time-convergence towards the fixed-point $x_*$ }

The time-convergence of the solution $x(t)$ of Eq. \ref{xquadsolu} towards the fixed-point $x_*$
can be analyzed via the series expansion
\begin{eqnarray}
x(t)=  \frac{x_*}{ 1   -  e^{- F^{[1]}t} \left( 1-  \frac{x_*}{x(0)}  \right)  }
&& = x_* \sum_{p=0}^{+\infty} e^{- p F^{[1]}t} \left( 1-  \frac{x_*}{x(0)}  \right)^p
\nonumber \\
&& = x_* \left[ 1+e^{-  F^{[1]}t} \left( 1-  \frac{x_*}{x(0)}  \right)+ e^{- 2 F^{[1]}t} \left( 1-  \frac{x_*}{x(0)}  \right)^2 +... \right]
\label{xquadsoluSeriesTime}
\end{eqnarray}
that involves the exponentially-decaying factors $e^{- p F^{[1]}t} $ in time 
with the infinite series of relaxation rates $[p F^{[1]}]$ with $p \in {\mathbb N}^*$.


\subsubsection{Powers $x^k(t) $ of the explicit solution $x(t)$ of Eq. \ref{xquadsolu} in terms of the powers $x^q(0)$ of the initial condition $x(0)$ }

To make the link with the Carleman approach described in the next subsection,
one wishes to write
the powers $x^k(t)$ of the solution $x(t)$ of Eq. \ref{xquadsolu} 
in terms of the powers $x^{q}(0) $ of the initial condition $x(0)$
\begin{eqnarray}
x^k(t) && =  \frac{x^k(0) e^{k F^{[1]}t}}{ \left[ 1+  \frac{x(0)}{x_*}(e^{ F^{[1]}t} -1) \right]^k }
= x^k(0) e^{ k F^{[1]}t} \sum_{p=0}^{+\infty} \frac{(k-1+p)!}{(k-1)! p! } \left( - \frac{x(0)}{x_*}\right)^p (e^{ F^{[1]}t} -1)^p
\nonumber \\
&& =  x^k(0) e^{ kF^{[1]}t} \sum_{p=0}^{+\infty} (-1)^p \frac{(k-1+p)!}{(k-1)! p! } \left(  \frac{x(0)}{x_*}\right)^p 
\sum_{j=0}^p e^{ j F^{[1]}t} (-1)^{p-j} \frac{ p!}{j! (p-j)! }
\nonumber \\
&& =   \sum_{p=0}^{+\infty} x^{p+k}(0)  \frac{(k-1+p)!}{(k-1)!  x_*^p}  
\sum_{j=0}^p e^{ (k+j) F^{[1]}t}  \frac{ (-1)^{j}}{j! (p-j)! }
\label{xquadsoluSeriesX0calcul}
\end{eqnarray}
where it is convenient to replace $p$ by $q=p+k$ and $j$ by $n=k+j$ in the two sums to obtain 
the final result
\begin{eqnarray}
x^k(t) && =   \sum_{q=k}^{+\infty} x^{q}(0)  \frac{(q-1)!}{(k-1)!  x_*^{q-k}}  
\sum_{n=k}^{q} e^{ n F^{[1]}t}  \frac{(-1)^{n-k} }{(n-k)! (q-n)! }
\label{xquadsoluSeriesX0}
\end{eqnarray}
that will be useful to compare with the Carleman approach described in the next subsection.


\subsection{ Alternative solution via the Carleman approach and the spectral decomposition of the Carleman matrix}

\label{subsec_deterCarleman}

In the Carleman approach, the one-dimensional dynamics of Eq. \ref{xquad} with quadratic non-linearity
translates in the infinite-dimensional space of all the integer powers $x^n(t)$ with $n \in {\mathbf N}$
into the linear dynamics
\begin{eqnarray}
\frac{ dx^n(t)}{dt} && = n x^{n-1}(t) \frac{ dx(t)}{dt} =n x^{n-1}(t) \left( F^{[1]}x(t) +F^{[2]}x^2(t) \right) 
\nonumber \\
&& = n F^{[1]} x^n(t) + n F^{[2]} x^{n+1}(t) \equiv \sum_{k=0}^{+\infty} U_{n,k} x^k(t)
\label{xquadn}
\end{eqnarray}
where the Carleman matrix ${\bold U}$ is upper-bidiagonal 
with the matrix elements
\begin{eqnarray}
U_{n,k} = n  F^{[1]} \delta_{k,n}   + n F^{[2]} \delta_{k,n+1}
\label{Udeterupper}
\end{eqnarray}

So the solution of the linear dynamics of Eq. \ref{xquadn}
\begin{eqnarray}
x^k(t) = \sum_{q=0}^{+\infty} \langle k \vert e^{t {\bold U} } \vert q \rangle  x^q(0)
\label{xquadnPropagator}
\end{eqnarray}
can be written in terms of the explicit spectral decomposition of Eq. \ref{Uupperbidiagspectral}
derived in subsection \ref{app_upperBidiag} of Appendix \ref{app_SpectralMatriceBiDiag}
\begin{eqnarray}
\langle k \vert e^{ t {\bold U} } \vert q \rangle
= \sum_{n=0}^{+\infty} e^{t e_n } \langle k  \vert r_{n} \rangle \langle l_n \vert q \rangle
\label{Uupperbidiagspectraldeter}
\end{eqnarray}
with the following notations :

(i) the eigenvalues $e_n$ of the upper-bidiagonal matrix ${\bold U}$ are directly given by the diagonal matrix elements $U_{n,k} = n  F^{[1]} $ of Eq. \ref{Udeterupper}
\begin{eqnarray}
e_n = U_{n,n} = n  F^{[1]} 
\label{Udeteruppereigen}
\end{eqnarray}

(ii) the right eigenvectors of Eq. \ref{RightUpperBi} read using Eq. \ref{Udeterupper}
and the notation $x_*  =  - \frac{F^{[1]}}{F^{[2]}}$ of Eq. \ref{fixedPoint} 
\begin{eqnarray}
\langle k \vert r_n \rangle =r_n(k) = \begin{cases}
\displaystyle 
\prod_{j=k}^{n-1} \frac{ U_{j,j+1} }{e_n-e_j }
=\prod_{j=k}^{n-1} \frac{  j F^{[2]} }{ (n-j) F^{[1]} }
= \left( - \frac{  1  }{  x_* } \right)^{n-k} \frac{ (n-1)!}{(k-1)! (n-k)! }
  \ \ \ \text{  for } \ \ 0 \leq k \leq n-1
\\
1   \text{ for } \ \ k=n
\\
0   \text{  for } \ \ k>n 
\end{cases}
\label{RightUpperBideter}
\end{eqnarray}

(iii) the left eigenvectors of Eq. \ref{xquadLrec} read using Eq. \ref{Udeterupper}
and the notation $x_*  =  - \frac{F^{[1]}}{F^{[2]}}$ of Eq. \ref{fixedPoint} 
\begin{eqnarray}
 \langle l_n \vert q \rangle = l_n^*(q) 
 = \begin{cases}
 0   \text{ for } \ \ q<n
\\
1   \text{  for } \ \ q=n 
\\
 \displaystyle   \prod_{j=n+1}^q \frac{ U_{j-1,j} }{e_n-e_j} 
 =  \prod_{j=n+1}^q \frac{  (j-1)  F^{[2]} }{  (n-j) F^{[1]} } 
= \left(  \frac{  1  }{  x_* } \right)^{q-n} \frac{ (q-1)!}{(n-1)! (q-n)! }
 \ \ \text{ for } \ \ q>n
 \end{cases}
\label{xquadLrecupper}
\end{eqnarray}

Putting everything together, Eq. \ref{xquadnPropagator} becomes
\begin{eqnarray}
x^k(t) && = \sum_{q=0}^{+\infty} 
\left( \sum_{n=0}^{+\infty} e^{t e_n } \langle k  \vert r_{n} \rangle \langle l_n \vert q \rangle  \right) x^q(0)
=  \sum_{q=k}^{+\infty} x^q(0)
 \sum_{n=k}^{q} e^{t e_n } \langle k  \vert r_{n} \rangle \langle l_n \vert q \rangle   
 \nonumber \\
 && =  \sum_{q=k}^{+\infty} x^q(0)
 \sum_{n=k}^{q} e^{t n F^{[1]} } \left[ \left( - \frac{  1  }{  x_* } \right)^{n-k} \frac{ (n-1)!}{(k-1)! (n-k)! } \right]
 \left[ \left(  \frac{  1  }{  x_* } \right)^{q-n} \frac{ (q-1)!}{(n-1)! (q-n)! } \right]
  \nonumber \\
 && =  \sum_{q=k}^{+\infty} x^q(0)  \frac{(q-1)!}{(k-1)! x_*^{q-k}}
 \sum_{n=k}^{q} e^{t n F^{[1]} }  
   \frac{(-1)^{n-k}  }{ (n-k)! (q-n)! } 
\label{xkdeter}
\end{eqnarray}
that coincides with the calculation of Eq. \ref{xquadsoluSeriesX0}
based on the direct exact solution.

\subsection{ Discussion }

This simple deterministic example is thus useful to see how the Carleman approach described in subsection \ref{subsec_deterCarleman}
is in agreement with the direct calculation described in subsection \ref{subsec_deterDirect}.


\section{ Explicit spectral decomposition of bidiagonal matrices }

\label{app_SpectralMatriceBiDiag}

In this Appendix, we write the explicit spectral decompositions of bidiagonal matrices
that are useful in various sections of the paper.

\subsection{ Case of upper-bidiagonal matrices ${\bold U}$    }

\label{app_upperBidiag}

In this subsection, we focus on the upper-bidiagonal matrix 
${\bold U} $ with matrix elements for $(n,k) \in {\mathbb N}^2$
\begin{eqnarray}
\langle n \vert {\bold U}\vert k \rangle =  U_{n,n} \delta_{k,n} + U_{n,n+1} \delta_{k,n+1} = e_n \delta_{k,n} + U_{n,n+1} \delta_{k,n+1}
\label{Uupperbidiag}
\end{eqnarray}
where the diagonal elements $U_{n,n} $ giving the possible eigenvalues $e_n = U_{n,n}  $
will be assumed to be all different
in order to write the spectral decomposition 
\begin{eqnarray}
 {\bold U} = \sum_{n=0}^{+\infty} e_n \vert r_{n} \rangle \langle l_n \vert
\label{Uupperbidiagspectral}
\end{eqnarray}
in terms of the right and left eigenvectors 
\begin{eqnarray}
{\bold U} \vert r_n \rangle && = e_n \vert r_n \rangle
\nonumber \\
\langle l_n \vert {\bold U}  && = e_n \langle l_n \vert
\label{UbidiagEigen}
\end{eqnarray}
that form a bi-orthogonal basis
\begin{eqnarray}
 \langle l_n \vert r_{n'} \rangle = \delta_{n,n'}
\nonumber \\
{\bold 1}=\sum_{n=0}^{+\infty}  \vert r_n \rangle \langle l_n \vert
\label{BiorthogBidiag}
\end{eqnarray}

Let us now focus on a given eigenvalue $e_n$
in order to compute the corresponding right and left eigenvectors 
in the two following subsections. 


\subsubsection{ Computation of the right eigenvector $\vert r_n \rangle$ associated to the eigenvalue $e_n = U_{n,n}  $ }

The eigenvalue Eq. \ref{UbidiagEigen}
 for  right eigenvector can be projected on the basis $\langle k \vert  $ to obtain
\begin{eqnarray}
e_n \langle k \vert r_n \rangle= \sum_{q=0}^{+\infty} \langle k \vert {\bold U}\vert q \rangle \langle q \vert r_n \rangle
= e_{k} \langle k \vert r_n \rangle + U_{k,k+1} \langle k+1 \vert r_n \rangle
\label{EigenRbi}
\end{eqnarray}
i.e. more explicitly for the components $r_n(k)=\langle k \vert r_n \rangle  $ with $k=0,1,2,..$
\begin{eqnarray}
k=0 : \ \ \ \ \ \ \ e_n r_n(0) && = e_0 r_n(0) + U_{0,1} r_n(1)
\nonumber \\
k=1 : \ \ \ \ \ \ \ e_n r_n(1) && = e_1 r_n(1) + U_{1,2} r_n(2)
\nonumber \\
...
\nonumber \\
k=n-1 : \ \ \ e_n r_n(n-1) &&= e_{n-1} r_n(n-1) + U_{n-1,n} r_n(n)
\nonumber \\
k=n : \ \ \ \ \ \ \ \ e_n r_n(n) &&= e_n r_n(n) + U_{n,n+1} r_n(n+1)
\nonumber \\
k=n+1 : \ \ \ e_n r_n(n+1) && = e_{n+1} r_n(n+1) + U_{n+1,n+2} r_n(n+2)
\nonumber \\
k=n+2 : \ \ \ e_n r_n(n+2) && = e_{n+2} r_n(n+2) + U_{n+2,n+3} r_n(n+3)
\nonumber \\
...
\label{EigenRbiExpli}
\end{eqnarray}

The equation for $k=n$ yields that $r_n(n+1)=0$, and the equations for $k>n$ then yield iteratively
that all the components for $k \geq n+1$ also vanish
\begin{eqnarray}
 r_n(k) =  0 \ \ \text{ for $k >n$ }
\label{Rnkvanish}
\end{eqnarray}
One can then choose the following normalization for the component $k=n$
\begin{eqnarray}
 r_n(k=n) =  1
\label{Rnnchoice}
\end{eqnarray}
and compute the other components for $k=n-1,n-2,..,0$ via the recurrence 
\begin{eqnarray}
 r_n(k) =  \frac{ U_{k,k+1} }{ e_n-e_k}r_n(k+1) 
\label{xquadRrec}
\end{eqnarray}
to obtain the final result 
\begin{eqnarray}
\langle k \vert r_n \rangle =r_n(k) = \begin{cases}
\displaystyle \prod_{j=k}^{n-1} \frac{ U_{j,j+1} }{e_n-e_j }  \ \ \ \text{  for } \ \ 0 \leq k \leq n-1
\\
1   \text{ for } \ \ k=n
\\
0   \text{  for } \ \ k>n 
\end{cases}
\label{RightUpperBi}
\end{eqnarray}


\subsubsection{ Computation of the left eigenvector $\langle l_n \vert
 $ associated to the eigenvalue $e_n = U_{n,n}  $ }

The eigenvalue Eq. \ref{UbidiagEigen} for left eigenvector $\langle l_n \vert $ 
an be projected onto the basis $\vert q \rangle$ to obtain
\begin{eqnarray}
e_n \langle l_n \vert q \rangle = \sum_{k=0}^{+\infty}  \langle l_n \vert k \rangle \langle k \vert {\bold U}\vert q \rangle  = \langle l_n \vert q \rangle e_q  + \langle l_n \vert q-1 \rangle U_{q-1,q} 
\label{EigenLbi}
\end{eqnarray}
i.e. more explicitly for the components $ \langle l_n \vert q \rangle = l_n^*(q)$ with $q=0,1,2,...$
\begin{eqnarray}
q=0 : \ \ \ \ \ \ \ e_n l_n^*(0) && = l_n^*(0) e_0  + 0
\nonumber \\
q=1 : \ \ \ \ \ \ \ e_n l_n^*(1) &&= l_n^*(1) e_1  + l_n^*(0) U_{0,1} 
\nonumber \\
...
\nonumber \\
q=n-1 : \ \ \ e_n l_n^*(n-1) &&= l_n^*(n-1) e_{n-1}  + l_n^*(n-2) U_{n-2,n-1} 
\nonumber \\
q=n : \ \ \ \ \ \ \ \ e_n l_n^*(n) &&= l_n^*(n) e_n  + l_n^*(n-1) U_{n-1,n} 
\nonumber \\
q=n+1 : \ \ \ e_n l_n^*(n+1) &&= l_n^*(n+1) e_{n+1}  + l_n^*(n) U_{n,n+1} 
\nonumber \\
q=n+2 : \ \ \ e_n l_n^*(n+2) &&= l_n^*(n+2) e_{n+2}  + l_n^*(n+1) U_{n+1,n+2} 
\nonumber \\
...
\label{EigenLbiExpli}
\end{eqnarray}

The equation for $q=n$ yields that $l_n^*(n-1)=0$, and the equations for $q<n$ then yield iteratively
that all components for $q \leq n-1$ also vanish
\begin{eqnarray}
 l_n^*(q) =  0 \ \ \text{ for $0 \leq q <n$ }
\label{Lnkvanish}
\end{eqnarray}
The previous choice $r_n(n)=1 $ of Eq. \ref{Rnnchoice} and the normalization of 
Eq. \ref{BiorthogBidiag} 
\begin{eqnarray}
1= \langle l_n \vert r_n \rangle = \sum_{q=0}^{+\infty}  \langle l_n \vert q \rangle \langle q \vert r_n \rangle
 = \langle l_n \vert n \rangle \langle n \vert r_n \rangle= l_n^*(n) r_n(n)= l_n^*(n)
 \label{BiorthogBidiagcheck}
\end{eqnarray}
determines the component $ l_n^*(q=n) =1$,
while the components for $q>n$ can be then computed via the recurrence
\begin{eqnarray}
 l_n^*(q) =  l_n^*(q-1) \frac{ U_{q-1,q} }{e_n-e_q}
\label{EigenLbiRec}
\end{eqnarray}
to obtain the final result
\begin{eqnarray}
 \langle l_n \vert q \rangle = l_n^*(q) 
 = \begin{cases}
 0   \text{ for } \ \ q<n
\\
1   \text{  for } \ \ q=n 
\\
 \displaystyle   \prod_{j=n+1}^q \frac{ U_{j-1,j} }{e_n-e_j} \ \ \text{ for } \ \ q>n
 \end{cases}
\label{xquadLrec}
\end{eqnarray}


\subsection{ Spectral properties of lower-bidiagonal matrices ${\bold M}$}

\label{app_lowerBidiag}

To obtain the spectral properties of lower-bidiagonal matrices ${\bold M}$,
one can remake the computations as in the previous subsection,
but one can also obtain directly the results by considering that 
the lower-bidiagonal matrix ${\bold M}$ is the adjoint ${\bold U}^{\dagger}$
of a upper-bidiagonal matrix ${\bold U}$
\begin{eqnarray}
{\bold M} = {\bold U}^{\dagger}
\label{TUtranspose}
\end{eqnarray}
with the following correspondence with the matrix elements of Eq. \ref{Uupperbidiag}
\begin{eqnarray}
\langle k \vert {\bold M}\vert n \rangle
 = \langle n \vert {\bold U}\vert k \rangle^* && =  U_{n,n}^* \delta_{k,n} + U_{n,n+1}^* \delta_{k,n+1} 
\nonumber \\
&& \equiv M_{k,k} \delta_{n,k} + M_{k,k-1} \delta_{n,k-1} \ \ \ \text{ with $M_{k,k}= U_{k,k}^*$ and $M_{k,k-1}=U_{k-1,k}^* $ }
\label{UupperbidiagT}
\end{eqnarray}
and with the spectral decomposition of Eq. \ref{Uupperbidiagspectral}
\begin{eqnarray}
{\bold M}  = {\bold U}^{\dagger} && = \sum_{n=0}^{+\infty} e_n^* \vert l_n \rangle \langle r_n \vert
\nonumber \\
&& \equiv  \sum_{n=0}^{+\infty} E_n \vert R_n \rangle \langle L_n \vert
\label{UupperbidiagspectralT}
\end{eqnarray}
with the following conclusions :

(i) the eigenvalues $E_n$ of the lower-bidiagonal matrix ${\bold M}$ are given by the diagonal elements $M_{n,n}$
as expected and correspond to the complex conjugate of the eigenvalues $e_n$ of ${\bold U}$
\begin{eqnarray}
E_n=e_n^*=U_{n,n}^* = M_{n,n}
\label{UupperbidiagspectralTen}
\end{eqnarray}

(ii) the right eigenvector $ \vert R_n \rangle$ of ${\bold M} $ associated to $E_n$ correspond to 
$\vert l_n \rangle $ with the components obtained from Eq \ref{xquadLrec}
\begin{eqnarray}
\langle q \vert R_n \rangle = \langle q \vert l_n \rangle
 = l_n(q) 
 = \begin{cases}
 0   \text{ for } \ \ q<n
\\
1   \text{  for } \ \ q=n 
\\
 \displaystyle   \prod_{j=n+1}^q \frac{ U^*_{j-1,j} }{e_n^*-e_j^*} =\prod_{j=n+1}^q \frac{ M_{j,j-1} }{E_n-E_j}
 \ \ \text{ for } \ \ q>n
 \end{cases}
\label{xquadLrecT}
\end{eqnarray}

(iii) the left eigenvector $\langle L_n \vert $ of ${\bold M} $ associated to $E_n$ correspond to 
$\langle r_n \vert  $ with the components obtained from Eq \ref{RightUpperBi}
\begin{eqnarray}
\langle L_n \vert k \rangle = \langle r_n \vert k \rangle
 =r_n^*(k) = \begin{cases}
\displaystyle \prod_{j=k}^{n-1} \frac{ U_{j,j+1}^* }{e_n^*-e_j^* } = \prod_{j=k}^{n-1} \frac{ M_{j+1,j} }{E_n-E_j }
 \ \ \ \text{  for } \ \ 0 \leq k \leq n-1
\\
1   \text{ for } \ \ k=n
\\
0   \text{  for } \ \ k>n 
\end{cases}
\label{RightUpperBiT}
\end{eqnarray}


\section{ Explicit spectral decomposition of block-bidiagonal matrices}

\label{app_SpectralMatriceBiBLOCK}

In this Appendix, the goal is to generalize the previous Appendix \ref{app_SpectralMatriceBiDiag} to the case of 
block-bidiagonal matrices.
The diagonal-block ${\bold U}^{[0]} $ can be decomposed into its blocks ${\bold U}^{[0]}_{[n,n]} $ 
associated to a given degree $n$
\begin{eqnarray}
 {\bold U}^{[0]} && = \sum_{n=0}^{+\infty} {\bold U}^{[0]}_{[n,n]}
\label{U0blocknn}
\end{eqnarray}
and the spectral decomposition of each block ${\bold U}^{[0]}_{[n,n]} $
\begin{eqnarray}
  {\bold U}^{[0]}_{[n,n]} && = \sum_{\alpha=0}^n 
  e_{n,\alpha} \vert r^{[0]}_{n,\alpha}\rangle \langle l^{[0]}_{n,\alpha}\vert
\label{U0blockspectral}
\end{eqnarray}
is assumed to be known in terms of its eigenvalues $e_{n,\alpha} $ 
and the bi-orthogonal basis of its left eigenvectors $\langle l^{[0]}_{n,\alpha}\vert $ 
and right eigenvectors $\vert r^{[0]}_{n,\alpha}\rangle $.
[Note that this spectral decomposition of diagonal-block $ {\bold U}^{[0]}  $
 replaces the orthonormalized basis $\vert k \vert$ of the previous Appendix  \ref{app_SpectralMatriceBiDiag}].


\subsection{ Case of block-upper-bidiagonal matrices ${\bold U}= {\bold U}^{[0]}  +  {\bold U}^{[1]} $    }

\label{app_upperBiBLOCK}

In this subsection, we focus on the case of block-upper-bidiagonal matrices 
${\bold U}= {\bold U}^{[0]}  +  {\bold U}^{[1]} $
where the off-diagonal block ${\bold U}^{[1]}$ can be decomposed into the blocks ${\bold U}^{[1]}_{[n,n+1]} $
associated to the consecutive degrees $n$ and $(n+1)$
\begin{eqnarray}
 {\bold U}^{[1]} && = \sum_{n=0}^{+\infty} {\bold U}^{[1]}_{[n,n+1]}
\label{U1blockupper}
\end{eqnarray}
The eigenvalues of the block-matrix  ${\bold U}= {\bold U}^{[0]}  +  {\bold U}^{[1]} $
are the same as the eigenvalues $ e_{n,\alpha}$ of the diagonal block ${\bold U}^{[0]} $ of Eq. \ref{U0blockspectral},
while the corresponding right and left eigenvectors
\begin{eqnarray}
({\bold U}^{[0]} + {\bold U}^{[1]}) \vert r_{n,\alpha} \rangle && = e_{n,\alpha} \vert r_{n,\alpha} \rangle
\nonumber \\
\langle l_{n,\alpha} \vert ({\bold U}^{[0]} + {\bold U}^{[1]})  && = e_{n,\alpha} \langle l_{n,\alpha} \vert
\label{UblockbidiagEigen}
\end{eqnarray}
can be computed as explained in the following subsections.

 
 \subsubsection{ Perturbation theory for ${\bold U}(\epsilon) = {\bold U}^{[0]} +\epsilon {\bold U}^{[1]} $ 
  at all orders in $\epsilon$ }

 It is technically convenient to introduce a perturbation parameter $\epsilon \in [0,1]$ and to define the matrices
that have exactly the same block-diagonal part ${\bold U}^{[0]} $ 
but have a reduced block-off-diagonal part $ {\bold U}^{[1]}$
\begin{eqnarray}
{\bold U}(\epsilon) = {\bold U}^{[0]} +\epsilon {\bold U}^{[1]}
  \label{mseries}
\end{eqnarray}
in order to interpolate continuously between the block-diagonal case ${\bold U}(\epsilon=0)={\bold U}^{[0]} $ 
and the full upper-bidiagonal matrix ${\bold U}(\epsilon=1)={\bold U}^{[0]}+{\bold U}^{[1]}={\bold U}$ one is interested in.
The eigenvalues $e_{n,\alpha}$ of $ {\bold U}(\epsilon)$ are independent of $\epsilon$
while the corresponding right eigenvector $\vert r_{n,\alpha}(\epsilon) \rangle $ and left eigenvector 
$\langle  l_{n,\alpha}(\epsilon) \vert $ depend continuously on $\epsilon$
and can be expanded in power-series
\begin{eqnarray}
\vert  r_{n,\alpha}(\epsilon)  \rangle && = \sum_{p=0}^{+\infty} \epsilon^p \vert r_{n,\alpha}^{(p)}\rangle
\nonumber \\
\langle   l_{n,\alpha}(\epsilon) \vert && = \sum_{p=0}^{+\infty} \epsilon^p \langle  l_{n,\alpha}^{(p)}  \vert
  \label{rl0series}
\end{eqnarray}
around the unperturbed eigenvectors for $\epsilon=0$

Let us plug the series expansions of Eqs \ref{rl0series}
into 
the eigenvalue equation for the right eigenvector $\vert r_{n,\alpha} (\epsilon) \rangle $
\begin{eqnarray}
0  = \left( {\bold U}(\epsilon) - e_{n,\alpha}   \right) \vert r_{n,\alpha} (\epsilon) \rangle
&& = \left( {\bold U}^{[0]}- e_{n,\alpha}  +\epsilon {\bold U}^{[1]}    \right)  \sum_{p=0}^{+\infty} \epsilon^p \vert r_{n,\alpha}^{(p)}\rangle
\nonumber \\
&& = 
  \sum_{p=0}^{+\infty} \epsilon^p \left( {\bold U}^{[0]} - e_{n,\alpha}  \right) \vert r_{n,\alpha}^{(p)}\rangle
+  \sum_{p=0}^{+\infty} \epsilon^{p+1} {\bold U}^{[1]} \vert r_{n,\alpha}^{(p)}\rangle
\nonumber \\
&& =  \left( {\bold U}^{[0]} - e_{n,\alpha}  \right) \vert r_{n,\alpha}^{(0)}\rangle + \sum_{p=1}^{+\infty} \epsilon^p
\left[\left( {\bold U}^{[0]} - e_{n,\alpha}  \right) \vert r_{n,\alpha}^{(p)} \rangle + {\bold U}^{[1]}    \vert r_{n,\alpha}^{(p-1)} \rangle
\right]
  \label{powerright}
\end{eqnarray}
and into 
the eigenvalue equation for the left eigenvector $\langle  l_{n,\alpha} (\epsilon)\vert $
\begin{eqnarray}
0  = \langle  l_{n,\alpha} (\epsilon)\vert\left( {\bold U}(\epsilon) - e_{n,\alpha}   \right) 
&& =  \sum_{p=0}^{+\infty} \epsilon^p \langle  l_{n,\alpha}^{(p)}  \vert\left( {\bold U}^{[0]}- e_{n,\alpha}  +\epsilon {\bold U}^{[1]} \right)
\nonumber \\
&& =  \sum_{p=0}^{+\infty} \epsilon^p \langle  l_{n,\alpha}^{(p)}  \vert\left( {\bold U}^{[0]}- e_{n,\alpha}   \right)
+  \sum_{p=0}^{+\infty} \epsilon^{p+1} \langle  l_{n,\alpha}^{(p)}  \vert {\bold U}^{[1]} 
\nonumber \\
&& =\langle  l_{n,\alpha}^{(0)}  \vert\left( {\bold U}^{[0]}- e_{n,\alpha}   \right)+ \sum_{p=1}^{+\infty} \epsilon^p
\left[ \langle  l_{n,\alpha}^{(p)} \vert \left( {\bold U}^{[0]} - e_{n,\alpha}  \right)  +\langle  l_{n,\alpha}^{(p-1)} \vert   {\bold U}^{[1]}  
\right]
  \label{powerleft}
\end{eqnarray}

These eigenvalues equations are satisfied for the unperturbed case $\epsilon=0$,
while the vanishing at all orders $\epsilon^p$ leads to the following simple recurrences 
between consecutive orders for any $p=1,2,..$ 
\begin{eqnarray}
0  && = \left( {\bold U}^{[0]} - e_{n,\alpha}  \right) \vert r_{n,\alpha}^{(p)} \rangle + {\bold U}^{[1]}    \vert r_{n,\alpha}^{(p-1)} \rangle
\nonumber \\
0 && =  \langle  l_{n,\alpha}^{(p)} \vert \left( {\bold U}^{[0]} - e_{n,\alpha}  \right)  +\langle  l_{n,\alpha}^{(p-1)} \vert   {\bold U}^{[1]}  
  \label{powerrightp}
\end{eqnarray}
Note that these simplifications with respect to the usual perturbation theory familiar from quantum mechanics 
come from the fact that the eigenvalues $e_{n,\alpha} $ do not depend on the perturbation parameter $\epsilon$ here.

It is useful to introduce the inverse 
${\bold G}_{n,\alpha}^{(0)} $ of the operator $( e_{n,\alpha} - {\bold U}^{[0]} )$
in the subspace orthogonal to $\vert r_{n,\alpha}^{(0)} \rangle  \langle l_{n,\alpha}^{(0)}  \vert$
\begin{eqnarray}
{\bold G}_{n,\alpha}^{(0)} \equiv \left( \mathbb{1} - \vert r_{n,\alpha}^{(0)} \rangle  \langle l_{n,\alpha}^{(0)}\vert \right)
\frac{1 }{  e_{n,\alpha}  - {\bold U}^{[0]}  } 
\left( \mathbb{1} - \vert r_{n,\alpha}^{(0)} \rangle  \langle l_{n,\alpha}^{(0)} \vert \right) 
= \sum_{(k,\beta) \ne (n,\alpha)} \frac{\vert r_{k,\beta}^{(0)} \rangle  \langle l_{k,\beta}^{(0)} \vert  }{  e_{n,\alpha} -e_{k,\beta}  } 
  \label{green}
\end{eqnarray}
in order to rewrite the recurrences of Eq. \ref{powerrightp} as
 \begin{eqnarray}
 \vert   r_{n,\alpha}^{(p)} \rangle && = {\bold G}_{n,\alpha}^{(0)} {\bold U}^{[1]}    \vert r_{n,\alpha}^{(p-1)} \rangle  
\nonumber \\
 \langle  l_{n,\alpha}^{(p)} \vert &&  = \langle  l_{n,\alpha}^{(p-1)} \vert   {\bold U}^{[1]} {\bold G}_{n,\alpha}^{(0)}
    \label{recRLop}
\end{eqnarray}
so that the iteration up to the unperturbed eigenvectors $\vert r_{n,\alpha}^{(p=0)} \rangle $
and $\langle  l_{n,\alpha}^{(p=0)} $
 yields
 \begin{eqnarray}
 \vert   r_{n,\alpha}^{(p)} \rangle &&  
 = {\bold G}_{n,\alpha}^{(0)} {\bold U}^{[1]}  {\bold G}_{n,\alpha}^{(0)} {\bold U}^{[1]}   \vert r_{n,\alpha}^{(p-2)} \rangle
 = ... = \left( {\bold G}_{n,\alpha}^{(0)} {\bold U}^{[1]} \right)^p \vert r_{n,\alpha}^{(0)} \rangle
\nonumber \\
 \langle  l_{n,\alpha}^{(p)} \vert &&  
 =  \langle  l_{n,\alpha}^{(p-2)} \vert   {\bold U}^{[1]} {\bold G}_{n,\alpha}^{(0)}{\bold U}^{[1]} {\bold G}_{n,\alpha}^{(0)}
 = .. =  \langle  l_{n,\alpha}^{(0)} \vert  \left({\bold U}^{[1]}  {\bold G}_{n,\alpha}^{(0)}  \right)^p
    \label{recRLopIter}
\end{eqnarray}

Up to now the analysis is actually valid for any block-upper-triangular matrix.
In the next subsection, we take into account the 
specific form of Eq. \ref{U1blockupper} for the off-diagonal part $ {\bold U}^{[1]}$
to obtain more explicit expressions.

 
 \subsubsection{ Taking into account the specific form of off-diagonal part 
 $  {\bold U}^{[1]}  = \displaystyle \sum_{n=0}^{+\infty} {\bold U}^{[1]}_{[n,n+1]}$  }

Let us now take into account that the off-diagonal part $ {\bold U}^{[1]}$ of Eq. \ref{U1blockupper}
has non-vanishing matrix elements only between consecutive degrees $[n,n+1]$,
while the Green function ${\bold G}_{n,\alpha}^{(0)} $ of Eq. \ref{green}
has non-vanishing matrix elements only between coinciding degrees $[k,k]$.

As a consequence, the recurrences of Eq \ref{recRLop}
for $p=1$ 
 \begin{eqnarray}
 \vert   r_{n,\alpha}^{(1)} \rangle && = {\bold G}_{n,\alpha}^{(0)}
 {\bold U}^{[1]}    \vert r_{n,\alpha}^{(0)} \rangle  
 =  \left[\sum_{(k,\beta) \ne (n,\alpha)} \frac{\vert r_{k,\beta}^{(0)} \rangle  \langle l_{k,\beta}^{(0)} \vert  }{  e_{n,\alpha} -e_{k,\beta}  } 
 \right] {\bold U}^{[1]}    \vert r_{n,\alpha}^{(0)} \rangle 
 =   \sum_{\alpha_1}   \vert r_{n-1,\alpha_1}^{(0)} \rangle
 \frac{ \langle  l_{n-1,\alpha_1}^{(0)} \vert  {\bold U}^{[1]}    \vert r_{n,\alpha}^{(0)} \rangle  }{  e_{n,\alpha} -e_{n-1,\alpha_1}  }   
\nonumber \\
 \langle  l_{n,\alpha}^{(1)} \vert &&  = \langle  l_{n,\alpha}^{(0)} \vert   {\bold U}^{[1]} {\bold G}_{n,\alpha}^{(0)}
 = \langle  l_{n,\alpha}^{(0)} \vert   {\bold U}^{[1]}
  \left[\sum_{(k,\beta) \ne (n,\alpha)} \frac{\vert r_{k,\beta}^{(0)} \rangle  \langle l_{k,\beta}^{(0)} \vert  }{  e_{n,\alpha} -e_{k,\beta}  } 
 \right]
 = \sum_{\alpha_1} 
 \frac{ \langle  l_{n,\alpha}^{(0)} \vert   {\bold U}^{[1]}\vert r_{n+1,\alpha_1}^{(0)} \rangle}{  e_{n,\alpha} -e_{n+1,\alpha_1}  } 
 \langle  l_{n+1,\alpha_1}^{(0)} \vert  
    \label{recRLop1}
\end{eqnarray}
yields that the first correction $\vert   r_{n,\alpha}^{(1)} \rangle $ of the right eigenvector
is a linear combination of the unperturbed right eigenvectors $ \vert r_{n-1,\alpha_1}^{(0)} \rangle$
of the lower degree $(n-1)$, 
while the first correction $ \langle  l_{n,\alpha}^{(1)} \vert $ of the left eigenvector
is a linear combination of the unperturbed left eigenvectors $ \langle  l_{n+1,\alpha_1}^{(0)} \vert$
of the bigger degree $(n+1)$.

Plugging Eqs \ref{recRLop1}
into the recurrences of Eq \ref{recRLop}
for $p=2$ 
 \begin{eqnarray}
 \vert   r_{n,\alpha}^{(2)} \rangle && = {\bold G}_{n,\alpha}^{(0)} {\bold U}^{[1]}    \vert r_{n,\alpha}^{(1)} \rangle  
 =   \sum_{\alpha_1} {\bold G}_{n,\alpha}^{(0)} {\bold U}^{[1]}   \vert r_{n-1,\alpha_1}^{(0)} \rangle
 \frac{ \langle  l_{n-1,\alpha_1}^{(0)} \vert  {\bold U}^{[1]}    \vert r_{n,\alpha}^{(0)} \rangle  }{  e_{n,\alpha} -e_{n-1,\alpha_1}  }   
 \nonumber \\
 && = 
 \sum_{\alpha_2} \sum_{\alpha_1} \vert r_{n-2,\alpha_2}^{(0)} \rangle
 \left( \frac{ \langle  l_{n-2,\alpha_2}^{(0)} \vert {\bold U}^{[1]}   \vert r_{n-1,\alpha_1}^{(0)} \rangle }
 {e_{n,\alpha} -e_{n-2,\alpha_2}} \right)
\left( \frac{ \langle  l_{n-1,\alpha_1}^{(0)} \vert  {\bold U}^{[1]}    \vert r_{n,\alpha}^{(0)} \rangle  }
 {  e_{n,\alpha} -e_{n-1,\alpha_1}  }    \right)
\nonumber \\
 \langle  l_{n,\alpha}^{(2)} \vert &&  = \langle  l_{n,\alpha}^{(1)} \vert   {\bold U}^{[1]} {\bold G}_{n,\alpha}^{(0)}
 =  \sum_{\alpha_1} 
 \frac{ \langle  l_{n,\alpha}^{(0)} \vert   {\bold U}^{[1]}\vert r_{n+1,\alpha_1}^{(0)} \rangle}{  e_{n,\alpha} -e_{n+1,\alpha_1}  } 
 \langle  l_{n+1,\alpha_1}^{(0)} \vert    {\bold U}^{[1]} {\bold G}_{n,\alpha}^{(0)}
 \nonumber \\
 && = \sum_{\alpha_2} \sum_{\alpha_1} 
\left( \frac{ \langle  l_{n,\alpha}^{(0)} \vert   {\bold U}^{[1]}\vert r_{n+1,\alpha_1}^{(0)} \rangle}{  e_{n,\alpha} -e_{n+1,\alpha_1}  } \right)
\left( \frac{ \langle  l_{n+1,\alpha_1}^{(0)} \vert   {\bold U}^{[1]}\vert r_{n+2,\alpha_2}^{(0)} \rangle}{  e_{n,\alpha} -e_{n+2,\alpha_2}  } \right)
 \langle  l_{n+2,\alpha_2}^{(0)} \vert   
    \label{recRLop2}
\end{eqnarray}
yields that the second correction $\vert   r_{n,\alpha}^{(2)} \rangle $ of the right eigenvector
is a linear combination of the unperturbed right eigenvectors $ \vert r_{n-2,\alpha_2}^{(0)} \rangle$
of the lower degree $(n-2)$, 
while the second correction $ \langle  l_{n,\alpha}^{(2)} \vert $ of the left eigenvector
is a linear combination of the unperturbed left eigenvectors $ \langle  l_{n+2,\alpha_2}^{(0)} \vert$
of the bigger degree $(n+2)$.

Via iteration, one obtains the following results in terms of the notation $\alpha_0=\alpha$ :

(i) For $1 \leq p \leq n$, the correction  
$\vert   r_{n,\alpha_0}^{(p)} \rangle $ of the right eigenvector
is a linear combination of the unperturbed right eigenvectors $ \vert r_{n-p,\alpha_p}^{(0)} \rangle$
of the lower degree $(n-p)$ with the following coefficients
 \begin{eqnarray}
 \vert   r_{n,\alpha_0}^{(p)} \rangle  =  \sum_{\alpha_p} ...
 \sum_{\alpha_2} \sum_{\alpha_1} \vert r_{n-p,\alpha_p}^{(0)} \rangle
\left[  \prod_{j=1}^p
 \left( \frac{ \langle  l_{n-j,\alpha_j}^{(0)} \vert {\bold U}^{[1]}   \vert r_{n-j+1,\alpha_{j-1}}^{(0)} \rangle }
 {e_{n,\alpha_0} -e_{n-j,\alpha_j}} \right) \right]
    \label{REigenOrderp}
\end{eqnarray}
while there are no corrections of order $p >n$.

(ii) For $p \geq 1$, the correction $ \langle  l_{n,\alpha_0}^{(p)} \vert $ of the left eigenvector
is a linear combination of the unperturbed left eigenvectors $ \langle  l_{n+p,\alpha_p}^{(0)} \vert$
of the bigger degree $(n+p)$ with the following coefficients
 \begin{eqnarray}
 \langle  l_{n,\alpha_0}^{(p)} \vert && 
  =  \sum_{\alpha_p} ... \sum_{\alpha_2} \sum_{\alpha_1} 
\left[ \prod_{j=1}^p
\left( \frac{ \langle  l_{n+j-1,\alpha_{j-1}}^{(0)} \vert   {\bold U}^{[1]}\vert r_{n+j,\alpha_j}^{(0)} \rangle}
{  e_{n,\alpha_0} -e_{n+j,\alpha_j}  } \right) \right] 
 \langle  l_{n+p,\alpha_p}^{(0)} \vert   
    \label{LEigenOrderp}
\end{eqnarray}

In summary, the corrections of various orders $p$ that appear in
perturbation series of Eqs \ref{rl0series} have for non-vanishing coefficients 
in the different blocks $(n-p)$ and $(n+p)$ respectively for right and left eigenvectors.


\subsubsection{ Conclusion for the right and left eigenvectors of the block-upper-bidiagonal matrices ${\bold U}= {\bold U}^{[0]}  +  {\bold U}^{[1]} $  }

For $\epsilon=1$, the perturbation series of Eqs \ref{rl0series}
\begin{eqnarray}
\vert r_{n,\alpha}  \rangle=\vert r_{n,\alpha} (\epsilon=1) \rangle && 
=\vert r_{n,\alpha}^{(0)}\rangle + \sum_{p=1}^{+\infty}  \vert r_{n,\alpha}^{(p)}\rangle
\nonumber \\
\langle  l_{n,\alpha} \vert = \langle  l_{n,\alpha} (\epsilon=1) \vert && 
=  \langle  l_{n,\alpha}^{(0)}  \vert + \sum_{p=1}^{+\infty}  \langle  l_{n,\alpha}^{(p)}  \vert
  \label{rlfinaltriangular}
\end{eqnarray}
with the explicit results of Eqs \ref{REigenOrderp}
and \ref{LEigenOrderp}
leads to the following coefficients in the bi-orthogonal basis of unperturbed eigenvectors :

(i) the right eigenvectors $ \vert r_{n,\alpha_0} \rangle $ have for coefficients
\begin{eqnarray}
\langle  l_{k,\beta}^{(0)}  \vert r_{n,\alpha_0} \rangle  = \begin{cases}
\displaystyle \delta_{\beta,\alpha_{n-k}} \sum_{\alpha_{n-k-1}} ...
 \sum_{\alpha_2} \sum_{\alpha_1} 
\left[  \prod_{j=1}^{n-k}
 \left( \frac{ \langle  l_{n-j,\alpha_j}^{(0)} \vert {\bold U}^{[1]}   \vert r_{n-j+1,\alpha_{j-1}}^{(0)} \rangle }
 {e_{n,\alpha_0} -e_{n-j,\alpha_j}} \right) \right]  \ \ \ \text{  for } \ \ 0 \leq k \leq n-1
\\
\delta_{\beta,\alpha_0}    \text{ for } \ \ k=n
\\
0   \text{  for } \ \ k>n 
\end{cases}
\label{RightUpperBlock}
\end{eqnarray}

(ii)  the left eigenvectors $\langle  l_{n,\alpha} \vert $
have for coefficients
 \begin{eqnarray}
 \langle l_{n,\alpha_0} \vert r_{q,\beta}^{(0)} \rangle 
 = \begin{cases}
 0   \text{ for } \ \ q<n
\\
\delta_{\beta,\alpha_0}    \text{  for } \ \ q=n 
\\
 \displaystyle   \delta_{\beta,\alpha_{q-n}} \sum_{\alpha_{q-n-1}} ... \sum_{\alpha_2} \sum_{\alpha_1} 
\left[ \prod_{j=1}^{q-n}
\left( \frac{ \langle  l_{n+j-1,\alpha_{j-1}}^{(0)} \vert   {\bold U}^{[1]}\vert r_{n+j,\alpha_j}^{(0)} \rangle}
{  e_{n,\alpha_0} -e_{n+j,\alpha_j}  } \right) \right] 
 \ \ \text{ for } \ \ q>n
 \end{cases}
\label{LeftUpperBlock}
\end{eqnarray}

These formulas \ref{RightUpperBlock}
and \ref{LeftUpperBlock} concerning the block-upper-bidiagonal case
generalize the results of Eqs \ref{RightUpperBi}
and \ref{xquadLrec}
concerning the case of block-upper-bidiagonal matrices that are recovered when there are no internal structures in the blocks.

 

\subsection{ Case of block-lower-bidiagonal matrices ${\bold M}= {\bold M}^{[-1]} + {\bold M}^{[0]}    $    }

\label{app_lowerBiBLOCK}

As in Eq. \ref{TUtranspose} of the previous Appendix,
the results for block-lower-bidiagonal matrices ${\bold M}= {\bold M}^{[-1]} + {\bold M}^{[0]}    $
can be directly translated from the case of block-upper-bidiagonal matrices
${\bold U}= {\bold U}^{[0]}  +  {\bold U}^{[1]} $ considered in the previous subsection \ref{app_upperBiBLOCK}
by considering that ${\bold M}$ is the adjoint ${\bold U}^{\dagger}$
of ${\bold U}$
\begin{eqnarray}
{\bold M} && = {\bold U}^{\dagger} 
\nonumber \\
{\bold M}^{[0]} && =[ {\bold U}^{[0]}]^{\dagger} 
\nonumber \\
{\bold M}^{[-1]} && =[ {\bold U}^{[1]}]^{\dagger} 
\label{TUtransposeBlock}
\end{eqnarray}
The spectral decomposition of Eq. \ref{U0blockspectral} for the diagonal block ${\bold U}^{[0]}_{[n,n]} $
\begin{eqnarray}
{\bold M}^{[0]}_{[n,n]} = [{\bold U}^{[0]} ]^{\dagger} 
= \sum_{\alpha=0}^n 
  e_{n,\alpha}^* \vert l^{[0]}_{n,\alpha}\rangle \langle r^{[0]}_{n,\alpha}\vert
  \equiv
  \sum_{\alpha=0}^n 
  E_{n,\alpha} \vert R^{[0]}_{n,\alpha}\rangle \langle L^{[0]}_{n,\alpha}\vert
\label{TUzerotransposeBlockSpectral}
\end{eqnarray}
and the spectral decomposition of the full matrix ${\bold U}$
\begin{eqnarray}
{\bold M} = [{\bold U} ]^{\dagger} 
= \sum_{n=0}^{+\infty} \sum_{\alpha=0}^n 
  e_{n,\alpha}^* \vert l_{n,\alpha}\rangle \langle r_{n,\alpha}\vert
  \equiv
 \sum_{n=0}^{+\infty}  \sum_{\alpha=0}^n 
  E_{n,\alpha} \vert R_{n,\alpha}\rangle \langle L_{n,\alpha}\vert
\label{TUtransposeBlockSpectral}
\end{eqnarray}
lead to the following conclusions :

(i) the eigenvalues $E_{{\bold U}^{[1]}\alpha}$ of the ${\bold M}^{[0]}$ and of ${\bold M} $
are the complex conjugate of the eigenvalues $e_{n,\alpha}$ of ${\bold U}^{[0]} $ and of ${\bold U}$
\begin{eqnarray}
E_{n,\alpha} = e_{n,\alpha}^*
\label{Ecce}
\end{eqnarray}

(ii) the scalar products of the right eigenvector $ \vert R_{n,\alpha_0}\rangle=\vert l_{n,\alpha_0}\rangle$ 
with the unperturbed left eigenvectors $\langle L^{[0]}_{q,\beta} \vert =\langle r^{[0]}_{q,\beta}\vert $ 
are obtained from Eq. \ref{LeftUpperBlock}
 \begin{eqnarray}
 \langle L^{[0]}_{q,\beta} \vert R_{n,\alpha_0}\rangle =
\left( \langle l_{n,\alpha_0} \vert r_{q,\beta}^{(0)} \rangle \right)^*
 = \begin{cases}
 0   \text{ for } \ \ q<n
\\
\delta_{\beta,\alpha_0}    \text{  for } \ \ q=n 
\\
 \displaystyle   \delta_{\beta,\alpha_{q-n}} \sum_{\alpha_{q-n-1}} ... \sum_{\alpha_2} \sum_{\alpha_1} 
\left[ \prod_{j=1}^{q-n}
\left( \frac{ \langle r_{n+j,\alpha_j}^{(0)} \vert   {\bold M}^{[-1]}\vert l_{n+j-1,\alpha_{j-1}}^{(0)}
 \rangle}
{  E_{n,\alpha_0} -E_{n+j,\alpha_j}  } \right) \right] 
 \ \ \text{ for } \ \ q>n
 \end{cases}
\label{RightLowerBlock}
\end{eqnarray}

(iii) the scalar products of the left eigenvector $\langle L_{n,\alpha_0} \vert =\langle r_{n,\alpha_0}\vert $ 
with the unperturbed left eigenvectors $ \vert R^{[0]}_{k,\beta}\rangle=\vert l^{[0]}_{k,\beta}\rangle$
are obtained from Eq \ref{RightUpperBi}
\begin{eqnarray}
\langle L_{n,\alpha_0} \vert R^{[0]}_{k,\beta}\rangle =
\left(\langle  l_{k,\beta}^{(0)}  \vert r_{n,\alpha_0} \rangle \right)^*
 = \begin{cases}
\displaystyle \delta_{\beta,\alpha_{n-k}} \sum_{\alpha_{n-k-1}} ...
 \sum_{\alpha_2} \sum_{\alpha_1} 
\left[  \prod_{j=1}^{n-k}
 \left( \frac{ \langle r_{n-j+1,\alpha_{j-1}}^{(0)}
  \vert {\bold M}^{[-1]}   \vert  l_{n-j,\alpha_j}^{(0)}
   \rangle }
 {E_{n,\alpha_0} -E_{n-j,\alpha_j}} \right) \right]  \ \ \ \text{  for } \ \ 0 \leq k \leq n-1
\\
\delta_{\beta,\alpha_0}    \text{ for } \ \ k=n
\\
0   \text{  for } \ \ k>n 
\end{cases}
\label{LeftLowerBlock}
\end{eqnarray}

These formulas \ref{RightLowerBlock}
and \ref{LeftLowerBlock} concerning the block-lower-bidiagonal case
generalize the results of Eqs \ref{xquadLrecT}
and \ref{RightUpperBiT}
concerning the case of block-lower-bidiagonal matrices that are recovered when there are no internal structures in the blocks.


\section{ Diffusions in dimension $d$ with block-diagonal Carleman matrices ${\bold M}={\bold M}^{[0]} $}

\label{app_dimensiond}

In this Appendix, we focus on diffusions in dimension $d$ 
with block-diagonal Carleman matrices ${\bold M}={\bold M}^{[0]} $
in order to see if
the strategy described in subsection \ref{subsec_Ricatti} 
for the dimension $d=2$ can be applied to the 
Stratonovich system of Eq. \ref{ItoSDEsinglegjmultiplicativediag} in dimension $d>2$.

\subsection{ Dynamics for the $(d-1)$ ratios $R_j(t)=\frac{x_j(t)}{x_1(t)}$ for $j=2,..,d$ }

As in Eq. \ref{Ricatti} concerning the dimension $d=2$,
it is useful to replace the $(d-1)$ variables $x_j(t)$ by the $(d-1)$ ratios
 \begin{eqnarray}
R_j(t) \equiv \frac{ x_j(t) }{x_1(t) }  \in ]-\infty,+\infty[ \ \ \text{ with periodic B.C. at $(\pm \infty)$ }
\label{Ricattij}
\end{eqnarray} 
They satisfy the following Stratonovich system
 obtained from Eq. \ref{ItoSDEsinglegjmultiplicativediag}
 that does not depend on $x_1(t)$
 \begin{eqnarray}
dR_j(t)   && = \frac{\displaystyle \left( \sum_{i=1}^d f^{[1]}_{ji} x_i(t)
\right) dt + \sqrt{ 2  D^{[2]}_j } \  x_j(t) dB_j(t)}{x_1(t)} 
- R_j(t) \frac{ \displaystyle \left( \sum_{i=1}^d f^{[1]}_{1i} x_i(t)
\right) dt + \sqrt{ 2  D^{[2]}_1 } \  x_1(t) dB_1(t)}{x_1(t)}
\nonumber \\
&& =  \left(f^{[1]}_{j1}  - f^{[1]}_{11}R_j(t)
+ \sum_{i=2}^d f^{[1]}_{ji} R_i(t) 
 - R_j(t) \sum_{i=2}^d f^{[1]}_{1i} R_i(t) \right) dt
+ R_j(t) \left[ \sqrt{ 2  D^{[2]}_j }  dB_j(t)
- \sqrt{ 2  D^{[2]}_1}  dB_1(t) \right]
\nonumber \\
&& \equiv {\cal F}_j[R_2(t),..,R_d(t) ] dt  
+ \sum_{\alpha=1}^{d} {\cal G}_{j \alpha}[R_2(t),..,R_d(t) ] dB_{\alpha}(t)
\label{ItoSDEsinglegjmultiplicativediagdratio}
\end{eqnarray}
where the Stratonovich forces ${\cal F}_j[R_2(t),..,R_d(t) ] $ are polynomial of degree 2 
with respect to its $(d-1)$ variables
\begin{eqnarray}
{\cal F}_j[R_2(t),..,R_d(t) ]  && \equiv  f^{[1]}_{j1}  - f^{[1]}_{11}R_j(t)
+ \sum_{i=2}^d f^{[1]}_{ji} R_i(t) 
 - R_j(t) \sum_{i=2}^d f^{[1]}_{1i} R_i(t) 
\label{ItoSDEsinglegjmultiplicativediagdratioCALFORCE}
\end{eqnarray}
while the amplitudes of the $d$ noises $\alpha=1,..,d$ that appear in the SDE of $R_j(t)$ with $2 \leq j \leq d$
 \begin{eqnarray}
 {\cal G}_{j \alpha}[R_2(t),..,R_d(t) ]  = R_j \left[ \sqrt{ 2  D^{[2]}_j } \delta_{j,\alpha} - \sqrt{ 2  D^{[2]}_1} \delta_{1,\alpha}\right]
\label{GjaR}
\end{eqnarray}
produce the following symmetric
diffusion matrix ${\cal D}_{ij}(R_2,..,R_d)$ of Eq. \ref{Dij}
for $2 \leq i \leq d$ and $2 \leq j \leq d$
\begin{eqnarray}
{\cal D}_{ji}[R_2,..,R_d ]
&& \equiv \frac{1}{2} \sum_{\alpha=1}^{d} {\cal G}_{j \alpha}[R_j ]       {\cal G}_{i \alpha}[R_i ]
 = \frac{1}{2} R_j R_i \sum_{\alpha=1}^{d} 
 \left[ \sqrt{ 2  D^{[2]}_j } \delta_{j,\alpha} - \sqrt{ 2  D^{[2]}_1} \delta_{1,\alpha}\right]
 \left[ \sqrt{ 2  D^{[2]}_i } \delta_{i,\alpha} - \sqrt{ 2  D^{[2]}_1} \delta_{1,\alpha}\right]
  \nonumber \\
&& =  R_j R_i  
 \left(    D^{[2]}_j  \delta_{j,i} +    D^{[2]}_1 \right)
\label{DijR}
\end{eqnarray}
with its diagonal elements for $j=i$
\begin{eqnarray}
{\cal D}_{jj}[R_2,..,R_d ]
=  R_j^2   \left(    D^{[2]}_j   +    D^{[2]}_1 \right)
\label{DijRdiag}
\end{eqnarray}
and its of--diagonal elements for $j\ne i$
\begin{eqnarray}
{\cal D}_{j\ne i}[R_2,..,R_d ]
 =  R_j R_i      D^{[2]}_1 
\label{DijRoff}
\end{eqnarray}

The Ito forces ${\cal F}_j^{Ito}[R_2,..,R_d ]  $ that can be computed from the Stratonovich forces ${\cal F}_j[R_2,..,R_d ] $ 
 via Eq. \ref{StratoItoCorrespondance} using Eq. \ref{GjaR}
 \begin{eqnarray}
{\cal F}_j^{Ito}[R_2,..,R_d ] 
&& = {\cal F}_j[R_2,..,R_d ]
+ \frac{1}{2} \sum_{i=2}^d \sum_{\alpha=1}^{d} {\cal G}_{i \alpha}(R_i) \frac{\partial {\cal G}_{j \alpha}(R_j)}{\partial R_i} 
\nonumber \\
&& = {\cal F}_j[R_2,..,R_d ]
+ R_j
\left(    D^{[2]}_1 
+  D^{[2]}_j 
\right)
\label{StratoItoCorrespondanceR}
\end{eqnarray}
are useful to write the generator of Eq. \ref{Generator} 
\begin{eqnarray}
{\cal L}
&&  = \sum_{j=2}^d  {\cal F}^{Ito}_j[R_2,..,R_d ]  \frac{ \partial }{\partial R_j}
+  \sum_{j=2}^d  \sum_{i=2}^d {\cal D}_{i,j}[R_2,..,R_d ]  \frac{ \partial^2 }{\partial R_i \partial R_j}
\label{GeneratorRj}
\end{eqnarray}

As discussed arounf Eq. \ref{SteadycurrentR} concerning the dimension $d=2$, one expects that this dynamics
  will converge towards a steady state $\rho_{st}(R_2,..,R_d)$
satisfying
\begin{eqnarray}
0= {\cal L}^{\dagger} \rho_{st}[R_2,..,R_d ]
&&  = - \sum_{j=2}^d  \frac{ \partial }{\partial R_j} 
\left[ {\cal F}^{Ito}_j[R_2,..,R_d ] \rho_{st}((R_2,..,R_d))
-  \sum_{i=2}^d  \frac{ \partial }{\partial R_i } \left(  {\cal D}_{i,j}[R_2,..,R_d ] \rho_{st}[R_2,..,R_d ] \right) \right]
\nonumber \\
&& \equiv  - \sum_{j=2}^d  \frac{ \partial }{\partial R_j} {\cal J}^{st}_j[R_2,..,R_d ]
\label{GeneratorRdaggerj}
\end{eqnarray}
i.e. the corresponding steady current with $(d-1)$ components for $j=2,..,d$
\begin{eqnarray}
{\cal J}^{st}_j[R_2,..,R_d ] && =  {\cal F}^{Ito}_j[R_2,..,R_d ] \rho_{st}[R_2,..,R_d ]
-  \sum_{i=2}^d  \frac{ \partial }{\partial R_i } \left(  {\cal D}_{i,j}[R_2,..,R_d ] \rho_{st}[R_2,..,R_d ] \right) 
\nonumber \\
 && = \left(  {\cal F}^{Ito}_j[R_2,..,R_d ] -  \sum_{i=2}^d  \frac{ \partial {\cal D}_{i,j}[R_2,..,R_d ]}{\partial R_i } 
 \right) \rho_{st}[R_2,..,R_d ]
-  \sum_{i=2}^d {\cal D}_{i,j}[R_2,..,R_d ]  \frac{ \partial \rho_{st}[R_2,..,R_d ]}{\partial R_i }   
\label{currentRj}
\end{eqnarray}
should have a vanishing divergence.


\subsection{ Dynamics of $x_1(t)$ and its finite-time Lyapunov exponent $\lambda_1(T)$ }

 The remaining coordinate $x_1(t)$ satisfies the dynamics obtained from Eq. \ref{ItoSDEsinglegjmultiplicativediag}
 \begin{eqnarray}
\frac{ dx_1(t) }{x_1(t) }  && =  \left( f^{[1]}_{11} + \sum_{i=2}^d f^{[1]}_{1i} R_i(t)
\right) dt + \sqrt{ 2  D^{[2]}_1 } \  dB_1(t) 
\label{ItoSDEsinglegjmultiplicativediagx1Rj}
\end{eqnarray}
It is then useful to introduce the normalized noise 
 \begin{eqnarray}
  dB^{\perp}(t) = \frac{1}{\sqrt{\displaystyle  \sum_{l=1}^d \frac{1}{D^{[2]}_l}}} 
   \sum_{i=1}^d \frac{ dB_i(t)}{\sqrt{  D^{[2]}_i } }
\label{Borthogd}
\end{eqnarray}
that is uncorrelated with the $(d-1)$ linear combinations $ \left( \sqrt{   D^{[2]}_j }  dB_j(t)
- \sqrt{   D^{[2]}_1}  dB_1(t) \right) $ that appear in the SDE of the ratios $R_j(t)$
for $j=2,..,d$
 \begin{eqnarray}
{\mathbb E} \left[ dB_R^{\perp}(t) \left( \sqrt{   D^{[2]}_j }  dB_j(t)
- \sqrt{   D^{[2]}_1}  dB_1(t) \right) \right] = 0 \ \ \text{ for } \ \ j=2,..,d
 \label{DReffexplicheckorthod}
\end{eqnarray} 
Then one can rewrite Eq. \ref{Borthogd} as
 \begin{eqnarray}
\sqrt{  \sum_{l=1}^d \frac{1}{D^{[2]}_l}}  dB^{\perp}(t) 
&& =  \frac{ dB_1(t)}{\sqrt{  D^{[2]}_1 } }
+  \sum_{j=2}^d \frac{ \left( \sqrt{  D^{[2]}_j } dB_j(t) - \sqrt{   D^{[2]}_1}  dB_1(t)\right) +\sqrt{   D^{[2]}_1}  dB_1(t) }{  D^{[2]}_j  }
\nonumber \\
&& = dB_1(t) \left[  \frac{ 1}{\sqrt{  D^{[2]}_1 } }  
+ \sum_{j=2}^d \frac{ \sqrt{   D^{[2]}_1}   }{  D^{[2]}_j  }
\right] 
+  \sum_{j=2}^d \frac{ \left( \sqrt{  D^{[2]}_j } dB_j(t) - \sqrt{   D^{[2]}_1}  dB_1(t)\right)  }{  D^{[2]}_j  }
\nonumber \\
&& = dB_1(t) \left[  \sum_{j=1}^d \frac{ \sqrt{   D^{[2]}_1}   }{  D^{[2]}_j  }\right] 
+  \sum_{j=2}^d \frac{ dR_j(t)   - {\cal F}_j[R_2(t),..,R_d(t) ] dt   }{  R_j(t)  \sqrt{ 2 } D^{[2]}_j  }
\label{Borthogdrewriting}
\end{eqnarray}
where we have used the SDE of Eq. \ref{ItoSDEsinglegjmultiplicativediagdratio}.
One can then plug
 \begin{eqnarray}
 \sqrt{ 2  D^{[2]}_1} dB_1(t) 
 && = \frac{ \displaystyle \sqrt{ 2 \sum_{l=1}^d \frac{1}{D^{[2]}_l}}  dB^{\perp}(t)  - \sum_{j=2}^d \frac{ dR_j(t)   - {\cal F}_j[R_2(t),..,R_d(t) ] dt   }{  R_j(t)   D^{[2]}_j  }}
{ \displaystyle \sum_{i=1}^d \frac{ 1  }{  D^{[2]}_i  }}
\nonumber \\
&& =   \sqrt{ \frac{2 }{\displaystyle \sum_{l=1}^d \frac{1}{D^{[2]}_l}}}  dB^{\perp}(t)  
- \frac{  1}
{ \displaystyle \left[\sum_{i=1}^d \frac{ 1  }{  D^{[2]}_i  } \right]} \sum_{j=2}^d \frac{ dR_j(t)      }{  R_j(t)   D^{[2]}_j  }
+ \frac{ dt }
{ \displaystyle \left[ \sum_{i=1}^d \frac{ 1  }{  D^{[2]}_i  } \right] } \sum_{j=2}^d \frac{  {\cal F}_j[R_2(t),..,R_d(t) ]    }{  R_j(t)   D^{[2]}_j  }
\label{dB1inverse}
\end{eqnarray}
into Eq. \ref{ItoSDEsinglegjmultiplicativediagx1Rj} to obtain
 \begin{eqnarray}
\frac{ dx_1(t) }{x_1(t) }  && =  \left( f^{[1]}_{11} + \sum_{i=2}^d f^{[1]}_{1i} R_i(t)
+ \frac{ 1 }
{ \displaystyle \left[ \sum_{l=1}^d \frac{ 1  }{  D^{[2]}_l  } \right] } \sum_{j=2}^d \frac{  {\cal F}_j[R_2(t),..,R_d(t) ]    }{  R_j(t)   D^{[2]}_j  }\right) dt 
\nonumber \\
&& +\sqrt{ \frac{2 }{\displaystyle \sum_{l=1}^d \frac{1}{D^{[2]}_l}}}  dB^{\perp}(t)  
- \frac{  1}
{ \displaystyle \left[\sum_{i=1}^d \frac{ 1  }{  D^{[2]}_i  } \right]} \sum_{j=2}^d \frac{ dR_j(t)      }{  R_j(t)   D^{[2]}_j  }
\label{ItoSDEsinglegjmultiplicativediagx1RjdRcalcul}
\end{eqnarray}

Let us now use the explicit form of the Stratonovich forces ${\cal F}_j[R_2(t),..,R_d(t) ] $ of Eq. \ref{ItoSDEsinglegjmultiplicativediagdratio} to simplify the term in $dt$
 \begin{eqnarray}
&& \left[ \sum_{l=1}^d \frac{ 1  }{  D^{[2]}_l  } \right] \left( f^{[1]}_{11} + \sum_{i=2}^d f^{[1]}_{1i} R_i(t) \right)
+\sum_{j=2}^d \frac{ {\cal F}_j[R_2(t),..,R_d(t) ]  }{  D^{[2]}_j  R_j(t)}   
  \nonumber \\
&& =   \sum_{j=1}^d \frac{  f^{[1]}_{jj}  }{  D^{[2]}_j } 
 + \sum_{i=2}^d  \frac{ f^{[1]}_{1i}  }{  D^{[2]}_1  }R_i(t) 
 +\sum_{j=2}^d \left( \frac{ f^{[1]}_{j1} }{ D^{[2]}_j } \right) \frac{1}{R_j(t)}
+ \sum_{j=2}^d \sum_{i\ne 1,j} \left( \frac{f^{[1]}_{ji}}{  D^{[2]}_j }    \right)
 \frac{R_i(t) }{R_j(t)}
\label{SimpliOrderdt}
\end{eqnarray}
so that Eq. \ref{ItoSDEsinglegjmultiplicativediagx1RjdRcalcul}
becomes
 \begin{eqnarray}
\frac{ dx_1(t) }{x_1(t) }  && = 
 \frac{ 1 }
{ \displaystyle \left[ \sum_{l=1}^d \frac{ 1  }{  D^{[2]}_l  } \right] } 
\left(   \sum_{j=1}^d \frac{  f^{[1]}_{jj}  }{  D^{[2]}_j } 
 + \sum_{i=2}^d  \frac{ f^{[1]}_{1i}  }{  D^{[2]}_1  }R_i(t) 
 +\sum_{j=2}^d \left( \frac{ f^{[1]}_{j1} }{ D^{[2]}_j } \right) \frac{1}{R_j(t)}
+ \sum_{j=2}^d \sum_{i\ne 1,j} \left( \frac{f^{[1]}_{ji}}{  D^{[2]}_j }    \right)
 \frac{R_i(t) }{R_j(t)}\right) dt 
\nonumber \\
&& +\sqrt{ \frac{2 }{\displaystyle \sum_{l=1}^d \frac{1}{D^{[2]}_l}}}  dB^{\perp}(t)  
- \frac{  1}
{ \displaystyle \left[\sum_{i=1}^d \frac{ 1  }{  D^{[2]}_i  } \right]} \sum_{j=2}^d \frac{ dR_j(t)      }{  R_j(t)   D^{[2]}_j  }
\label{ItoSDEsinglegjmultiplicativediagx1RjdR}
\end{eqnarray}

The integration over the time interval $t \in [0,T]$ (with the same remark as after Eq. \ref{StratoSDE2DF1D2dynx1RBRetorthogRinteg})
yields that the finite-time Lyapunov exponent $\lambda_1(T)$ 
of Eq. \ref{lambdajT}
 \begin{eqnarray}
\lambda_1(T) && =  \frac{1}{T} \ln \left\vert \frac{x_1(T)} {x_1(0) } \right\vert   =  \frac{1}{T} \int_0^T \frac{ dx_1(t)}{x_1(t) }
\nonumber \\
&&  =  \frac{ \displaystyle \sum_{j=1}^d \frac{  f^{[1]}_{jj}  }{  D^{[2]}_j }  }
{ \displaystyle \left[ \sum_{l=1}^d \frac{ 1  }{  D^{[2]}_l  } \right] } 
  +\sqrt{ \frac{2 }{\displaystyle \sum_{l=1}^d \frac{1}{D^{[2]}_l}}}  B^{\perp}(t)  
- \frac{  1}
{ \displaystyle \left[\sum_{i=1}^d \frac{ 1  }{  D^{[2]}_i  } \right]} \sum_{j=2}^d \frac{ \ln \left\vert \frac{R(T)} {R(0) } \right\vert    }{     D^{[2]}_j  }
+ A_T  [R_2(0 \leq t \leq T);..;R_d(0 \leq t \leq T)]
\label{LyapunovDimd}
\end{eqnarray}
involves the following functional $A_T  [R_2(0 \leq t \leq T);..;R_d(0 \leq t \leq T)]$
of the stochastic trajectories of the ratios $ R_j(0 \leq t \leq T)$
 \begin{eqnarray}
&& A_T  [R_2(0 \leq t \leq T);..;R_d(0 \leq t \leq T)]
\nonumber \\
&& \equiv  \frac{1}{T} \int_0^T dt 
     \frac{ 1 }
{ \displaystyle \left[ \sum_{l=1}^d \frac{ 1  }{  D^{[2]}_l  } \right] } 
\left(   \sum_{i=2}^d  \frac{ f^{[1]}_{1i}  }{  D^{[2]}_1  }R_i(t) 
 +\sum_{j=2}^d \left( \frac{ f^{[1]}_{j1} }{ D^{[2]}_j } \right) \frac{1}{R_j(t)}
+ \sum_{j=2}^d \sum_{i\ne 1,j} \left( \frac{f^{[1]}_{ji}}{  D^{[2]}_j }    \right)
 \frac{R_i(t) }{R_j(t)}\right)   
\label{defTimeAveraged}
\end{eqnarray}

As in Eq. \ref{ergodicForLyapunov},
this time-averaging over the time-window $[0,T]$
can be thus evaluated for large $T \to + \infty$ by an average over the steady state $\rho_{st}(R_2,..,R_d) $
 \begin{eqnarray}
&& A_T  [R_2(0 \leq t \leq T);..;R_d(0 \leq t \leq T)] 
\nonumber \\
&& \opsimeq_{T \to + \infty}
\int_{-\infty}^{+\infty} dR_2 .. \int_{-\infty}^{+\infty} dR_d \rho_{st}[R_2,..,R_d]   
 \frac{ 1 }
{ \displaystyle \left[ \sum_{l=1}^d \frac{ 1  }{  D^{[2]}_l  } \right] } 
\left(   \sum_{i=2}^d  \frac{ f^{[1]}_{1i}  }{  D^{[2]}_1  }R_i 
 +\sum_{j=2}^d \left( \frac{ f^{[1]}_{j1} }{ D^{[2]}_j } \right) \frac{1}{R_j}
+ \sum_{j=2}^d \sum_{i\ne 1,j} \left( \frac{f^{[1]}_{ji}}{  D^{[2]}_j }    \right)
 \frac{R_i }{R_j}\right)   
 \nonumber \\
&&  \equiv  A(T=\infty)
\label{ergodicForLyapunovd}
\end{eqnarray}

As a consequence, the asymptotic value for $T=+\infty$ of the Lyapunov exponent of Eq. \ref{LyapunovDimd}
 \begin{eqnarray}
\lambda_1(T=\infty) &&   =  \frac{ \displaystyle \sum_{j=1}^d \frac{  f^{[1]}_{jj}  }{  D^{[2]}_j }  }
{ \displaystyle \left[ \sum_{l=1}^d \frac{ 1  }{  D^{[2]}_l  } \right] } 
+ A(T=\infty)
\label{LyapunovDimdinfty}
\end{eqnarray}
while the Carleman moment
\begin{eqnarray}
m_T(n,0) && = {\mathbb E} \left( x_1^{n}(T) \right) 
 = {\mathbb E} \left( x_1^{n}(0) e^{ T n \lambda_1(T) } \right)
\nonumber \\
&&  \opsimeq_{T \to + \infty}   e^{ T n
  \frac{  \sum_{j=1}^d \frac{  f^{[1]}_{jj}  }{  D^{[2]}_j }  }
{  \left[ \sum_{l=1}^d \frac{ 1  }{  D^{[2]}_l  } \right] }     }
  {\mathbb E} \left(  e^{  n \sqrt{ \frac{2 }{ \sum_{l=1}^d \frac{1}{D^{[2]}_l}}}   B^{\perp} (t) } \right)
  {\mathbb E} \left(  e^{ T n A_T  [R_2(0 \leq t \leq T);..;R_d(0 \leq t \leq T)]  } \right)  
  \nonumber \\
&&  \opsimeq_{T \to + \infty}  
e^{ T \left[ n
  \frac{  \sum_{j=1}^d \frac{  f^{[1]}_{jj}  }{  D^{[2]}_j }  }
{  \left[ \sum_{l=1}^d \frac{ 1  }{  D^{[2]}_l  } \right] }    +  n^2  \frac{2 }{ \sum_{l=1}^d \frac{1}{D^{[2]}_l}} 
+  \Phi(n) \right] }
\label{LyapunovTgeneratinglargedevnExplid}
\end{eqnarray}
where $\Phi(n)$ represents the Scaled Cumulant Generating Function 
of the additive observable $A_T[R_2(0 \leq t \leq T);..;R_d(0 \leq t \leq T)] $ as in Eq. \ref{SCGFA} 
and corresponds to the highest eigenvalue of the following tilted version
of the generator ${\cal L}  $ of Eq. \ref{GeneratorRj}
\begin{eqnarray}
{\cal L}_n && = {\cal L} 
+   \frac{ n }
{ \displaystyle \left[ \sum_{l=1}^d \frac{ 1  }{  D^{[2]}_l  } \right] } 
\left(   \sum_{i=2}^d  \frac{ f^{[1]}_{1i}  }{  D^{[2]}_1  }R_i 
 +\sum_{j=2}^d \left( \frac{ f^{[1]}_{j1} }{ D^{[2]}_j } \right) \frac{1}{R_j}
+ \sum_{j=2}^d \sum_{i\ne 1,j} \left( \frac{f^{[1]}_{ji}}{  D^{[2]}_j }    \right)
 \frac{R_i }{R_j}\right)   
\label{GeneratorStratoRnd}
\end{eqnarray}
as in Eq. \ref{GeneratorStratoRn} concerning $d=2$.


\subsection{ Example for the dimension $d=3$  }

To be more concrete, let us write explicitly the case of the simplest dimension $d=3$ 
beyond the case $d=2$ analyzed in subsection \ref{subsec_Ricatti} of the main text.

\subsubsection{ Dynamics of the two ratios $R_2(t)=\frac{x_2(t)}{x_1(t)}$ and $R_3(t)=\frac{x_3(t)}{x_1(t)}$ }

In dimension $d=3$, the Stratonovich system of Eq. \ref{ItoSDEsinglegjmultiplicativediagdratio}
for the two ratios $R_2(t)=\frac{x_2(t)}{x_1(t)}$ and $R_3(t)=\frac{x_3(t)}{x_1(t)}$ read
\begin{eqnarray}
dR_2(t)   && =  \left[f^{[1]}_{21} +( f^{[1]}_{22}- f^{[1]}_{11}) R_2(t) +  f^{[1]}_{23} R_3(t)
  -   f^{[1]}_{12} R^2_2(t) -  f^{[1]}_{13} R_2(t)  R_3(t) \right] dt
\nonumber \\
&& + R_2(t) \left[ \sqrt{ 2  D^{[2]}_2 }  dB_2(t)
- \sqrt{ 2  D^{[2]}_1}  dB_1(t) \right]
\nonumber \\
dR_3(t)   && =  \left[f^{[1]}_{31}  +  f^{[1]}_{32} R_2(t) +  (f^{[1]}_{33} - f^{[1]}_{11}) R_3(t)
  -   f^{[1]}_{12} R_2(t)R_3(t) -   f^{[1]}_{13} R^2_3(t) \right] dt
\nonumber \\
&& + R_3(t) \left[ \sqrt{ 2  D^{[2]}_3 }  dB_3(t)
- \sqrt{ 2  D^{[2]}_1}  dB_1(t) \right]
\label{ItoSDEsinglegjmultiplicativediagdratiod3}
\end{eqnarray}

The corresponding $2 \times 2$ diffusion matrix has the two diagonal elements of Eqs \ref{DijRdiag} and 
the of--diagonal element of Eq. \ref{DijRoff}
\begin{eqnarray}
{\cal D}_{22}[R_2,R_3 ] && =  R_2^2   \left(    D^{[2]}_2   +    D^{[2]}_1 \right)
\nonumber \\
{\cal D}_{33}[R_2,R_3 ] && =  R_3^2   \left(    D^{[2]}_3   +    D^{[2]}_1 \right)
\nonumber \\
{\cal D}_{23}[R_2,R_3 ]&& =  R_2R_3    D^{[2]}_1 
\label{DijRd3}
\end{eqnarray}

The Ito forces of Eq. \ref{StratoItoCorrespondanceR}
 \begin{eqnarray}
{\cal F}_2^{Ito}[R_2,R_3 ] 
&&  = {\cal F}_2[R_2,..,R_d ]+ R_2\left(    D^{[2]}_1 +  D^{[2]}_2 \right)
\nonumber \\
&& = f^{[1]}_{21} +( f^{[1]}_{22}- f^{[1]}_{11} +  D^{[2]}_1 +  D^{[2]}_2 ) R_2 +  f^{[1]}_{23} R_3
  -   f^{[1]}_{12} R^2_2 -  f^{[1]}_{13} R_2  R_3
\nonumber \\
{\cal F}_3^{Ito}[R_2,R_3 ] 
&&  = {\cal F}_3[R_2,..,R_d ]+ R_3\left(    D^{[2]}_1 +  D^{[2]}_3 \right)
\nonumber \\
&&= f^{[1]}_{31}  +  f^{[1]}_{32} R_2 +  (f^{[1]}_{33} - f^{[1]}_{11} + D^{[2]}_1 +  D^{[2]}_3 ) R_3
  -   f^{[1]}_{12} R_2R_3 -   f^{[1]}_{13} R^2_3
\label{StratoItoCorrespondanceRd3}
\end{eqnarray}
are useful to write the generator of Eq. \ref{GeneratorRj} 
\begin{eqnarray}
{\cal L}
&&  =   {\cal F}^{Ito}_2[R_2,R_3]  \frac{ \partial }{\partial R_2}
+ {\cal F}^{Ito}_3[R_2,R_3]  \frac{ \partial }{\partial R_3}
\nonumber \\
&& + {\cal D}_{22}[R_2,R_3 ]\frac{ \partial^2 }{\partial R_2^2 }
+ {\cal D}_{33}[R_2,R_3 ]\frac{ \partial^2 }{\partial R_3^2 }
+  2 {\cal D}_{23}[R_2,R_3 ]  \frac{ \partial^2 }{\partial R_2 \partial R_3}
\label{GeneratorRj3d}
\end{eqnarray}
whose adjoint ${\cal L}^{\dagger} $ appears in Eq. \ref{GeneratorRdaggerj}
satified by the steady state $\rho_{st}[R_2,R_3 ] $
\begin{eqnarray}
0&& = {\cal L}^{\dagger} \rho_{st}[R_2,R_3 ]
\nonumber \\
&&  = -   \frac{ \partial }{\partial R_2} 
\left[ {\cal F}^{Ito}_2[R_2,R_3 ] \rho_{st}(R_2,R_3)
-    \frac{ \partial }{\partial R_2 } \left(  {\cal D}_{22}[R_2,R_3 ] \rho_{st}[R_2,R_3 ] \right) 
-    \frac{ \partial }{\partial R_3 } \left(  {\cal D}_{23}[R_2,R_3 ] \rho_{st}[R_2,R_3 ] \right) 
\right]
\nonumber \\
&& -   \frac{ \partial }{\partial R_3} 
\left[ {\cal F}^{Ito}_3[R_2,R_3 ] \rho_{st}(R_2,R_3)
-    \frac{ \partial }{\partial R_3 } \left(  {\cal D}_{33}[R_2,R_3 ] \rho_{st}[R_2,R_3 ] \right) 
-    \frac{ \partial }{\partial R_2 } \left(  {\cal D}_{23}[R_2,R_3 ] \rho_{st}[R_2,R_3 ] \right) 
\right]
\label{GeneratorRdaggerd3}
\end{eqnarray}


\subsection{ Dynamics of $x_1(t)$ and its finite-time Lyapunov exponent $\lambda_1(T)$ }

For $d=3$, the additive functional of Eq. \ref{defTimeAveraged} of the stochastic trajectories of the two ratios
$R_2(0 \leq t \leq T) $ and $R_3(0 \leq t \leq T) $
reads
 \begin{eqnarray}
&& A_T  [R_2(0 \leq t \leq T);R_3(0 \leq t \leq T)]
\label{defTimeAveraged3} \\
&& =  \frac{1}{T} \int_0^T dt 
     \frac{ \left(  \frac{ f^{[1]}_{12}  }{  D^{[2]}_1  }R_2(t)  +  \frac{ f^{[1]}_{13}  }{  D^{[2]}_1  }R_3(t) 
+\left( \frac{ f^{[1]}_{21} }{ D^{[2]}_2 } \right) \frac{1}{R_2(t)}
 + \left( \frac{ f^{[1]}_{31} }{ D^{[2]}_3 } \right) \frac{1}{R_3(t)}
+   \left( \frac{f^{[1]}_{23}}{  D^{[2]}_2 }    \right) \frac{R_3(t) }{R_2(t)}
+   \left( \frac{f^{[1]}_{32}}{  D^{[2]}_3 }    \right) \frac{R_2(t) }{R_3(t)}
\right)    }
{ \displaystyle \left[ \frac{ 1  }{  D^{[2]}_1  } +  \frac{ 1  }{  D^{[2]}_2  }  +  \frac{ 1  }{  D^{[2]}_3  } \right] } 
\nonumber
\end{eqnarray}
with its asymptotic value for $T=+\infty$ of Eq. \ref{ergodicForLyapunovd}
 \begin{eqnarray}
&& A_T  [R_2(0 \leq t \leq T);R_3(0 \leq t \leq T)] 
\nonumber \\
&& \opsimeq_{T \to + \infty}
\int_{-\infty}^{+\infty} dR_2  \int_{-\infty}^{+\infty} dR_3 \rho_{st}[R_2,R_3]   
     \frac{ \left[  \frac{ f^{[1]}_{12}  }{  D^{[2]}_1  }R_2  +  \frac{ f^{[1]}_{13}  }{  D^{[2]}_1  }R_3 
+\left( \frac{ f^{[1]}_{21} }{ D^{[2]}_2 } \right) \frac{1}{R_2}
 + \left( \frac{ f^{[1]}_{31} }{ D^{[2]}_3 } \right) \frac{1}{R_3}
+   \left( \frac{f^{[1]}_{23}}{  D^{[2]}_2 }    \right) \frac{R_3 }{R_2}
+   \left( \frac{f^{[1]}_{32}}{  D^{[2]}_3 }    \right) \frac{R_2 }{R_3}
\right]    }
{ \displaystyle \left[ \frac{ 1  }{  D^{[2]}_1  } +  \frac{ 1  }{  D^{[2]}_2  }  +  \frac{ 1  }{  D^{[2]}_3  } \right] } 
 \nonumber \\
&&  \equiv  A(T=\infty)
\label{ergodicForLyapunovd3}
\end{eqnarray}
that appears in the 
asymptotic value of Eq. \ref{LyapunovDimdinfty} of the Lyapunov exponent 
 \begin{eqnarray}
\lambda_1(T=\infty) &&   =  \frac{\displaystyle \left( \frac{  f^{[1]}_{11}  }{  D^{[2]}_1  } +  \frac{  f^{[1]}_{22}  }{  D^{[2]}_2  }  +  \frac{  f^{[1]}_{22}  }{  D^{[2]}_3  } \right)  }
{ \displaystyle \left( \frac{ 1  }{  D^{[2]}_1  } +  \frac{ 1  }{  D^{[2]}_2  }  +  \frac{ 1  }{  D^{[2]}_3  } \right) } 
+ A(T=\infty)
\label{LyapunovDimd3}
\end{eqnarray}
while the Carleman moment of Eq. \ref{LyapunovTgeneratinglargedevnExplid} becomes
\begin{eqnarray}
m_T(n,0) && = {\mathbb E} \left( x_1^{n}(T) \right) 
 = {\mathbb E} \left( x_1^{n}(0) e^{ T n \lambda_1(T) } \right)
\nonumber \\
&&  \opsimeq_{T \to + \infty}  
e^{ T \left[  \frac{\displaystyle n \left( \frac{  f^{[1]}_{11}  }{  D^{[2]}_1  } +  \frac{  f^{[1]}_{22}  }{  D^{[2]}_2  }  +  \frac{  f^{[1]}_{22}  }{  D^{[2]}_3  } \right) + 2 n^2 }
{ \displaystyle \left( \frac{ 1  }{  D^{[2]}_1  } +  \frac{ 1  }{  D^{[2]}_2  }  +  \frac{ 1  }{  D^{[2]}_3  } \right) }      
+  \Phi(n) \right] }
\label{LyapunovTgeneratinglargedevnExplid3}
\end{eqnarray}
where $\Phi(n)$ represents the Scaled Cumulant Generating Function 
of the additive observable $A_T[R_2(0 \leq t \leq T);R_3(0 \leq t \leq T)] $ as in Eq. \ref{SCGFA} 
and corresponds to the highest eigenvalue of the following tilted version
of the generator ${\cal L}  $ of Eq. \ref{GeneratorRj3d}
\begin{eqnarray}
{\cal L}_n && = {\cal L} 
+  n  \frac{ \left[  \frac{ f^{[1]}_{12}  }{  D^{[2]}_1  }R_2  +  \frac{ f^{[1]}_{13}  }{  D^{[2]}_1  }R_3 
+\left( \frac{ f^{[1]}_{21} }{ D^{[2]}_2 } \right) \frac{1}{R_2}
 + \left( \frac{ f^{[1]}_{31} }{ D^{[2]}_3 } \right) \frac{1}{R_3}
+   \left( \frac{f^{[1]}_{23}}{  D^{[2]}_2 }    \right) \frac{R_3 }{R_2}
+   \left( \frac{f^{[1]}_{32}}{  D^{[2]}_3 }    \right) \frac{R_2 }{R_3}
\right]    }
{ \displaystyle \left[ \frac{ 1  }{  D^{[2]}_1  } +  \frac{ 1  }{  D^{[2]}_2  }  +  \frac{ 1  }{  D^{[2]}_3  } \right] } 
\label{GeneratorStratoRn3d}
\end{eqnarray}
as in Eq. \ref{GeneratorStratoRn} concerning $d=2$.


\subsection{ Discussion  }

In conclusion, the main ideas described in subsection \ref{subsec_Ricatti} 
for the dimension $d=2$ can be extended to analyze the
Stratonovich system of Eq. \ref{ItoSDEsinglegjmultiplicativediag} in dimension $d>2$,
but it is much more difficult to obtain explicit results,
in particular for the steady state $\rho_{st}(R_2,..,R_d)$ involving
$(d-1)$ variables in Eq. \ref{GeneratorRdaggerj}
and even for the special case $d=3$ for the steady state 
$\rho_{st}(R_2,R_3)$ involving
two variables in Eq. \ref{GeneratorRdaggerd3}.
 Further work is needed to see if one can make progress in this analysis.


\end{document}